\pdfoutput=1

\documentclass[twoside,a4paper,BCOR=18mm,12pt,DIV=11,bibliography=totoc]{scrbook}

\usepackage[ngerman, english]{babel}		
\usepackage{hyphenat}
\usepackage{tabularx}
\usepackage{booktabs}

\usepackage{color}

\usepackage{graphicx}
\usepackage{caption}
\usepackage{subcaption}

\usepackage{cite}

\usepackage{amsmath}
\usepackage{amssymb}

\usepackage{textcomp}
\usepackage{hyperref}

\usepackage{cleveref} 

\addto\captionsenglish{}

\newcommand{\g}{g_\mathrm{1 D}}
\newcommand{\gtd}{g_\mathrm{3 D}}
\newcommand{\nod}{n_\mathrm{1 D}}
\newcommand{\as}{a_\mathrm{s}}
\newcommand{\cohfact}{\langle \cos(\varphi) \rangle}

\newcommand{\me}{\mathrm{e}}
\newcommand{\re}{\mathrm{Re}}

\newcommand{\rcite}[1]{ref.~\cite{#1}}

\newcommand{\secref}[1]{section~\ref{#1}}

\newcommand{\tss}[1]{\textsuperscript{#1}}
\newcommand{\tsub}[1]{\textsubscript{#1}}

\newcommand{\Hz}{\kern 0.2em Hz}
\newcommand{\kHz}{\kern 0.2em kHz}
\newcommand{\MHz}{\kern 0.2em MHz}
\newcommand{\G}{\kern 0.2em G}
\newcommand{\ms}{\kern 0.2em ms}
\newcommand{\ns}{\kern 0.2em ns}
\newcommand{\us}{\kern 0.2em{\textmu}s}

\newcommand{\um}{\kern 0.2em{\textmu}m}
\newcommand{\mm}{\kern 0.2em mm}
\newcommand{\nm}{\kern 0.2em nm}

\newcommand{\nK}{\kern 0.2em nK}
\newcommand{\uK}{\kern 0.2em{\textmu}K}

\makeatletter
\newcommand{\vast}{\bBigg@{4}}
\newcommand{\Vast}{\bBigg@{5}}
\makeatother

\setcapindent{0pt}
\addtokomafont{caption}{\small}
\addtokomafont{captionlabel}{\sffamily\bfseries}

\begin{document}
		
\pagenumbering{roman}		
\begin{titlepage}
{\centering

\includegraphics[height=0.16\textwidth]{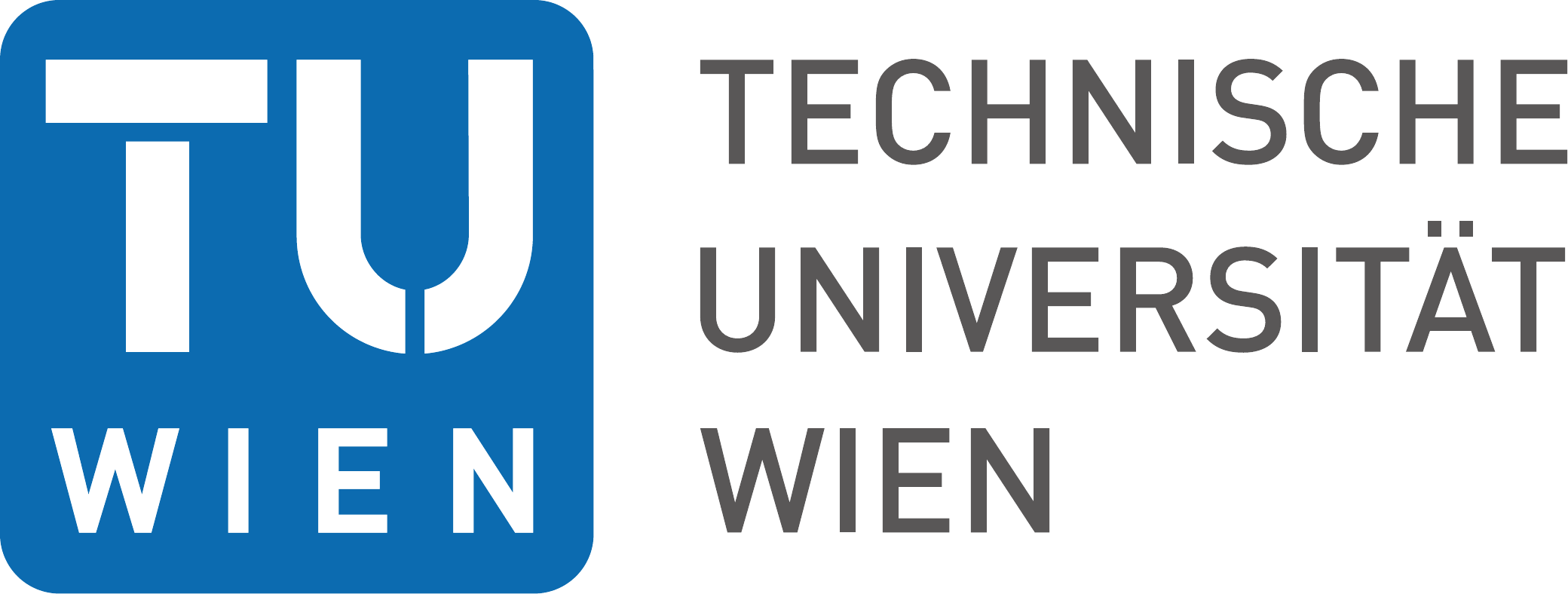}

\vspace{1.7em}
 
\large DISSERTATION
\vspace{1.5em}
 
\Large
\textbf{Correlations and dynamics of tunnel-coupled one-dimensional Bose gases} \\
\normalsize
\vspace{3em}

ausgef\"uhrt zum Zwecke der Erlangung des akademischen Grades eines\\
Doktors der technischen Wissenschaften\\
\vspace{2em}
 
unter Leitung von \\
\textbf {Univ. Prof. Dipl.-Ing. Dr. techn. Hannes-J\"org Schmiedmayer}\\
E141 Atominstitut\\
\vspace{2em}

eingereicht an der\\
Technischen Universit\"at Wien\\ 
Fakult\"at f\"ur Physik\\
\vspace{2em}

von \\
\textbf{Dipl.-Ing. Thomas Schweigler} \\
\vspace{2em}

Wien, M\"arz 2019

} 
%

\vspace*{\fill}

\noindent Revised version. \today

\newpage
\thispagestyle{empty}
\mbox{}

\newpage
\vspace*{41em}
\begin{flushleft}
\hspace{0.5em}\textbf{Referees:}\\
\vspace{1em}
\begin{tabular}{p{6cm} p{7.2cm}}
Hannes-J\"org Schmiedmayer & (Technische Universit\"at Wien, Austria) \\
Selim Jochim & (Universit\"at Heidelberg, Germany) \\
Ian Spielman & (National Institute of Standards and
Technology and the University of Maryland, USA)
\end{tabular}
\end{flushleft}

\newpage
\thispagestyle{empty}
\mbox{}

\end{titlepage}


\addchap*{Zusammenfassung}

\selectlanguage{ngerman}

Das Verständnis von Quanten\hyp{}Vielteilchensystemen ist von großer Bedeutung für viele Teilbereiche der Physik.
Während man die grundlegenden mikroskopischen Bewegungsgleichungen solcher Systeme oft einfach aufstellen kann ist deren exakte Lösung nur in den seltensten Fällen möglich.
Für das Verständnis der Eigenschaften von Quanten\hyp{}Vielteilchensystemen benötigt man deshalb effektive theoretische Beschreibungen als auch experimentelle Modellsysteme um deren Vorhersagen zu testen.

In dieser Dissertation präsentieren wir eine Reihe von Experimenten mit ultrakalten Bosegasen.
Ein solches System stellt ein gut isoliertes, flexibles und robustes Modellsystem für Quanten\hyp{}Vielteilchenphysik dar.
Zahlreiche erprobte Methoden zur Manipulation und Vermessung des Systems stehen zur Verfügung.
In unserem speziellen Fall beschäftigen wir uns mit ultrakalten eindimensionalen Bosegasen bestehend aus Rubidium Atomen.

Wir erzeugen zwei solcher ultrakalten Gase in einem Doppeltopf-Potential.
Die Atome können von einem Topf in den anderen tunneln, was je nach Höhe der Tunnelrate zu einer mehr oder weniger starken Phasenkohärenz zwischen den beiden Teilsystemen führt.
Der ortsaufgelöste Phasenunterschied zwischen beiden Gasen kann mit Hilfe von Materiewellen-Interferenz gemessen werden.
Damit lassen sich die räumlichen Korrelationen dieses Phasenunterschieds untersuchen.

Untersucht man ob sich Korrelationsfunktionen höherer Ordnung in Korrelationen niedrigerer Ordnung zerlegen lassen, so kann man daraus Rückschlüsse auf die Wechselwirkungen im Systems ziehen.
Kann man alle Korrelationsfunktionen mit Ordnung größer als zwei zerlegen, so handelt es sich um ein nicht-wechselwirkendes System und die Fluktuationen folgen einer Gauß-Verteilung.
In dieser Arbeit präsentieren wir die Messung nicht-zerlegbarer Korrelationsfunktionen vierter Ordnung und untersuchen damit die Wechselwirkung zwischen den kollektiven Anregungen unseres Quanten\hyp{}Vielteilchensystems.
Durch Einstellen der Tunnelrate zwischen den beiden Teilsystemen im Doppeltopf\hyp{}Potential können wir unterschiedliche Grade von Nicht-Zerlegbarkeit der Korrelationsfunktionen beobachten.

Ausgehend von einem solchen nicht-Gaußschen Zustand beobachten wir außerdem eine dynamische Entwicklung zu einem Zustand mit zerlegbaren Korrelationsfunktionen (Gaußsche Fluktuationen).
Wir starten in einem Doppeltopf mit Tunneln und fahren dann die Potential-Barriere zwischen den Teilsystemen hoch um jenes zu unterbinden.
Anschließend beobachten wir wie die anfänglich nicht-Gaußschen Phasen-Fluktuationen in Gaußsche Fluktuationen übergehen.
Hierbei handelt es sich um die erste experimentelle Beobachtung eines solchen Prozesses in  Quanten\hyp{}Vielteilchensystemen.
Untersuchungen dieser Art von Dynamik sind wichtig um zu verstehen wie Gaußsche Gleichgewichtszustände erreicht werden können.

Zu guter Letzt diskutieren wir die dynamische Entstehung von Phasenkohärenz in einem Doppeltopf-Potential mit Tunneln.
Wir beobachten diese Entwicklung im Experiment ausgehend von zwei unterschiedlichen Anfangszuständen.
Einerseits teilen wir eine Wolke in zwei Teilsysteme und regen globale Oszillationen in deren Phasendifferenz an.
Diese Oszillationen werden daraufhin gedämpft und ein Zustand mit Phasenkohärenz stellt sich ein.
Der zweite Anfangszustand besteht aus zwei unabhängigen Teilsystemen welche schlagartig durch Tunneln gekoppelt werden.
Auch hier sehen wir die Entstehung von Phasenkohärenz zwischen beiden Teilsystemen.

\selectlanguage{english}

\addchap*{Abstract}

Understanding quantum many-body systems is of great importance for many branches of physics.
While it is often easy to state the basic equations for the microscopic motion of countless particles, their exact solution is only possible in rare cases.
To understand the properties of quantum many-body systems, one therefore needs effective theoretical descriptions as well as experimental model systems to test their predictions.

In this thesis, we present a series of experiments with ultracold Bose gases.
Such gases represent a well-isolated, flexible and robust model system for quantum many-body physics.
Numerous proven methods for manipulating and measuring such systems are available.
In our particular case, we are working with ultracold one-dimensional Bose gases consisting of rubidium atoms.

We create two such ultracold gases in a double well potential. 
The atoms can tunnel from one well into the other, which leads,  depending on the strength of the tunneling, to various degrees of phase locking between the two subsystems.
Employing matter-wave interference, we can measure the spatially resolved phase difference between the two gases.
This makes it possible to investigate spatial correlations of this phase difference.

By investigating whether correlation functions of higher order can be factorized into correlations of lower order, we can investigate the interaction properties of the system.
For a non-interacting system, all correlation functions with orders greater than two factorize and one observes Gaussian fluctuations.
In this thesis, we present the measurement of non-factorizing fourth-order correlation functions, leading to an experimental characterization of the interactions between the collective excitations of the quantum many-body system.
The degree of non-factorizibility, i.e., the degree of non-Gaussianity of the phase fluctuations, depends on the tunneling strength, which is tuneable in the experiment.

Starting from such a non-Gaussian state, we are able to observe the dynamical evolution towards a state with factorizing correlation functions (Gaussian fluctuations).
More precisely, we start in a double well with tunneling and then 
abruptly decouple the two subsystems.
Subsequently, we observe how the initially non-Gaussian phase fluctuations become Gaussian.
This represents the first experimental demonstration of `Gaussification' in quantum many-body systems.
Investigating this type of dynamics is important to understand how Gaussian equilibrium states can be reached by quantum mechanical evolution.

Last but not least, we discuss the dynamical emergence of phase coherence in a double well potential with tunneling.
We experimentally investigate the evolution starting from two different initial states.
In the first case, we split a cloud of atoms into two subsystems and trigger global oscillations in their relative phase.
The oscillations subsequently damp and phase coherence sets in.
In the second case, two independent subsystems are suddenly coupled by tunneling.
Again, we see the emergence of phase coherence between the two subsystems.
	
\tableofcontents



\let\oldchapmark=\chaptermarkformat
\renewcommand*{\chaptermarkformat}{}
\chapter*{Introduction}
\chaptermark{Introduction}
\addcontentsline{toc}{chapter}{Introduction}

\setcounter{page}{1}		
\pagenumbering{arabic}	

%
%
%
%

The understanding of quantum many-body systems is of great importance for many different fields, ranging from solid state to high energy physics.
Cold atom experiments have proven to be a valuable tool for investigating quantum many-body physics~\cite{bloch2008many,Bloch2012,Goldman_2014,proukakis2017universal}.
They are very well isolated from the environment, can be manipulated with a number of different techniques and are easily accessible for measurements. 
A large number of different geometries are possible, enabling the study of one-, two- and three-dimensional as well as lattice systems.
 
In this thesis we study ultracold one-dimensional (1D) Bose gases. 
Such 1D systems are of great interest because their dynamics is strongly affected by the restricted phase space available for scattering~\cite{Giamarchi04,Cazalilla11}, which may lead to many intriguing properties like, for example, integrability~\cite{Caux_2011}. 
Moreover, the reduced number of dimensions often makes 1D systems theoretically tractable.

A common method to characterize cold atom systems is to measure correlations.
While the measurement of second-order correlation functions has been part of the experimental toolbox for many years, higher-order correlation functions have rarely been used to analyze experiments so far~\cite{Hodgman2011,Dall2013,Endres2013,Langen15,Feng2019}.
However, the measurement of correlation functions of order bigger than two is of particular interest for investigating the interaction properties of quantum many-body systems.
Analyzing whether higher-order correlation functions factorize into correlations of lower order is a direct test whether the system is interacting or not~\cite{ZinnJustin}.
In the absence of interactions, all higher-order correlation functions factorize, the fluctuations follow a Gaussian distribution.
If interactions are present, the fluctuations follow a non-Gaussian distribution and non-factorizing higher-order correlations appear. 
In \cref{chap:corr}, we present measurements of non-factorizing fourth-order correlation functions.
The correlations are calculated for the measured relative phase fluctuations of two tunnel-coupled 1D Bose gases.
Note that we published the results in \rcite{Schweigler17}.

Another very active topic where cold atom systems are used is the investigation of equilibration and thermalization~\cite{Polkovnikov2011,Eisert2015,Langen2015b}.
It is still an open question how an isolated quantum many-body system can reach thermal equilibrium through the unitary evolution inherent to quantum mechanics. 
One possible scenario to investigate is the `quench' from an interacting to a free system.
In this case, thermalization leads to an equilibrium state with Gaussian fluctuations, starting from initially non-Gaussian fluctuations.
In general, an evolution from a non-Gaussian to a Gaussian state is often termed `Gaussification' in the literature; several theoretical studies exist~\cite{Gogolin_2016,Gluza2016,Sotiriadis_2016,Sotiriadis_2017,murthy2018relaxation,gluza2018equilibration}.
However, an experimental demonstration has been lacking so far.   
In \cref{chap:gaussification}, we give such a demonstration.
The experimental procedure consists of suddenly switching off the tunneling between two 1D Bose gases and subsequently investigating the evolution of the relative phase fluctuations.
A publication about the results is currently in preparation.

Investigating the dynamics of two tunnel-coupled 1D Bose gases is of great interest as well. 
While the result for the Gaussification as well as many previous experimental results of our group~\cite{Gring12,Kuhnert2013,Langen13b,Langen15,Rauer16,Rauereaan7938} can, at least approximately, be explained by a non-interacting effective field theory, one has a genuinely interacting system here.
The occurrence and damping of Josephson oscillations for two tunnel-coupled 1D Bose gases was recently publish by a different experiment from our group in \rcite{Pigneur18}.
We present similar measurements in \cref{chap:non_equi}.
With our experimental apparatus, we have access to different observables and parameter regimes.
Our results should therefore be seen as being complementary to the ones presented in \rcite{Pigneur18}.
Moreover, we investigate the rephasing of two independent clouds after abruptly switching on the tunnel coupling, a situation theoretically discussed in \rcite{dalla_torre_13}.

\section*{Outline of this thesis}

\begin{description}
	\item[\Cref{chap:exp_setup}] gives a brief overview of the experimental apparatus used in this thesis. 
	Moreover, it contains a brief discussion of the experimental cycle. 
	\item[\Cref{chap:theo}] introduces the basic theoretical concepts and models used in this thesis.
	\item[\Cref{chap:imaging}] discusses the used absorption imaging systems in detail. 
	We focus on the processes preventing a perfect imaging and discuss how to simulate absorption images in order to estimate the influence of the imaging process onto measured data.
	\item[\Cref{chap:extracting}] discusses how we analyze the absorption images taken after time of flight in order to extract information about the in-situ properties of the system.
	\item[\Cref{chap:corr}] contains the results for the phase correlation function of two tunnel-coupled 1D Bose gases.
	We experimentally measure the second and fourth-order correlation functions.
	By analyzing whether the fourth-order correlation function factorizes into lower-order correlations, we will investigate the interactions between phononic excitations.
	\item[\Cref{chap:gaussification}] presents an experimental demonstration of Gaussification by discussing the dynamics after decoupling two initially tunnel-coupled gases.
	%
	\item[\Cref{chap:non_equi}] shows the results for the dynamics in the double well trap with tunneling.
	%
\end{description}

\renewcommand*{\chaptermarkformat}{\oldchapmark}

%
%
%
%
%

\chapter{Experimental setup}
\label{chap:exp_setup}

In this chapter, we will give a brief overview of the experimental apparatus used to obtain the results presented in this thesis.
For a detailed description of the experimental apparatus see the preceding PhD and master's theses \cite{goebelthesis,gring2012thesis,rauer2012mastersthesis,kuhnert2013thesis,langen2013thesis,rauer2018thesis}.
Note that this chapter is meant as a recapitulation for readers who are already familiar with our setup or similar cold atom experiments.
Unfamiliar readers might want to start by reading the theses cited above.

For a general introduction about cold atom experiments \rcite{pethick_smith_2008} can be consulted.
The thesis \cite{schummthesis} also contains a good introduction which is more focused towards atomchip experiments.
To get an overview about atomchips, see the book \cite{Reichel11}.
Note that throughout this thesis, we will work with $^{87}$Rb atoms.
For its physical properties see \rcite{steck2015}.

\begin{figure}
	\centering
	\includegraphics[width=3.54in]{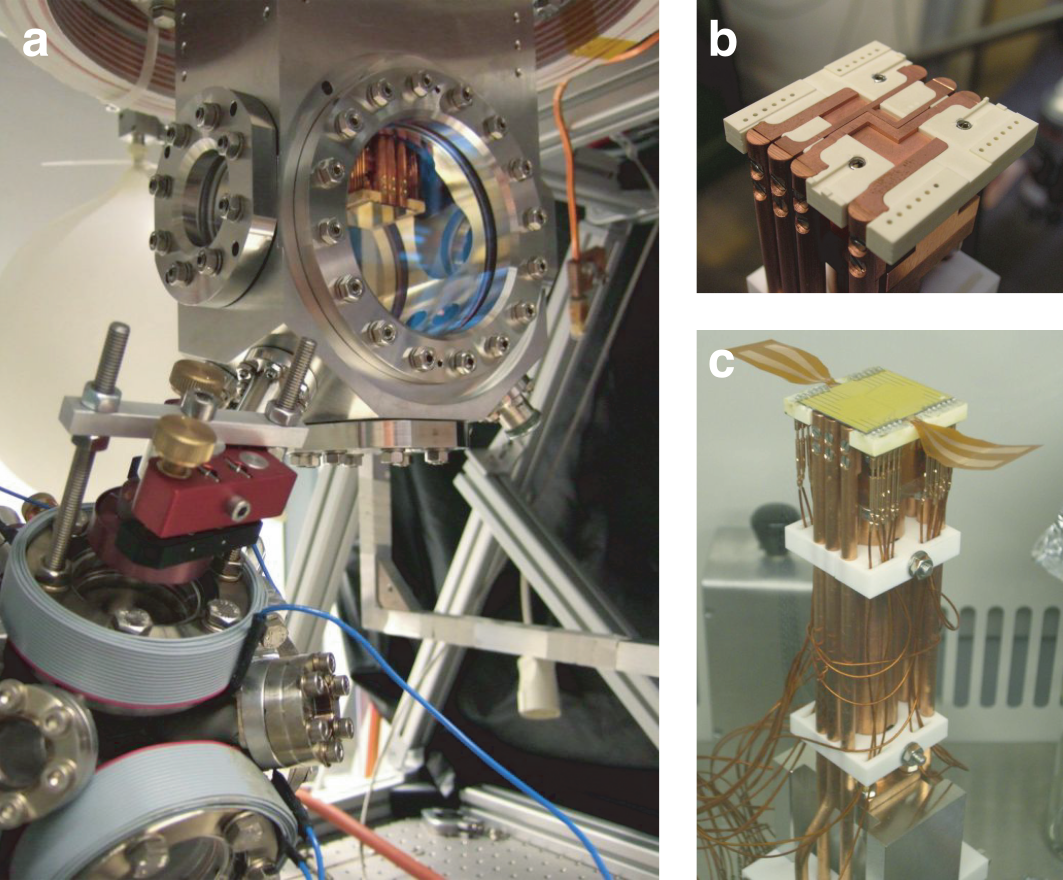}
	\caption{\textbf{Experimental setup.}		
		\textbf{(a)} The two vacuum chambers of our experiment are shown. 
		In the upper chamber one can see the atomchip being mounted up-side down. Note that the picture was taken before the chambers have been surrounded with coils and optics.
		\textbf{(b)} Chip-mount without atomchip. 
		One sees the macroscopic copper structures.
		\textbf{(c)} Completed mount with the atomchip on top.		
		Figure adapted from \cite{rauer2018thesis}. 
		}
	\label{fig:expsetup}
\end{figure}

\section{The experimental cycle in brief}

In this section we will give a brief overview of the experimental cycle used for this thesis.
Our experiment consist of two vacuum chambers as can be seen in \cref{fig:expsetup}a.
The lower chamber contains Rubidium dispensers. 
The $^{87}$Rb atoms evaporated by the dispensers are collected in a six beam magneto optical trap (MOT).
We will refer to it as the `lower MOT'.
With the help of a near resonant laser beam, internally referred to as `push beam', the atoms are pushed from the lower MOT trough a narrow tube connection into the upper vacuum chamber.
There they are collected in another MOT, which we will call the `upper MOT'.

The upper vacuum chamber contains the atomchip mounted up-side-down from above (see \cref{fig:expsetup}a). 
Of course, this obstructs the optical access from above.
Therefore, we are using two beams reflected on the atomchip's gold surface in addition to four incoming beams in order to create the upper MOT.
Both the lower and upper MOT are on for approximately 10{\,s}. 
We will often refer to this time-span as the `MOT phase'.
During this time (until about a second before the MOTs are switched off) the atoms are continuously pushed up from the lower MOT and collected in the upper MOT.
While the size of the atom cloud in the lower MOT stays more or less the same after a quick equilibration, the atom cloud in the upper MOT continuously grows by collecting and cooling more and more atoms.
This growth is rather fast at the beginning and slows down considerably towards the end.

At the end of the MOT phase, the magnetic fields as well as the laser intensities and frequencies are ramped to compress and shift the upper MOT.
The purpose of this is to give the atom cloud the right size and position for being transferred into the magnetic trap later on.
Before the transfer, immediately following the MOT phase, we switch all magnetic fields off and perform polarization gradient cooling \cite{Dalibard89}.
Afterwards, a bias field is switched on to provide a quantization axis for the atoms.
This is necessary in order to transfer all $^{87}$Rb atoms into the low field seeking~\cite{steck2015} $F = 2$, $m_F = 2$ state via optical pumping.
Subsequently the atoms are loaded into a cigar shaped macroscopic magnetic trap.

The necessary fields for this macroscopic trap are generated by a Helmholtz pair located outside of the vacuum chamber and a copper structure mounted below the atomchip (see \cref{fig:expsetup}b).
Note that it is a bit confusing what is meant by below/above here as the atomchip is mounted up-side down.
With below, we mean here that the gold surface of the atomchip is on the outside and the copper structures are sandwiched between atomchip and mount.
Note that the wires on the atomchip are not used for creating the macroscopic trap.

The atom cloud is held in the macroscopic trap for about 6\,s.
During this time, we perform forced evaporative cooling using radio frequency (390{\kHz} to 14.9{\MHz}) magnetic fields.
We will often refer to this fields as the `cooling (radio frequency) fields'.
They are also generated by one of the copper structures located below the atomchip.


After the first stage of evaporation, the cooling fields are switched off again.
At this point, we have around $2 \times 10^6$ atoms at a temperature of a few {\textmu}K.
The atom cloud is subsequently transferred into a microscopic magnetic trap produced by the wires on the atomchip (in combination with external bias fields).
In the microscopic magnetic trap we continue with the forced evaporative cooling, preparing the one-dimensional (1D) quasicondensates used for the experimental investigations.
Note that the cooling mechanism is quite different from the standard three-dimensional (3D) case after the cloud becomes effectively one-dimensional \cite{Rauer16,rauer2018thesis}.
At the end, the cloud contains around 10000 atoms at a temperature of a few tens of nanokelvin. 
The features of the atomchip and the possible trapping geometries will be discussed in the next \cref{sec:intro_atomchip}.

The experimental cycle ends with releasing the cloud from the atomchip trap and taking absorption images after time of flight (TOF) expansion.
The properties and limitations of the imaging systems will play a crucial role for the results presented in this thesis.
We will therefore devote the whole \cref{chap:imaging} to discussing the imaging process.

\section{The atomchip and possible trapping geometries}
\label{sec:intro_atomchip}

\begin{figure}
	\centering
	\includegraphics[width=0.85\linewidth]{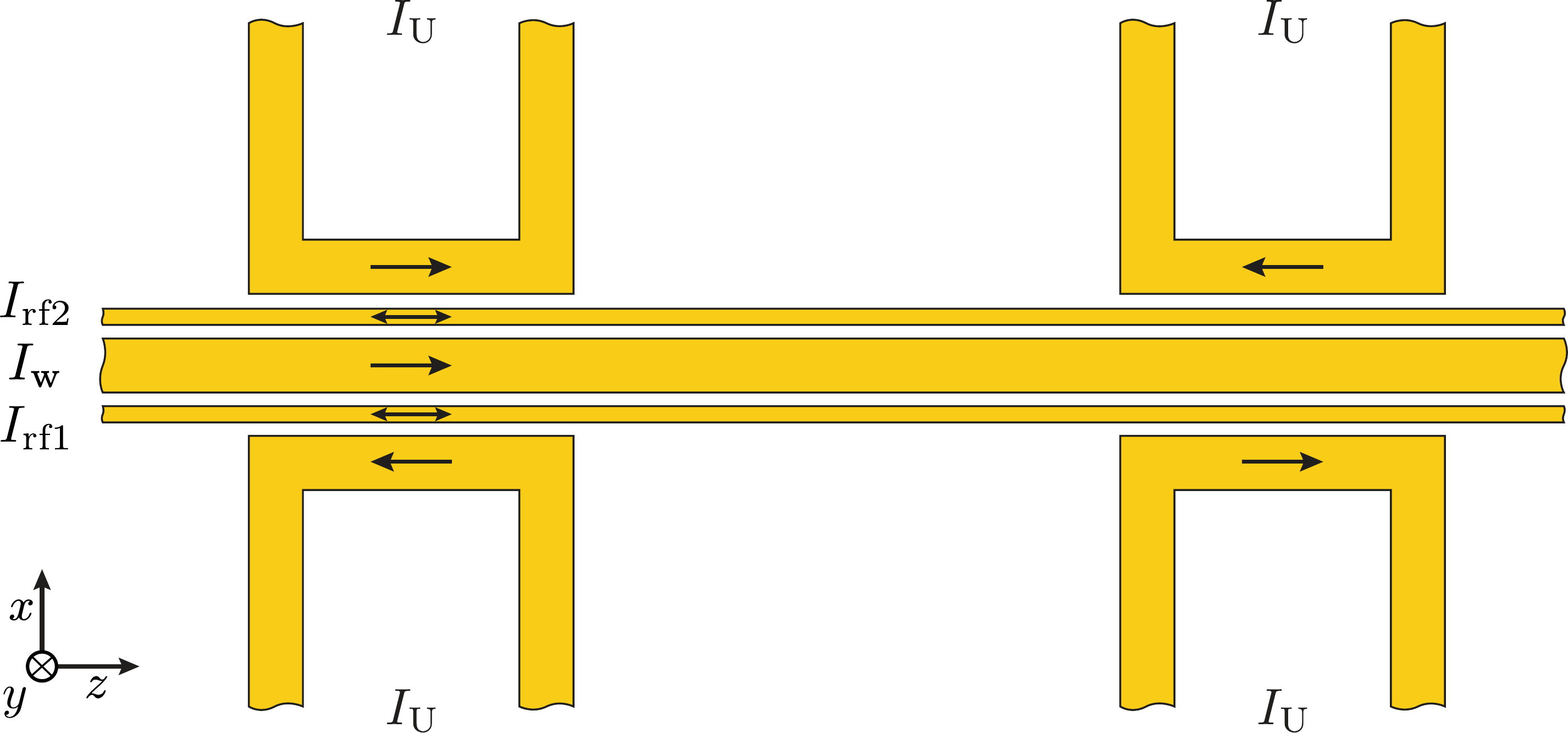}
	\caption{\textbf{Wires on the atomchip.}
		A schematic illustration showing only the wires of the atomchip which were actually used in this thesis.
		Figure reproduced with permission from \cite{rauer2018thesis}.
	}
	\label{fig:figchipwires}
\end{figure}

The microscopic magnetic trap used for the experiments is produced by the wires of an atomchip~\cite{Reichel11}.
Our atomchip consists of a single gold layer isolated from a silicon substrate by a thin layer of silicon dioxide.
There were several wires structured into the gold layer.
\Cref{fig:figchipwires} gives a schematic illustration of selected chip wires (the ones used in this thesis).
For more details about the chip used in this experiment see \rcite{goebelthesis,kuhnert2013thesis}.

Let us now discuss the different structures and currents in \cref{fig:figchipwires}.
The current $I_\mathrm{W}$ flowing in the $z$ direction in the central trapping wire creates a circular magnetic field.
Together with homogeneous bias fields in the $x$ and $z$ direction, this gives a tight harmonic confinement in $x$ and $y$~\cite{Reichel11}.
The homogeneous bias fields are produced by Helmholtz pairs located outside of the vacuum chamber.
The bias field in the $z$ direction is commonly referred to as Ioffe-field.

To confine the atoms also in the $z$ direction, we need the currents $I_\mathrm{U}$ in the U shaped chip wires.
Sending the currents in the appropriate directions (as marked by the arrows in the figure) leads to a harmonic trap in the $z$ direction.
While the confinement in $x$ and $y$ is very strong (typically around 2.1{\kHz}), the confinement in $z$ is quite weak (typically around 12{\Hz}).
This highly elongated trap, therefore, leads to cigar shaped clouds.
In the following, we will often refer to the elongated direction (the $z$ direction) as the `longitudinal' direction.
The perpendicular tightly confined directions will be the `transverse' directions.
The trapping frequencies in the longitudinal and transverse direction will be denoted by $\omega_\parallel/2\pi $ and $\omega_\perp/2\pi $ respectively.
Note that in the experimental setup $z$ and $x$ are horizontal and $y$ vertical.


Besides the central trapping wire, two smaller wires (currents $I_\mathrm{rf1,2}$) exist.
They are used for generating radio frequency magnetic fields in order to `dress' the atoms, i.e., off-resonantly couple the different Zeeman states leading to a new eigenbasis of `dressed' state.
We will often refer to this magnetic fields as the `dressing (radio frequency) fields'.
With the help of the dressing fields, we can realize, among others, a double well geometry in the $x$ or $y$ direction \cite{Schumm05,schummthesis,Lesanovsky2006}.
By independently controlling the amplitude and phase of $I_\mathrm{rf1}$ and $I_\mathrm{rf2}$, we can change the polarization and amplitude of the radio frequency magnetic field at the position of the atoms.
To get a double well separation in the horizontal $x$ direction, we need a linear polarization in the $y$ direction.
This is achieved with equal amplitudes\footnote{Due to technical imperfection we actually have to use slightly different amplitudes for the two wires.} and a pi phase-shift between the currents.
Note that also the frequency of the dressing field influences the resulting shape of the dressed state potential.
For the results of this thesis, we used a frequency ($360${\kHz}) about 30{\kHz} lower than the Zeeman splitting ($\approx 390${\kHz}) in the center of the trap.
   

Let us conclude by discussing the longitudinal confinement.
Using the U wires on the atomchip, we unfortunately can only generate harmonic traps. 
In order to realize other trapping geometries in the longitudinal $z$ direction, we recently started superimposing optical dipole potentials.
In this thesis, we used only an optical box trap, for which the light was shaped by a simple mask \cite{rauer2018thesis,Rauereaan7938}, in some of the measurements.
In the future, we will be able to generate a large variety of optical dipole traps with the help of a digital micromirror device (DMD) \cite{tajik2017mastersthesis}.


\chapter{Theoretical basics}
\label{chap:theo}

In this chapter we will introduce the theoretical models used in this thesis. Additional discussions of the specific theoretical calculations done for comparison with the experimental results can be found in the respective \cref{chap:corr,chap:gaussification,chap:non_equi}.

We will start by discussing the Lieb-Liniger model for Bosons in one dimension (1D) in \cref{sec:lieb_liniger}.
From the Lieb-Liniger Hamiltonian, we will then derive low energy effective field theories (Bogoliubov theory and Luttinger liquid model) for the quasicondensate regime in \cref{sec:low_energy}.
How to connect the three dimensional (3D) theory for a very elongated system to the 1D models will be the topic of \cref{sec:3D_to_1D}. 
This includes also a discussion about corrections to the purely 1D models.   
In \cref{sec:DW_theo}, we will introduce some theoretical models describing two clouds in a double well trap.
A comparison of the thermal fluctuations following from the different models is given in \cref{sec:theo_comparison}.
We conclude the chapter with \cref{sec:theo_tof}, where we discuss the expansion after the atom clouds are released from the trap.

\section{The Lieb-Liniger model}
\label{sec:lieb_liniger}

The Hamiltonian for bosonic fields $\hat { \psi } (z)$ in 1D, interacting via a delta-function potential, is given by
\begin{equation}
\hat { H } = \int \mathrm { d } z \, \hat { \psi } ^ { \dagger } ( z ) \left[ - \frac { \hbar ^ { 2 } } { 2 m } \partial _ { z } ^ { 2 } + U ( z ) - \mu + \frac { \g } { 2 } \hat { \psi } ^ { \dagger } ( z ) \hat { \psi } ( z ) \right] \hat { \psi } ( z ). \label{eq:trapped_lieb_liniger}
\end{equation}
Here, the strength of the delta-function interaction is characterized by the 1D coupling constant $\g$. In our case the interactions are repulsive, i.e., $\g > 0$. In addition to the kinetic energy term (first term) and the interaction energy term (last term), the Hamiltonian also includes the trapping potential $U ( z )$ and the chemical potential $\mu$, which sets the average number $N$ of Bosons (atoms) in the system. 
The Hamiltonian \labelcref{eq:trapped_lieb_liniger} (without the trapping potential) is usually known as Lieb-Liniger model.
A solution for the ground state~\cite{lieb_liniger_63} and excitation spectrum~\cite{lieb_63} of that model are known.

Depending on the temperature $T$, the interaction strength $\g$, the total number $N$ of particles (or equivalently the chemical potential $\mu$) and the trapping potential $U ( z )$, the system can be in several distinctive regimes~\cite{petrov_00,Kheruntsyan03,petrov2004low}. 
Detailed discussions following various approaches can be found in the previously written PhD theses \cite{rauer2018thesis,kuhnert2013thesis,langen2013thesis}. 
Here we will restrict ourselves to a short discussion following \rcite{Kheruntsyan03}, which treats the case of a uniform gas. 
In this case, the relevant parameters are the Lieb-Liniger parameter
\begin{equation}
\gamma = \frac{m \g}{\hbar ^ { 2 } \nod} \label{lieb_liniger_param}
\end{equation}
and the reduced temperature $\tau = T / T _ { d }$, where the degeneracy temperature is given by 
\begin{equation}
	T_d = \frac{\hbar ^ { 2 } \nod  ^ { 2 }}{2 m k_\mathrm{B}} .
\end{equation} 
Here the uniform average atomic density is denoted by $\nod $ and the atomic mass by $m$. 
The formula for $T_d$ can be understood from $\nod \Lambda_{T_d} \approx 1$, meaning that the de Broglie wavelength $\Lambda_{T_d}$ at the degeneracy temperature $T_d$ is approximately equal to the mean interparticle spacing $1/\nod $.
The Lieb-Liniger parameter $\gamma$ quantifies the amount of interactions in the system.

In the experiment, we typically have $T_d \approx 10${\uK} and $\gamma \approx 2 \times 10^{-3}$ for the atomic density in the center of the cloud.
This follows from  a transverse trapping frequency of $\omega_\perp \approx 2 \pi \times 1.4 ${\kHz} and an atomic density of $\nod \approx 60${\um}\tss{-1}.
The given values are typical for the double well traps. 
The transverse trapping frequency determines the interaction strength as discussed in \cref{sec:3D_to_1D}.
To be more precise, we have used \cref{eq:g1d_approx} in order to obtain $\g$. 

We usually work with temperatures $T$ between 10 and 100{\nK}. 
The latter gives a reduces temperature of $\tau \approx 10^{-2}$. 
This places us well in the quasicondensate regime.
In this regime, many of the local properties look similar to a true condensate.
However, quasicondensates lack the global coherence that true condensates exhibit.
The criterion to be in the quasicondensate regime is~\cite{Kheruntsyan03} 
\begin{equation}
	\tau^2  \lesssim \gamma \lesssim 1. \label{eq:quasicond_criterion}
\end{equation} 
In our case it is well fulfilled ($  10^{-4}  <  2 \times 10^{-3} < 1$). 

As already mentioned, a uniform gas was assumed in the calculations leading to the criteria \labelcref{eq:quasicond_criterion}. However the authors of \rcite{Kheruntsyan03} state that they expect their conclusion to remain valid when using the local density approximation for sufficiently slowly varying systems. Note that the condition~\labelcref{eq:quasicond_criterion} will not be fulfilled at the edges of the trapped cloud. For example, for a temperature of $T = 100${\nK} the first inequality is not fulfilled for $\nod  \lesssim 20${\um}\tss{-1}.

In the harmonically trapped case treated in \rcite{petrov_00,petrov2004low} additional complications arise due to the discrete level-structure. 
However, for the parameters in our experiment, the discrete level structure is not so important as the interaction energy is much bigger then the level-spacing. 

\section{Low energy effective field theories}
\label{sec:low_energy}

\subsection{Bogoliubov theory for quasicondensates}
\label{sub_sec:bogo_single_cond}

Mora and Castin developed a Bogoliubov theory for the quasicondensate regime in \rcite{Mora2003}.
Their starting point is the discretized version of \labelcref{eq:trapped_lieb_liniger},
\begin{equation}
\hat { H } = \sum_z l  \ \hat { \psi } ^ { \dagger } ( z ) \left[ - \frac { \hbar ^ { 2 } } { 2 m } \Delta  + U ( z ) - \mu + \frac { \g } { 2 } \hat { \psi } ^ { \dagger } ( z ) \hat { \psi } ( z ) \right] \hat { \psi } ( z ). \label{eq:discrete_lieb_liniger}
\end{equation}
Here, $l$ denotes the distance between the discrete space points. 
The differential operators are replaced by the finite differences
\begin{equation}
\Delta f (z ) =  \frac { f( z + l ) + f (z - l  ) - 2 f (z ) } { l ^ { 2 } }, \quad \nabla f ( z ) = \frac { f ( z + l ) - f ( z - l ) } { 2 l }.
\end{equation}
Note that it is crucial to work with the discretized version \cref{eq:discrete_lieb_liniger} instead of its continuum equivalent \labelcref{eq:trapped_lieb_liniger}.
The reasons for this as well as appropriate values for $l$ are discussed in \rcite{Mora2003}.

The bosonic operator $\hat\psi(z)$ ($\hat\psi^\dagger(z)$) of \cref{eq:discrete_lieb_liniger} annihilates (creates) a particle in a box of size $l$ at the position $z$.
Their commutation relation is therefore
\begin{equation}
\left[ \hat { \psi } ( z ) , \hat { \psi } ^ { \dagger } \left( z ^ { \prime } \right) \right] = \frac { \delta _ { z , z ^ { \prime } } } { l},
\end{equation}
where $\delta _ { z , z ^ { \prime } }$ is the Kronecker delta.
To derive the Bogoliubov theory, the phase density representation
\begin{equation}
\hat \psi  ( z ) \approx e ^ { i \hat { \theta } (  z  ) } \sqrt { \hat { \rho } (  z  ) }. \label{eq:phase_dens}
\end{equation}
is used.
Here, $\hat { \theta } (  z  ) $ and $\hat { \rho } (  z  )$ are the Hermitian phase and density operators respectively. 
They fulfill the commutation relations
\begin{equation}
{ \left[ \hat { \rho } ( z ) , \hat { \theta } \left( z ^ { \prime } \right) \right] \approx \frac { i \delta _ { \mathrm { zz } ^ { \prime } } } { l } , \quad \left[ \hat { \rho } ( z ) , \hat { \rho } \left( z ^ { \prime } \right) \right] = 0 },  \quad { \left[ \hat { \theta } ( z ) , \hat { \theta } \left( z ^ { \prime } \right) \right] = 0 }
 \label{eq:phase_dens_comm_discr}
\end{equation}
for conjugate variables.
Note that the operator giving the number of atoms in the box of length $l$ at position $z$ is given by
\begin{equation}
	l \,  \hat \psi^\dagger (z) \hat { \psi } ( z )  = l \, \hat { \rho } ( z ) .
\end{equation}  
Also note that defining a hermitian phase operator is problematic and can only be done approximately in certain limits. 
Therefore the approximate equality in equation \labelcref{eq:phase_dens} and \labelcref{eq:phase_dens_comm_discr}. For a detailed discussion of the phase operator see \rcite{Mora2003,gerry_knight_2004}. 

To continue with the derivation of the Bogoliubov theory, the density operator $\hat { \rho } $ is split into a background density $\rho_0$ and fluctuations $\delta\hat\rho$ on top of this background density, i.e.
\begin{equation}
\hat { \rho } ( z ) = \rho _ { 0 } ( z ) + \delta \hat { \rho } ( z ).
\end{equation}
The operator $\delta\hat\rho$ fulfills the same commutation relations \labelcref{eq:phase_dens_comm_discr} as $\hat { \rho } $ itself. Subsequently the Hamiltonian is expanded in the small parameters 
\begin{equation}
\frac { | \delta \hat { \rho } | } { \rho _ { 0 } } \ll 1 , \quad  | l \nabla \hat { \theta } | \leqslant 1. \label{eq:small_parameters_bogo}
\end{equation}
In other words, one assumes the density fluctuations and the phase gradient to be small. 
Furthermore, it is assumed that both small parameters are of the same order.
One can then sort the terms in the Hamiltonian according to their order in the small parameters.

The zeroth order only depends on the background density $\rho_0 (z)$.
Minimizing it with respect to $\rho_0 (z)$ gives the Gross-Pitaevskii equation
\begin{equation}
	\left[ - \frac { \hbar ^ { 2 } } { 2 m } \Delta + U ( z ) - \mu + \g \, \rho_0 (z) \right] \sqrt { \rho_0 (z) } = 0. \label{eq:gpe_mora}
\end{equation}
For a density profile $\rho_0 (z)$ fulfilling this equation the first order contribution to the Hamiltonian always gives 0. The leading order contribution to the fluctuations is therefore given by the second-order Hamiltonian
\begin{align}
H^{(2)}_\mathrm{Bogo} = \sum _ { z } l  \vast[ &- \frac { \hbar ^ { 2 } } { 2 m } \frac { \delta \hat { \rho } } { 2 \sqrt { \rho _ { 0 } } } \Delta \left( \frac { \delta \hat { \rho } } { 2 \sqrt { \rho _ { 0 } } } \right) + \frac { \hbar ^ { 2 } \delta \hat { \rho } ^ { 2 } } { 8 m \rho _ { 0 } ^ { 3 / 2 } } \Delta \sqrt { \rho _ { 0 } } \nonumber \\ &\quad + \frac { \g } { 2 } \delta \hat { \rho } ^ { 2 } + \frac { \hbar ^ { 2 } } { 2 m }  \sqrt { \rho _ { 0 } ( z ) \rho _ { 0 } \left( z + l  \right) } \frac { \left[ \hat { \theta } \left( z + l \right) - \hat { \theta } ( z ) \right] ^ { 2 } } { l ^ { 2 } } \vast]. \label{eq:H2_mora}
\end{align}
Here we omitted a complex number contribution as it does not affect the dynamics of the phase or density fluctuations.

Starting from \labelcref{eq:gpe_mora} and \labelcref{eq:H2_mora} we can do further approximations. One is the Thomas-Fermi approximation \cite{castin2001simple} consisting of neglecting the derivative of the density profile, i.e., $\Delta \rho_0 = 0$. One neglects the corresponding terms in \cref{eq:gpe_mora} as well as \cref{eq:H2_mora}. A second approximation is to assume density fluctuations to be of short wavelength, i.e., neglecting the first term in \labelcref{eq:H2_mora}. Employing both approximations leads to the the Luttinger liquid Hamiltonian
\begin{equation}
H^{(2)}_\mathrm{ LL } = \sum _ { z } l  \left\{ \frac { \g } { 2 } \delta \hat { \rho } ^ { 2 } ( z ) + \frac { \hbar ^ { 2 } } { 2 m }  \sqrt { \rho _ { 0 } ( z ) \rho _ { 0 } \left( z + l  \right) } \frac { \left[ \hat { \theta } \left( z + l \right) - \hat { \theta } ( z ) \right] ^ { 2 } } { l ^ { 2 } } \right\}. \label{eq:H2LL_mora}
\end{equation}
For simplicity we will often use the continuum version ($\lim l \to 0$) of this Hamiltonian:
\begin{equation}
H^{(2)}_\mathrm{ LL } =  \int \mathrm { d } z \,  \left[ \frac { \g } { 2 } \delta \hat { \rho } ^ { 2 } ( z ) + \frac { \hbar ^ { 2 } } { 2 m }   \rho _ { 0 } ( z ) \left( \frac { \partial \hat \theta (z)  } { \partial z } \right) ^ { 2 }  \right]. \label{eq:H2LL_cont}
\end{equation}
Note that in general, we will always use the same symbols for the fields in the continuum as well as the discretized models.
This includes the bosonic fields $\hat \psi$ as well as the phase ($\hat \theta$) and density ($\delta \hat \rho$) fields.
Also, the letter $z$ denotes the spatial coordinate in the continuum as well as the sites in the discretized space.


\subsection{Derivation in classical fields approximation}
\label{sec:bogo_classical_field_derivation}

Note that the derivation in \rcite{Mora2003}, which we discussed above was done for the quantum fields. 
However, as discussed in \rcite{Stimming_2010} and \cref{sec:val_classical_fields}, the fluctuations observable in the experiment are dominated by thermal noise. 
We therefore can also simply start form the classical fields approximation of \labelcref{eq:trapped_lieb_liniger}, i.e., simply omit the hats in the Hamiltonian. 
Using the phase density representation for the classical fields, we can write \cref{eq:trapped_lieb_liniger} as 
\begin{equation}
H = \int d z \left[ \frac { \hbar ^ { 2 } } { 2 m } \rho  \left( \frac { \partial \theta  } { \partial z } \right) ^ { 2 } + \frac { \hbar ^ { 2 } } { 2 m } \frac { 1 } { 4 \rho  } \left( \frac { \partial \rho  } { \partial z } \right) ^ { 2 } + ( U ( z ) - \mu ) \, \rho  + \frac { \g } { 2 } \rho^2 \right]. \label{eq:lieb_liniger_classical_field_dens}
\end{equation}
Here we have assumed that the field $\psi (z)$ vanishes at integration boundaries.
This is fulfilled for a finite number of particles in an appropriate trapping potential with the integration boundaries far enough away. 

Starting from \labelcref{eq:lieb_liniger_classical_field_dens} we can then derive the Bogoliubov theory by doing an expansion around the minimum solutions $\theta_0(z)$ and $\rho_0 (z)$. To find this minimum solutions one simply sets the first functional derivative of \labelcref{eq:lieb_liniger_classical_field_dens} to 0. The functional derivative with respect to $\theta(z)$ is zero for the phase field being a constant (along $z$). Inserting this into the functional derivative with respect to  $\rho (z)$ and setting it to zero gives the Gross-Pitaevskii equation \labelcref{eq:gpe_mora}. The second-order term in the Taylor series around this minimum solutions gives then exactly the continuum and classical field version of \labelcref{eq:H2_mora}.

\subsection{Numerical calculations}
\label{sec:bogo_matrix_vector}

The quadratic Hamiltonians \labelcref{eq:H2_mora} and \labelcref{eq:H2LL_mora} in discretized space are convenient for numerical calculations, while their respective continuum limits are often more suitable for analytical calculations (\cref{sec:homo_bogo}).

In order to do numerical calculations, we write the discretized quadratic Hamiltonians \labelcref{eq:H2_mora} and \labelcref{eq:H2LL_mora} in matrix vector form. We collect the fields on the $N$ different discretized space points into vectors, i.e.,
\begin{equation}
\vec{\hat { \theta }} = \left( \begin{array} { c } { \hat { \theta } ( 1 ) } \\ { \vdots } \\ { \hat { \theta } ( N ) } \end{array} \right) , \quad { \vec {\hat { \theta }}} ^ {\  T } = \left( \hat { \theta } ( 1 ) \dots \hat { \theta } ( N ) \right) 
\end{equation}
for phase fields and analogous for the density fluctuations $\delta \rho (z)$. The Hamiltonians can then be written in the form
\begin{equation}
\hat { H }^{(2)} = \frac { 1 } { 2 }  { \vec {\hat { \theta }}} ^ {\  T} K \vec{\hat { \theta }}  + \frac { 1 } { 2 } \delta {\vec { \hat { p } }} ^ {\ T } L \delta \vec { \hat { \rho } }, \label{eq:H2_matrix_vector}
\end{equation}
where the matrices $K$ and $L$ depend on what particular quadratic Hamiltonian we are looking at.
Note that the models discussed so far do not have any cross-terms between phase and density, therefore we also have no cross-terms in \cref{eq:H2_matrix_vector}.
Note that a general quadratic Hamiltonian could have such cross-terms~\cite{BROADBRIDGE1979}. 

Starting from \labelcref{eq:H2_matrix_vector}, it is easy to calculate thermal expectation values as discussed in \cref{sec:bogo_thermal_exp_val}, \cref{eq:thermal_matrix_vector}.
The Heisenberg equations of motion are also easily derived as 
\begin{equation}
\partial_t \, \delta \vec { \hat { \rho } } = \frac { 1 } { \hbar l } K   \vec {\hat { \theta }}, \quad 
\partial_t \, \vec {\hat { \theta }} = - \frac { 1 } { \hbar l } L \delta \vec { \hat { \rho } } . \label{eq:matrix_vector_eq_motion}
\end{equation}
Note that these are basically the same as the Hamilton equations of motion, as expected for conjugate variables.
The only difference is the factor $1/\hbar l$.
It stems from the fact that \cref{eq:phase_dens_comm_discr} differs from the usual commutation relation for conjugate variables by exactly that factor.

Starting from \cref{eq:H2_matrix_vector,eq:matrix_vector_eq_motion,eq:thermal_matrix_vector} a set of codes was developed as part of this PhD and applied in \cref{chap:corr,chap:gaussification,chap:non_equi} as well as \rcite{Rauereaan7938,rauer2018thesis,gluza2018}. 
Let us conclude by noting that similar equations like \cref{eq:H2_matrix_vector,eq:matrix_vector_eq_motion,eq:thermal_matrix_vector} also exist for quadratic Hamiltonians in continuous space~\cite{ZinnJustin}.

\subsection{Thermal expectation values}
\label{sec:bogo_thermal_exp_val}

Let us start by discussing the thermal expectation values in classical fields approximation.
There, we simply omit the hats in the Hamiltonian.
The thermal probability density for a certain field configuration is then given by
\begin{equation}
	\rho = \exp(-\beta H)/\mathcal{Z}, \label{eq:thermal_classical_dens_func}
\end{equation}
where $H$ is the Hamiltonian depending on the classical fields, $\mathcal{Z}$ is the partition function and $\beta = 1/k_\mathrm{B} T$.
If the Hamiltonian, for example, depends on the phase $\theta(z)$ and density fluctuations $\delta\rho(z)$, the partition function is written as the functional integral~\cite{ZinnJustin} 
\begin{equation}
\mathcal{Z} = \int \mathcal{D} \theta \int \mathcal{D} \delta\rho \ e ^ { - \beta H }.
\end{equation} 

It is easy to see that \cref{eq:thermal_classical_dens_func} with a quadratic Hamiltonian leads to Gaussian fluctuations for the fields.
Gaussian fluctuations are fully determined by their mean values and covariance matrix~\cite{kardar_2007}.
The covariance matrix can be easily obtained by kernel inversion~\cite{ZinnJustin}. 
For example, starting from \labelcref{eq:H2_matrix_vector} the thermal expectation values of the conjugate fields are
\begin{align}
&\langle \rho ( n ) \rangle = \langle \theta ( n ) \rangle = \langle \rho ( m ) \theta ( n ) \rangle = 0  \nonumber \\
&\langle \rho ( m ) \rho ( n ) \rangle = k_\mathrm{B} T \left(  L ^ { - 1 } \right) _ { m n }, \quad\langle \theta ( m ) \theta ( n ) \rangle = k_\mathrm{B} T \left( K  ^ { - 1 } \right) _ { m n } \label{eq:thermal_matrix_vector}
\end{align}
Be aware that the expectation values of the cross-terms only vanish because there are no cross-terms in the Hamiltonian \labelcref{eq:H2_matrix_vector}.
Note that the matrix vector form \labelcref{eq:H2_matrix_vector} for the Hamiltonian was introduced for the discretized space.
However, also the Hamiltonians for a finite continous system will have a similar form when represented in their eigenbasis. 
Therefore, \cref{eq:thermal_matrix_vector} will also be useful in this case as can be seen from \cref{sec:homo_bogo}.

Also in the quantum case, a quadratic Hamiltonian leads to a Gaussian thermal state.
One can see this from the fact that a quadratic Hamiltonian leads to a quadratic euclidean action~\cite{ZinnJustin}.

\subsection{Homogeneous background density}
\label{sec:homo_bogo}

In this section, we will discuss the case of a homogeneous background density $\rho_0 (z) = \nod  $.
Note that we will in general denote the background density by $\rho_0 (z)$ while $\nod$ is always only used for the value of the atomic density in the homogeneous case.  
Using $\rho_0 (z) = \nod  $ in \labelcref{eq:H2_mora} and taking the continuum limit gives
\begin{equation}
\hat H^{(2)} = \int d z \left[  \frac {\hbar^2}{8 m \nod } \, \left(\frac { \partial \delta \hat { \rho } } { \partial z }  \right)^2 + \frac { \g } { 2 } \delta \hat { \rho } ^ { 2 } + \frac { \hbar ^ { 2 } \nod  } { 2 m }  \left( \frac { \partial \hat \theta  } { \partial z } \right) ^ { 2 } \right]. \label{eq:H2_homo}
\end{equation}
Here we also rewrote the first term containing derivations of the density fluctuations.
For this we assumed the surface term to vanish, which requires appropriate boundary conditions, like Neumann boundary conditions, periodic boundary conditions or vanishing density fluctuations on the boundary. 
Remember that neglecting the first term in \cref{eq:H2_homo} gives the Luttinger liquid Hamiltonian \labelcref{eq:H2LL_cont}.

In the following, we want to discuss the case of Neumann boundary conditions:
\begin{equation}
\left. \frac{ \partial \hat \theta  }{ \partial z }	\right\rvert_{z \, = \, 0, \, L} = \left. \frac{ \partial \delta \hat { \rho } }{ \partial z }	\right\rvert_{z \, = \, 0, \, L} = 0 \label{eq:box_BC_bogo}
\end{equation}
Having Neumann boundary conditions for the phase implies that the particle current on the boundaries vanishes. 
This corresponds to the physical situation of having a hard-walled box.
For the density fluctuations, there is no similar argument.
However, the numerically obtained eigenmodes for a box with finite wall-width (not shown) look very similar to the ones obtained with the Neumann boundary conditions.
We therefore conclude that the Neumann boundary conditions are also a sensible choice for the density fluctuations when investigating the case of a hard-walled box.

Keeping \cref{eq:box_BC_bogo} in mind we can expand the phase fluctuations as
\begin{equation}
	\hat { \theta } ( z ) = \sqrt { \frac { 2 } { L } } \sum _ { n = 1 } ^ { \infty } \cos \left( n \frac { \pi } { L } z \right) \hat { \theta } _ { n } + \frac { \hat { \theta } _ { 0 } } { \sqrt { L } } .
\end{equation}
The same we do for the density fluctuations.
Inserting the expansions into \cref{eq:H2_homo} and using the orthonormality of the cosine modes gives
\begin{equation}
\hat H^{(2)} = \sum _ { n = 1 } ^ { \infty } \left[ \left( \frac { \g } { 2 } +  \frac { \hbar ^ { 2 } k _ { n } ^ { 2 } } { 8 m \nod } \right)  \delta \hat { \rho } _ { n } ^ { 2 } + \frac { \hbar ^ { 2 } k _ { n } ^ { 2 } } { 2 m } \nod \, \hat { \theta } _ { n } ^ { 2 } \right] +  \frac { \g } { 2 } \, \delta \hat { \rho } _ { 0 } ^ { 2 } . \label{eq:H2_box_bogo}
\end{equation}
Here we denoted the wavenumber of the cosine modes as
\begin{equation}
	k _ { n } = n \frac { \pi } { L }. \label{eq:k_box}
\end{equation}

By using \cref{eq:thermal_matrix_vector}, we can directly see from \cref{eq:H2_box_bogo} that the thermal expectation values in classical fields approximation are
\begin{align}
\left\langle  \theta_n \right\rangle &= \left\langle  \delta\rho_n  \right\rangle = 0 \label{eq:homo_theo_first_moments} \\
\left\langle  \theta_n \theta_m \right\rangle_\mathrm{CF} &= \delta_{n,m} \frac{k_\mathrm{B} T}{2} \frac{2 m}{\hbar ^ { 2 } k _ { n } ^ { 2 } \nod } = \delta_{n,m} \, \frac{2} {\lambda_T}  \, \frac {1}{k_n^2} \label{eq:cf_phase_exp_box_bogo} \\
\left\langle  \delta\rho_n  \delta\rho_m \right\rangle_\mathrm{CF} &= \delta_{n,m} \frac{k_\mathrm{B} T}{2} \left( \frac { \g } { 2 } +  \frac { \hbar ^ { 2 } k _ { n } ^ { 2 } } { 8 m \nod } \right)^{-1} = \delta_{n,m} \frac{k_\mathrm{B} T}{\g} \left( 1 +  \frac { \xi_\mathrm{h}^2 k_n^2 }{4} \right)^{-1}. \label{eq:cf_dens_exp_box_bogo}
\end{align}
Here the indices $n,\ m$ run from $0$ to $\infty$ for the density fluctuations, and from $1$ to $\infty$ for the phase fluctuations. 
Note that the expectation values for the zero mode $\theta_0$ of the phase are not defined, which does not lead to any complications since a global phase offset of a single condensate has no physical meaning.
Furthermore, we have used the thermal coherence length given by
\begin{equation}
\lambda _ { T } = \frac{2 \hbar ^ { 2 } \nod}{ m k _ { B } T } \label{eq:def_lambda_T}
\end{equation}
in \cref{eq:cf_phase_exp_box_bogo} and the healing length
\begin{equation}
	\xi _ { \mathrm { h } } = \frac{\hbar}{ \sqrt { \g \nod m }} \label{eq:healing_length}
\end{equation}
in \cref{eq:cf_dens_exp_box_bogo}.

As discussed in \cref{sec:bogo_thermal_exp_val}, the expectation values
\cref{eq:homo_theo_first_moments,eq:cf_phase_exp_box_bogo,eq:cf_dens_exp_box_bogo}
are all that is necessary to characterize the Gaussian thermal state following from the quadratic Hamiltonian \labelcref{eq:H2_box_bogo}.
Note that \cref{eq:homo_theo_first_moments,eq:cf_phase_exp_box_bogo,eq:cf_dens_exp_box_bogo} are the result for the full Bogoliubov theory \labelcref{eq:H2_homo}.
In case of the first moments \labelcref{eq:homo_theo_first_moments} and the second moments \labelcref{eq:cf_phase_exp_box_bogo} of the phase fluctuations, the results coincide with the ones for the Luttinger liquid model.
For the second moments of the density fluctuations on the other hand, we have to neglect the second term in the brackets in \cref{eq:cf_dens_exp_box_bogo} in order to get the results for the Luttinger liquid Hamiltonian.
 
Of course, one can also define bosonic creation and annihilation operators in order to bring the Hamiltonian \labelcref{eq:H2_homo} into the form 
\begin{equation}
\hat H^{(2)} = \sum _ { n = 1 } ^ { \infty } \epsilon_n \left( \hat { b } _ { n } ^ { \dagger } \hat { b } _ { n } + \frac{1}{2}\right).
\end{equation}
Starting from this, one can calculate quantum expectation values. 
We will here not go into details but only state the mode energies.
For the full Bogoliubov theory, they are given by
\begin{equation}
	\epsilon_n = \hbar k_n c \ \sqrt{ 1 +  \frac { \xi_\mathrm{h}^2 k_n^2 }{4} } . \label{eq:mode_energy_box_bogo}
\end{equation}
By neglecting the term $	\xi_\mathrm{h}^2 k_n^2 / 4$ under the square root we get the results for the Luttinger liquid theory:
\begin{equation}
\epsilon_n = \hbar k_n c  \label{eq:mode_energy_box_LL}
\end{equation}
From this, we see that $c$ is the speed of sound for the Luttinger liquid theory (long wavelength excitations).
Its value can be calculated with
\begin{equation}
	c = \sqrt{ \frac{ \g \nod }{m}}. \label{eq:speed_of_sound}
\end{equation}

In general, we obtain the Luttinger liquid theory from the Bogoliubov theory by neglecting the term $	\xi_\mathrm{h}^2 k_n^2 / 4$.
This is true for \cref{eq:H2_box_bogo,eq:cf_dens_exp_box_bogo,eq:mode_energy_box_bogo}.
We can therefore argue that Luttinger liquid and Bogoliubov theory give the same results when
\begin{equation}
	\frac { \xi_\mathrm{h}^2 k^2 }{4} \ll 1
\end{equation} 
leading to the criteria
\begin{equation}
k \ll \frac{2}{\xi _ \mathrm { h } }.
\end{equation}
For typical experimental parameters we have $\xi _ \mathrm { h } \approx 0.35${\um}. This means that the limiting length scale below which the difference between Luttinger liquid and full Bogoliubov theory becomes important is way below the imaging resolution (see discussion in \cref{sec:res_vandor,sec:width_of_Gaussian_psf}).

So far we discussed the homogeneous case with Neumann boundary conditions.
Note that for the homogeneous system with periodic boundary conditions we can do an expansion into plane waves and get basically the same result.
To be more precise, the form of \cref{eq:H2_box_bogo,eq:cf_phase_exp_box_bogo,eq:cf_dens_exp_box_bogo,eq:mode_energy_box_bogo} stays the same, however, the allowed values of the wavenumber $k$ are different:
\begin{equation}
\tilde k _ { n } = n \frac { 2 \pi } { L } \quad \text{with} \quad  n = -\infty \dots \infty . \label{eq:k_per}
\end{equation}
We see that the spacing of the k-modes is twice as big as in \cref{eq:k_box}. 
On the other hand also negative k-values exist, which leads to the same mode density with respect to $|k|$.

Using \cref{eq:cf_phase_exp_box_bogo} with \cref{eq:k_per} and considering the limit of infinite system size, we can calculate the variance of the phase difference between two spatial points as
\begin{equation}
\left\langle \left[ \theta( z + \Delta z ) - \theta( z ) \right] ^ { 2 } \right\rangle = \frac{2}{\lambda_T} \left| \Delta z \right| \label{eq:phase_fluct_single_cond}.
\end{equation}

\subsubsection{Validity of the classical fields approximation}
\label{sec:val_classical_fields}

Doing the classical fields approximation, i.e., leaving out the hats, corresponds to making two approximations: 
Firstly, the quantum (zero temperature) fluctuations are neglected. 
Secondly, the Bose-Einstein distribution is replaced by the Rayleigh-Jeans distribution. 
Both approximations are good when the occupation of the modes under consideration is sufficiently large.
This leads to the criterion
\begin{equation}
	\beta \, \epsilon_n \ll 1 , \label{eq:classical_crit}
\end{equation}
where $\epsilon_n$ are the mode energies and $\beta = 1 / k_\mathrm{B} T$.
Using the mode energies \labelcref{eq:mode_energy_box_LL} for the Luttinger liquid Hamiltonian, \cref{eq:classical_crit} leads to the criterion
\begin{equation}
	k \ll \frac{k_\mathrm{B} T}{\hbar c} \equiv k_\mathrm{c} . \label{eq:classical_crit_LL}
\end{equation}
In other words, the investigated or relevant length scales must be bigger than $\lambda_\mathrm{c} = 2\pi / k_\mathrm{c}$.

%

For a typical experimental transverse trap frequency and atomic density, the speed of sound \labelcref{eq:speed_of_sound} is $c \approx 2 \times 10^{-3}\,\mathrm{m}/\mathrm{s}$. 
The lowest temperature achieved so far is about 10{\nK}, leading to $k_\mathrm{c} = 0.65 \ 1/$\textmu{m}. 
This is on the limit of what we can resolve with our imaging systems (see \cref{chap:imaging,sec:width_of_Gaussian_psf} as well as \rcite{gring2012thesis}). 
Therefore, it is very challenging to resolve quantum fluctuations with the current setup (see also the discussion in \rcite{Stimming_2010}).
Even if we would have the necessary imaging resolution, there would still be the problem of technical noise.
In short, we could not yet observe the quantum fluctuations of the system in thermal equilibrium.

In most measurements, the experimental temperatures are quite a bit higher (around $30${\nK}) than the lowest achieved 10{\nK}.
Therefore, we will apply the classical fields approximation in the theoretical modeling of the experimental investigation described in \cref{chap:corr,chap:gaussification,chap:non_equi}. 

\section{From 3D to 1D}
\label{sec:3D_to_1D}

We introduced the Hamiltonian for the 1D case in \labelcref{eq:trapped_lieb_liniger} but so far did not discuss how it is connected to the 3D world and its scattering properties.
As already discussed in \cref{sec:intro_atomchip}, we work with highly elongated traps with two tightly confined transverse directions ($x$ and $y$) and the longitudinal $z$ direction.
This means that dynamics in the transverse direction is basically frozen out and we are left with an effectively 1D system with coordinate $z$.
  
In \rcite{Olshanii98}, a relation between the 3D scattering length $\as$ and the 1D interaction constant $\g$ was developed:
\begin{equation}
\g = 2 \hbar \omega_\perp \as \left( 1 - 1.03 \frac { \as } { a _ { \perp } } \right) ^ { - 1 } \label{eq:g1d_olshanii}
\end{equation}
Here, we assume that the trap frequency in both transverse directions is the same and given by $\omega_\perp$.
The harmonic oscillator length in the transverse direction is
\begin{equation}
	a_\perp = \sqrt{\hbar/(m \omega_{\perp})} \label{eq:harm_osc_len}.
\end{equation}
For our typical experimental parameters the ratio $ { \as }/ { a _ { \perp } } < 0.02$.
We will therefore in the following always use the approximate formula 
\begin{equation}
\g \approx 2 \hbar \omega _ { \perp } \as \quad \mathrm{for} \quad  { \as }/ { a _ { \perp } } \ll 1. \label{eq:g1d_approx}
\end{equation}

Note that the condition  $ { \as }/ { a _ { \perp } } \ll 1$ means that the scattering has 3D character.
In this case we can also start from the 3D Hamiltonian of interacting Bosons and assume that the atoms are in the ground state of the transverse harmonic potential. 
Integrating out the transverse direction then leads to \labelcref{eq:g1d_approx}.

Note that \cref{eq:trapped_lieb_liniger} is only valid in the 1D regime.
In other words, the thermal energy $k_\mathrm{B} T$ and the mean interaction energy $\g \rho_0/2$ ($\rho_0$ is the 1D density of atoms) per particle should be much smaller than the transverse level spacing $\hbar \omega_{ \perp }$.
In the experiment, we typically have a temperature of $T = 40${\nK} leading to $k_\mathrm{B} T/(2 \pi \hbar) = 833${\Hz} compared with the trap frequency of typically $1.4${\kHz}.
The ratio between the mean interaction energy per particle and the level spacing is simply given by
\begin{equation}
	\frac{\g \rho_0(z)}{2 \hbar \omega_\perp} = \as  \rho_0(z),
\end{equation} 
where we have used \cref{eq:g1d_approx}.
With a typical 1D atomic density of $\rho_0(z=0) = 60${\um}\tss{-1} for the center of the trap (maximum density) and $\as = 5.2${\nm} \cite{van_Kempen_2002} this gives $\as  \rho_0 = 0.31$.
The inequalities $\hbar \omega_{ \perp } > \g \rho_0/2$ and $\hbar \omega_{ \perp } > k_\mathrm{B} T$ are therefore only approximately fulfilled.


Some of the consequences following from the only approximate fulfillment of $\hbar \omega_{ \perp } > \g \rho_0/2$ can be considered using the approach discussed in \cite{Salasnich02}\footnote{Note that there is a vast body of literature discussing similar approaches like \rcite{Salasnich02} or extending its findings. In particular, we want to mention \cite{Salasnich2004,Massignan03,Kamchatnov2004,Gerbier_2004}.}.
It is based on an approximate consideration of the broadening of the transverse wavefunction.
Starting point is the Hamiltonian for Bosons in 3D, interacting via a delta functional potential. 
For an arbitrary potential in the longitudinal $z$ direction and a harmonic potential in the transverse direction, the Hamiltonian is given by
\begin{equation}
\hat H = \int \mathrm { d } \vec r \  \hat \Psi^\dagger ( \vec r ) \left[ - \frac { \hbar ^ { 2 } } { 2 m } \partial _ { \vec r } ^ { 2 } + \frac { m \omega _ { \perp } ^ { 2 } \left( x ^ { 2 } + y ^ { 2 } \right) } { 2 } + U ( z ) - \mu +  \gtd | \hat \Psi ( \vec r ) | ^ { 2 } \right] \hat \Psi ( \vec r ). \label{eq:H3D}
\end{equation}
Here $\partial _ { \vec r } ^ { 2 } = \frac{\partial^2}{{\partial x}^2} + \frac{\partial^2}{{\partial y}^2} + \frac{\partial^2}{{\partial z}^2}$ and the 3D coupling constant is given by $ \gtd = 4 \pi \hbar ^ { 2 } \as / m$.

We will now use the classical fields approximation and make the ansatz
\begin{equation}
\Psi ( \vec r ) = \psi ( z ) \times \frac { 1 } { \sigma(z) \sqrt { \pi } } e ^ { - \frac {  x ^ { 2 } + y ^ { 2 }  } { 2 \sigma^2(z) } }
\label{eq:H_salasnich_not_mini}
\end{equation}
of having a Gaussian wavefunction in the transverse direction. 
Furthermore, we will assume the width of the Gaussian to fluctuate slowly in the transverse direction, i.e., we will neglect $\partial_{ z } \sigma(z)$. 
After integrating out the transverse direction, one gets a Hamiltonian depending on $\sigma(z)$ and $\psi(z)$:
\begin{align} 
	H =   \int d z \  \psi^*(z) \Bigg[ - \frac { \hbar ^ { 2 } } { 2 m } \frac { \partial ^ { 2 } } { \partial z ^ { 2 } } + U(z) &- \mu + \frac { 1 } { 2 } \, \gtd \, \frac { \sigma ^ { - 2 } (z)} { 2 \pi } \, | \psi (z) | ^ { 2 }  \notag \\ &  + \frac { \hbar ^ { 2 } } { 2 m } \sigma ^ { - 2 } (z) + \frac { m \omega _ { \perp } ^ { 2 } } { 2 } \, \sigma ^ { 2 } (z) \Bigg]  \psi(z) .
\end{align}
Minimizing it with respect to $\sigma(z)$ gives
\begin{equation}
\sigma ^ { 2 } (z) = a _ { \perp } ^ { 2 } \sqrt { 1 + 2 \as  | \psi (z) | ^ { 2 } }. \label{eq:sigma_wf_broadened}
\end{equation}
Inserting this into \cref{eq:H_salasnich_not_mini,eq:H3D} gives
\begin{equation} 
H =  { \int d z \ \psi ^ { * }(z) \left[ - \frac { \hbar ^ { 2 } } { 2 m } \frac { \partial ^ { 2 } } { \partial z ^ { 2 } } + U(z) - \mu +  \hbar \omega_{ \perp } \sqrt { 1 + 2 \as  | \psi (z) | ^ { 2 } } \right] \psi (z) }. \label{eq:broadened_lieb_liniger}
\end{equation}
In the limit of $\as  | \psi (z) | ^ { 2 } \ll 1 $, i.e., the mean interaction energy per particle is much smaller than the transverse level spacing, this reduces to \cref{eq:trapped_lieb_liniger} with the 1D coupling strength given by \cref{eq:g1d_approx}.
Note that in \cref{sec:theo_comparison}, the implications of the density broadening for the phase fluctuations are discussed.

Starting from \cref{eq:broadened_lieb_liniger} and using functional derivatives along the lines of \cref{sec:bogo_classical_field_derivation}, we can find the modified Gross-Pitaevskii equation
\begin{equation}
\left[ - \frac { \hbar ^ { 2 } } { 2 m } \Delta + U ( z ) - \mu + \hbar \omega _ { \perp } \frac { 1 + 3 \as \rho _ { 0 } } { \sqrt { 1 + 2 \as \rho _ { 0 } } }\right] \sqrt { \rho _ { 0 } } = 0. \label{eq:gpe_broadened}
\end{equation}
Moreover we can obtain a modified second-order Hamiltonian. 
It is given by replacing
\begin{equation}
 \frac { \g } { 2 } \delta { \rho } ^ { 2 } (z) \quad \rightarrow \quad \frac{1}{2} \hbar \omega _ { \perp } \as \frac { 2 + 3 \as \rho_0 (z) } { \left( 1 + 2 \as \rho_0 (z) \right) ^ { 3 / 2 } } \delta { \rho } ^ { 2 } (z) \label{eq:g1D_broadened}
\end{equation}
in \labelcref{eq:H2_mora}. 
Similarly one can proceed with the other low energy effective field theories discussed in \cref{sec:low_energy,sec:DW_theo}.
Note that we only sketched the derivation in the continuum using the classical fields approximation.
However, one can do a calculation similar to the one in \rcite{Mora2003} in the discretized space starting from a quantum version of \cref{eq:broadened_lieb_liniger} \cite{rauer2018thesis}.

\section{Theory for the double well}
\label{sec:DW_theo}

As discussed in \cref{chap:exp_setup}, we can create a double well potential where we trap two clouds separated along one of the strongly confined directions.
Tunneling of atoms between the two wells can occur if the separating barrier is finite. 
In this section we will therefore extend the discussions of the previous sections to models for the tunnel-coupled double well.

The starting point for the different effective models will again be the Lieb-Liniger Hamiltonian. 
To get a description for the double well, the Hamiltonian for a single cloud \labelcref{eq:trapped_lieb_liniger} is simply duplicated to describe the left/right cloud respectively, and a linear tunneling term is added \cite{Whitlock2003}: 
\begin{equation}
	\hat H_\mathrm{DW} = \hat H_1 + \hat H_2 + \hat H_t
		\label{eq:coupled_lieb_liniger} 
\end{equation}
with
\begin{align}
	\hat H_{1,2} &=  \int \mathrm { d } z \,  { \hat \psi } ^ { \dagger } _{1,2} ( z ) \left[ - \frac { \hbar ^ { 2 } } { 2 m } \partial _ { z } ^ { 2 } + U ( z ) - \mu + \frac { \g } { 2 } { \hat  \psi } ^ { \dagger } _{1,2} ( z )  {\hat \psi_{1,2} } ( z ) \right]  { \hat \psi_{1,2} } ( z )  \\
	\hat H_t &= - \hbar J \int d z \left[ \hat \psi _ { 1 } ^ { \dagger } ( z ) \hat \psi _ { 2 } ( z ) + \hat \psi _ { 2 } ^ { \dagger } ( z ) \hat \psi _ { 1 } ( z ) \right]. \label{eq:Ht_full}
\end{align}
Here the single particle tunneling rate $J$ sets the strength of the tunnel coupling.
The first term in $\hat H_t$ describes a particle being annihilated in well 2 at longitudinal position $z$ and one particle being created in well 1 also at longitudinal position $z$, i.e., tunneling from well 2 into well 1. 
The second term describes tunneling in the opposite direction.

We will now again assume to be in the quasicondensate regime (see discussion in \cref{sec:lieb_liniger}) and discuss the corresponding low energy effective field theories for the two clouds in \cref{sub_sec:uncoupled_theo,sec:coupled_DW,sec:therm_sG}.


\subsection{Uncoupled double well}
\label{sub_sec:uncoupled_theo}

In the case of zero tunnel coupling ($ \hat H_t = 0$), we get the low energy description simply by replacing each of the Hamiltonians $\hat H_{1,2}$ in \cref{eq:coupled_lieb_liniger} by one of the quadratic Hamiltonians discussed in \cref{sub_sec:bogo_single_cond}. 
Each of the condensates is then described by independent Hamiltonians (\cref{eq:H2_mora,eq:H2LL_mora,eq:H2LL_cont}), which are quadratic in the respective phase and density fluctuations of the individual condensates. For example, in case of the Luttinger liquid Hamiltonian \labelcref{eq:H2LL_cont} we have
\begin{equation}
H^{(2)}_{\mathrm{ LL } \, 1,2} =  \int \mathrm { d } z \,  \left[ \frac { \g } { 2 } \delta \hat { \rho } ^ { 2 }_{1,2} ( z ) + \frac { \hbar ^ { 2 } } { 2 m }   \rho _ { 0 } ( z ) \left( \frac { \partial \hat \theta_{1,2} (z)  } { \partial z } \right) ^ { 2 }  \right]. 
\end{equation}
Here we have assumed the same density profile $\rho_0 (z)$ for both clouds.

Starting from such a description, it is often convenient to introduce relative and common density and phase fluctuations: 
\begin{equation}
	\delta \hat \rho_\pm ( z ) = \frac { 1 } { 2 } \left[ \delta \hat \rho _ { 1 } ( z ) \pm \delta  \hat \rho _ { 2 } ( z ) \right] , \quad \hat \varphi_\pm ( z ) = \hat \theta _ { 1 } ( z ) \pm \hat \theta _ { 2 } ( z ) . \label{eq:def_rel_com}
\end{equation}
Here the common fluctuations are marked by the subscript $+$ and the relative fluctuations by the subscript $-$. 
The commutation relations are
\begin{align}
	\left[\delta \hat \rho_+ ( z ),\hat \varphi_+ ( z^\prime )\right] = \left[\delta \hat \rho_- ( z ),\hat \varphi_- ( z^\prime )\right] &= \left[\delta \hat \rho_1 ( z ),\hat \varphi_1 ( z^\prime )\right] = \left[\delta \hat \rho_2 ( z ),\hat \varphi_2 ( z^\prime )\right] \notag \\
	\left[\delta \hat \rho_+ ( z ),\hat \varphi_- ( z^\prime )\right] = \left[\delta \hat \rho_- ( z ),\hat \varphi_+ ( z^\prime )\right] &= \left[\delta \hat \rho_1 ( z ),\hat \varphi_2 ( z^\prime )\right] = \left[\delta \hat \rho_2 ( z ),\hat \varphi_1 ( z^\prime )\right] = 0 .
\end{align}
Note that \cref{chap:corr,chap:gaussification,chap:non_equi} are mostly concerned with the relative phase fluctuation. 
In these chapters, we will therefore, for brevity, drop the minus sign when denoting the relative degrees of freedom.
However, for now, we will keep it.

Using the definitions of \cref{eq:def_rel_com}, we can also write the low energy effective Hamiltonian for the uncoupled double well system as sum of two independent Hamiltonians depending only on the common or relative fluctuations respectively (instead of the fluctuations of the individual condensates):
\begin{equation}
\hat H^{(2)}_\mathrm{DW} = \hat H^{(2)}_1 + \hat H^{(2)}_2 = \hat H^{(2)}_+  + \hat H^{(2)}_- .
\end{equation}
Note that for this step it is important to assume a symmetric double well leading to the same density profile $\rho_0 (z)$ for both wells. The case of an asymmetry in the density profile and its effect on non-equilibrium dynamics is extensively discussed in \rcite{Langen18}. 

Due to our particular definition \labelcref{eq:def_rel_com}, the pre-factors in $\hat H^{(2)}_\pm $ are different compared to $\hat H^{(2)}_{1,2} $.  
For example, in case of the continuous Luttinger liquid Hamiltonian we have
\begin{equation}
H^{(2)}_{\mathrm{ LL } \, \pm} =  \int \mathrm { d } z \,  \left[ \g \, \delta \hat { \rho } ^ { 2 }_\pm ( z ) + \frac { \hbar ^ { 2 } } { 4 m }   \rho _ { 0 } ( z ) \left( \frac { \partial \hat \varphi_\pm (z)  } { \partial z } \right) ^ { 2 }  \right]. \label{eq:H_LL_com_rel}
\end{equation}

\subsection{Non-vanishing tunnel coupling}
\label{sec:coupled_DW}

Deriving a low energy description for the case of non-vanishing tunnel coupling is not so simple. One way would be to completely neglect the density fluctuations in the tunnel coupling term, i.e.,
\begin{equation}
\hat \psi_{1,2}  ( z ) \approx e ^ { i \hat { \theta }_{1,2} (  z  ) } \sqrt { { \rho_0 } (  z  ) }. 
\end{equation} 
leading to
\begin{equation}
\hat H_t = -  2 \hbar J \int d z \rho_0 (z) \cos\left( \hat \varphi_-(z) \right). \label{eq:Ht_sG}
\end{equation}
Together with the Luttinger liquid model for $H_{1,2}$ this leads to the sine-Gordon Hamiltonian for the relative degrees of freedom:
\begin{equation}
	\hat H _ { \mathrm { SG } } = \int d z \left[ \g \, \delta \hat \rho ^ { 2 }_- (z)  + \frac { \hbar ^ { 2 } \rho_0 (z)} { 4 m } \left( \frac { \partial \hat \varphi_- (z) } { \partial z } \right) ^ { 2 } - 2 \hbar J \rho_0 (z) \cos \left( \hat \varphi_- (z) \right) \right] \label{eq:H_SG} .
\end{equation}
The common degrees of freedom are described by the Luttinger liquid Hamiltonian \labelcref{eq:H_LL_com_rel}.

It was was proposed in \rcite{Gritsev2007} that \cref{eq:H_SG} often describes the relevant physics for the coupled double well. 
The validity of \cref{eq:H_SG} as an approximation for \cref{eq:coupled_lieb_liniger} is discussed in \cref{sec:theo_comparison} and \rcite{Beck18}.
The thermal fluctuations following from $H _ { \mathrm { SG } }$ will be the topic of the next \cref{{sec:therm_sG}}.

%

Within the classical fields approximation, we can also expand the tunneling term up to second order in the density fluctuations following the procedure outlined in \cref{sec:bogo_classical_field_derivation}. Expanding up to second order leads to
\begin{equation}
H_t = -  2 \hbar J \int d z \left[ \rho_0 (z) + \delta \rho_+ (z) - \frac{1}{2}\frac{\delta \rho^2_- (z)}{\rho_0 (z)}\right] \cos( \varphi_- (z) ). \label{eq:extended_SG}
\end{equation}
Note that the first order term does not vanish for $\rho_0 (z)$ being a solution of the Gross-Pitaevskii equation \labelcref{eq:gpe_mora}, it is therefore unclear whether $\delta \rho$ is still a small parameter.
The usefulness and validity of \labelcref{eq:extended_SG} is therefore questionable.
It is only mentioned for completeness and to argue that the tunnel coupling term can lead to coupling between common and relative degrees of freedom, which might potentially be important for explaining the results presented in \cref{chap:non_equi}.


In the case of strong tunnel coupling one can assume the relative phase $\varphi_- (z)$ itself to be small and of the same order as the other small parameters \cref{eq:small_parameters_bogo}. 
With this a Bogoliubov theory for the strong tunneling was developed in~\rcite{Whitlock2003}.
Starting from \cref{eq:extended_SG}, we can replace $\cos( \varphi_- (z) )$ with $1 - \varphi_-^2 (z)/2 $. 
We then can account for the zeroth and first order terms by rescaling the chemical potential.
The leading order term is therefore the second-order term 
\begin{equation}
H_t^{(2)} = \hbar J \int d z  \ \rho_0 (z) \, \varphi_-^2 (z). \label{eq:Ht_2}
\end{equation}


\subsection{Thermal fluctuations within the sine-Gordon model}
\label{sec:therm_sG}

Let us start by noting that the sine-Gordon Hamiltonian \cref{eq:H_SG} is non-quadratic, which leads to non-Gaussian phase fluctuations.
For the classical fields approximation, this connection is apparent from \cref{eq:thermal_classical_dens_func}.
Investigating the non-Gaussianity will be a big part of \cref{chap:corr}.  

Let us continue by discussing some theoretical predictions following from the sine-Gordon Hamiltonian \cref{eq:H_SG} in classical fields approximation and thermal equilibrium.
Moreover, we assume an infinite system with homogeneous background density, i.e., $\rho_0 (z) = \nod$.
One can then obtained some analytical results via the transfer-matrix method.
We will here not get into details about the used method, but just present some relevant results.
For details about the calculations and additional results see \rcite{Grisins2013,Schweigler17,Beck18}.

In the classical fields approximation, the thermal phase fluctuations following from the sine-Gordon model \cref{eq:H_SG} are characterized by two length scales \cite{Grisins2013}:
The phase coherence length 
\[
\lambda_T= \frac{2\hbar^2\nod}{m k_\mathrm{B} T}
\]
describing the randomization of the phase due to the temperature $T$, and the healing length of the relative phase 
\[
l_J=\sqrt{\frac{\hbar}{4mJ}}
\]
determining the restoration of the phase coherence through the tunnel coupling $J$. 

The dimensionless ratio 
\begin{equation}
q = \frac{\lambda_T}{l_J} \label{eq:q}
\end{equation}
is directly related to the observable quantity $\langle \cos(\varphi_-) \rangle$, which we usually call `coherence factor'.
It determines the relevance of non-quadratic contributions to the Hamiltonian \labelcref{eq:H_SG} and is used to quantify the `phase locking' in the experiment (see discussion in \cref{sec:phase_locking}).
The dependence of $\langle \cos(\varphi_-) \rangle$ on $q$ is shown in \cref{fig:cohvsq}.
Note that the ratio $q$ is the actual physically relevant parameter.
Changing the thermal coherence length $\lambda_T$ just simply leads to a rescaling of the spatial coordinate $z$.
For the same value of $q$, but different values of $\lambda_T$, one gets the same results when using the normalized coordinate $z/\lambda_T$.

Starting from the transfer matrix method, an efficient numerical method was developed in \rcite{Beck18}.
It consists of a stochastic evolution along the spatial coordinate $z$ and delivers numerical realizations for classical fields, following the thermal distribution for a particular Hamiltonian.
The method can be applied to a large class of different Hamiltonians.
However, certain criteria have to be fulfilled.
For example, we can generally only apply it to infinite homogeneous systems.
None of the parameters in the Hamiltonian may depend on $z$.
The numerical realizations represent a part of the homogeneous infinite system. 

For the sine-Gordon Hamiltonian \cref{eq:H_SG} with constant background density $\rho_0 (z) = \nod$ all criteria for the applicability of the method are fulfilled.
Note that in classical fields approximation the thermal distribution \cref{eq:thermal_classical_dens_func} for the sine-Gordon Hamiltonian will factorize in a part concerning phase fluctuations and a part concerning density fluctuations only.
The thermal phase fluctuations are therefore completely independent from the thermal density fluctuations.
This means that we can have a stochastic process giving us numerical realizations for the relative phase $\varphi_-(z)$ only.
In the following, we will often refer to this stochastic process as the `sine-Gordon stochastic process'.
Note that the phase profiles produced by it have a global ambiguity of $2 \pi n$, where $n$ is an integer. 
Therefore, only periodic functions of the phase $\varphi_-(z)$ or functions of phase differences $\varphi_-(z) - \varphi_-(z')$ between two spatial points should be calculated from them (see also the discussion in \cref{sec:phase_extraction,fig:phase_fitting}).

In \cref{chap:corr}, the sine-Gordon stochastic process will be used for a comparison of experimentally obtained phase correlation functions with the predictions following from the thermal sine-Gordon model in classical fields approximation.
More theoretical results can therefore be found in that chapter, especially in \cref{sec:compare_to_SG}.


\section{Comparison of the different models}
\label{sec:theo_comparison}


In the experiment, we routinely probe the phase fluctuations. 
We will therefore mostly focus on them in this section. 
In classical fields approximation and thermal equilibrium, the Bogoliubov Hamiltonian \labelcref{eq:H2_mora} and the Luttinger liquid Hamiltonian \labelcref{eq:H2LL_mora} give the same phase fluctuations, as can be seen from the discussion in \cref{sec:bogo_thermal_exp_val}. 
For the homogeneous system the variance of phase fluctuations between two different points was calculated in \cref{sec:homo_bogo}, \cref{eq:phase_fluct_single_cond} as
\begin{equation}
\left\langle \left[ \theta( z + \Delta z ) - \theta( z ) \right] ^ { 2 } \right\rangle = \frac{2}{\lambda_T} \left| \Delta z \right|.
\end{equation}
The validity of the classical fields approximation was discussed in \cref{sec:val_classical_fields}.

\begin{figure}
	\centering
	\begin{subfigure}{0.49\textwidth}
		\centering
		\includegraphics[width=\linewidth]{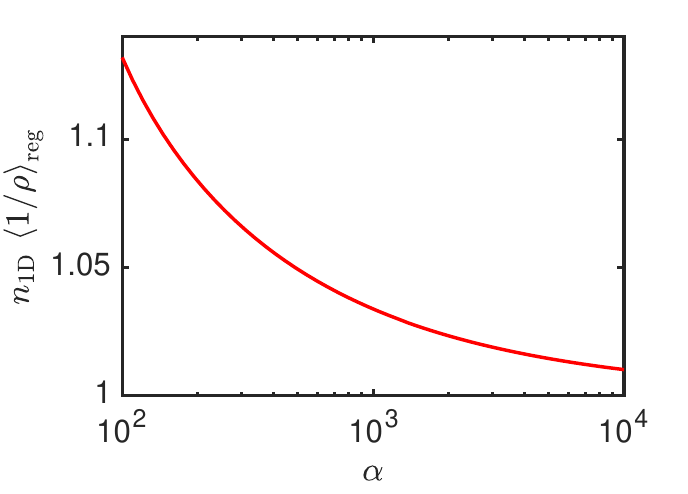}
		\caption{}
		\label{fig:rescalefactvsalpha}
	\end{subfigure}
	\begin{subfigure}{0.49\textwidth}
		\centering
		\includegraphics[width=\linewidth]{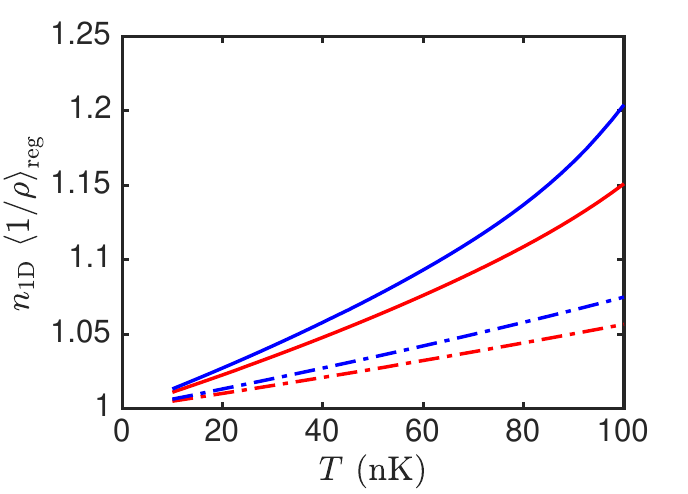}
		\caption{}
		\label{fig:rescalefactvst}
	\end{subfigure}
	\caption{
		\textbf{Rescale factors for \boldmath{$\lambda_{ T }$}.} \textbf{(\subref{fig:rescalefactvsalpha})} Shows the rescaling factor $\nod \left\langle 1 / \rho \right\rangle_\mathrm{reg}$ (according to \cref{eq:rescaled_lT}), for comparing the phase fluctuations of the full Hamiltonian \labelcref{eq:lieb_liniger_classical_field_dens} to the ones of the Luttinger liquid theory \labelcref{eq:H2LL_cont}, as a function of $\alpha$ \labelcref{eq:alpha_def}. 
		\textbf{(\subref{fig:rescalefactvst})} The red curves represent the same as is shown in (\subref{fig:rescalefactvsalpha}) as a function of temperature. The solid/dash-dotted curve is for $\nod  = 60  $ and 100{\um}\tss{-1} respectively. 
		The blue lines represent the corresponding results for the broadened Hamiltonian \labelcref{eq:broadened_lieb_liniger}.
		The plot was made for a transverse trap frequency of $\omega_{ \perp } = 2 \pi \times 1.4 ${\kHz}.
		The values for $\alpha$ range from 84 to 8400 for the solid curves and from 390 to $39000$ for the dash-dotted ones.
	}
\end{figure}

As discussed in \cite{Beck18}, the full Hamiltonian \labelcref{eq:trapped_lieb_liniger} in classical field approximation \labelcref{eq:lieb_liniger_classical_field_dens} basically leads to the same phase fluctuation, but with the rescaled thermal coherence length
\begin{equation}
\tilde { \lambda } _ { T } = \frac{\lambda _ { T }}{\nod \left\langle 1 / \rho \right\rangle _ { \mathrm { reg } }} . \label{eq:rescaled_lT}
\end{equation}
Here $\rho$ stands for the local one dimensional atomic density. 
Note that one has to regularize $\left\langle 1 / \rho \right\rangle$ in order to avoid a division by zero (see discussion in \cite{Beck18}).
The relevant parameter determining the rescaling factor in the case \labelcref{eq:lieb_liniger_classical_field_dens} without density broadening is given by
\begin{equation}
\alpha = \frac { \lambda _ { T } ^ { 2 } } { 4 \xi _ { \mathrm { h } } ^ { 2 } } \label{eq:alpha_def}
\end{equation}
with the healing length $\xi_\mathrm{h}$ \labelcref{eq:healing_length}. 
This means that we have the proportionality $\alpha \propto \nod ^3 / T^2$.
The dependence of the rescaling factor on $\alpha$ is shown in \cref{fig:rescalefactvsalpha} . 
One sees that the fluctuations are enhanced in comparison with the Luttinger liquid theory.

In the case of density broadening \labelcref{eq:broadened_lieb_liniger}, the rescaling factor also depends explicitly on the 1D density $\nod$. 
The temperature dependence is shown for a few values of the density in \cref{fig:rescalefactvst}.
One sees that the phase fluctuations are even more enhanced than in the case without density broadening.

The validity of the rescaling \cref{eq:rescaled_lT} was rechecked in the process of writing this thesis. 
For the variance of the cosine-transformed phases (see discussion in \cref{sec:homo_bogo,sub_sec:calc_corr_momentum}), good agreement was found up to $k_\mathrm{c}$ \labelcref{eq:classical_crit_LL} for $\alpha = 85$ and $525$ (not shown). 
Above $k_\mathrm{c}$ deviations are visible.
However, for these $k$-values one does not expect the classical fields approximation to work anyway.

Note that no simple rescaling law was found for the density fluctuations. 
However, it seems that the deviations from Bogoliubov theory are fairly small. 
For the variances of the cosine transformed densities the deviations are on the order of a few percent for $k < k_\mathrm{c}$ \labelcref{eq:classical_crit_LL}.
Again this was checked for $\alpha = 85$ and $525$ (not shown). 

Comparing the phase fluctuations following from the coupled Lieb-Liniger model \labelcref{eq:coupled_lieb_liniger} with the ones following from the sine-Gordon model \labelcref{eq:H_SG} we can get agreement in a certain parameter range when rescaling $q$ \labelcref{eq:q} in addition to $\lambda_T$.
The rescaled value is given by \cite{Beck18}
\begin{equation}
	\tilde q = q \ \sqrt{\frac{\left\langle \sqrt{\rho_1 \rho_2} \right\rangle}{\left\langle 1 / \rho_{1,2} \right\rangle_\mathrm{reg} }},
\end{equation}
where $\rho_{1,2}$ are the densities in the left or right cloud respectively.
Note that it doesn't matter whether we use $\rho_{1}$ or $\rho_{2}$ in order to evaluate $\left\langle 1 / \rho \right\rangle_\mathrm{reg}$ as we assume a symmetric double well trap.
As discussed in \rcite{Beck18}, the rescaling works quite well for the parameter range typically used in the experiment.


\section{Time of flight expansion}
\label{sec:theo_tof}

For all measurements presented in this thesis the atomic density after time of flight (TOF) expansion is probed via absorption imaging (see \cref{chap:imaging}). 
By measuring this atomic density, we want to infer quantities of the system before expansion. 
This connection is not completely trivial and needs a theoretical modeling of the expansion in TOF, which will be the topic of this section.

\subsection{Expansion without interactions}
\label{sec:theo_tof_no_inter}

Let us start by assuming that the quasicondensates in well 1 and 2 before expansion ($t=0$) are described by the classical fields
\begin{equation}
\Psi_{1,2}(\vec{r},t=0) = \phi_{1,2}(x,y,t=0) \, \psi_{1,2}(z,t=0) \label{eq:before_expansion}
\end{equation} 
respectively.
The fields $\psi_{1,2}(z,t=0)$ are stochastic, i.e., they change from shot to shot. 
Their statistics can be given by thermal and/or quantum fluctuations, or the fluctuations of a non-equilibrium situation. 
For the transverse part $\phi_{1,2}$ we will assume the harmonic oscillator ground state, i.e.,
\begin{equation}
	\phi_{1,2}(x,y,t=0) = \frac{1}{\sqrt{\pi a_\perp^2}} \, \me^{-((x \pm d)^2 + y^2)/2 a_\perp^2}, \label{eq:trans_wf_harm_osc}
\end{equation}
with $a_\perp $ being the harmonic oscillator length \labelcref{eq:harm_osc_len}.
Note that the two wells are separated by a distance of $2d$ in the transverse $x$ direction. 
We will assume no separation in the transverse $y$ direction.

By using \cref{eq:trans_wf_harm_osc} we neglect the interaction induced broadening of the transverse wavefunction (see discussion in \cref{sec:3D_to_1D}).
Further, we will also neglect interactions during the expansion. 
This is quite well justified for the parameters in our experiment \cite{rauer2018thesis}. 
The free expansion is independent for the different spatial directions, meaning that $\Psi_{1,2}$ will keep its factorizable form.
It will be the product of a freely expanding harmonic oscillator ground state and a freely expanding longitudinal field $\psi_{1,2}(z,t)$, where $t$ is the TOF.

Free expansion of the harmonic oscillator ground state leads to a growing Gaussian wavefunction.
The wavefunction stays normalized, its standard deviation evolves like
\begin{equation}
\sigma_\mathrm{wf} = a_\perp \sqrt{1 + \omega_\perp^2 t^2}. \label{eq:harm_osc_width_expansion}
\end{equation}
Moreover, a fluctuating phase factor appears, which will be the cause of interference fringes.

With this assumptions, the atomic density after expansion is~\cite{schummthesis,langen2013thesis}
\begin{align}
\left|\Psi_1 + \Psi_2\right|^2 =  \frac{1}{\pi \sigma_\mathrm{wf}^2} & \, \me^{-y^2/\sigma_\mathrm{wf}^2} \ \Bigg\{  \me^{-(x + d)^2/\sigma_\mathrm{wf}^2} \psi_1^*(z) \psi_1(z)  + \me^{-(x - d)^2/\sigma_\mathrm{wf}^2} \psi_2^*(z) \psi_2(z) \notag \\ &+ 2 \, \me^{-(x^2 + d^2)/\sigma_\mathrm{wf}^2}\, \re \left[ \exp\left(i \ 2 \pi \frac{x}{\lambda_\mathrm{F}}\right) \psi_1^*(z) \psi_2(z)\right]  \Bigg\}.
\label{eq:dens_after_TOF}
\end{align}
Here we omitted the time arguments in $\psi_{1,2}(z,t)$ and $\sigma_\mathrm{wf}(t)$ for brevity; we will continue omitting it in the following. 
The last term in \cref{eq:dens_after_TOF} leads to interference fringes with the fringe spacing given by
\begin{equation}
\lambda_\mathrm{F} = \frac{\pi m a_\perp^2 \sigma_\mathrm{wf}^2}{\hbar t d}. \label{eq:fringe_spacing}
\end{equation}
Writing $\psi_{1,2}$ in phase/density representation \labelcref{eq:phase_dens}
\begin{equation}
	\psi_{1,2}(z) = \sqrt{\rho_{1,2}(z)} \, \mathrm{e}^{i \theta_{1,2}(z)},
\end{equation} 
we can write \cref{eq:dens_after_TOF} as
\begin{align}
\left|\Psi_1 + \Psi_2\right|^2 =  &\frac{1}{\pi \sigma_\mathrm{wf}^2}  \, \me^{-y^2/\sigma_\mathrm{wf}^2} \ \Bigg\{  \me^{-(x + d)^2/\sigma_\mathrm{wf}^2} \rho_1(z)  + \me^{-(x - d)^2/\sigma_\mathrm{wf}^2} \rho_2(z) \notag \\ &+ 2 \, \me^{-(x^2 + d^2)/\sigma_\mathrm{wf}^2}\,  \sqrt{\rho_1(z) \rho_2(z)}  \, \cos \left[ \ 2 \pi \frac{x}{\lambda_\mathrm{F}} - \left(\theta_1(z) - \theta_2(z)\right)\right]  \Bigg\}.
\label{eq:dens_after_TOF_phase_dens}
\end{align}
For $x \gg d$, this can be written in the form 
\begin{equation}
\left|\Psi_1 + \Psi_2\right|^2 \approx A(z) \, \mathrm{e}^{-\frac{x^2 + y^2}{ \sigma_\mathrm{wf}^2}}  \left\{ 1 + C(z) \, \cos \left[ 2 \pi \frac{x}{\lambda_\mathrm{F}} - \left(\theta_1(z) - \theta_2(z)\right) \right] \right\},
\label{eq:dens_after_TOF_simple}
\end{equation}
which is used to fit the interference pattern in order to extract the relative phase $\varphi_- (z) = \theta_1(z) - \theta_2(z)$ from the measured atomic densities after TOF (see \cref{sec:phase_extraction}).

Note that a similar treatment of the expansion not making the assumption of having stochastic classical fields, but working with operators can be found in \rcite{van2018projective}.  

\subsection{Expansion with interactions}
\label{sec:theo_tof_inter}

If one wants to consider the density broadening and the interaction during expansion, the formalism presented in \rcite{Massignan03} might be an efficient way. 
However, here we will only talk about how to get a better guess for the width of the transverse wavefunction in TOF.
We will use this for the discussion of the imaging resolution in \cref{chap:imaging}.

For the in-situ clouds, we discussed in \cref{sec:3D_to_1D} how to approximately consider the influence of the interactions with a Gaussian ansatz.
This lead to the broadened transverse width for the in-situ clouds given in \cref{eq:sigma_wf_broadened}. 
Let us denote it here by $\sigma_\mathrm{b}$.
Using a Gaussian ansatz~\cite{Massignan03} also for the transverse expansion and neglecting the dynamics in the longitudinal direction one can derive an analytic expression for the time evolution of the transverse width $\sigma_\mathrm{wf}$. 
It is simply given by~\cite{rauer2018thesis}
\begin{equation}
\sigma_\mathrm{wf} = \sigma_\mathrm{b} \sqrt{1+\omega^2_\perp t^2}, \label{eq:exp_inter_gauss}
\end{equation}
i.e., by replacing $a_{\perp}$ in~\cref{eq:harm_osc_width_expansion} with the broadened width $\sigma_\mathrm{b}$.

\chapter{Probing the atom cloud through absorption imaging}
\label{chap:imaging}
	
The aim of an absorption imaging system~\cite{Ketterle1999,smith2011absorption} is to measure the atomic density (integrated along the imaging direction) as well as possible. 
A big part of this chapter will be concerned with the unwanted effects preventing a perfect image. 
These effects include the optical resolution (\cref{sec:opt_imag_res}), the `recoil blurring' due to absorption and re-emission of photons by the atoms (\cref{sec:recoil_blurring}) as well as coarse graining through the finite pixel size of the camera.


\begin{figure}
	\centering
	\includegraphics[width=1\linewidth]{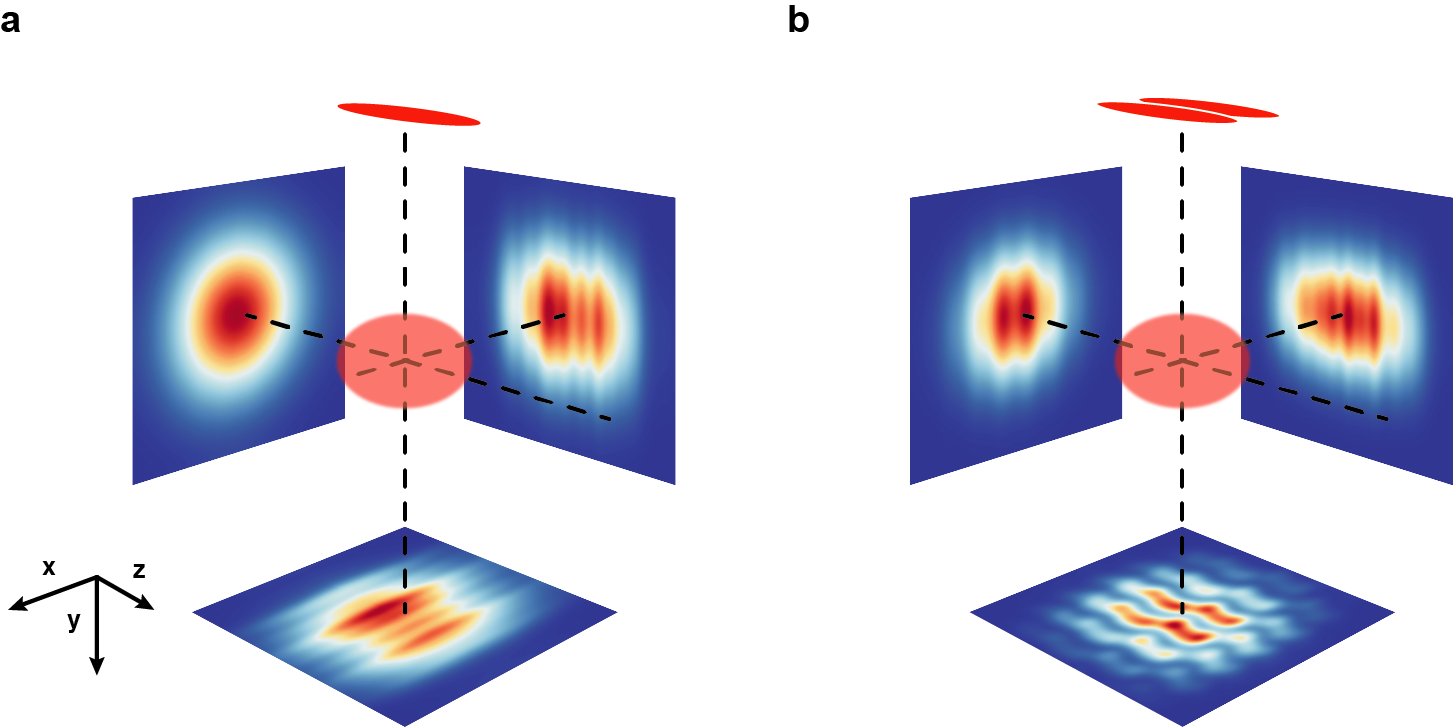}
	\caption{\textbf{Possible imaging directions.}
		The red cigars in the figure illustrate the in situ atom clouds in a single well \textbf{(a)} and a double well \textbf{(b)} trap.
		After being released from the trap, the clouds expand in time of flight (TOF), as illustrated by the red blob.
		Subsequently absorption images are taken.
		The pictures show typical atomic densities (integrated along the imaging direction) as recorded from the different directions.
		The example images in the $x$ direction (`transverse' imaging system) show speckle patterns, which can be used for thermometry as discussed in \cref{sec:DR_therm}.
		The example image in the $y$ direction (`vertical' imaging system) in \textbf{(b)} shows the interference pattern following from the expansion and overlap of the two clouds released from the double well trap. 
		From such images, the relative phase $\varphi_-(z)$ between the two condensates can be extracted (see \cref{sec:phase_extraction}). 
		The imaging system in the $z$ direction (`longitudinal' imaging system) can be used to record integrated interference pictures (example image in \textbf{(b)}) or to measure the imbalance in the atom number between the two wells~\cite{rauer2018thesis}.
		Figure reproduced with permission from~\cite{rauer2018thesis}.}
	\label{fig:figimagsys}
\end{figure}

In our experimental setup, we have absorption imaging systems for all spatial directions $x$, $y$ and $z$. 
See \cref{fig:figimagsys} for a schematic of the alignment of the different imaging systems.
Remember that $z$ is the direction along the weakly confined (the longitudinal) direction of the cloud, while $y$ is aligned with the direction of gravity.
For details about the imaging systems see~\cite{gring2012thesis,rauer2012mastersthesis,langen2013thesis}.
Some quantities for the different imaging systems are summarized at the end of the chapter in \cref{tab:imaging_param}.

In absorption imaging the atom cloud is probed with a resonant laser beam. The incoming laser beam is attenuated as photons get absorbed by the atom cloud. For a two level system, the relation between the intensity $I_0$ of the incoming beam and the intensity $I$ after absorption is given by
\begin{equation}
\ln\left(\frac{I_0(x,y)}{I(x,y)}\right) + \frac{I_0(x,y)  - I(x,y)}{I_{\mathrm{sat}}} = \sigma \tilde{\rho}(x,y). \label{eq:abs_imag_basic}
\end{equation}
Here we have assumed that equilibrium has been reached and the populations of the ground and excited state do not change anymore. 
The equation is therefore not valid shortly after switching the imaging light on or changing its intensity. 
Furthermore, without loss of generality, we have assumed to image in the $z$ direction. The saturation intensity is denoted by $I_{\mathrm{sat}}$, $\sigma$ represents the absorption cross-section and $\tilde{\rho}(x,y)$ is the atomic density integrated in imaging direction $z$. 
Measuring $I(x,y)$ and $I_0(x,y)$ (measured by taking a second picture without any atoms present) one can determine $\tilde{\rho}(x,y)$. 
Note that in \cref{eq:abs_imag_basic} the possible re-absorption of spontaneously emitted photons is neglected. 
No literature discussing the importance of this effect is known to us. 

\section{Scattering cross section for the different imaging configurations}
\label{sec:im_cross_section}

We image the cloud of \tss{87}Rb by using the D\tsub{2} (5\tss{2}S\tsub{1/2} $\rightarrow$ 5\tss{2}P\tsub{3/2}) transition, which is also used for laser cooling. The total angular momentum quantum number is $F=2$ for the ground state and $F'=3$ for the excited state. One therefore has five ground $m_F$-states and seven excited $m_{F'}$-states. 
This means that the simple formula \labelcref{eq:abs_imag_basic} is not valid anymore. However, for some particular combinations of polarization and atomic quantization axis (determined by the applied bias field) one gets an effective two-level system. Typically one chooses such a configuration for absorption imaging. 

One example would be the case of circularly polarized light and quantization axis aligned with the imaging direction. 
When the imaging light is switched on, the populations of the different ground $m_F$ and excited $m_{F'}$ states will rearrange and ultimately reach an equilibrium. 
For the discussed situation, one always ends up in the $|F=2, m_F = \pm 2 \rangle \rightarrow |F'=3, m_{F'} = \pm 3 \rangle$ cyclic transition where the sign of $m_F$ and $m_{F'}$ depends on whether we have left or right circularly polarized light.

\Cref{eq:abs_imag_basic} is then valid with the on-resonance cross-section being given by~\cite{steck2015}
\begin{equation}
\sigma_0 = \frac{3\lambda^2}{2\pi}, \label{eq:sigma_0}
\end{equation}
where $\lambda$ is the wavelength of the imaging light. 
With this $\sigma_0$, we can then calculate the saturation intensity using   
\begin{equation}
I_{\mathrm{sat}}^0 = \frac{\hbar \omega}{2 \sigma_0 \tau}, \label{eq:i_sat_0}
\end{equation}
where $\hbar \omega$ is the photon energy and $\tau$ is the lifetime of the excited state (see \cite{steck2015} for a numerical value). 
For off resonant frequencies, the cross-section follows a Lorentzian function with the natural linewidth $1/\tau$. 
Note that the resonance frequency changes with the magnetic bias field as discussed in \cref{eq:landor_detuning_B_dep,eq:landor_detuning_B_dep_num} in \cref{sec:LAndor_obe}.
This configuration is used for the longitudinal and the vertical imaging systems in our experiment. 

A more complicated effective two-level system is achieved with linear polarization along the quantization axis of the atoms~\cite{steck2015}. 
The effective two-level system is again reached after initial equilibration. 
In the stationary state all the different $m_F$ states are occupied. 
The scattering cross-section is reduced with respect to \cref{eq:sigma_0} by a factor $\alpha$, i.e.,
\begin{equation}
\sigma = \frac{\sigma_0}{\alpha}, \label{eq:sigma_rescaled}
\end{equation}
which leads to
\begin{equation}
I_{\mathrm{sat}}  = I_{\mathrm{sat}}^0 \alpha. \label{eq:isat_rescaled}
\end{equation}
The numerical value is $\alpha = 1.83$ \cite{steck2015} for small magnitudes of the magnetic bias field fixing the atomic quantization axis.
It increases with field-strength as discussed in \cref{sec:TAndor_obe} and presented in \cref{fig:tandoralphavsb}. 
Also the linewidth increases with field strength (\cref{fig:linewidthvsbtandor}), while the lineshape stays Lorentzian (\cref{fig:linewidthvsbtandor}). 
That the latter follows from the observation that we have an effective two-level transition. 
This configuration is used for the transverse imaging system in our experiment.

To investigate the influence of the initial equilibration as well as perturbations to the configurations (e.g., stray fields), it is useful to simulate the absorption process with optical Bloch equations.
This is done in \cref{chap:obe}, where calculations using the parameters of the different imaging systems are presented.
Some of the results show substantial deviations from the quantities for the idealized situations.
Unfortunately it would be rather involved to consider these deviations when analyzing the experimental absorption images.
We will therefore not do that and simply use \cref{eq:sigma_0} for the longitudinal and the vertical imaging systems and \cref{eq:sigma_rescaled} with $\alpha = 1.83$ for the transverse imaging system.


Another effect to keep in mind is the Doppler shift due to photon absorption during the imaging process. When a photon gets absorbed, it transfers its momentum to the atom which therefore gets accelerated. The difference in velocity is given by the recoil-velocity, its value for \tss{87}Rb is $v_r = 5.8845${\mm}/s \cite{steck2015}. 
The photon might then be re-emitted by spontaneous or stimulated emission. 
The first happens in a random direction, it therefore does not lead to an average acceleration. 
The latter happens in the direction of the imaging light, leading to an acceleration of the atom counteracting the one of the initial absorption. 
The number $n$ of photons absorbed and not re-emitted by stimulated emission therefore leads to an average acceleration of $\Delta v = n \, v_r$ in the imaging direction.
This leads to a Doppler shift of the resonance frequency. 

For small velocities, the frequency shift is approximately given by $\Delta f = - v/c \times f_0$.
Here, $v$ is the atomic velocity, $c$ the speed of light and $f_0$ the unshifted resonance frequency. 
In our case, this lead to a shift of 7.54{\kHz} per scattered photon.
Typically around 200 photons per atom are absorbed during the imaging process. 
This leads to a Doppler shift of 1.51{\MHz}, which is still rather small compared to the natural linewidth of 6.07{\MHz}. 
In the experiment one will set the light frequency approximately to the mean of unshifted and maximally shifted frequency, i.e., the relevant value for the detuning is half the Doppler shift. 
For this value, the scattering cross section is reduced to about 95\%.  

However, in the experiment, the imaging intensity is quite inhomogeneous.
There are regions where up to twice as many photons are scattered. 
For half of this Doppler shift, the scattering cross section is reduced to about 80\%. 
The Doppler shift is therefore not necessarily negligible. 
Despite this, we will ignore it when analyzing experimental absorption images. 

\section{Optical imaging resolution for a Gaussian cloud}
\label{sec:opt_imag_res}

All optical imaging systems are fundamentally limited by diffraction. As we are using laser-light for imaging, we will only discuss coherent image formation. Coherent image formation deals with the complex amplitudes $U$ of the light-field rather than with intensities as in incoherent image formation. 

Let us again assume that we image in the $z$ direction.
For a thin object, the light field amplitude $U_3(x_3,y_3)$ in the imaging plane is given by the convolution of the light field amplitude $U_1(x_1,y_1)$ in the object plane (immediately after the thin object) with the 2D amplitude point spread function (APSF) $h(x,y)$~\cite{gu2000advanced}. We have
\begin{equation}
U_3(x_3,y_3) \propto \int \int_{-\infty}^{\infty} dx_1 dy_1 \  U_1(x_1,y_1) \; h(x_1 + M x_3,y_1 + M y_3), \label{eq:psf_thin_obj}
\end{equation}   
where $M$ is the magnification of the imaging system. In the expression we omitted a phase factor depending on $x_3$ and $y_3$ as it is of no importance for our discussion. We will also omit this phase factor in all other expressions of this chapter.
Note that for \cref{eq:psf_thin_obj} to be valid, we have to assume that the lens law is fulfilled and $h(x,y)$ falls off quickly.
If the light field $U_1(x_1,y_1)$ consists only of a single point at $x_1 = y_1 = 0$, i.e., is given by the Dirac delta function $\delta(x_1,y_1)$, then $U_3(x_3,y_3)$ is proportional to the amplitude point spread function $h(M x_3,M y_3)$, hence its name.
The form of $h(x,y)$ depends on the details of the imaging system and whether the object is in focus, or better said on the distance between the object plane and the plane in focus. For a circular aperture with numerical aperture NA and object in focus, one has~\cite{gu2000advanced}
\begin{equation}
h(x,y) = \tilde{h}(\tilde{r}(x,y)) = 2 \pi \frac{J_1(\tilde{r})}{\tilde{r}} \label{eq:airy_amp}
\end{equation}
where
\begin{equation}
\tilde{r} = \frac{2\pi}{\lambda} \, \sqrt{x^2+y^2} \, \mathrm{NA}.
\end{equation}
Here $J_n(x)$ represent the Bessel functions and $\lambda$ is the wavelength of the light. Squaring the expression in \cref{eq:airy_amp} gives the well known airy pattern. 

Often it is more convenient to work with the image in the object plane 
\begin{equation}
\tilde{U_3}(\tilde{x}_3,\tilde{y}_3) = U_3\left(-\frac{\tilde{x}_3}{M},-\frac{\tilde{y}_3}{M}\right).
\end{equation}
Using this quantity, \cref{eq:psf_thin_obj} turns into 
\begin{equation}
\tilde{U_3}(\tilde{x}_3,\tilde{y}_3) \propto \int \int_{-\infty}^{\infty} dx_1 dy_1 \  U_1(x_1,y_1) \; h(x_1 -  \tilde{x}_3,y_1 -\tilde{y}_3), \label{eq:psf_thin_obj_onj_plane}
\end{equation}
which can be written as 
\begin{equation}
\tilde{U_3}(\tilde{x}_3,\tilde{y}_3) \propto \mathcal{F}^{-1} \left(\mathcal{F}( U_1(x_1,y_1) ) \times c(k_x,k_y) \right). \label{eq:apply_ctf}
\end{equation}
Here $\mathcal{F}$ and $\mathcal{F}^{-1}$ represent the 2D Fourier transform and its inverse respectively, and we have defined the 2D coherent transfer function (CTF) as
\begin{equation}
c(k_x,k_y) = \mathcal{F}^{-1} ( h(x,y) ).
\end{equation}
Note that a narrow APSF corresponds to a wide CTF and good imaging resolution.

The APSF in \cref{eq:airy_amp} becomes narrower, i.e., the resolution becomes better, for higher values of NA. 
However, with increasing NA the depth of focus decreases, meaning that the resolution worsens more and more rapidly with the defocus distance. Therefore, the optical resolution of imaging systems with a high numerical aperture can be substantially reduced when imaging extended objects (extended in the imaging direction). This is the case when measuring an atom cloud after time of flight (TOF) expansion. 



In the following, we will discuss the effective coherent transfer function for a cloud with Gaussian atomic density in the imaging direction $z$ and arbitrary dependence on $x$ and $y$. Note that this case was discussed in~\cite{buecker2009}. 
Therefore, we will only state and discuss the result and the assumptions made in its derivation here. 

We assume that the atomic density after TOF has the form
\begin{equation}
\rho(x,y,z) = \frac{1}{\sqrt{\pi} \sigma_\mathrm{wf}}  \mathrm{e}^{-\frac{(z-z_0)^2}{\sigma_\mathrm{wf}^2}} \tilde{\rho}(x,y),
\end{equation}
where $z$ represents the imaging direction, and $\tilde{\rho}(x,y)$ is an arbitrary function. 
The standard deviation of the atomic wavefunction after TOF is denoted by $\sigma_\mathrm{wf}$ leading to a standard deviation of $\sigma_\mathrm{wf}/\sqrt{2}$ for the Gaussian atomic density.
For information about how to calculate $\sigma_\mathrm{wf}$ see the discussion in \cref{sec:theo_tof} as well as \cref{eq:harm_osc_width_expansion,eq:exp_inter_gauss}.
As the origin for the $z$ coordinate we choose the in-focus plane (fulfilling the lens law), which means that $z_0$ represents the defocus distance for the center of the Gaussian cloud.
Further we assume that each infinitesimal slice along the $z$ direction of this extended cloud is imaged independently.
For each slice we apply \cref{eq:psf_thin_obj_onj_plane} and subsequently obtain the total image as a superposition. 
By doing so, we neglect that the incoming/outgoing light is modified through the atoms before/after the slice under consideration. One speaks of a semi-transparent cloud and calls the assumption the first Born approximation~\cite{gu2000advanced}.


We can then model the imaging of the extended cloud like a thin cloud being imaged with an effective coherent transfer function, i.e., we use \cref{eq:apply_ctf} with an effective $c(k_x,k_y)$ and
\begin{equation}
\label{eq:U1_gauss_cloud}
U_1(x_1,y_1) = U_0 \ \mathrm{e}^{-\sigma \tilde{\rho}(x_1,y_1)/2}.\footnote{Note that using the exponential in the formula rather than the linear approximation is somewhat inconsistent with the assumption of semi-transparency used to calculate the effective CTF.} 
\end{equation}
Here $\sigma$ represents the atomic scattering cross section and we have assumed a homogeneous incoming beam with amplitude $U_0$. Moreover, we assumed an imaging intensity far smaller than the saturation intensity.
Note that the factor $1/2$ in the exponential of \cref{eq:U1_gauss_cloud} appears because we deal with amplitudes and not intensities.

In case of a circular aperture, we get for the effective CTF
\begin{equation}
c_{eff}(k_x,k_y)  \propto \Theta\left(\frac{\mathrm{NA}}{\lambda} - k_t\right) \  \mathrm{e}^{-\left(\pi \frac{\lambda}{2} k_t^2 \sigma_\mathrm{wf} \right)^2} \ \mathrm{e}^{-i \pi \lambda \, k_t^2 z_0}, \label{eq:ctf_gauss_cloud}
\end{equation}
where 
\begin{equation}
k_t^2 = k_x^2+ k_y^2.
\end{equation}
The first term in \cref{eq:ctf_gauss_cloud} represents the fundamental diffraction limit of the optical system. The second term represents the decrease in resolution due to the finite extend of the Gaussian cloud. As soon as the second term dominates (is much narrower than the fist term), one cannot improve the imaging resolution by improving the optical system anymore. The last term appears if the center of the cloud is not in focus.
Note that the proportionality factor in \cref{eq:ctf_gauss_cloud} does not depend on the cloud position or extend, this will become important for a discussion later on.

For some reason only the real part of \cref{eq:ctf_gauss_cloud} was stated in \rcite{buecker2009}. 
For examples how the effective point spread function looks for various imaging situations used in the experiment see \cref{sec:res_tandor,sec:res_vandor}.

\section{Recoil blurring}
\label{sec:recoil_blurring}

During the imaging process photons are absorbed by the atoms and subsequently re-emitted. The absorption only gives a push in the imaging direction. 
It therefore does not influence the absorption image per se but can lead to the cloud being pushed out of focus during imaging. During the spontaneous emission on the other hand photons are emitted into a completely random direction. This leads to a random walk and a blurring of the image as discussed in \rcite{Ketterle1999}. 
In the following we will treat this subject a bit more carefully and discuss some aspects usually overlooked. 

The position of an atom absorbing $N$ photons during the imaging time $\Delta t$ is given by
\begin{equation}
\vec{r} = \vec{e}_N \  v_{\mathrm{rec}} \  \frac{\Delta t}{N} + \vec{e}_{N-1} \  v_{\mathrm{rec}} \ 2 \frac{\Delta t}{N} + \dots + \vec{e}_1 \  v_{\mathrm{rec}} \ \Delta t . \label{eq:r_vec_recoil_1}
\end{equation}   
Here $v_{\mathrm{rec}}$ represents the recoil velocity and the unit vectors $\vec{e}_j$ represent the directions in which the recoil from the $j$-th photon kicks the atom. 
Note that the lifetime of the 5\tss{2}P\tsub{3/2} state of \tss{87}Rb (the excited state of the D\tsub{2} line used for imaging) is 26{\ns}, which is much shorter then typical imaging durations (50 and 75{\us} are currently used in our experiment).
We can therefore neglect the time lag between absorption and spontaneous emission of the photons.

What we are interested in is the traveled distance squared: 
\begin{equation}
\left|\vec{r}\right|^2  = v_{\mathrm{rec}}^2  \  \frac{{\Delta t}^2}{N^2} \ \sum_{j=1}^{N} j^2 + \mathrm{mixed \ terms \ containing \ } \vec{e}_j \cdot \vec{e}_k \mathrm{\ with\ } j \neq  k .
\end{equation}
The sum can be evaluated to 
\begin{equation}
\sum_{j=1}^{N} j^2 = \frac{1}{6} N (N+1) (2N + 1).
\end{equation}
For the expectation value of the squared distance all the mixed terms drop out and we get
\begin{equation}
\left\langle \left|\vec{r}\right|^2 \right\rangle  = v_{\mathrm{rec}}^2  \  {\Delta t}^2 \  \frac{1}{6 N} (N+1) (2N + 1) .
\end{equation}
For a large number of scattered photons $N$ we can use the approximate formula
\begin{equation}
\left\langle \left|\vec{r}\right|^2 \right\rangle  \approx v_{\mathrm{rec}}^2  \  {\Delta t}^2 \  \frac{N}{3}, \label{eq:recoil_dist_usual}
\end{equation}
which is the expression given in~\cite{Ketterle1999}.

Note that as \cref{eq:r_vec_recoil_1} represents a sum of independent random variables the distribution of $\vec{r}$ will converge to a Gaussian for many absorbed photons, i.e.,
\begin{equation}
P(\vec{r}) \ dx \, dy \, dz = \frac{1}{({2 \pi \sigma_\mathrm{rc}^2})^{3/2}} \ e^{-\frac{|\vec{r}|^2}{2 \sigma_\mathrm{rc}^2}} \ dx \, dy \, dz \label{eq:recoil_endpoint_dist_3D}
\end{equation}
with 
\begin{equation}
 \left\langle \left|\vec{r}\right|^2 \right\rangle = 3 \, \sigma_\mathrm{rc}^2  . \label{eq:recoil_dist_sigma_connection}
\end{equation}


We now have expressions for the blurring at the end of the imaging procedure.
However, what really counts for absorption imaging is not the position of the atoms at the end of the exposure time, but the position of the atoms when the photons get absorbed. 
For the expectation value of the squared distance for this absorption points we get
\begin{align}
\begin{split}
\left\langle \left|\vec{r}\right|^2 \right\rangle_{\mathrm{ab}} &= \frac{1}{N} \sum_{n=1}^{N} \left\langle \left|\vec{r}\right|^2 \right\rangle_n = \frac{1}{N} \sum_{n=1}^{N} \left[ v_{\mathrm{rec}}^2    \left(\frac{\Delta t}{N}\right)^2   \frac{1}{6} n (n+1) (2n + 1) \right] \\
&=  \frac{1}{N}  v_{\mathrm{rec}}^2    \left(\frac{\Delta t}{N}\right)^2   \frac{1}{12}  N  (1+N)^2  (2+N) \approx  v_{\mathrm{rec}}^2  \,  {\Delta t}^2 \,  \frac{N}{12} .
\end{split}
\end{align}
Note that this is a factor 4 smaller than \cref{eq:recoil_dist_usual} which is normally used to estimate the effect of recoil blurring. 

The distribution of the points where the photons are absorbed is not a Gaussian anymore.
Instead, it is an average over Gaussian functions with different variance. 
We will now again assume that we image in the $z$ direction and only discuss the distribution for the 2D plane perpendicular to it. 
We have
\begin{equation}
P_\mathrm{ab}(x,y) = \frac{1}{N} \sum_{n=1}^{N} \frac{1}{2 \pi \sigma_{\mathrm{rc} \, n}^2} \mathrm{e}^{-\frac{x^2+y^2}{2 \sigma_{\mathrm{rc} \, n}^2}}, \label{eq:2D_dist_abs}
\end{equation}
with 
\begin{equation}
 \sigma_{\mathrm{rc} \, n}  = \frac{1}{3} \left\langle \left|\vec{r}\right|^2 \right\rangle_n = \frac{1}{3}  v_{\mathrm{rec}}^2 \left(\frac{\Delta t}{N}\right)^2 \frac{n^3}{3} = C \ n^3. \label{eq:def_sigma_n}
\end{equation}
\Cref{eq:2D_dist_abs} can be written as
\begin{equation}
P_\mathrm{ab}(x,y) = C_1 \sum_{n=1}^{N} n^{-3} \mathrm{e}^{-\frac{C_2}{n^3}}, 
\end{equation}
where
\begin{align}
C_1 &=\frac{1}{N} \frac{1}{2 \pi C} \\
C_2 &= \frac{x^2 + y^2}{2 C} 
\end{align}
with $C$ defined in \cref{eq:def_sigma_n}. 
The expression can be further simplified when approximating the sum by an integral, setting the lower integration bound to zero and making the substitution $t=C_2/n^3$. 
We get
\begin{align}
P_\mathrm{ab}(x,y) \approx \frac{C_1}{3 C_2^\frac{2}{3}} \int_{\frac{C_2}{N^3}}^{\infty} dt \ t^{\frac{2}{3} - 1} \mathrm{e}^{-t} = \frac{C_1 \ \Gamma(\frac{2}{3})}{3 \ C_2^\frac{2}{3}} \ \Gamma\left(\frac{C_2}{N^3},\frac{2}{3}\right),
\end{align}
where $\Gamma(s)$ represents the Gamma function, and  
\begin{equation}
\Gamma(x,s) = \frac{1}{\Gamma(s)} \int_{x}^{\infty} t^{s-1} \mathrm{e}^{-t}
\end{equation}
is the upper incomplete gamma function\footnote{Note that there are different definitions for the incomplete gamma functions, this is the definition implemented in MATLAB.}.

\section{Combining the different effects and simulating pictures}
\label{sec:imag_sim_pics}

We will now discuss how to combine the different effects. 
We will start by assuming that the effective 2D coherent transfer function is approximately given by~\cref{eq:ctf_gauss_cloud} for a Gaussian cloud. The Gaussian shape is a good approximation for the transverse wavefunction in-situ and after TOF, when having only a single cloud of atoms~\cite{rauer2018thesis}. 

In the case of two clouds in a double well potential, one gets interference patterns in the direction of the separation. The patterns will have an envelope of approximately Gaussian shape. If the direction of separation coincides with the imaging direction, one can strictly speaking not apply~\cref{eq:ctf_gauss_cloud} anymore. Without real justification, except the Gaussian envelope, we will still apply the formula in this case to get a rough idea about the effect of the cloud extension. This situation occurs when measuring two clouds separated in  horizontal $x$ direction with the transverse imaging system. 

Note that the spatial resolution of the longitudinal imaging system is not particularly important for the measurements presented in this thesis.
We will therefore not explicitly discuss it in this chapter.
Therefore we will also not discuss the shape of the longitudinal density profile or the validity of \cref{eq:ctf_gauss_cloud} here.  


Neglecting any effects but the optical resolution, the field in the imaging plane is simply calculated by convolving the field in the object plane with the effective APSF, i.e., by using  \cref{eq:apply_ctf} with \cref{eq:U1_gauss_cloud,eq:ctf_gauss_cloud}.

As already discussed in \cref{sec:recoil_blurring}, the photon recoil during the imaging process modifies the atomic density. 
During the imaging process the atoms get pushed in imaging direction and perform a random walk in all three dimensions. 
The push in the imaging direction leads to different defocus distances 
for the different times during imaging. 
The random walk leads to different blurring for the different times. 
To consider the effect of the photon recoil, one therefore has to incoherently sum up (integrate) the intensities for the different times in the imaging process. 

For simulating pictures, we will divide the imaging duration into several smaller time intervals (typically 50).
We then calculate the defocus distances $z_0$, the cloud extensions in imaging direction $\sigma_\mathrm{wf}$ (possibly considering the modification by the random walk) and the blurred atomic densities $\tilde \rho (x,y)$ (integrated in imaging direction) for the mid-point of each time interval. 
Subsequently, we will assume that these quantities are constant over the duration of one interval and apply \cref{eq:U1_gauss_cloud,eq:ctf_gauss_cloud,eq:apply_ctf} for each time interval. 
The resulting fields $\tilde{U_3}(\tilde{x}_3,\tilde{y}_3) $ are then squared and summed up. 
For the summing up, it is important to note that the prefactor in \cref{eq:ctf_gauss_cloud} does not depend on the cloud position or width as already discussed above.

In addition to the effects already discussed, there is another source limiting the resolution of the longitudinal and transverse imaging system. 
The CCDs in the cameras shift during the exposure leading to a smearing of the image. 
The shift only happens in one spatial direction which coincides with the $z$ direction (the longitudinal direction of the atom cloud) for the transverse imaging system and the vertical $y$ direction for the longitudinal imaging system.
For simulating pictures, the shift is simply considered by shifting the sub-pictures generated to consider the imaging push and recoil blurring (see the discussion above).
The exposure time in the experiment is $75${\us} for the longitudinal and transverse imaging system and one row of the CCD is shifted every $16${\us}.
This leads to a smearing of the recorded picture over $4.7$ pixels.
Note that this shift is just an insufficiency of the cameras currently used and could easily (but costly) be fixed by simply exchanging the camera. 

After the simulated pictures have been calculated considering all the effects discussed so far, we add photon shot noise.
Before adding the noise, we have a certain number of detected photons at every pixel.
To add the shot noise, we then assume a Poisson distribution at every pixel, which has that photon count as its mean value.
From this distribution different at every pixel, we then draw a random variable which  represents the pixel of the simulated picture.  


In the following, we will discuss the imaging resolution for the transverse and the vertical imaging system in detail. The longitudinal imaging system will not be discussed as it is mainly used for checking the overall atom number difference between the two clouds of a double well potential \cite{rauer2018thesis}, an application for which the imaging resolution is not very important.


\section{Resolution of the transverse imaging}
\label{sec:res_tandor}

The transverse imaging system images from the horizontal $x$ direction (see \cref{fig:figimagsys}).
The imaging light is switched on for 75{\us} and has an intensity of 30\%  of the on resonance saturation intensity $I_\mathrm{sat}^0$ \labelcref{eq:i_sat_0}.
During this time, every atom typically absorbs around 200 photons. 
We usually take images after 2 or 11.2{\ms} TOF. 
The short TOF is for example used to measure the approximate in-situ density profile~\cite{Rauereaan7938,rauer2018thesis}. 
The long TOF is used for thermometry via `density ripples' (see discussion in \cref{sec:DR_therm} and \cite{manz10}). 
A detailed discussion about how the imaging system influences this thermometry can be found in \cref{sec:imag_DR}. 
In the following, we will discuss the imaging resolution for the two cases separately. 
We will start by discussing the case of the large TOF as this is also used for focusing the imaging system in the experiment.

\subsection{Resolution for large time of flight}

Let us start by noting that the imaging system has a high resolution objective with $\mathrm{NA} \approx 0.20$~\cite{rauer2018thesis}. 
Also note that for long TOF, the imaging light is not reflected off the atomchip but passes straight through the chamber \cite{gring2012thesis}. 
This is in contrast to the case of having small TOF discussed in the next subsection.

\begin{figure}
	\centering
	\includegraphics{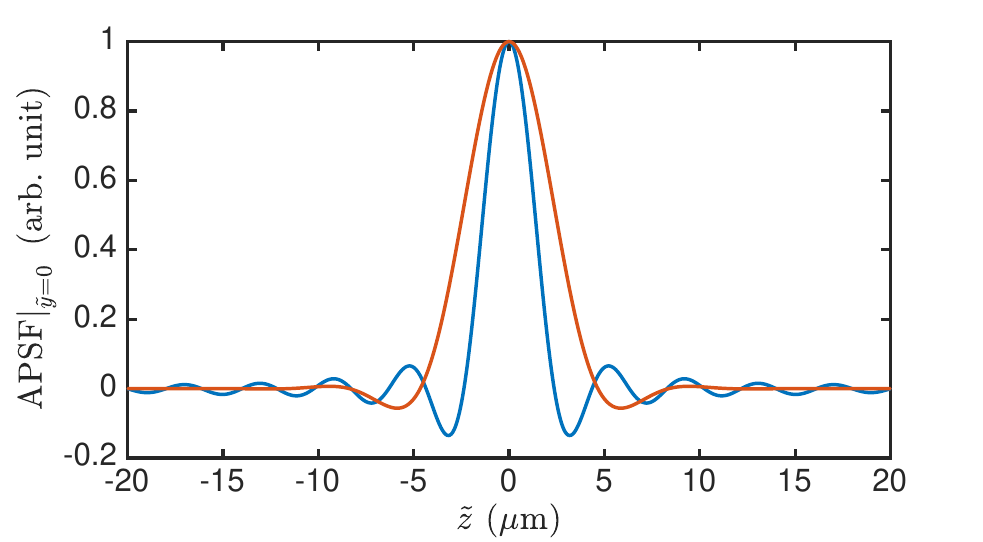}
	\caption{\textbf{APSF with and without cloud extension for the transverse imaging.} 
		The amplitude point spread function (APSF) following from \cref{eq:ctf_gauss_cloud} is shown.
		Note that the APSF is two dimensional, depending on the two coordinates $\tilde y$ and $\tilde z$, which are perpendicular to the imaging direction $x$.
		The tildes should emphasize that we are looking at the coordinates in the object plane.
		One coordinate is fixed to zero ($\tilde y = 0$) for plotting. 
		Note that the APSF is rotationally symmetric around the imaging direction $x$.
		Therefore, it does not matter whether we fix $\tilde y$ or $\tilde z$ to zero.
		The blue line presents the result without cloud extension ($\sigma_\mathrm{wf} = 0$), the red line with cloud extension after 11.2{\ms} TOF ($\sigma_\mathrm{wf} = 42.6${\um}). 
		Both functions have been normalized so that they have a maximum value of one. 
	}
	\label{fig:apsfdroptvsext}
\end{figure}

The amplitude point spread function (APSF) of a thin object is therefore very narrow. For a TOF of 11.2{\ms} the transverse width of the atom cloud substantially decreases the resolution. 
For the static trap, a typical transverse trap frequency would be $\omega_\perp = 2 \, \pi \times 2.1${\kHz}, a typical 1D density $\rho_0(z=0) = 120${\um}\tss{-1} in the center of the cloud. 
Using \cref{eq:exp_inter_gauss}\footnote{ 
	Here one has to use $\rho_0$ for $| \psi (z) | ^ { 2 } $ in \cref{eq:sigma_wf_broadened} to get $\sigma_\mathrm{b}$ used in the equation.
} with this values gives $\sigma_\mathrm{wf} = 42.6${\um}.
In \cref{fig:apsfdroptvsext} a comparison between the APSF for this $\sigma_\mathrm{wf}$ and for the case without cloud extension is shown. 
Remember that $\sigma_\mathrm{wf}$ is the value for the standard deviation of the wavefunction, not the density. 

Neglecting interactions, one gets $\sigma_\mathrm{wf} = 34.8${\um}.
This should be a better value for the edges of the cloud, where densities are low and interactions therefore are not that important. 
This means that the transverse width of the cloud varies quite a bit with the longitudinal $z$ direction in case of a non-homogeneous background density. 
Considering this variation when simulating pictures according to \cref{sec:imag_sim_pics} would be quite involved and is therefore not done in this thesis.



Note that the recoil blurring leads to a further broadening of the transverse width. 
Remember that typically around 200 photons get scattered per atom during the 75{\us} exposure time. From \cref{eq:recoil_dist_usual,eq:recoil_endpoint_dist_3D,eq:recoil_dist_sigma_connection} we see that at the end of the imaging process this leads to a blurring of $2.1${\um} (standard deviation of the density). 
The convolution of two Gaussian function yields a Gaussian where the variances are summed up. 
Due to the quadratic summing up, the standard deviation hardly changes.
We therefore just ignore the recoil blurring in the imaging direction here. 

Unfortunately, it is not straight forward to quantify and visualize the importance of all the different effects preventing a perfect image.
In particular, this is true for the imaging push, the recoil blurring and the CCD shift. 
As one has to sum the intensities of sub-pictures to consider this effects (see \cref{sec:imag_sim_pics}), one cannot define an overall APSF. 
In general, the importance of the effects will depend on the atomic density being imaged and the quantity one is interested in. 

However, we can still compare some numbers. As discussed in \cref{sec:recoil_blurring}, the photon recoil leads to a Gaussian smearing of the atomic density.
At the end of the imaging process, the standard deviation for the Gaussian smearing typically is $\sigma_\mathrm{rc} = 2.1${\um}. 
This $\sigma_\mathrm{rc}$ corresponds to a full width at half maximum (FWHM) of $2 \sigma_\mathrm{rc} \sqrt{2 \ln2} = 4.9${\um}. 
As a comparison, the FWHM of the squared APSF with cloud extension as shown in \cref{fig:apsfdroptvsext} is 3.5{\um}. 
Of course that does not mean that the recoil blurring dominates, as the point of photon absorption and not the final atom position after the imaging process is relevant (see discussion in \cref{sec:recoil_blurring}). 

As already mentioned in \cref{sec:imag_sim_pics}, the CCD shift of the cameras leads to a smearing over $4.7$ pixel.
For the transverse imaging system, this $4.7$ pixels correspond to $4.9${\um} in the object plane which is again on the same order of magnitude as the recoil blurring or the optical resolution.


The imaging push leads to a defocusing which decreases the optical resolution.
To discuss the magnitude of this decrease, we first need to know the defocus distance caused by the push. 
This is not so straightforward as discussed in the following.
 
\begin{figure}
	\centering
	\includegraphics{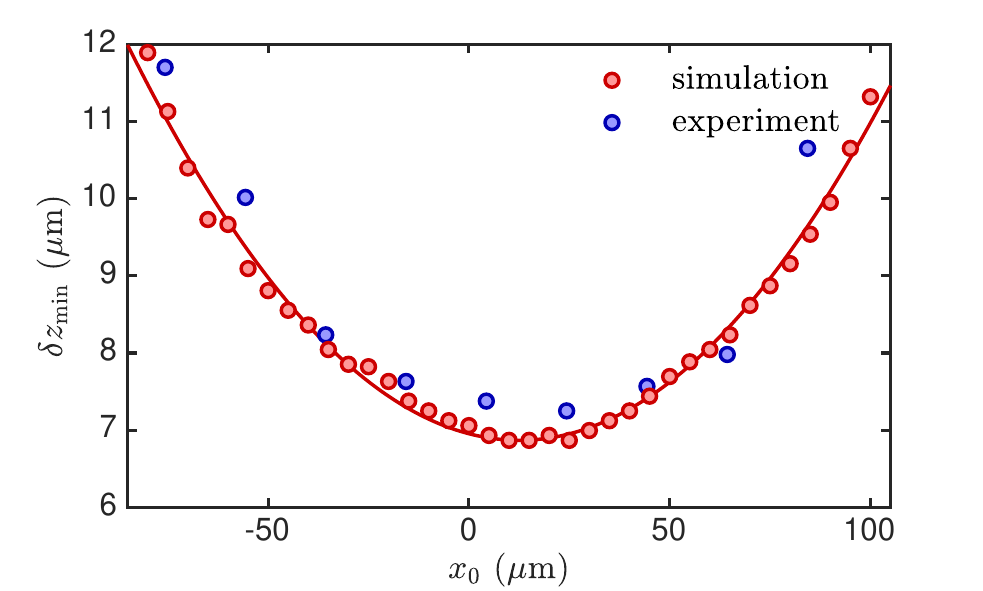}
	\caption{\textbf{Focusing the transverse imaging system:} 
		The minimum position $\delta z_\mathrm{min}$ of the $g_2$ function \labelcref{eq:g2_def} is plotted as a function of the defocus distance $x_0$.
		Remember that we are imaging in the $x$ direction here. 
		Therefore we also denote the defocus distance with $x_0$ instead of $z_0$ which was used in \cref{eq:ctf_gauss_cloud}.
		The red bullets represent the results obtained from 200 simulated pictures (see discussion in main text).
		Due to the imaging push, the value for the defocus distance $x_0$ changes during the exposure time.
		The values used for the plot are given by the mean of the initial and final values for $x_0$.
		The solid red line is a quadratic fit, it's minimum position is at $x_0 = 12.6${\um}. 
		The blue bullets represent data from experimental measurements, agreeing with the results from the simulated pictures quite well. 
		The values for $x_0$ are given by the position of the objective.
		As the position of the focal plane is not a priori known for the experimental data, the values for the horizontal axis have been shifted so that the minimum of the experimental data and the simulated data agree. 
	}
	\label{fig:focusingplot}
\end{figure}

In the experiment, the imaging system is focused by adjusting the position of the objective.
The focusing procedure consists of measuring density ripples (see \cref{sec:DR_therm}) with different positions for the objective.
We then find the ideal setting by minimizing the minimum position $\delta z_\mathrm{min}$ (see \cref{sec:imag_DR}) of the $g_2$ function \labelcref{eq:g2_def}.
As the cloud is pushed through the focus, it is unclear what initial defocus distance in the simulations this ideal configuration corresponds to. 
Therefore, we simulated pictures with density ripples according to \cref{sec:imag_sim_pics} for different initial defocus distances.
With this we can simulate the experimental focusing procedure. 
The results are shown in \cref{fig:focusingplot}.

The experimental data shown for comparison have been obtained for a double well trap with tunnel coupling.
The longitudinal confinement was harmonic.
Temperature and atom number were not analyzed. 
The TOF was $11.2${\ms}.

Note that the value for $\delta z_\mathrm{min}$ generally depends rather little on the physical parameters like temperature or atom number \cite{rauer2018thesis}, as long as they don't influence the optical imaging resolution.
We therefore simply chose typical values for the simulation.


We simulated the pictures for two clouds with temperature $T = 40${\nK} in a double well potential without tunnel coupling.
The longitudinal density profile was chosen to be homogeneous with $\nod = 60 ${\um}\tss{-1} (in one well) on a length of $62.94${\um}. 
Beyond this range, it falls to zero with sharp edges.
The transverse trap-frequency was chosen as $\omega_\perp = 2\pi \times 1.4${\kHz}, which is a typical value for the double well trap used in the experiment.
Note that the 1D atomic density and especially the transverse trap frequency determine the transverse width $\sigma_\mathrm{wf}$ after TOF expansion.
This in turn influences the optical imaging resolution.
However, the used values for $\nod$ and $\omega_\perp$ should be pretty good guesses for the true values in the experiment.
Using \cref{eq:exp_inter_gauss}, they lead to a transverse width of $\sigma_\mathrm{wf} = 32.1${\um} after TOF.

Moreover, note that the classical stochastic $\psi_{1,2} (z)$ fields used to calculate the atomic density after TOF (see \cref{sec:theo_tof_no_inter}) follow the thermal statistics for the Hamiltonian \labelcref{eq:coupled_lieb_liniger} without any tunnel coupling.
The stochastic process discussed in \rcite{Beck18} was used to obtain the numerical shots. 

During the imaging, the atoms are pushed for a total distance of $45${\um} in imaging direction. It turns out that the ideal distance for the midpoint between initial and final position is shifted approx.\ 12.5{\um} in imaging (=push) direction when compared to the focal plane (see \cref{fig:focusingplot}). This means that for the focused imaging system the atom cloud is pushed from 10{\um} before the focus to 35{\um} after the focus. This makes sense as the acceleration is constant and one wants to spent as much time as possible close to the focus. Note that the minimum position of the $g_2$ functions for the simulated pictures and measurements agree quite well. For a more general discussions about focusing imaging systems see the references \cite{Langen13comment,putra2014optimally}.

Let us now come back to the discussion about the influence of the imaging push.
For a the maximum defocus distance of 35{\um}, the FWHM of the squared APSF with cloud extension increases to 4.0{\um} as compared to 3.5{\um} for the in-focus case.

Let us conclude this section by noting that during the imaging time of 75{\us} the cloud falls 8.2{\um} due to gravity. 
This is however not of great importance as the direction of the fall coincides with one of the transverse trapping directions. 
We are usually interested in the fluctuations along the longitudinal direction for which the fall due to gravity does not decrease the imaging resolution. 
One however has to keep it in mind when measuring the transverse cloud width.

\subsection{Resolution for small time of flight}

\begin{figure}
	\centering
	\includegraphics{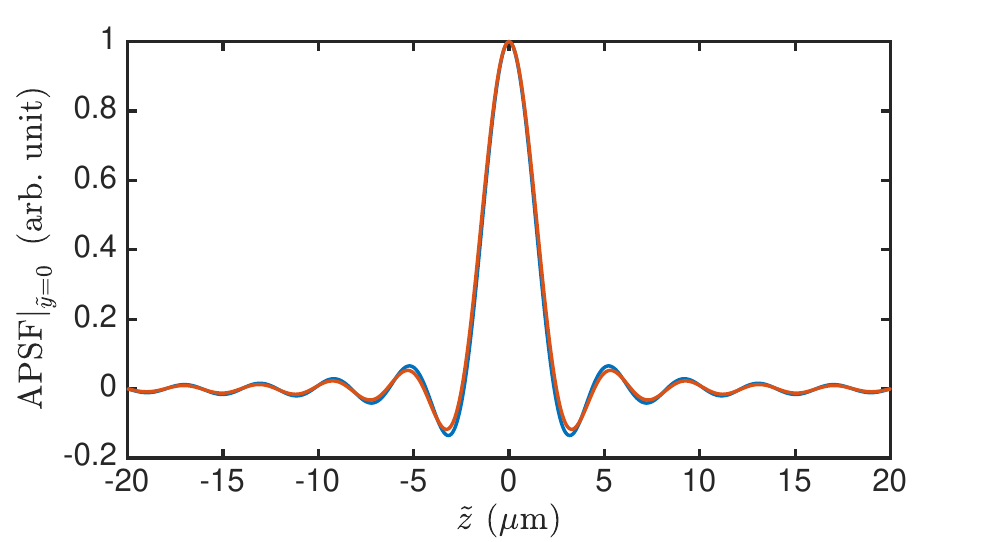}
	\caption{\textbf{APSF with and without cloud extension for the transverse imaging.} 
		Same as \cref{fig:apsfdroptvsext} but with cloud extension $\sigma_\mathrm{wf} = 7.6${\um} after 2{\ms} TOF for the red curve.
		}
	\label{fig:apsf2mstofoptvsext}	
\end{figure}

For a short TOF, the transverse width of the atom cloud is still quite small. For the static trap, a typical transverse trap frequency would be $\omega_\perp = 2 \, \pi \times 2.1${\kHz}, a typical 1D density $\rho_0 (z=0) = 120${\um}\tss{-1} in the center of the cloud. Using \cref{eq:exp_inter_gauss} with this values gives $\sigma_\mathrm{wf} = 7.6${\um}. 
Without interactions one gets $\sigma_\mathrm{wf} = 6.2${\um}. 
Both values are the standard deviation of the wavefunction, not the density. 
As discussed above, the recoil blurring leads to a further broadening of this width, which is however rather small and will therefore be neglected here.
We will simply use $\sigma_\mathrm{wf} = 7.6${\um} for the amplitude point spread function shown in \cref{fig:apsf2mstofoptvsext}. 
As one can see, this transverse width does not substantially reduce the optical resolution.

For short TOFs the imaging light is reflected on the atomchip~\cite{gring2012thesis} under an angle of approximately two degrees.
This leads to a formation of a standing wave pattern below the atomchip.
The atom cloud will be within this standing wave pattern which creates many problems when analyzing the recorded images.
It also makes a detailed discussions of the imaging push and the recoil-blurring very difficult. 
As short TOFs have not really been used for obtaining the physical results in this thesis, we will not discuss this here. 

Note that the atoms fall about 1.5{\um} due to gravity during the exposure time.
Also, note that for $\rho_0 = 120${\um}\tss{-1} and $\sigma_\mathrm{wf} = 7.6${\um} the 3D peak density after TOF is about $0.66${\um}\tss{-3}, corresponding to a mean distance of $d = 1.15${\um} between the atoms. This distance $d = 1.5 \lambda_0$ is comparable to the wavelength $\lambda_0$ of the imaging light. Therefore resonant van der Waals interactions between the atoms might play a role~\cite{Chomaz12}. However, treating this topic goes beyond of what we want to discuss here.

\section{Resolution of the vertical imaging system}
\label{sec:res_vandor}

The vertical imaging system images from the vertical $y$ direction (see \cref{fig:figimagsys}).
The numerical aperture is $\mathrm{NA}\approx 0.079$~\cite{rauer2018thesis}.
One complication is that due to the unusual imaging method~\cite{rauer2012mastersthesis} with the imaging beam traversing the optical system at an angle, there is an astigmatism in the imaging system. This leads to the the focus in the longitudinal $z$ direction being about 120{\um} different from the focus in the transverse $x$ direction. It is found that the ideal point for measuring interference fringes lies in the middle between the two focal points~\cite{rauer2018thesis}. Some of the data presented in this thesis was obtained using this point, for other data the transverse focal point was used. 

The imaging system is focused by adjusting the TOF. 
For the longitudinal focus we have 15.2{\ms} TOF and for the transverse focus it is 16{\ms}. 
The ideal TOF for recording interference patterns was measured as TOF$=15.6${\ms}~\cite{rauer2018thesis}, i.e., it lies right in the middle between the two focus points. 
Using~\cref{eq:exp_inter_gauss} with this TOF gives $\sigma_\mathrm{wf} = 59.4${\um} for the width of the cloud after expansion, when using the parameters $\omega_\perp = 2 \, \pi \times 2.1${\kHz} and $\rho_0 (z = 0) = 120${\um}\tss{-1} typical for the static trap. 
As discussed in \cref{sec:res_tandor} the recoil blurring does not substantially alter the transverse width. 

The imaging light is switched on for 50{\us} with an intensity of typically $0.22 \, I^0_\mathrm{sat}$. For that intensity, every atom scatters 171 photons on average. During the exposure time, the atoms fall for 7.7{\um} in the imaging direction and are pushed for 25{\um} in the same direction. 
It is unclear how to treat the astigmatism with the formalism discussed in this chapter. For the discussion here and the simulation of pictures in \cref{chap:corr}, we will therefore simply assume to be 100{\um} defocused in both directions and ignore the effect of the imaging push.
Note that contrary to the cameras used for the transverse and longitudinal imaging system, the CCD of the camera used for the vertical imaging system does not shift during the exposure.

\begin{figure}
	\centering
	\includegraphics{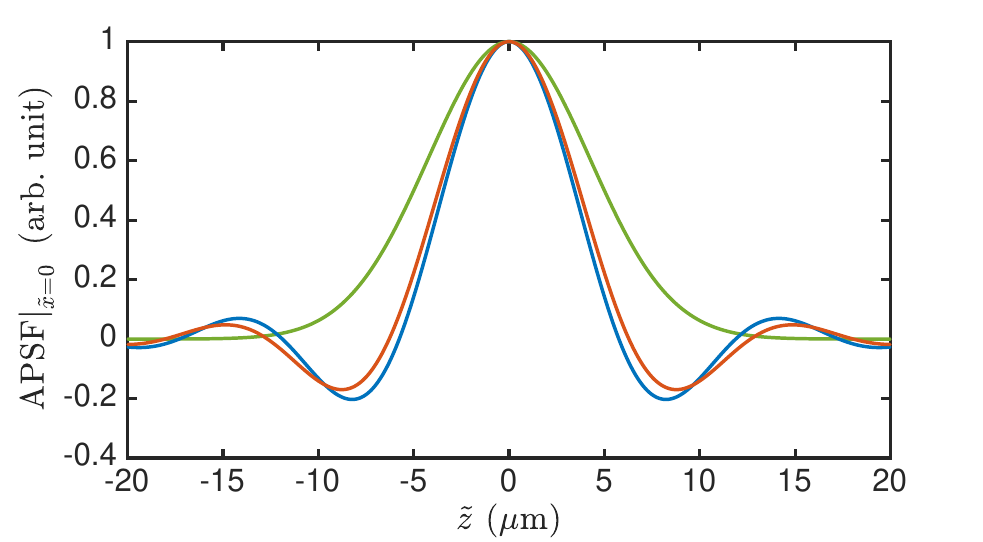}
	\caption{\textbf{APSF with and without cloud extension for the vertical imaging.}
	Similar to \cref{fig:apsfdroptvsext} but for the vertical imaging system.	 
	The blue line shows the APSF without cloud extension, the red line with cloud extension $\sigma_\mathrm{wf} = 59.4${\um} after 15.6{\ms} TOF. Both functions are calculated for a defocus distance $y_0 = 100${\um} and have been normalized so that they have a maximum value of one. 
	For comparison, a Gaussian with standard deviation $\sigma^\prime_\mathrm{PSF} = \sqrt{2} \times 3${\um} is shown in green (see discussion in the main text).}
	\label{fig:apsfVAndoroptvsext}
\end{figure}

The APSF for a  defocus distance $y_0 = 100${\um}\footnote{
	Remember that we are imaging in the $y$ direction here. 
	Therefore we also denote the defocus distance with $y_0$ instead of $z_0$ which was used in \cref{eq:ctf_gauss_cloud}.
} 
and a cloud width of $\sigma_\mathrm{wf} = 59.4${\um} is plotted in \cref{fig:apsfVAndoroptvsext}. As one can see, the cloud extension has only a minor influence on the imaging resolution. 
As a comparison, a Gaussian with $\sigma^\prime_\mathrm{PSF} = \sqrt{2} \times 3${\um} is plotted. This `corresponds' to a Gaussian with $\sigma_\mathrm{PSF} = 3${\um} for the atomic density (squaring reduces the standard deviation by a factor $\sqrt{2}$). 
Such a Gaussian is used to smear out theoretical phase profiles in order to approximately consider the imaging resolution (see discussion in \cref{sub_sec:eff_psf_fringes,sec:width_of_Gaussian_psf}).

Note that the optical resolution is quite a bit worse than for the transverse imaging system (see \cref{sec:res_tandor}).
The FWHM of the squared APSF without cloud extension is 5.2{\um}.
In comparison, the FWHM for the recoil blurring is around 3.6{\um} at the end of the imaging procedure (see \cref{eq:recoil_dist_usual,eq:recoil_endpoint_dist_3D,eq:recoil_dist_sigma_connection}).
Considering that the blurring at the time of photon absorption and not at the end of the imaging process is relevant, we expect the recoil blurring to be of only minor influence.

\begin{table}[h]
	  \renewcommand{\arraystretch}{1.3}
	\centering
	\begin{tabular}{lrr} 
		\toprule
		\textbf{parameter} & \multicolumn{2}{c}{\textbf{imaging system}} \\
		\cmidrule{2-3}
		& transverse  & vertical  \\ 
		\midrule 
		NA & 0.2 & 0.079 \\
		typical TOF (ms) & 11.2 & 15.6 \\
		typical $\sigma_\mathrm{wf}$ after TOF ({\textmu}m) & 42.6 & 59.4 \\
		imaging duration ({\textmu}s) & 75 & 50 \\
		photons scattered per atom & 200 & 171 \\
		FWHM recoil blurring ({\textmu}m) & 4.9 & 3.6 \\
		imaging push distance ({\textmu}m) & 45 & 25 \\
		fall distance ({\textmu}m) & 8.2 & 7.7 \\
		CCD shift ({\textmu}m) & 4.9 & 0 \\
		\midrule
		\textbf{optical resolution} & & \\
		FWHM w/o cloud extension ({\textmu}m) & 2.0 & 5.2  \\
		FWHM w/ cloud extension ({\textmu}m) & 3.5  & 5.6 \\ 
		\bottomrule
	\end{tabular}
	\caption{\textbf{Parameters for the transverse and vertical imaging systems.}
	Note that the stated values for the photons scattered per atom and the values for the quantities depending on it (FWHM of the recoil blurring and imaging push distance) should only be understood as an order of magnitude.
	The FWHM of the recoil blurring is given for the end of the imaging duration.
	For the optical resolution, the FWHM of the squared APSF is given.
	The values for the transverse imaging system were calculated with zero defocus distance, for the vertical imaging system a defocus distance of 100{\um} was assumed.
	In order to avoid misinterpreting the given numbers, please consult \cref{sec:res_tandor,sec:res_vandor} for details.
	}
	\label{tab:imaging_param}
\end{table}

\chapter{Extracting information from the absorption images}
\label{chap:extracting}

\section{Thermometry via density ripples}
\label{sec:DR_therm}

In-situ phase fluctuations lead to pronounced density fluctuations (density ripples) in time of flight (TOF). 
This can be used for thermometry.
For a detailed discussion and illustrations of how the density ripples emerge, see \rcite{Imambekov09,manz10,rauer2018thesis}.
We will here only shortly recapitulate some basics and then focus on the influence of the imaging system and the case of density ripples for two clouds in a double well potential.

To get a value for the temperature, we will analyze the $g_2$ function
\begin{equation}
g _ { 2 } ( \delta z ) = \frac { \int \mathrm { d } z \langle n ( z + \delta z ) n ( z ) \rangle } { \int \mathrm { d } z \langle n ( z + \delta z ) \rangle \langle n ( z ) \rangle }, \label{eq:g2_def}
\end{equation}
where $n(z)$ is the density after TOF integrated over the transverse $x$ and $y$ directions. 
The averaging is done over many experimental realizations.

\subsection{Effect off the imaging system onto the density ripples}
\label{sec:imag_DR}

For typical TOFs, the density ripple pattern is dominated by rather short wavelength fluctuations~\cite{Imambekov09}. 
The imaging resolution therefore heavily influences the measured pattern. 
Note that we can in principle measure the density ripple patterns with the transverse as well as the vertical imaging system.
However, we will always use the transverse imaging system, because its spatial resolution is quite a bit better.

\begin{figure}
	\centering
	\begin{subfigure}{0.49\textwidth}
		\centering
		\includegraphics[width=\linewidth]{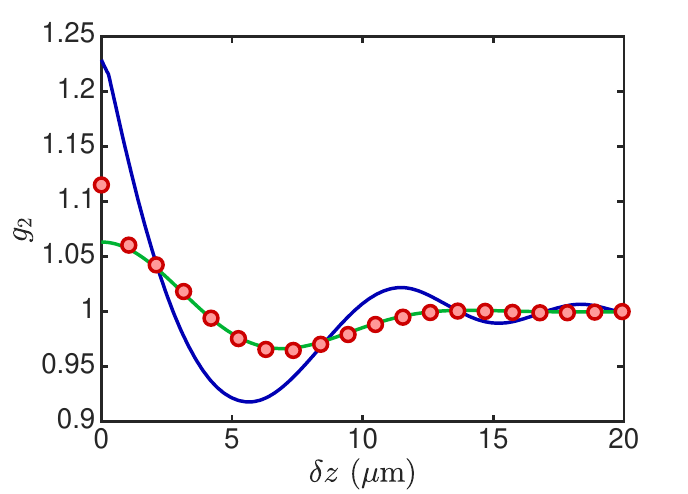}
		\caption{$\lambda_T = 16.7${\um}}
		\label{fig:t40g2}
	\end{subfigure}
	\begin{subfigure}{0.49\textwidth}
		\centering
		\includegraphics[width=\linewidth]{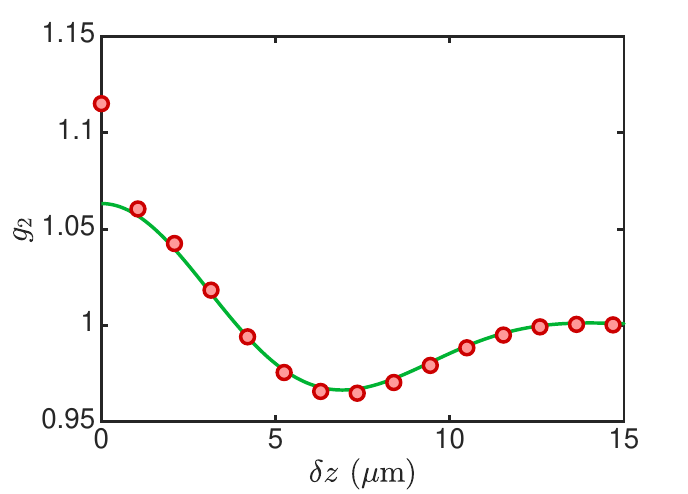}
		\caption{$\lambda_T = 16.7${\um}}
		\label{fig:t40g2detail}
	\end{subfigure} \\
	\begin{subfigure}{0.49\textwidth}
		\centering
		\includegraphics[width=\linewidth]{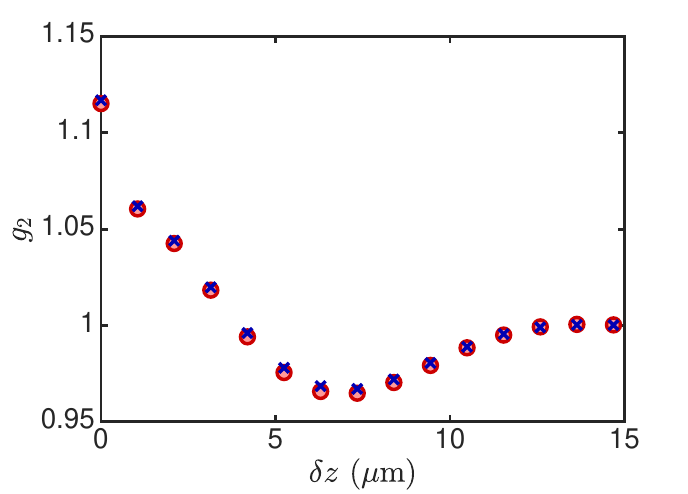}
		\caption{$\lambda_T = 16.7${\um}}
		\label{fig:t40g2wwodensfluct}
	\end{subfigure}
	\begin{subfigure}{0.49\textwidth}
		\centering
		\includegraphics[width=\linewidth]{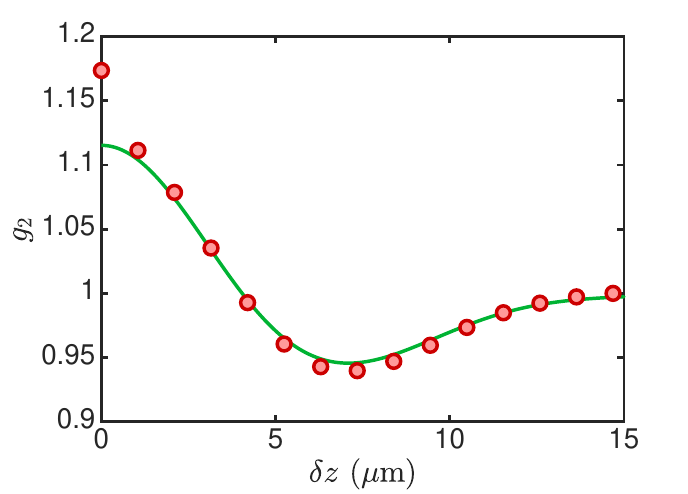}
		\caption{$\lambda_T = 6.7${\um}}
		\label{fig:t100g2detail}
	\end{subfigure} \\
	\caption{
		\textbf{Influence of the imaging system onto the density ripples.} \textbf{(\subref{fig:t40g2})} The $g_2$ function \labelcref{eq:g2_def} is shown without any consideration of the imaging resolution (blue line), extracted from simulated pictures (red bullets) and for the effective consideration of the imaging process through convolution with a Gaussian as discussed in the main text (green line). 
		The width of the Gaussian was chosen so that the minimum for the $g_2$ function agrees with the minimum following from the simulated pictures. \textbf{(\subref{fig:t40g2detail})} Represents a detailed view of (\subref{fig:t40g2}). 
		One sees that the influence of the imaging is fairly well, but not perfectly modeled by the Gaussian point spread function (PSF) for the densities. 
		It seems like the $g_2$ contrast for the simulated pictures is increased compared to the $g_2$ function with Gaussian PSF. 
		The deviation of the value for $\delta z = 0$ is due to photon shot noise.
		\textbf{(\subref{fig:t100g2detail})} Presents the same as (\subref{fig:t40g2detail}), but for a higher temperature. 
		\textbf{(\subref{fig:t40g2wwodensfluct})} Considering the in-situ density fluctuations (blue crosses) also leads to a small difference in the $g_2$ function. 
		The results have been calculated from 1000 realizations for the Bogoliubov theory \labelcref{eq:H2_mora} in classical field approximation. 
		A box-like profile ($\nod = 60 ${\um}\tss{-1}) of a length of 62{\um} with smoothed edges was used.
		The thermal coherence length $\lambda_{T}$ is stated below the pictures.  
	}
	\label{fig:DR_imaging_influence}
\end{figure}

In the following, we will investigate the influence of the imaging process by analyzing simulated pictures (see \cref{sec:imag_sim_pics}).
As we use the transverse imaging system in the experiment, we will also use the appropriate imaging parameters (see \cref{sec:res_tandor}) in the simulation.
The results presented in \cref{fig:t40g2} show that the $g_2$ function is greatly modified by the simulated imaging process.
Previously~\cite{rauer2018thesis,Rauer16,Rohringer2015} it was always assumed that the influence of the imaging system can be modeled by convolving the numerical realizations for the 1D density profiles after TOF with a Gaussian of a certain width.
The minimum position $\delta z_\mathrm{min}$ of the $g_2$ function is then used to determine the width of this Gaussian point spread function (PSF) and the contrast (the difference between the minimum and the maximum) of the $g_2$ function determines the measured temperature.
Note that under the minimum position $\delta z_\mathrm{min}$ we understand the first local minimum of the $g_2$ function as shown in \cref{fig:DR_imaging_influence}. 
As can be seen from \cref{fig:t40g2detail,fig:t100g2detail}, the assumption of an effective Gaussian PSF seems to be fulfilled rather well, but not perfectly, leading to a small systematic error (see \cref{fig:tfittedfromsimu}).

It can be seen from \cref{fig:t40g2wwodensfluct} that the in-situ density fluctuations\footnote{
Note that we used a transverse trapping frequency of $\omega_\perp = 2 \pi \times 1.4${\kHz} (typical for the double well) to calculate the density fluctuations.
} 
only have a minor influence on the $g_2$ function. 
In the following we will therefore neglect them when fitting the $g_2$ function to get the temperature.

\subsection{Extracting a temperature}
\label{sec:DR_fit}

The temperature is extracted by checking which $g_{2}$ function from a number of guesses for the temperature and the width of the Gaussian PSF fits the experimental result best.
In the fit, the point for $\delta z = 0$ is neglected as it contains a large contribution coming from the shot noise during imaging.
The fit is performed considering the first point with $\delta z > 0$ up to the second point after the minimum of the $g_2$ function, all with the same weight.

\begin{figure}
	\centering
	\includegraphics{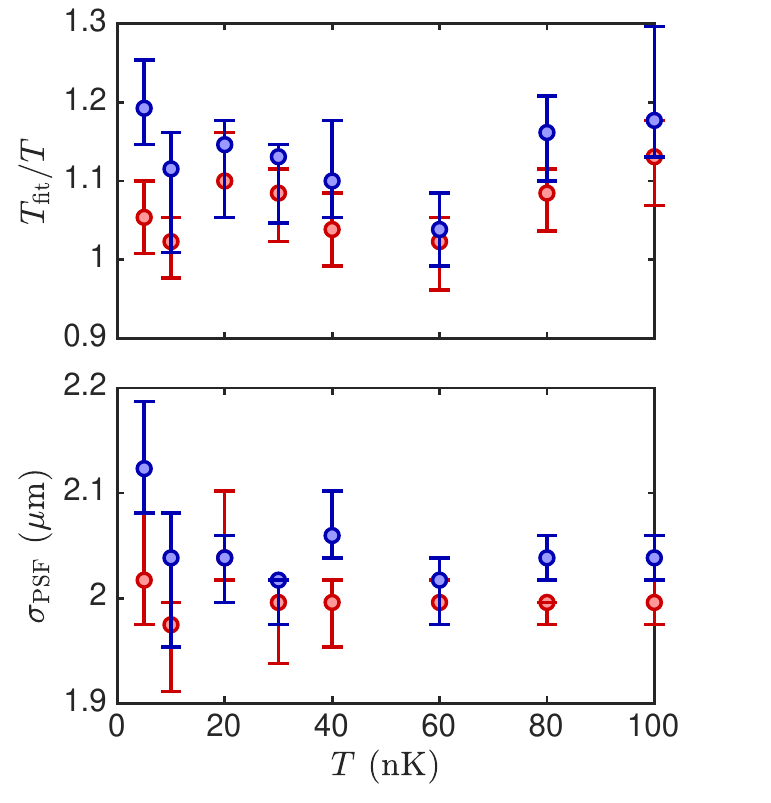}
	\caption{\textbf{Testing density ripple thermometry on the simulated pictures:} 1000 simulated pictures have been produced for various temperatures. The $g_2$ functions \labelcref{eq:g2_def} extracted from the simulated pictures were then fitted with $g_2$ functions calculated from the Luttinger liquid model in classical field approximation where  the initial density fluctuations have been neglected. 
	For this, $10^4$ numerical realizations have been used. For a discussion of the fitting procedure see the main text. 
	In the upper plot, the ratios of the extracted temperatures $T_\mathrm{fit}$ and the input temperatures $T$ for the simulated pictures are shown.
	The red markers show the results when the pictures have been simulated from classical thermal fluctuations following the Luttinger liquid model without initial density fluctuations.
	The extraction of the temperature works fairly well.
	However, there is a small systematic deviation, stemming from the fact that a Gaussian PSF does not perfectly model the imaging process. 
	For the blue markers, initial density fluctuations following the Bogoliubov model in classical fields approximation have been considered. 
	One sees an additional systematic error, which at least partly is due to fitting the wrong width $\sigma_\mathrm{PSF}$ of the effective Gaussian PSF (lower plot). 
	A box-like density profile ($\nod = 60 ${\um}\tss{-1}) of a length of 62{\um} with smoothed edges was used.
	The transverse trapping frequency was $\omega_\perp = 2 \pi \times 1.4${\kHz}, which is a typical value for the double well.
	The error-bars represent the 80\% confidence intervals for the fitted quantities obtained by using bootstrapping. 
	}
	\label{fig:tfittedfromsimu}
\end{figure}

As can be seen from \cref{fig:tfittedfromsimu}, the extraction of the temperature from the simulated pictures works fairly well. 
However there is a small systematic deviation, stemming from the fact that a Gaussian PSF does not perfectly model the imaging process.
Also neglecting the density fluctuations in the fitting process seems to lead to an error stemming at least partly from fitting the wrong width of the Gaussian PSF.
Considering the density fluctuations in the theory used for fitting will probably reduce this error.
However it is not completely clear what theory best describes the density fluctuations on the relevant length scales.
Therefore only phase fluctuations were considered when calculating the theory used to fit the experimental data with.  

\subsection{Density ripples in the double well}
\label{sec:DR_in_DW}

From absorption imaging, one gets the atomic density integrated in the direction of the imaging. 
Using the transverse imaging system, we get the atomic density integrated along the transverse $x$ direction. 
Integrating along the other transverse direction ($y$) during data analysis gives the one dimensional density profiles for the single experimental realizations.

In the framework of classical fields we can write the one dimensional density profile as
\begin{equation}
n(z) = \int d x d y \left| \Psi _ { 1 } + \Psi _ { 2 } \right| ^ { 2 } . \label{eq:long_dens_prof}
\end{equation}
Here $\Psi_{1,2}(\vec r,t)$ represent the fields for the two wells after TOF. 
As in \cref{sec:theo_tof_no_inter}, we will then assume them to be separable into a function of the transverse coordinates $x$ and $y$ times a function of the longitudinal coordinate $z$:
\begin{equation}
\Psi_{1,2}(\vec{r},t) = \phi_{1,2}(x,y,t) \, \psi_{1,2}(z,t) \label{eq:separabel_TOF}
\end{equation} 
To be more precise, we assume separability for the initial fields before TOF expansion and we assume an independent free expansion for the transverse and longitudinal directions.

We will choose the transverse functions $\phi_{1,2}(x,y,t=0)$ to be normalized before TOF expansion. 
Due to the assumed separable unitary evolution, they will stay normalized in our model during the TOF expansion, i.e.,
\begin{equation}
\int d x \, d y \left| \phi _ { 1, 2 } \right| ^ { 2 } = 1
\end{equation}
for all times $t$. 

Inserting \labelcref{eq:separabel_TOF} into \labelcref{eq:long_dens_prof} gives
\begin{equation}
n(z,t) = \left| \psi_1 (z,t) \right| ^ { 2 } + \left| \psi_2 (z,t) \right| ^ { 2 } + 2 \re\left[ \psi_1 (z,t) \psi_2^* (z,t) \int{ d x \, d y \,  \phi_1  \phi_2^*} \right]
\end{equation}
If we assume zero initial overlap between the two clouds in the double well, the last term vanishes. 
Due to the assumed separable unitary evolution, this implies that it vanished for all times $t$, even if the fields overlap and interfere in TOF.
In the following we will therefore always use
\begin{equation}
n(z,t) \approx \left| \psi_1 (z,t) \right| ^ { 2 } + \left| \psi_2 (z,t) \right| ^ { 2 }. \label{eq:add_dens_profiles}
\end{equation}
Here we used the approximately equal sign as the two clouds will also initially  overlap slightly.

For the expanding harmonic oscillator ground state we have (see \cref{sec:theo_tof_no_inter})
\begin{equation*}
\phi_{1,2}(x,y,t) = \frac{1}{\sqrt{\pi \sigma_\mathrm{wf}^2(t)}} \me^{-((x \pm d)^2 + y^2)/2  \sigma_\mathrm{wf}^2(t)} \quad \text{with} \quad  \sigma_\mathrm{wf}(t) = a_\perp \sqrt{1 + \omega_\perp^2 t^2},
\end{equation*}
which gives 
\begin{equation}
	\int{ d x \, d y \,  \phi_1  \phi_2^*} = \me^{-\left[d/(2 a_\perp)\right]^2}
\end{equation}
for all times $t$. For typical experimental parameters the double well separation $d$ is much larger than the harmonic oscillator length $a_\perp$ \labelcref{eq:harm_osc_len}, meaning that the use of \cref{eq:add_dens_profiles} is justified.

Let us use the short-hand notation 
\begin{equation}
	n_{1,2} = \left| \psi_{1,2} (z,t) \right| ^ { 2 },
\end{equation}
and make the assumptions
\begin{align}
	\left\langle n_1 ( z ) \right\rangle &= \left\langle n_2 ( z ) \right\rangle \notag \\
	\left\langle n_1 ( z ) \, n_1 \left( z ^ { \prime } \right) \right\rangle = \left\langle  n_2 ( z ) \, n_2 \left( z ^ { \prime } \right) \right\rangle \quad &, \quad
	\left\langle n_1 ( z ) \, n_2 \left( z ^ { \prime } \right) \right\rangle = \left\langle  n_2 ( z ) \, n_1 \left( z ^ { \prime } \right) \right\rangle
\end{align}
valid for the symmetric double well.
Inserting \labelcref{eq:add_dens_profiles} into \labelcref{eq:g2_def} we then get
\begin{equation}
g _ { 2 } = \frac { 1 } { 2 } \cdot 
\frac { \int \mathrm { d } z  \left[ \left\langle n_1 ( z + \delta z ) \, n_1 ( z) \right\rangle +  \left\langle n_1 ( z + \delta z) \, n_2 ( z) \right\rangle \right]} 
{ \int \mathrm { d } z \left\langle n_1 ( z + \delta z ) \right\rangle \left\langle n_1 ( z)  \right\rangle } .
\end{equation}

In the case of two independent condensates we have
\begin{equation}
	\left\langle n_1 ( z ) \, n_2 \left( z ^ { \prime } \right) \right\rangle = \left\langle n_1 ( z ) \right\rangle \left\langle n_2 \left( z ^ { \prime } \right) \right\rangle
\end{equation}
leading to 
\begin{equation}
	g_2 = \frac { 1 } { 2 } \left( 1 + \frac { \int \mathrm { d } z \langle n_1 ( z + \delta z ) \, n_1 ( z ) \rangle } { \int \mathrm { d } z \langle n_1 ( z + \delta z ) \, \rangle \langle n_1 ( z ) \rangle } \right).
\end{equation}
This means that for two independent condensates the $g_2$ contrast is reduced by one half compared to the single condensate.

\begin{figure}
	\centering
	\begin{subfigure}{0.53\textwidth}
		\centering
		\includegraphics[width=\linewidth]{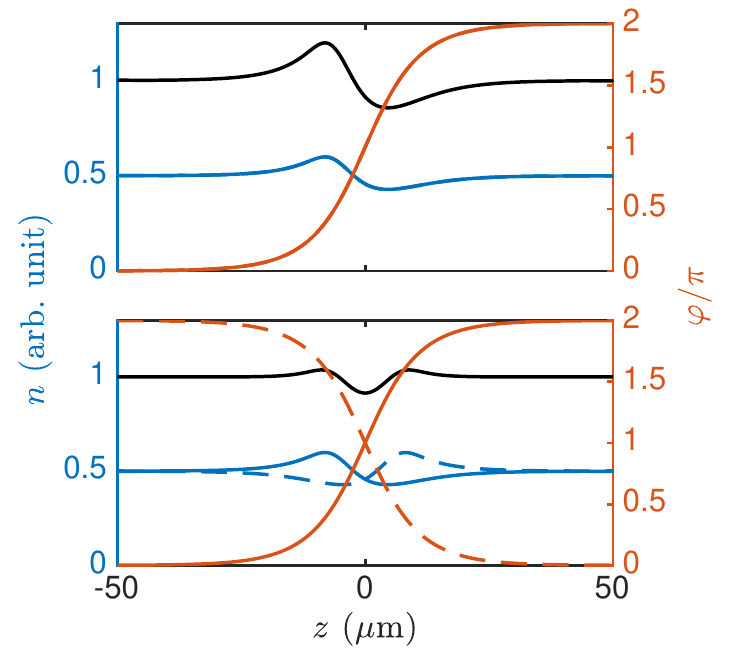}
		\caption{}
		\label{fig:relcompropagation}
	\end{subfigure}
	\begin{subfigure}{0.46\textwidth}
		\begin{subfigure}{\textwidth}
			\centering
			\includegraphics[width=\linewidth]{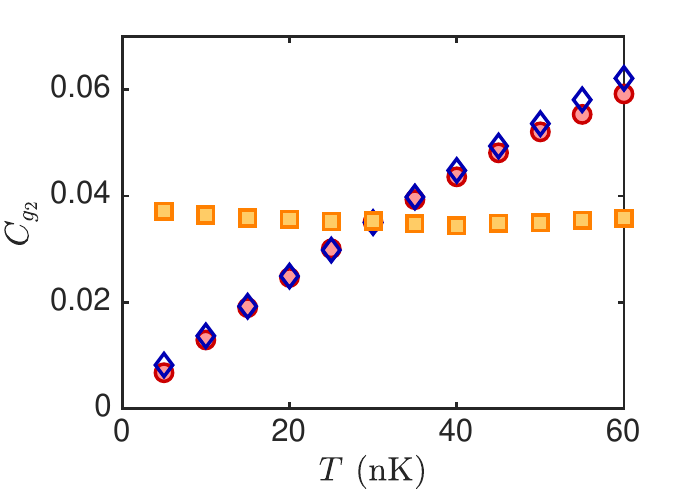}
			\caption{}
			\label{fig:drcontrastrelcom}
		\end{subfigure} \\
		\begin{subfigure}{\textwidth}
			\centering
			\includegraphics[width=\linewidth]{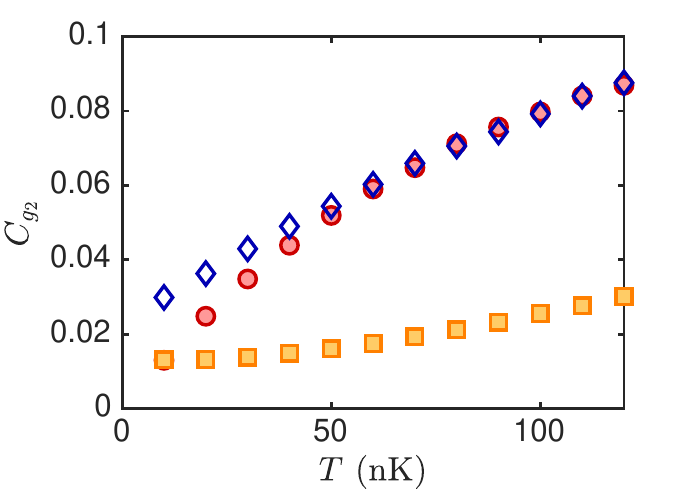}
			\caption{}
			\label{fig:drcontrastrelcomwiderange}
		\end{subfigure}
	\end{subfigure}
	\caption{
		\textbf{Influence of the common and relative phase fluctuations onto the density ripples.} \textbf{(\subref{fig:relcompropagation})} Illustrates how common (upper plot) and relative (lower plot) in-situ phase fluctuations lead to density ripples in TOF. 
		The red solid and dashed lines show the in-situ phase profiles of the left and right condensates. 
		The blue lines the resulting density profiles after 11.2{\ms} TOF. The black line the sum of both. 
		No in-situ density fluctuations have been assumed. 
		One sees that the relative in-situ phase fluctuations (lower plot) lead to much smaller density fluctuations than common phase fluctuations (upper plot).
		\textbf{(\subref{fig:drcontrastrelcom})} Shows the scaling of the $g_2$ contrast $C_{g_2}$ with the common and relative temperature. The red bullets show the case where both temperatures are varied with the horizontal axis. For the blue diamonds the relative temperature is fixed to 30{\nK} and only the common temperature is varied. For the orange squares it is the other way around. \textbf{(\subref{fig:drcontrastrelcomwiderange})} Shows the same with the relative temperature fixed to 120{\nK} (blue diamonds) and the common temperature fixed to 10{\nK} (orange squares). Again, only in-situ phase fluctuations and no in-situ density fluctuations have been assumed and a box-like profile ($\nod = 60 ${\um}\tss{-1}) of a length of 62{\um} with smoothed edges was used. One sees that the relative phase fluctuations only substantially influence the $g_2$ contrast for big imbalances between the relative and common temperature.
	}
	\label{fig:com_rel_g2}
\end{figure}

When looking at the double well system in terms of common and relative degrees of freedom \labelcref{eq:def_rel_com}, we see (as illustrated in \cref{fig:com_rel_g2}) that the density ripples are dominated by the common phase fluctuations. 
The influence of the relative phase fluctuations on the $g_2$ contrast $C_{g_2}$ seems to be only minor unless the magnitude of the relative fluctuations is much larger than the one of the common fluctuations.
Note that apart from the discussed small influence on the $g_2$ contrast the relative phase fluctuations also slightly influence the minimum positions of the $g_2$ function (not shown), which will also slightly change the extracted temperature.
The minor influence of the relative fluctuation onto the $g_2$ function also means that it is rather insensitive to a possible tunnel coupling between the wells as this, in the low energy effective theory, only influences the relative phase fluctuations.

\section{Measuring the relative phase}
\label{sec:phase_extraction}

Many of the experimental results presented in this thesis have been obtained for the horizontally ($x$ direction) split double well. 
As discussed in \cref{sec:theo_tof}, TOF expansion leads to interference fringes in the direction of separation. Imaging in $y$ direction (vertical imaging system) gives the atomic density integrated in this direction. 
We therefore get the integrated density $\tilde \rho (z,x)$ showing interference fringes in the $x$  direction (see \cref{fig:figimagsys}).
The position of the interference fringes depends on the longitudinal coordinate $z$ and is connected to the in-situ phase difference $\varphi_-(z)$ \labelcref{eq:def_rel_com} between the two condensates (see discussion in \cref{sec:theo_tof}). 

\begin{figure}
	\begin{center}		
		\includegraphics{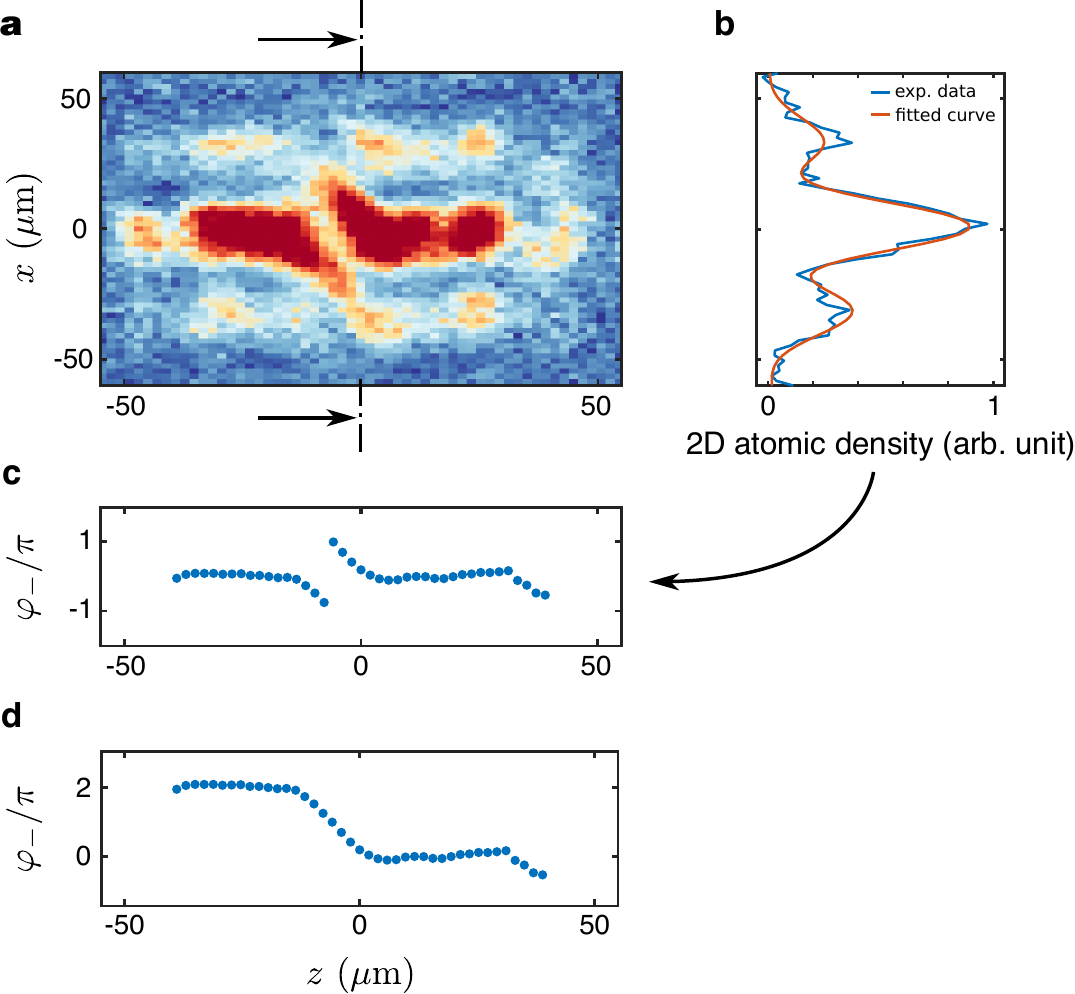}		
	\end{center}	
	\caption{ \textbf{Extraction of the relative phase.} 		
		\textbf{(a)} The two condensates interfere in 15.6{\ms} TOF. 
		The picture shows the resulting interference pattern recorded through the vertical imaging system. 
		The color encodes the atomic density, red corresponding to high density and blue to low density. 
		\textbf{(b)} For each $z$-value, the atomic density depending on $x$ is fitted by the function given in \cref{eq:fr_fit_func}, as an example the fit for $z = 0$ is shown. 
		The fitted phase shift of the cosine function represents (at least approximately) the in-situ phase difference $\varphi_-(z) \ \mathrm{mod} \ 2 \pi$ between the two condensates (plotted in \textbf{(c)}). 
		\textbf{(d)} Imposing continuity onto the phase profile (by shifting the local value of the phase by multiples of $2 \pi$) gives phase differences not restricted to the interval $[-\pi,\pi)$. 
		A global ambiguity of $2 \pi n$ (with $n$ being an integer) remains. 
	}	
	\label{fig:phase_fitting}	
\end{figure}

The goal is now to extract a guess for $\varphi_-(z)$ at a certain position $z$ from the phase of the interference patter at that point $z$. The extraction happens by fitting the function \cite{langen2013thesis}
\begin{equation}
\label{eq:fr_fit_func}
f(x) = A \, \mathrm{e}^{-\frac{(x-x_0)^2}{ \sigma^2}}  \left[1+C \cos\left(2 \pi \frac{x - x_0}{\lambda_\mathrm{F}} + \varphi_-\right)\right] + B 
\end{equation}
to the 2D atomic density $\tilde \rho (z,x)$ at every point $z$. 
The form of \cref{eq:fr_fit_func} comes from integrating \cref{eq:dens_after_TOF_simple} over $y$ and accounting for a possible noise-floor with the parameter $B$.
The parameters to fit are the center $x_0$ and width $\sigma$ of the Gaussian envelope, the fringe contrast $C$, the fringe spacing $\lambda_\mathrm{F}$, the relative phase $\varphi_-$, the overal amplitude $A$ and the overall offset $B$. 
See \cref{fig:phase_fitting} for an illustration of the fitting procedure.


%
%



Clearly, by fitting with \cref{eq:fr_fit_func} one can only determine the phase $\varphi_-$ up to a multiple of $2\pi$.
However, as illustrated in \cref{fig:phase_fitting}, one can get a continuous phase profile with unambiguous phase differences $\varphi_-(z) - \varphi_-(z')$ by imposing continuity.
To be more precise, we assume that the phases fitted from neighboring pixel-slices do not differ by more than $\pi$ and add the appropriate multiples of $2\pi$ to the fitted phases to ensure this.
Clearly, a global ambiguity of $2 \pi n$ (with $n$ being an integer) remains.

If the relative phase field fluctuate too strongly, this phase unwrapping procedure doesn't work reliably anymore.
Even more, the phase fitting with \cref{eq:fr_fit_func} becomes unreliable.
This will be discussed in more detail in \cref{sec:ph_err}.

\section{Investigating time evolution}
\label{sec:extracting_time_evo}

In \cref{chap:gaussification,chap:non_equi} we will investigate the time evolution of the system after a change in the trapping geometry.
Note that the absorption imaging in TOF used in the experiment is destructive.
Therefore, one can only measure the system once.
However, we can still get the time evolution of observables by running the experimental cycle with the same parameters multiple times and simply measuring at different times.
Usually one also needs several realization for every measurement time in order to calculate expectation values.
This is a standard procedure used by most cold atom experiments investigating time evolution.
The idea is that for the same experimental parameters, the physical system is prepared in the same, or at least a similar state.
This should be understood in the context of quantities like temperature or atom number.
The thermal fluctuations, for example, will be different from shot to shot.
Therefore only the measurement of equal-time expectation values is possible.
This prevents, for example, the measurement on non-equal time correlation functions.

\chapter{Phase correlations in the double well}
\label{chap:corr}

In this chapter we will discuss the physical system of two 1D quasicondensates in a double well potential with varying tunnel coupling.
The condensates are prepared by evaporative cooling in the double well potential. 
Only the state at the end of the preparation process is investigated, not the process of evaporative cooling.

The discussions will focus on the relative phase $\varphi_-(z)$ between the condensates, which is extracted from interference fringes (\cref{sec:phase_extraction}). 
We will calculate correlations of the relative phase between different points in position and momentum space.
The discussion will not be restricted to second-order correlations only, but also treat higher-order correlations and investigate their factorization.

We will start with \cref{{sec:calc_corr}}, where we define the particular types of phase correlation functions which we will be investigating.
Moreover, we will discuss the motivation behind using this type of correlation functions as well as their physical significance.
In \cref{sec:preparation}, the details about the preparation process are given.
\Cref{sec:phase_locking} is about the observation of `phase locking' between the condensates and how to quantify it.
The sine-Gordon model \cref{eq:H_SG} is discussed as a theoretical description for the physical situation in \cref{sec:compare_to_SG}.
This includes a discussion of how to obtain the model parameters from experimental measurements.
Moreover, we discuss how the influence of the imaging system onto the measurement of the phase profiles can be efficiently considered.
In \cref{sec:exp_res_correlations} we present the experimental results in comparison to theory predictions.
In \cref{sec:corr_interpretation} we interpret the results and discuss their robustness. 
Finally, we give some concluding remarks in \cref{sec:corr_conclusion} 

Note that, in this chapter, we will drop the minus sign for denoting the relative degrees of freedom.
The relative phase and density fluctuations will simply be denoted by $\varphi (z)$ and $\delta \rho (z)$, respectively.
For the common phase and density fluctuations will continue to write $\varphi_+ (z)$ and $\delta \rho_+ (z)$, respectively.


\section{Calculating phase correlation functions}
\label{sec:calc_corr}


\subsection{Correlations in position space}
\label{sub_sec:calc_corr_pos}

Let us define the $N^{\mathrm{th}}$-order correlation function of the relative phase $\varphi (z)$ as  
\begin{equation}
G^{(N)}({\boldsymbol{z}},{\boldsymbol{z}}') = \left\langle [\varphi(z_1)- \varphi(z'_1)]\dots[\varphi(z_N) - \varphi(z'_N)] \right\rangle.
\label{eq:CorrelationFunction}
\end{equation}
Here we denote the coordinates along the length of the system by ${\boldsymbol{z}}=(z_1, \dots, z_N)$ and ${\boldsymbol{z}}'=(z'_1,\dots,z'_N)$.
Note that we wrote \cref{eq:CorrelationFunction} for the classical phase fields $\varphi$.
However, it also stays valid when using the phase operator $\hat \varphi (z)$.
For the experimental results, the brackets $\langle\ldots\rangle$ in \cref{eq:CorrelationFunction} denote averaging over many experimental realizations.
Similarly, we can average over many numerical realizations to get theory predictions.


Note that we always calculate correlations between phase differences $\varphi(z_i)- \varphi(z'_i)$, as the overall phase profile has a global ambiguity of $2 \pi n$ (see discussion in \cref{sec:therm_sG,sec:phase_extraction}).
By taking phase differences, this global ambiguity drops out.
In the following we will often choose ${\boldsymbol{z}}'= \boldsymbol{0}$, i.e., we are usually looking at phase differences with respect to the middle of the cloud ($z = 0$).  

In the absence of interactions, all information about the state of the system is contained in the correlation functions up to second-order~\cite{ZinnJustin}. 
Higher-order correlations $G^{(N)}$ with $N > 2$ fully factorize, i.e., they can be calculated by the Wick decomposition, a sum containing only products of $G^{(N)}$ with $N \leq 2$. In this case the quantum many-body states are Gaussian, i.e., fully characterized by their first and second moments (mean and variance).

Note that a quantum many-body state is generally fully characterized when all correlation functions are known.
This includes the correlations of the field variables as well as correlations of their conjugate and mixed correlations.
In the case of our system, we not only need to know the phase correlations, but also the correlation functions of the density fluctuations and cross-correlations between phase and density fluctuations.
Unfortunately, the fluctuations of the relative density are at least not directly accessible in our experiment.
For an indirect measurement of the second-order density correlations see \rcite{gluza2018}.

More generally, in the presence of interactions, 
$G^{(N)}$ can be decomposed into~\cite{ZinnJustin} 
\begin{equation}
G^{(N)}({\boldsymbol{z}},{\boldsymbol{z}}') = G_{\mathrm{dis}}^{(N)}({\boldsymbol{z}},{\boldsymbol{z}}') + G_{\mathrm{con}}^{(N)}({\boldsymbol{z}},{\boldsymbol{z}}') .
\label{eq:DecomposeCorrFunc}
\end{equation}
The first term $G_{\mathrm{dis}}^{(N)}$ is the {\em disconnected} part of the correlation function. 
It is fully determined by \emph{all} the lower-order correlation functions $G^{(N')}$ with $N'<N$ and therefore does not contain new information at order $N$.  

The second term, $G_{\mathrm{con}}^{(N)}$, is the {\em connected} part of the correlation function, and contains genuine new information about the system at order $N$. 
Factorization for all higher-order correlation functions is therefore equivalent to $G_{\mathrm{con}}^{(N)}=0$ for all $N>2$. 
In a diagrammatic expansion, $G^{(N)}_{\mathrm{con}}$ is given by a sum of fully connected diagrams with $N$ external lines~\cite{ZinnJustin,erne2018far}.

We calculate the connected part using~\cite{shiryaev2016probability}
\begin{align}
\begin{split}
G_{\mathrm{con}}^{(N)}({\boldsymbol{z}},{\boldsymbol{z}}') =  \sum_{\pi} \Bigg[ \ &(|\pi|-1)! \ (-1)^{|\pi|-1}  
\prod_{B \in \pi} \left\langle \prod_{i \in B} [\varphi(z_i)- \varphi(z'_i)] \right\rangle \Bigg]  . \label{general_formula_connected}
\end{split}
\end{align}
Here the sum runs over all possible partitions $\pi$ of $\{1,\dots,N\}$, the first (left) product runs over all blocks $B$ of the partition and the second (right) product runs over all elements $i$ of the block. 
$|\pi|$ is the number of blocks in the partition. The number of partitions is given by the Bell number $B_N$, which quickly grows with $N$. 
For $N=10$ we get $B_{10} = 115975$, for the next even order it would already be $B_{12} = 4213597$. 
Using the central moments in \labelcref{general_formula_connected}, all partitions containing blocks of size one do not contribute, this substantially reduces the number of terms in the sum. 
However, the computational effort still rises quickly with the order $N$. 
Note that using \labelcref{general_formula_connected} with the sample moments does not represent an unbiased estimator of the connected correlation function. 
However, for the large sample sizes used in the experiment or numerical calculations, the bias should be rather small.


Note that the correlation functions in \cref{eq:CorrelationFunction} reflect the correlations in the collective degrees of freedoms constructed from the conjugate fields $ \delta \rho $ and $\varphi$. 
The connected correlation function $G_{\mathrm{con}}^{(N)}$ for $N>2$ is therefore a direct measure of their interactions. 
This makes them very useful for investigating the validity of the various low energy effective field theories discussed in \cref{chap:theo}.

In contrast, the more commonly used correlation functions $\langle \mathrm{e}^{i[\varphi(z)-\varphi(z')]} \rangle$ and their higher-order generalizations \cite{Langen15} contain $G_{\mathrm{con}}^{(N)}$ up to arbitrary order even for the second-order function. 
One can still use them for investigating the (non-)Gaussianity of the phase fluctuations by applying a different kind of factorization formula \cite{Langen15}.
However, the results seem to be dominated by experimental imperfections like the finite imaging resolution or fluctuations of experimental parameters.
The same is true for quantities like the circular kurtosis.
However, such quantities might still be useful for theoretical calculations \cite{Beck18}.

\subsection{Correlations in momentum space}
\label{sub_sec:calc_corr_momentum}

To get the correlations in momentum space, we simply have to Fourier transform the correlations in position space, or equivalently simply apply the formulas to the Fourier transformed phase profiles.
As already discussed, the measured or simulated phase profiles have a global ambiguity of $2 \pi n$.
After Fourier transformation this translates to an undefined value for $k = 0$, the non-zero $k$-values are of course not affected.

As we have only a finite system, there is the question what Fourier-like transform to use.
As we want to transform unbound phase profiles, the use of a fast Fourier transform (FFT) is not ideal as it assumes periodicity at the boundaries of the spatial interval under consideration.   
Clearly, this is not even approximately fulfilled for unbound phase profiles.
It therefore seems more sensible to use the discrete cosine transform (DCT).
The cosine transform assumes vanishing derivative at the boundaries, an assumption that we think is much less restrictive for our application.
In many other respects it behaves equivalently to the Fourier transform.
In particular, the global $2 \pi n$ ambiguity of the phase profiles still only affects the result for $k = 0$.
The cosine transformed phases will be denoted by $\varphi_k$.

\section{Preparation of the system}
\label{sec:preparation}

We start our discussion from a thermal cloud in a single elongated magnetic trap on the atomchip. 
How to get to this point is discussed in \cref{chap:exp_setup} and the various previous theses \cite{gring2012thesis,kuhnert2013thesis,langen2013thesis,rauer2018thesis}.
We then ramp up the dressing fields to deform the single well into a double well potential.
Note that the cooling fields are switched off before the switch on of the dressing fields.
After the ramping up of the dressing fields, the cooling fields are switched on again.

The cooling in the double well trap consists of a frequency ramp with a duration of 470{\ms}, followed by a period of 400{\ms} where the frequency is held constant at the final value of the ramp.
After this, the cooling amplitude is ramped down in 500{\us} and the trap switched off immediately after to allow for TOF expansion and subsequent absorption imaging.
A similar procedure was used in \rcite{Betz2011,betz2011thesis}.

The segment of 400{\ms} of constant cooling frequency is intended to give the system time to thermalize.
However during this time one cannot avoid that atoms are removed from the trap.
Note that (at least for the tunnel-coupled double well) it is not possible to simply switch of the cooling and give the system time to thermalize with switched off cooling fields. 
The reason for this is that the cooling fields also lead to a dressing of the atoms, i.e., they deform the trap.
Switching off the cooling therefore introduces some dynamics.
This is especially critical in the coupled double well as, for example, a tilt in the double well can be introduced, leading to strong tunneling dynamics and subsequently strong heating. 
For this reason, we measure immediately after switching off the cooling fields, instead of waiting for some time without any cooling.
For the uncoupled double well this effect is not so important and one could also have a waiting time between cooling and measuring.
However, for consistency, also the uncoupled systems in this chapter have been prepared and measured in the same way, unless explicitly stated otherwise. 

\begin{figure}
	\centering
	\begin{subfigure}{0.49\textwidth}
		\centering
		\includegraphics[width=1\linewidth]{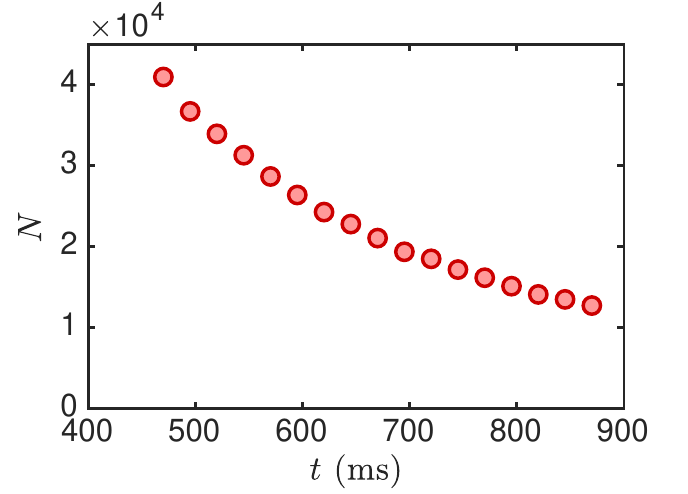}
		\caption{}
		\label{fig:scan4971slowcoolingramp}
	\end{subfigure}
	\begin{subfigure}{0.49\textwidth}
		\centering
		\includegraphics[width=1\linewidth]{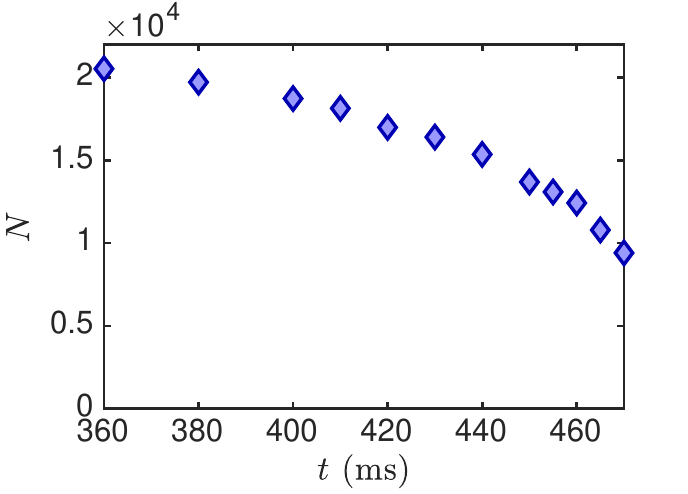}
		\caption{}
		\label{fig:scan5276fastcoolingramp}
	\end{subfigure}
	\caption{
		\textbf{Atom number evolution during system preparation.}
		The evolution of the atom number $N$ is shown as a function of the duration of evaporative cooling in a double well potential with intermediate tunnel coupling. 	
		\textbf{(\subref{fig:scan4971slowcoolingramp}) Slow cooling.}
		Only the last 400{\ms} are shown, during which the frequency of the evaporative cooling is kept constant. 
		Note that during the last 25{\ms} (interval between the last two data points), 763 atoms are lost on average, which corresponds to 30.5 atoms per ms.
		\textbf{(\subref{fig:scan5276fastcoolingramp}) Fast cooling.} The last 110{\ms} of the cooling procedure are shown.
		The cooling frequency is ramped during this time. 
		In the last 5{\ms} (interval between the last two data points), 1383 atoms are lost on average, corresponding to 277 atoms per ms. 
		For the last 30{\ms}, the numbers are 5946 atoms and 198 atoms/ms respectively.
	}
	\label{fig:cooling_ramp}
\end{figure}

The atom loss during the preparation process has been measured for a few prototypical cases.
One of it is shown in \cref{fig:scan4971slowcoolingramp}. 
We can see that, despite keeping the frequency of the cooling fields constant, there are still atoms removed from the trap at the end of the cooling sequence. 

For some of the results presented in this chapter, we also produced a non-equilibrium state on purpose by simply skipping the 400{\ms} segment with constant cooling frequency\footnote{
	Note that in this case the end-frequency of the cooling ramp has to be lower than for the longer cooling sequence, in order to get to similar atom numbers.}.
This results in an atom loss at the end of the cooling procedure which is about an order of magnitude faster (\cref{fig:scan5276fastcoolingramp}). 
This difference in preparation leads to qualitative differences in the obtained correlation functions as discussed in \cref{sec:exp_res_correlations}. 




\section{Phase locking between the condensates}
\label{sec:phase_locking}

In the double well with tunneling, phase coherence, i.e., a vanishing relative phase $\varphi(z)$ is energetically preferred.
This can also be seen from the theoretical models \cref{eq:Ht_full,eq:Ht_sG} when assuming a positive $J$.
In this thesis $J$ will always be positive, negative values for the effective $J$ might be achieved by techniques like shaking the double well \cite{Struck996} or with Raman assisted tunneling~\cite{Aidelsburger2011}.

The strength of this `phase locking' due to tunneling can be characterized by the coherence factor $\langle \cos(\varphi) \rangle$, a quantity that is zero for completely random phases (no phase locking) and approaches unity in the limit of strong phase locking (relative phase vanishes).
In thermal equilibrium  $\langle \cos(\varphi) \rangle$ increases with increasing tunneling strength and decreases with increasing temperature.
See \cref{sec:therm_sG,sec:get_sG_param,fig:cohvsq}  for the dependence of $\langle \cos(\varphi) \rangle$ on the parameters of the thermal sine-Gordon model.

\begin{figure}
	\centering
	\begin{subfigure}{0.49\textwidth}
		\centering
		\includegraphics[width=\linewidth]{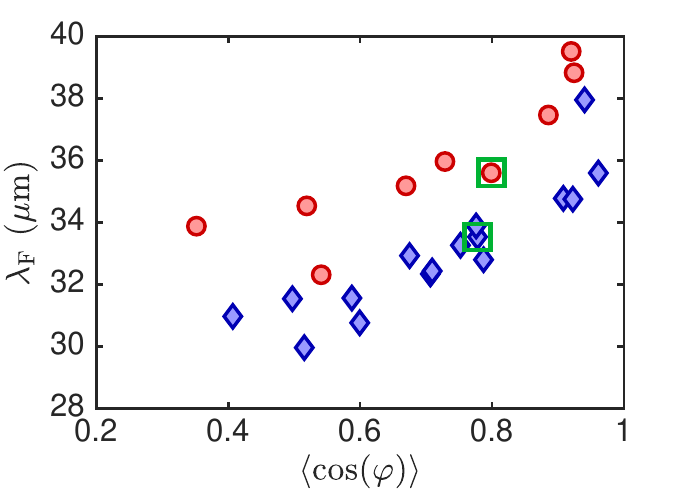}
		\caption{}
		\label{fig:fringespacingvscoh}
	\end{subfigure}
	\begin{subfigure}{0.49\textwidth}
		\centering
		\includegraphics[width=\linewidth]{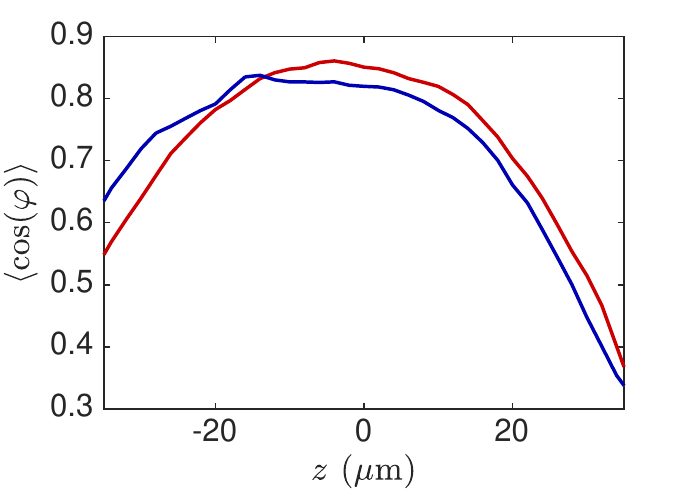}
		\caption{}
		\label{fig:cohspatialdependence}
	\end{subfigure}
	\caption{
		\textbf{Phase locking between two clouds in a double well with finite barrier.}
		\textbf{(\subref{fig:fringespacingvscoh})}~Shows the relation between fringe spacing and phase locking.
		The fringe spacing $\lambda_\mathrm{F}$ is plotted as function of the coherence factor $\langle \cos(\varphi) \rangle$.
		The coherence factor is calculated by averaging over the different experimental realizations as well as the central 50{\um} of the cloud.
		The red bullets show the results for the slowly cooled data (cooled in 870 ms), while the blue diamonds represent the fast cooled data (cooled in 470 ms).
		See \cref{sec:preparation} for details of the preparation process.
		\textbf{(\subref{fig:cohspatialdependence})}~Shows the spatial dependence of the coherence factor for two experimental scans.
		Again red corresponds to slow cooling and blue to fast cooling.
		The corresponding points in (\subref{fig:fringespacingvscoh}) are marked by the green squares.
	}
	\label{fig:phase_locking}
\end{figure}

In \cref{fig:fringespacingvscoh}, one sees that for similarly prepared states, the fringe spacing $\lambda_\mathrm{F}$ of the interference fringes increases with increasing phase coherence.
With similarly prepared, we mean that everything is approximately the same except for the double well separation determining the tunneling strength\footnote{
In the experiment, it is difficult to change one thing without also affecting something else.
}.
Remembering \cref{eq:fringe_spacing}, we know that increasing fringe spacing corresponds to decreasing double well separation, which in turn leads to increased tunneling.
This means that \cref{fig:fringespacingvscoh} just shows that, as expected, the phase locking increases with the tunneling strength.

Note that there is a different dependence between fringe spacing and coherence factor for the slow and the fast cooling procedure discussed in \cref{sec:preparation}.
This is not surprising.
However, we don't have an intuition why it changes in the way it does.
From the results of \cref{sec:exp_res_correlations}, it seems like the relative phase fluctuations for the fast cooled case have a higher effective temperature.
This can explain a difference in the dependence between $\lambda_\mathrm{F}$ and $\cohfact$.
However, one expects lower phase coherence for the hotter (fast cooled) data, which is the exact opposite of what \cref{fig:fringespacingvscoh} shows.
Another possible explanation might be the different dressing of the atoms due to the different cooling fields.
To get similar atom numbers, the frequency of the cooling fields will in general be closer to the trap bottom for fast cooling than for slow cooling. 
This might lead to different double well shapes for the two cases, i.e., to different barrier heights for the same double well separation.

\Cref{fig:cohspatialdependence} shows the spatial dependence of the coherence factor.
Note that for the presented experimental data, we have a harmonic longitudinal confinement leading to a spatially varying 1D background density $\rho_0 (z)$.
From the 1D theoretical models discussed in \cref{sec:DW_theo}, we therefore also expect the coherence factor to be non-constant, even for a spatially homogeneous single particle tunneling rate $J$.
However the observed spatial variation is stronger than what one expects just from the variation of the density.
This was tested by comparison with 1D stochastic Gross-Pitaevskii calculations in the harmonic trap with constant $J$ (not shown).
It is unclear whether 3D Gross-Pitaevskii calculations could, at least partly, explain the position dependence of the tunneling.


\section{Theoretic modeling}
\label{sec:compare_to_SG}

We will compare the experimentally obtained phase correlation functions to the predictions for the thermal fluctuations of the sine-Gordon model \cref{eq:H_SG} in classical fields approximation.
The results are obtained by producing numerical realizations using the sine-Gordon stochastic process discussed in \cref{sec:therm_sG} (see also \rcite{Schweigler17,Beck18}).
The quantities under consideration are then calculated by averaging over these numerical realizations.
Note that some theoretical results for the classical thermal sine-Gordon model were already discussed in \cref{sec:therm_sG,sec:theo_comparison}.

In order to make theoretical predictions, we first have to infer the model parameters from the experimental measurements.
This will be discussed in \cref{sec:get_sG_param}.
First, however, we will investigate the influence of the imaging procedure by simulating pictures.
This is important in order to know how to consider the finite resolution in the calculations of the theoretical predictions.

\subsection{Simulating pictures for the sine-Gordon theory}
\label{sec:corr_sG_simu_pic}

For many of the analysis in this chapter, it is crucial to understand the influence of the imaging system.
We therefore produced a number of simulated interference images following the procedure in \cref{sec:imag_sim_pics}.
In the experiment, we record the images with the vertical imaging system.
Therefore, we use the optical parameters of this imaging system for simulating the pictures (see \cref{chap:imaging}, particularly \cref{sec:res_vandor}).
From these simulated pictures, we then extract the phase profiles according to \cref{sec:phase_extraction}, i.e., the exact same way as for the experimentally recorded interference images.
We produced simulated pictures from fluctuations following the sine-Gordon Hamiltonian as well as the Luttinger liquid Hamiltonian with a quadratic tunneling term \cref{eq:Ht_2}.
In the following we will discuss the details of how the pictures were simulated for the sine-Gordon theory.
The details about the simulated pictures for the Gaussian fluctuations following from the tunnel-coupled Luttinger liquid model are discussed in \cref{sec:corr_Gauss_simu_pic}.

We simulate the pictures from numerically obtained phase profiles following the thermal statistics of the classical sine-Gordon model.
The in-situ density fluctuations are assumed to be zero.
The sine-Gordon stochastic process (see \cref{sec:therm_sG}) is used to generate the numerical realizations of the phase profiles.
They therefore represent a part of an infinite homogeneous system with the relevant physical parameters being given by $\lambda_T$ and $q$.
Simulated images are produces for a range of different values of this parameters.
Note that the background density was only assumed to be homogeneous for producing the phase fluctuations.
For getting the atomic density in TOF, a background density according to \rcite{Gerbier_2004} for a harmonic trap was assumed.
For the parameters defining the background density we chose the transverse trapping frequency $\omega_\perp = 2 \pi \times 1.35${\kHz}, the longitudinal trapping frequency $\omega_\parallel = 2 \pi \times 6.7${\Hz} and an atom number of $5000$ in each well.
These are typical values for the double well trap, leading to a longitudinal cloud extension of approximately 120{\um}.

When simulating the pictures, no imaging push was assumed.
The recoil blurring was considered.
The atomic density in 16{\ms} TOF was obtained according to \cref{eq:dens_after_TOF_phase_dens}.
The fringe spacing was chosen as $\lambda_\mathrm{F} = 22${\um} for the case of $q = 0$ and $\lambda_\mathrm{F} = 35${\um} for when $q \neq 0$.
For the coherent transfer function \cref{eq:ctf_gauss_cloud} the transverse width according to \cref{eq:harm_osc_width_expansion}, i.e., the width of the expanding harmonic oscillator ground state, was used.

\subsection{Efficient consideration of the finite imaging resolution}
\label{sub_sec:eff_psf_fringes}
\label{sec:width_of_Gaussian_psf}

Simulating and subsequently analyzing pictures to consider the influence of the imaging process on the measured phase profiles is computationally costly. 
We will therefore only do it for some prototypical cases.
Most of the time, we will approximately consider the finite imaging resolution by convolving the theoretical phases in position space with a Gaussian with a certain standard deviation $\sigma_\mathrm{PSF}$.
The convolution might, for example, be applied to numerically obtained phase profiles or the phase correlation functions \labelcref{eq:CorrelationFunction}.


In $k$-space the convolution translates into a multiplication with the Gaussian
\begin{equation}
\exp( - k ^ { 2 } \sigma_\mathrm{PSF} ^ { 2 } /2 ). \label{eq:psf_in_k_space}
\end{equation}
To get the best guess for $\sigma_\mathrm{PSF}$ and check the validity of this simplified consideration, we compare the results following from the smeared phase profiles with the ones calculated from the phase profiles extracted from the simulated pictures.
The pictures simulated from the fluctuations of the sine-Gordon model according to the discussion in \cref{sec:corr_sG_simu_pic} have been used.

%

\begin{figure}
	\centering
	\begin{subfigure}{0.49\textwidth}
		\centering
		\includegraphics[width=\linewidth]{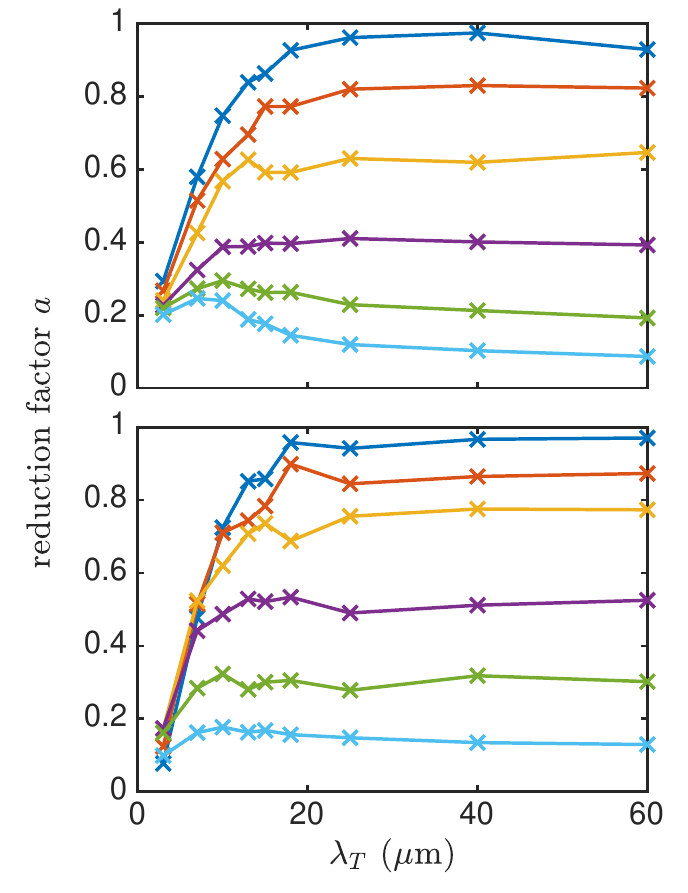}
		\caption{}
		\label{fig:imagingreductionvslt}
	\end{subfigure}
	\begin{subfigure}{0.49\textwidth}
		\centering
		\includegraphics[width=\linewidth]{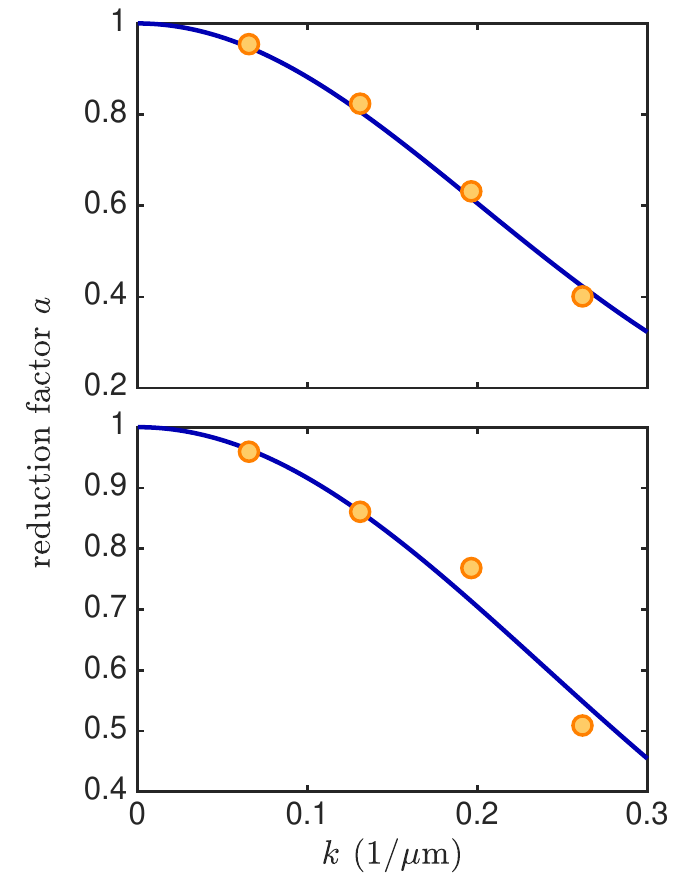}
		\caption{}
		\label{fig:imagingreductionfit}
	\end{subfigure}
	%
	%
	\caption{
		\textbf{Simulated influence of the imaging system onto the cosine transformed phase fluctuations.} 
		Plotted is the simulated reduction factor for the variance of the cosine transformed phase fluctuations (see \cref{eq:reduction_fact}). 
		The cosine transform was performed for the central 50{\um} of the cloud.
		The upper plots represent the results for $q=0$, the lower plots for $q = 2.9$.
		\textbf{(\subref{fig:imagingreductionvslt})} Shows the reduction factor depending on the thermal coherence length $\lambda_T$. 
		The different colors represent the reduction factors for the different modes.
		The first six modes are plotted, with the mode number increasing from top to bottom.
		The crosses represent the data, the lines are a guide to the eye.
		\textbf{(\subref{fig:imagingreductionfit})} The orange bullets represent the reduction factor as a function of the wavenumber $k$.
		The mean value over $\lambda = 25, 40$ and $60${\um} is shown.
		To this mean value the Gaussian function $\exp(- k^2 \sigma_\mathrm{PSF}^2 )$ is fitted, giving $\sigma_\mathrm{PSF} = 3.54${\um} for $q=0$ and $\sigma_\mathrm{PSF} = 2.96${\um} for $q=2.9$.
		Note that this difference for $\sigma_\mathrm{PSF}$ stems from a different fringe spacing used in the simulated pictures (see discussion in the main text).
	}
	\label{fig:imaging_reduction_cos_var}
\end{figure}

We will have a look at the variance $\langle \varphi_k^2 \rangle$ of the cosine transformed phase profiles.
Experimental results for this quantity will be discussed in \cref{sec:2p_cos_corr}.
We will take the central 25 pixels (50{\um}) as the input of the cosine transform, as will be done for the experimental results.
Comparing the results calculated from the phase profiles extracted from the simulated images with the results calculated from the original phase profiles (without any smearing) used as input of the simulation, one can calculate a imaging reduction factor 
\begin{equation}
a = \frac{\langle \varphi_k^2 \rangle_\text{simulated images}}{\langle \varphi_k^2 \rangle_\text{original data}} \label{eq:reduction_fact}
\end{equation}
for each mode independently.
In our simple effective model, this reduction factors should have a Gaussian $k$-dependence (the dependence should be \cref{eq:psf_in_k_space} squared) and be independent of temperature. 
However, as can be seen from \cref{fig:imagingreductionvslt}, the independence from the temperature is only true below a certain temperature and only for the lowest modes.

Fitting a Gaussian to the reduction factor (see \cref{fig:imagingreductionfit}) for the first four modes gives a $\sigma_\mathrm{PSF} \approx 3.5${\um} for the uncoupled scans ($q = 0$) and $\sigma_\mathrm{PSF} \approx 3${\um} for the coupled ones ($q \neq 0$).
The reason for the difference in $\sigma_\mathrm{PSF}$ lies solely in the fact that when simulating the pictures, we used a fringe spacing of $22${\um} for the uncoupled scans and a spacing of $35${\um} for the coupled scans.
These are typical values, which one observes in the experiment for the two cases (see \cref{sec:phase_locking}).
For the same fringe spacing, no substantial variation with $q$ could be detected (not shown).     

%

\subsection{Finding the model parameters}
\label{sec:get_sG_param}

\begin{figure}
	\centering
	\includegraphics{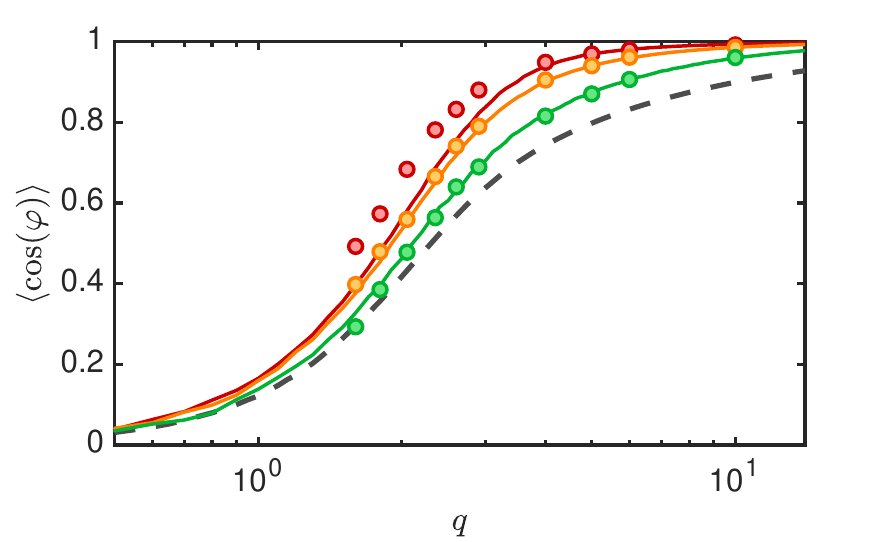}
	\caption{\textbf{Coherence factor for the sine-Gordon model.}
		The coherence factor $\cohfact$ is plotted as a function of the parameter $q = \lambda_T / l_J$ for the thermal phase fluctuations of the classical sine-Gordon model (see \cref{sec:therm_sG}).
		The dashed gray line represents the results without considering the finite imaging resolution.
		The solid lines are obtained when approximately considering the imaging resolution by convolving the numerically obtained phase profiles with a Gaussian of standard deviation $\sigma_\mathrm{PSF} = 3${\um} (see \cref{sub_sec:eff_psf_fringes,sec:width_of_Gaussian_psf}).
		The different colors represent the results for the different thermal coherence lengths $\lambda_T = 3${\um} (red), $\lambda_T = 7${\um} (orange) and $\lambda_T = 25${\um} (green).
		The bullets represent the corresponding results calculated from the simulated pictures (see \cref{sec:corr_sG_simu_pic}).
		One sees that the effective consideration of the imaging resolutions via Gaussian smearing works fairly well except for the case of $\lambda_T = 3${\um} (high Temperature, $T = 260${\nK} for a typical $\nod = 70${\um}\tss{-1}).  
	}
	\label{fig:cohvsq}
\end{figure}

As discussed in \cref{sec:therm_sG} we need the two parameters $q$ and $\lambda_{ T }$ to make theoretical predictions from the thermal sine-Gordon model for the homogeneous system (in classical fields approximation). 
Remember that the coherence factor $\cohfact$ depends only on $q$ and not explicitly on $\lambda_T$.
One would therefore think that $q$ can directly be fitted from the coherence factor independently from the value for $\lambda_T$.
However we still have to consider the finite imaging resolution.
As discussed in \cref{sec:therm_sG}, the classical sine-Gordon theory depends on the rescaled coordinate $z/\lambda_T$. 
This rescaling is the only way in which $\lambda_T$ enters the calculation.
Therefore, the extension of the PSF in the rescaled coordinate depends on $\lambda_T$, which in turn leads to a slight dependence of the predicted coherence factor on $\lambda_T$ when considering the finite imaging resolution.
This is presented in \cref{fig:cohvsq}.

\subsubsection{Using density ripple thermometry}

Assuming that common and relative degrees of freedom (\cref{eq:def_rel_com}) are in thermal equilibrium, we can determine the temperature $T$ through density ripple thermometry (\cref{sec:DR_therm}).
Subsequently we can get the thermal coherence length via
\[
\lambda _ { T } = \frac{2 \hbar ^ { 2 } \nod}{ m k _ { B } T }.
\]
Note that the background density is not homogeneous in the experiment.
We, therefore, use the average over the central $50${\um} of the cloud as the value for $\nod$\footnote{We take the average over the central $50${\um} as we are basically always using the central $50${\um} to calculate the different quantities.}.
Having obtained a value for $\lambda_{ T }$, we can then fit $q = \lambda_T/l_J$ from the experimentally obtained value of $\cohfact$.
Again, be aware that the coherence factor is not spatially uniform.
As for the atomic density, we will use the average over the central $50${\um} of the cloud for the fit.

\subsubsection{Fitting the temperature from the relative phase}
\label{sec:rel_T_fitting}

It seems like the temperature in the relative and common degrees of freedom doesn't always coincide (see \cref{sec:2p_corr_sum,sec:2p_cos_corr}). 
We can therefore try to fit both parameters $\lambda_T$ and $q$ from the measured relative phases.
One possibility is to self consistently fit $\lambda_T$ and $q$ from the coherence factor $\cohfact$ as well as the summed absolute value of the 2p function 
\begin{equation}
S^{(2)} = \sum_{\boldsymbol{z}}{\left|G^{(2)}_{\mathrm{con}}({\boldsymbol{z}},0)\right|}, \label{eq:2p_sum}
\end{equation} 
which is a measure for how strongly the relative phase fluctuates.
The fit is then done by a comparison with a table calculated for a large number of possible values of $q$ and $\lambda_T$.
The weighted sum 
\begin{equation}
\left| \frac{ S^{(2)}_\mathrm{exp} - S^{(2)}_\mathrm{theo} }{ S^{(2)}_\mathrm{exp} } \right| + 5 \, \left| \frac{\cohfact_\mathrm{exp} - \cohfact_\mathrm{theo}}{ \cohfact_\mathrm{exp} } \right|
\end{equation}
of the relative errors for the two quantities is minimized to find the ideal parameters.

\begin{figure}
	\centering
	\includegraphics{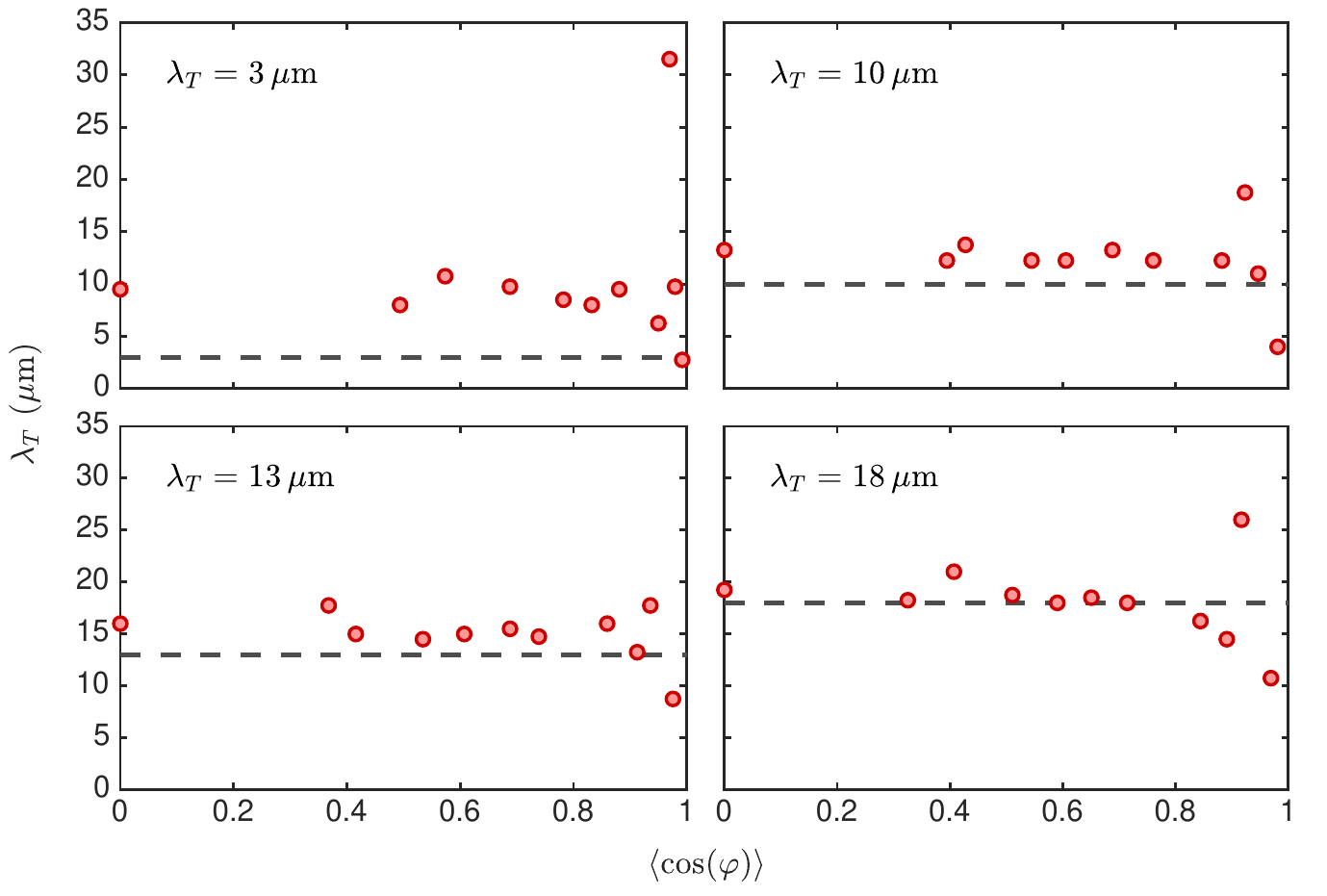}
	\caption{
		\textbf{Testing the temperature fit with simulated pictures.}
		The extraction of the thermal coherence length $\lambda_T$ from the relative phases according to the procedure described in \cref{sec:get_sG_param} is tested by applying it to pictures simulated from the thermal sine-Gordon model (see \cref{sec:corr_sG_simu_pic}).	
		The different subplots show the results for the different $\lambda_T$ used as input for simulating the pictures.
		This input value is indicated in the upper left corner and by the dashed gray lines.
		The red bullets present the results extracted from the simulated pictures, which should ideally coincide with the input values for $\lambda_T$.
	}
	\label{fig:ltfitsimu}
\end{figure}

Note that when testing the procedure with pictures simulated from fluctuations following the sine-Gordon model, the fitting doesn't always work well.
As can be seen from \cref{fig:ltfitsimu}, it often fails for strong phase locking and high temperatures.
We therefore also looked at the full distribution functions for the contrasts of the interference fringes in order to fit the parameters.
As discussed in \cref{chap:contrast_dist}, this shows promise but still needs work.

\section{Experimental results}
\label{sec:exp_res_correlations}

\subsection{Magnitude of the second-order correlation function}
\label{sec:2p_corr_sum}

As discussed in the last \cref{sec:get_sG_param}, we can independently measure $\lambda_T$ with density ripple thermometry and then get the the value for $q$ from a fit of $\cohfact$.
From the sine-Gordon theory for the parameters obtained in this way, we can then calculate a prediction for the quantity $S^{(2)}$ defined in \cref{eq:2p_sum}, which indicates the overall magnitude of the phase fluctuations.

\begin{figure}
	\centering
	\includegraphics{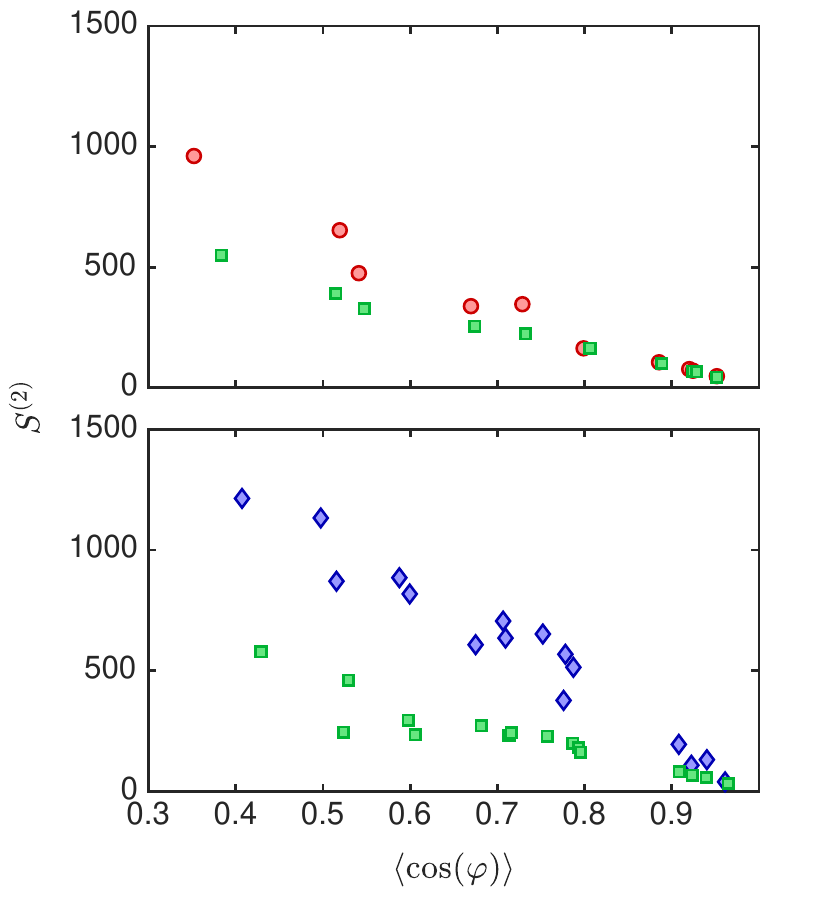}
	\caption{
		\textbf{Magnitude of the relative phase fluctuations.}
		The measure $S^{(2)}$ \labelcref{eq:2p_sum} for the magnitude of the relative phase fluctuations is plotted as a function of the coherence factor.
		In the upper subplot, the experimental results for the coupled slow cooled data is represented by the red bullets.
		The lower one shows the results for the fast cooled data marked by the blue diamonds.
		In both subplots the theory prediction for the temperature from density ripple thermometry is given by the green squares.
		Note that in the sum of \cref{eq:2p_sum} for $S^{(2)}$, the central 25 pixels (50{\um}) were used.
		For the theory values, the influence of the imaging was modeled via a convolution with a Gaussian with $\sigma_\mathrm{PSF} = 3${\um}.
	}
	\label{fig:sum2p}
\end{figure}

In \cref{fig:sum2p}, one sees a comparison between the theory prediction and the experimental results for $S^{(2)}$.
Note that the experimental results have been obtained for the harmonic trap, while the theory predictions have been calculated for the homogeneous sine-Gordon theory. 
One can see that the experimental results for $S^{(2)}$ are generally much bigger than the theory predictions, especially for the fast cooled scans.
While we wouldn't really expect agreement for the fast cooled scans, we hoped to see it for the slow cooled ones.

The reason for the discrepancy in the slow cooled case is still an open question.
One suspicion was that it's due to the non-homogeneity of the background density and tunneling strength (see \cref{sec:phase_locking,fig:cohspatialdependence}).
However, we were not able to confirm this suspicion by comparing to theoretical results obtained from stochastic Gross-Pitaevskii calculation for spatially varying tunnel coupling $J(z)$.
In these calculations, we chose $J(z)$ such that the experimentally seen spatial variation of $\cohfact$ is reproduced.

Another explanation would be that the relative and common degrees of freedom have different effective temperatures.
Fitting the thermal coherence length from the relative phase profiles according to the procedure described in \cref{sec:rel_T_fitting} gives decreased values (higher temperature) when compared to $\lambda_T$ measured via density ripple thermometry.
However, due to the shortcomings of the fitting procedure as discussed in \cref{sec:rel_T_fitting}, we decided not to present the results in form of a figure.

\subsection{Second order correlations in momentum space}
\label{sec:2p_cos_corr}

We will start by discussing the case of having two independent condensates, i.e., the double well barrier separating the two wells is high enough so that there is no tunneling possible.
For this case, we took data in the harmonic as well as in the box-like longitudinal trap \cite{Rauereaan7938,rauer2018thesis}.
With box-like trap we mean the superposition of the magnetic harmonic trap and a box shaped optical dipole potential.
We took absorption images with the vertical imaging system to record the interference patterns as well as with transverse imaging system in order to perform density ripple thermometry.

As discussed in \cref{sec:homo_bogo}, the Bogoliubov theory predicts 
\begin{equation}
\langle \varphi_k ^2 \rangle = \frac{4} {\lambda_T}  \frac {1}{k^2}  \label{eq:phi_k_var}
\end{equation} 
for the cosine transformed phases $\varphi_k$, in the case of a homogeneous system with Neumann boundary conditions.
Note that this is the result for the relative phase, whereas \cref{eq:cf_phase_exp_box_bogo} shows the results for the phase of a single condensate.
Comparing the two equations, one notices the difference in prefactor which is due to the particular definition \labelcref{eq:def_rel_com} of the relative and common degrees of freedom.

\begin{figure}
	\centering
	\begin{subfigure}{0.47\textwidth}
		\centering
		\includegraphics[width=\linewidth]{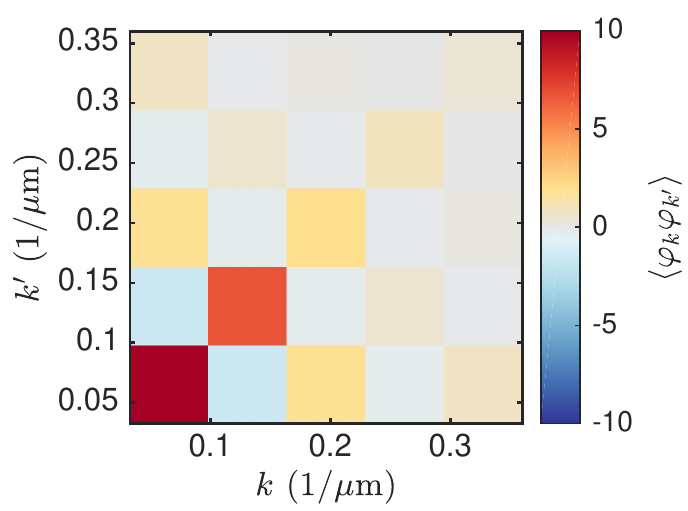}
		\caption{}
		\label{fig:scan54832p2dcoscorr}
	\end{subfigure}
	\ \ \ %
	\begin{subfigure}{0.47\textwidth}
		\centering
		\includegraphics[width=\linewidth]{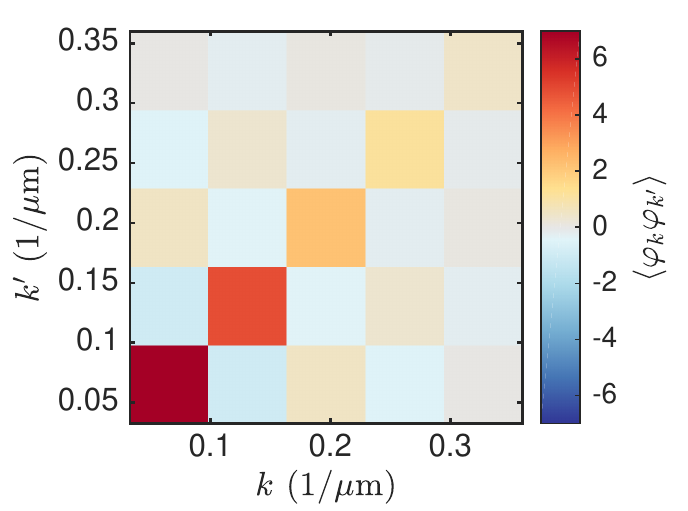}
		\caption{}
		\label{fig:scan5073b2p2dcoscorr}
	\end{subfigure}
	\caption{\textbf{Cosine transformed second-order correlation function.}
	Results for an uncoupled double well \textbf{(\subref{fig:scan54832p2dcoscorr})} and for a coupled double well \textbf{(\subref{fig:scan5073b2p2dcoscorr})} leading to intermediate phase locking ($\langle \cos (\varphi) \rangle  = 0.80$). 		
	Both measurements have been performed in a harmonic trap.
	The color represents the values for the covariance matrices of the cosine transformed phase profiles $\varphi_k$.
	The central 50{\um} have been used for the cosine transformation.
	Note that the value (44.33 for (\subref{fig:scan54832p2dcoscorr}) and 13.09 for (\subref{fig:scan5073b2p2dcoscorr})) for $k = k^\prime = k_1$, i.e., the lower leftmost data point, lies outside the color-range.
	The color-range was chosen like this to get better visibility.
	}
	\label{fig:2D_2p_cos_corr}
\end{figure}

It was checked that \cref{eq:phi_k_var} approximately holds true for the Luttinger liquid theory of a harmonically trapped system when cosine transforming only the central part. 
We can therefore directly compare the experimental results to \cref{eq:phi_k_var}.
The validity of this approach is also supported by the more or less diagonal form for the cosine transformed second-order correlation functions shown in \cref{fig:2D_2p_cos_corr}.


\begin{figure}
	\centering
	\begin{subfigure}{0.49\textwidth}
		\centering
		\includegraphics[width=\linewidth]{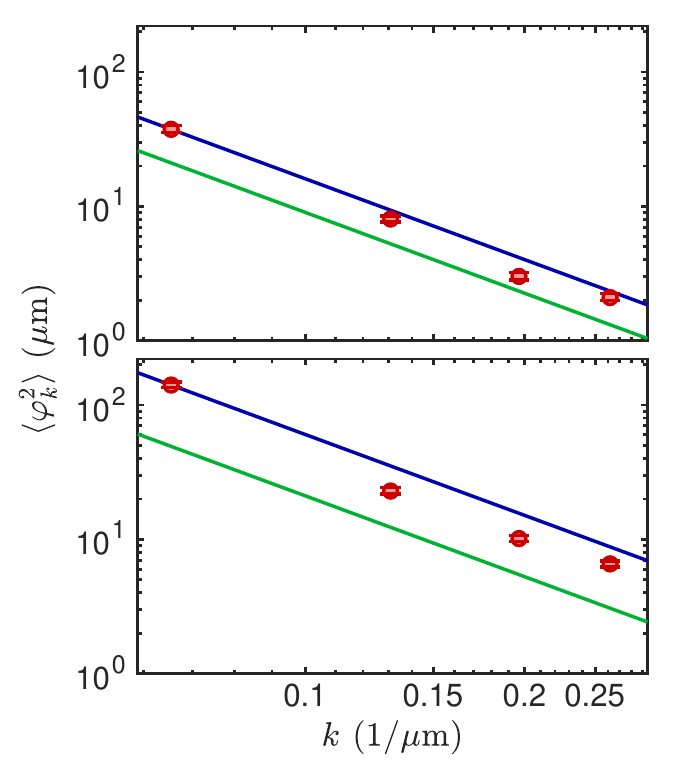}
		\caption{Harmonic trap}
		\label{fig:harmonictrapcoscorr2ploglog}
	\end{subfigure}
	\begin{subfigure}{0.49\textwidth}
		\centering
		\includegraphics[width=\linewidth]{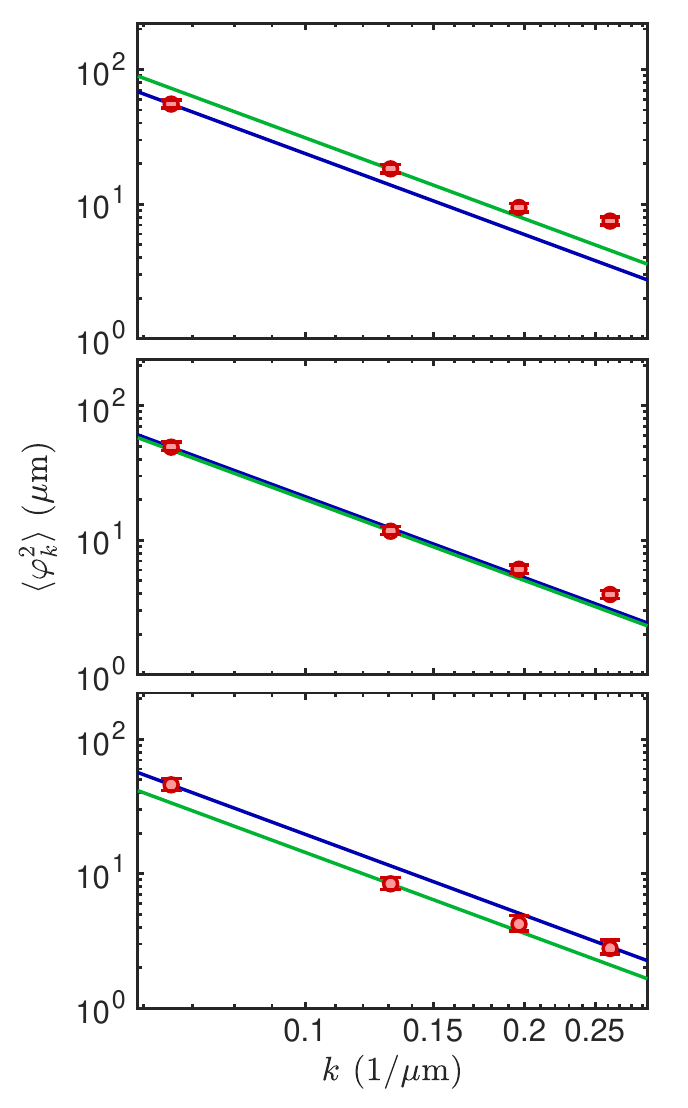}
		\caption{Box-like trap}
		\label{fig:boxtrapcoscorr2ploglog}
	\end{subfigure}
	%
	%
	\caption{
		\textbf{Variances of the cosine transformed relative phases for the uncoupled double well.} 
		The results for the cosine transformation of the central 50{\um} are shown.
		The experimental results are corrected by the expected influence of the imaging system as discussed in the main text. 
		\textbf{(\subref{fig:harmonictrapcoscorr2ploglog})} The upper/lower plots show the result for two different measurements in the harmonic trap. 
		The red bullets represent the corrected experimental results with the errorbars representing 80\% confidence intervals calculated using bootstrapping.
		The green solid line shows the variances expected for a thermal state with the coherence length $\lambda_{T \, \mathrm{DR}} = 44.3${\um} / $19.0${\um} (upper/lower plot) measured via density ripple thermometry. The blue solid line shows the same for the coherence length $\lambda_{T \, 1} = 24.8${\um} / $6.6${\um} (upper/lower plot) extracted from the first mode. 
		\textbf{(\subref{fig:boxtrapcoscorr2ploglog})} Shows the same when a box-trap is superimposed onto the harmonic trap. The length of the box trap is $65${\um} (upper plot), $98${\um} (middle plot) and $130${\um} (lower plot).
		The thermal coherence lengths are given by   $\lambda_{T \, \mathrm{DR}} = 12.9${\um} / $20.0${\um} / $27.8${\um} and $\lambda_{T \, 1} = 16.8${\um} / $18.9${\um} / $20.3${\um} for the upper/middle/lower plot.
	}
	\label{fig:cos_var_uncoupled}
\end{figure}

In \cref{fig:cos_var_uncoupled}, the experimental results for the variances of the cosine transformed phases are shown.
All presented results were obtained for double well traps without tunneling. 
This is confirmed by checking that the coherence factor $\langle \cos(\varphi) \rangle$ is indeed zero. 

Remember that the influence of the finite imaging resolution can approximately be considers by multiplying the theory prediction with the factor $e ^ { - k ^ { 2 } \sigma_\mathrm{PSF} ^ { 2 } }$  (see discussion in \cref{sec:width_of_Gaussian_psf}).
Here we follow the reverse approach, we correct the experimental results by the inverse of this factor.
The reason for this is, that the theoretically expected $1/k^2$ dependence leads to straight lines in the log-log plots.
For the results presented in \cref{fig:cos_var_uncoupled}, we used $\sigma_\mathrm{PSF} = 3.5${\um}.

\begin{figure}
	\begin{subfigure}{\textwidth}
		\centering
		\includegraphics{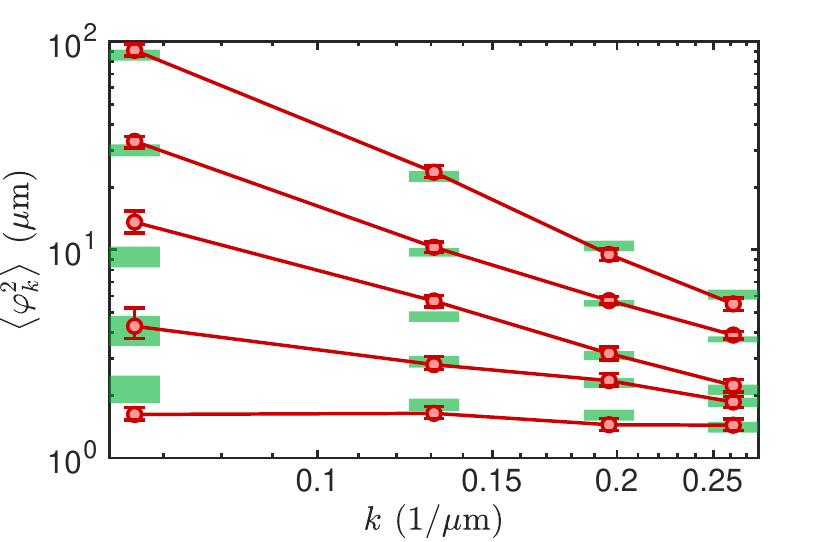}
		\caption{}
		\label{fig:scan5067scan5291scan5073bscan5266scan53102ploglog}
	\end{subfigure}
	\\
	\begin{subfigure}{\textwidth}
		\centering
		\includegraphics{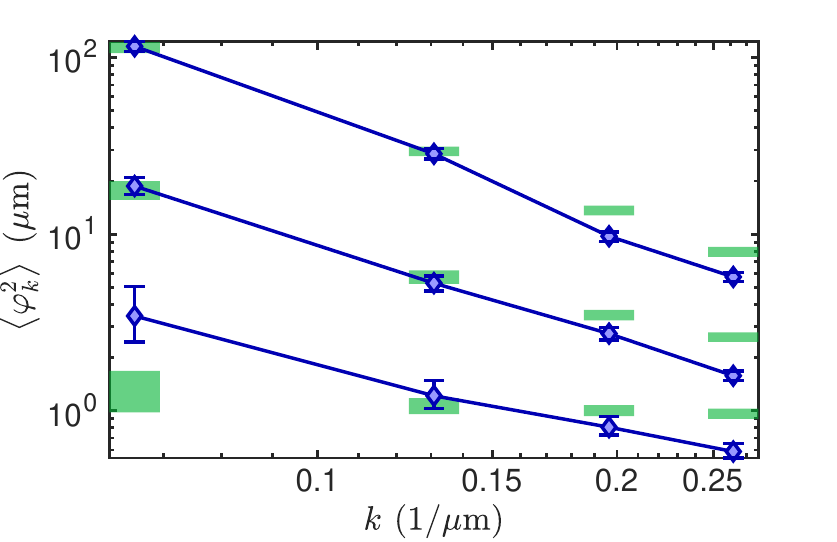}
		\caption{}
		\label{fig:scan5249scan5281scan50842ploglogwtheo}
	\end{subfigure}
	\caption{\textbf{Variances of the cosine transformed relative phases for different phase locking.}
	\textbf{(\subref{fig:scan5067scan5291scan5073bscan5266scan53102ploglog})} The red bullets represent the experimental results when cosine-transforming the central 50{\um} and correcting for the imaging resolution ($\sigma_\mathrm{PSF} = 3${\um} was used). 
	The errorbars represent 80\% confidence intervals obtained using bootstrapping.
	The red lines act as a guide to the eye, connecting data points that belong to the same measurement.
	The coherence factor quantifying the phase locking strength (see \cref{sec:phase_locking}) is (from top to bottom) $\langle \cos (\varphi) \rangle  = 0.35, \ 0.73, \ 0.80, \  0.89, \ 0.92$.  
	Note that for all measurements the system has been prepared in a similar way, with the slow cooling procedure described in \cref{sec:preparation}.
	The theory prediction from the sine-Gordon model in thermal equilibrium is given by the green bars.
	Both parameters ($\lambda_T$ and $q$) for the sine-Gordon theory are fitted from the relative phases according to the procedure discussed in \cref{sec:get_sG_param}.
	The height of the bars indicates the 80\% confidence interval for the theory predictions considering the finite experimental sample size.
	Note that all uncertainty comes from the finite sample size, no uncertainty in $\lambda_T$ and $q$ was assumed.
	The width of the bars was chosen arbitrarily.
	\textbf{(\subref{fig:scan5249scan5281scan50842ploglogwtheo})} Same as (\subref{fig:scan5067scan5291scan5073bscan5266scan53102ploglog}), but for the system prepared by the fast cooling procedure.
	The coherence factor is (from top to bottom) $\cohfact  = 0.41, \ 0.91, \ 0.96$.
	}
	\label{fig:cos_corr_coupled}
\end{figure}

Note that we checked by analyzing simulated pictures whether correcting the measured results in the way discussed above works well.
Consistent with the discussion in \cref{sec:width_of_Gaussian_psf}, we got good agreement for the first few modes (no shown), but saw deviations for higher modes.
Therefore, we are showing only the first four to five modes in \cref{fig:2D_2p_cos_corr,fig:cos_var_uncoupled,fig:cos_corr_coupled}.


For the harmonic trap, we see that the relative phase often fluctuates much stronger than is expected from the temperature measured with density ripple thermometry (\cref{fig:harmonictrapcoscorr2ploglog}).
Note that this is similar to what we discussed for double well traps with tunneling in \cref{sec:2p_corr_sum}.
Also here, the reason for this discrepancy is still unclear.
The magnitude of the discrepancy is different for different measurements; it generally seems to be smaller for lower temperatures.
However, for some of the lower temperature measurements, there seems to be a general gradient in the phase profiles.
Also note that the k-dependence never follows the $1/k^2$ dependence expected from Luttinger liquid theory. 
Having a waiting time between switching off the cooling fields and measuring the interference pattern doesn't resolve the difference between the magnitude of the relative phase-fluctuations and the expectations from the density ripple temperature (not shown).
It is still unclear if a waiting time leads to a more thermal k-dependence.
The experimental results are ambiguous in this respect and more data needs to be taken.
Note that for the hotter of the two measurements shown in \cref{fig:harmonictrapcoscorr2ploglog}, the phase fluctuations might be to fast to be resolved correctly (see discussion in \cref{sec:width_of_Gaussian_psf}).

\Cref{fig:boxtrapcoscorr2ploglog} shows the results when a box-trap is superimposed onto the harmonic trap.
In this box-like trap, the magnitude of the relative phase fluctuations fits quite well with the expectations from the temperature measured with density ripple thermometry.
Again one sees that the k-dependence doesn't follow the expected $1/k^2$ exactly.
Note that for the upper subplot (representing the results for a rather box-like trap) the higher modes are overpopulated in comparison to the thermal state. 
For the lowest plot, the trap is basically harmonic and the lowest mode is overpopulated as also observed for the fully harmonic trap.

A question that naturally comes to mind is whether the k-dependence is only non-thermal for the relative phases or also for the common phase fluctuations.
As discussed in \cref{sec:DR_in_DW,fig:com_rel_g2}, the density ripple patterns are dominated by common phase fluctuations.
However looking at the spectrum of the density ripple pattern, we couldn't really tell whether we see deviations from the thermal prediction or just noise.
More investigations concerning the influence of possible noise sources have to be done before drawing any conclusions in this respect.

Let us now discuss the coupled case in more detail.
As can be seen from \cref{fig:scan5067scan5291scan5073bscan5266scan53102ploglog}, the k-dependence of the variance for the cosine transformed relative phase becomes more flat with increased phase locking.
This is expected from the sine-Gordon theory.
Qualitative agreement between the experimental results and the prediction from sine-Gordon equilibrium theory can be seen.

For the case of the fast cooled scans on the other hand, the k-dependence doesn't flatten out that much with increasing phase locking (see \cref{fig:scan5249scan5281scan50842ploglogwtheo}).
There is not even qualitative agreement between the experimental results and the sine-Gordon equilibrium theory.
One sees a stronger decay with $k$ than expected from theory.


\subsection{Fourth-order connected correlation functions}
\label{sec:exp_res_higher_order}

\begin{figure}
	\centering
	\includegraphics{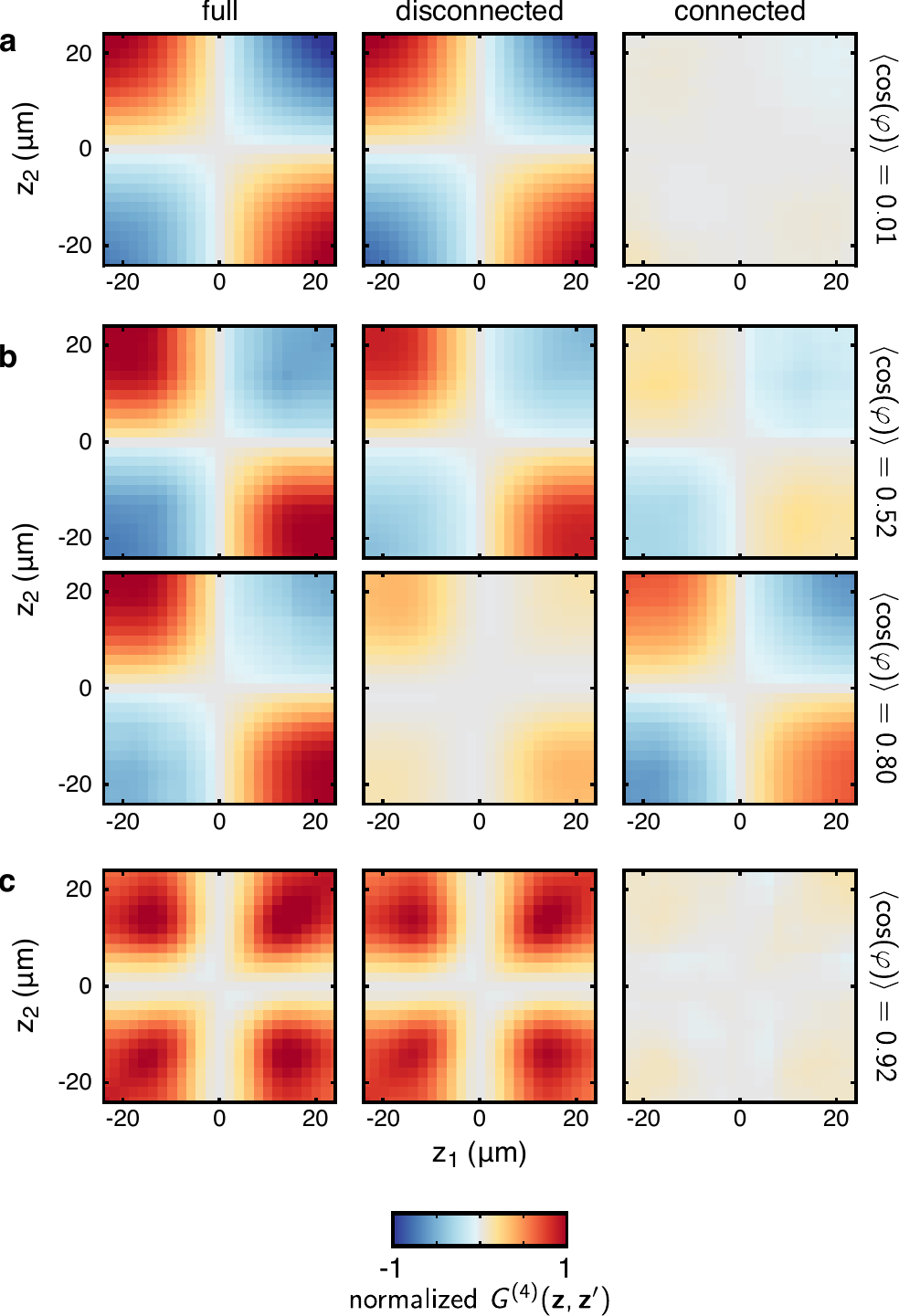}
	\caption{
		\textbf{  Decomposition of the fourth-order phase correlation functions $G^{(4)}(\boldsymbol{z},\boldsymbol{z}')$.}  
		Uncoupled ($\langle \cos(\varphi) \rangle \approx 0$; \textbf{a}), intermediate (\textbf{b}) and strongly phase-locked ($\langle \cos(\varphi) \rangle \approx 1$; (\textbf{c})) regimes. 
		To visualize the high-dimensional data, we choose $z_3 = -z_4 = 14${\um} and $\boldsymbol{z}' = 0$, which results in the observed symmetric crosses where the correlation function vanishes.
		The color marks the value of the full, disconnected and connected correlation functions, with each row normalized to its maximum value such that the color encodes the interval from $-1$ to $1$. 
	}
	\label{fig:wick_visual}
\end{figure}

As already argued in \cref{sec:calc_corr}, higher-order correlation functions are a useful tool to characterize the interaction properties of a system.
Here we will discuss the fourth-order phase correlations of systems prepared by slow and fast evaporative cooling as described in \cref{sec:preparation}.
The results presented in this section have been published in \rcite{Schweigler17}.

\Cref{fig:wick_visual} shows the experimental data for the full fourth-order correlation function $G^{(4)}({\boldsymbol{z}},{\boldsymbol{z}}')$, its disconnected, and connected parts, for different strengths of the phase locking between the condensates.
The system has been prepared with the slow cooling procedure for all results presented in the figure. 
In both limits, $\langle \cos(\varphi) \rangle  \approx 0$ (uncoupled condensates) and $\langle \cos(\varphi) \rangle  \approx 1$ (strongly coupled condensates), the connected part vanishes (\cref{fig:wick_visual}a and c).
The full fourth-order correlation function is given by its disconnected part, calculated from the second-order correlation function.
In other words, the fourth-order correlation function factorizes. 
For intermediate phase locking (\cref{fig:wick_visual}b) the fourth-order function cannot be described by second-order functions alone, and a substantial connected part remains.


Note that the sine-Gordon Hamiltonian \labelcref{eq:H_SG}
\begin{equation}
\hat H _ { \mathrm { SG } } = \int d z \left[ \g \, \delta \hat \rho ^ { 2 } (z)  + \frac { \hbar ^ { 2 } \rho_0 (z)} { 4 m } \left( \frac { \partial \hat \varphi (z) } { \partial z } \right) ^ { 2 } - 2 \hbar J \rho_0 (z) \cos \left( \hat \varphi (z) \right) \right], \label{eq:H_SG_corr}
\end{equation}
nicely reflects the observations in \cref{fig:wick_visual}. 
For $\langle \cos(\varphi) \rangle  \approx 0$, corresponding to $J \approx 0$, only the first part of $\hat H_{\mathrm{SG}}$, the quadratic Luttinger liquid Hamiltonian, remains, leading to Gaussian thermal states characterized by a vanishing connected correlation function $G^{(N)}_{\mathrm{con}}$ for $N>2$. For $\langle \cos(\varphi) \rangle  \approx 1$ we can replace the cosine in the Hamiltonian by its harmonic approximation (see \cref{eq:Ht_2}) leading to a quadratic Hamiltonian and Gaussian fluctuations as well. 

For intermediate phase locking (intermediate $\langle \cos(\varphi) \rangle $) we have to consider the full cosine potential leading to a non-vanishing fourth-order connected correlation function.

\begin{figure}
	\centering
	\includegraphics{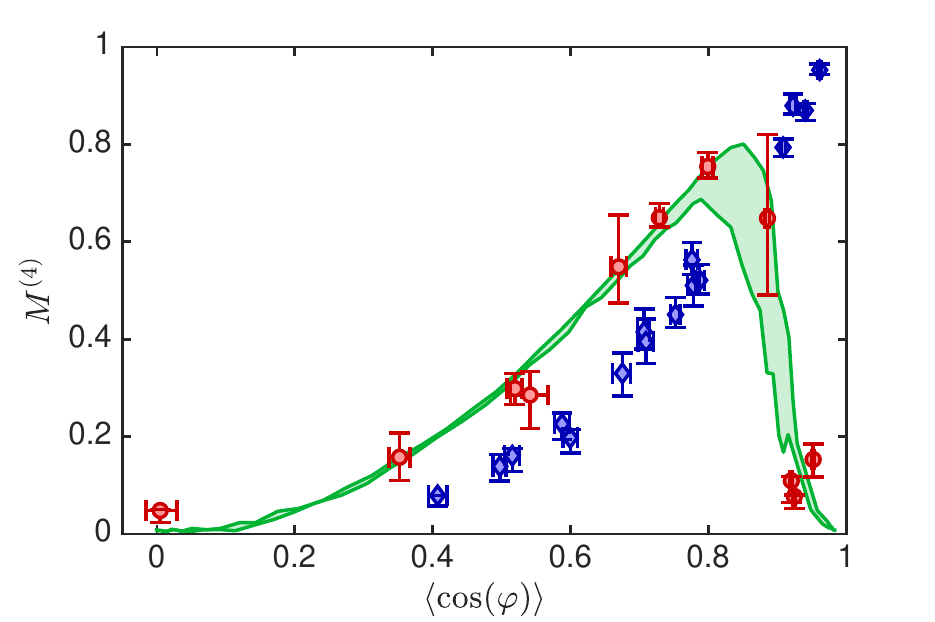}
	\caption{
		\textbf{ Relative size of the fourth-order connected correlation function.} 
		The results following from slow evaporative cooling (red bullets) and from the fast cooling procedure (blue diamond) are shown. 
		We plot the measure $M^{(4)}$ (\cref{eq:int_measure}) as a function of the phase locking strength quantified by $\cohfact$. 
		The error bars represent 80\% confidence intervals calculated using bootstrapping.
		One sees good agreement between the experimental results for slow cooling and the thermal sine-Gordon theory (green shaded region). 
		The fast-cooling data clearly deviates from the equilibrium theory prediction. 
		The theory prediction was calculated for the spread of $\lambda_T$ for the slow-cooled result as measured with density ripple thermometry.
		The borders of the green shaded area represent the predictions for $\lambda_T = 20${\um} (lower border) and $\lambda_T = 15${\um} (upper border). 
	}
	\label{fig:integrated_thermal}
\end{figure}

For a quantitative comparison between experiment and equilibrium sine-Gordon theory we calculate the measure
\begin{equation}
M^{(N)}=\frac{\sum_{\boldsymbol{z}}{\left|G^{(N)}_{\mathrm{con}}({\boldsymbol{z}},0)\right|}}{\sum_{\boldsymbol{z}}{\left|G^{(N)}({\boldsymbol{z}},0)\right|}}. \label{eq:int_measure}
\end{equation}
\Cref{fig:integrated_thermal} shows $M^{(4)}$ as a function of the phase locking strength quantified by the coherence factor $\langle \cos(\varphi) \rangle $.

Note that the thermal coherence length $\lambda_T$ as measured by density ripple thermometry is slightly different for each experimental data point.
We therefore calculated two theory curves, one for the minimum and one for the maximum value of the measured $\lambda_T$ for the slow cooled data.
These curves represent the borders of the green shaded area shown in \cref{fig:integrated_thermal}.
The plotted dependence onto the coherence factor is obtained by varying the parameter $q$ while keeping $\lambda_T$ constant at one of the two values.
The theory predictions have been calculated from $10^5$ numerical realizations obtained from the sine-Gordon stochastic process (see \cref{sec:therm_sG}).
The imaging resolution was considered by convolving the numerically obtained phase profiles with a Gaussian with $\sigma_\mathrm{PSF} = 3${\um}.

\begin{figure}
	\centering
	\includegraphics{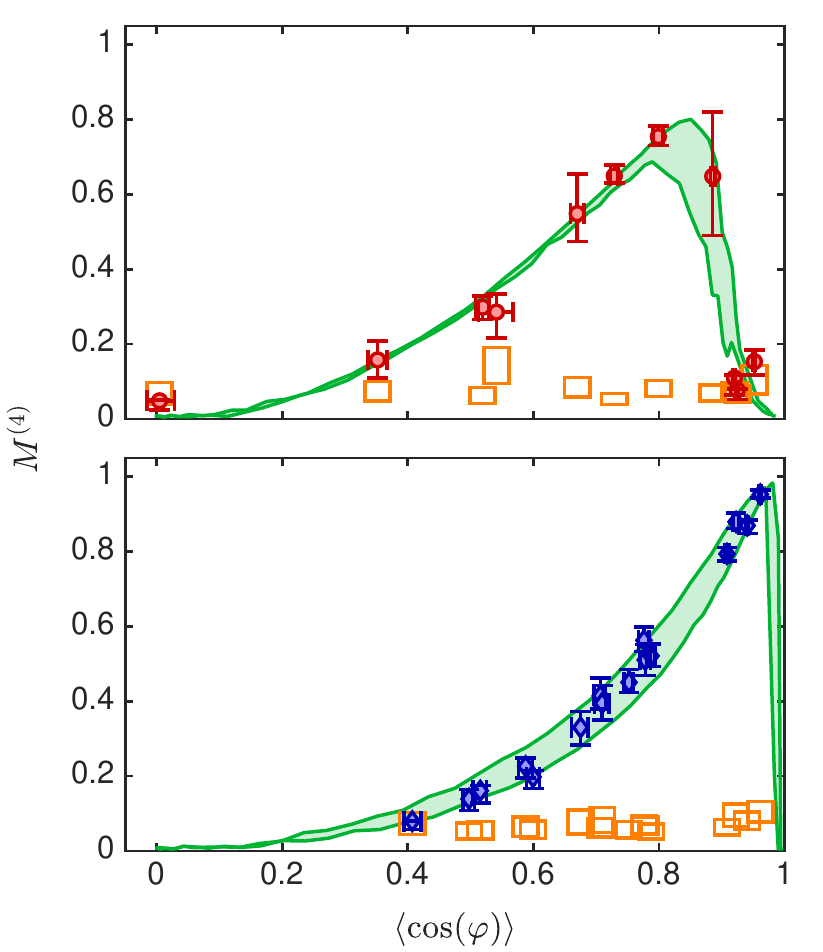}
	\caption{\textbf{ Relative size of the fourth-order connected correlation function.} 
		The same experimental results as in \cref{fig:integrated_thermal} are shown.
		In the upper subplot, the experimental results for slow cooling are marked by the red bullets.
		In the lower subplot, the blue diamonds mark the results for fast cooling.	
		The predictions from the thermal sine-Gordon theory are again given by the green shaded area. 
		In the upper subplot, the borders represent the results for $\lambda_T = 15$ and $20${\um} respectively (same values as in \cref{fig:integrated_thermal}).		
		In the lower subplot, we have $\lambda_T = 5${\um} for the left border and $\lambda_T = 3${\um} for the right border.
		However, the shown curves (in the lower subplot) are not what we expect for the experimental results for hot thermal systems (see discussion in main text).
		The orange rectangles represent the theory predictions for $M^{(4)}$ following from Gaussian fluctuation and considering the finite experimental sample size (see the main text for details).
		The height of the rectangles represents the 80\% confidence intervals, the width was chosen arbitrarily.
	}
	\label{fig:m4wgauss}
\end{figure}

The experimental results for the slowly cooled case agree well with the sine-Gordon equilibrium theory.
States prepared by the fast cooling procedure show a different behavior. 
The results are only consistent with thermal sine-Gordon theory when assuming much higher temperatures than have been measured with density ripple thermometry.
In the lower subplot of \cref{fig:m4wgauss}, the theoretical predictions for $\lambda_T = 3$ and $5${\um} are shown.
The imaging resolution was again considered by convolving the numerically obtained phase profiles with a Gaussian with $\sigma_\mathrm{PSF} = 3${\um}.
However, we don't expect this effective description for the influence of the imaging to stay valid for such high temperatures (see \cref{sec:width_of_Gaussian_psf}).
As discussed later in \cref{sec:corr_imaging}, the results obtained from simulating pictures for hot sine-Gordon phase profiles don't coincide with the shown theory curves, but are similar to the theory curves for much lower temperatures.

In \cref{fig:m4wgauss}, we also check the significance of the observed non-Gaussian fluctuations.
Due to the finite experimental sample size, we would get (small) non-zero values for the measures $M^{(N)}$ even for Gaussian phase fluctuations.
We therefore want to compare the experimental results to the predictions from Gaussian fluctuations.
For this, we first calculate the sample mean and covariance of the phase differences $\varphi(z) - \varphi(0)$ from the experimental data. 
The mean and covariance define a multivariate Gaussian distribution from which we draw $n$ samples (where $n$ is the experimental sample size, different for each point in the diagrams%
\footnote{
The experimental sample size for the results presented in \cref{fig:m4wgauss} varies from $n = 290$ to $n = 2800$, typically it is around $n = 1000$.
}%
) and calculate the measures. 
To get an estimated distribution for the measures, we repeat the procedure 999 times. 
From this, we can calculate the confidence intervals for the Gaussian predictions which are shown in \cref{fig:m4wgauss}.

\begin{figure}
	\centering
	\quad \includegraphics{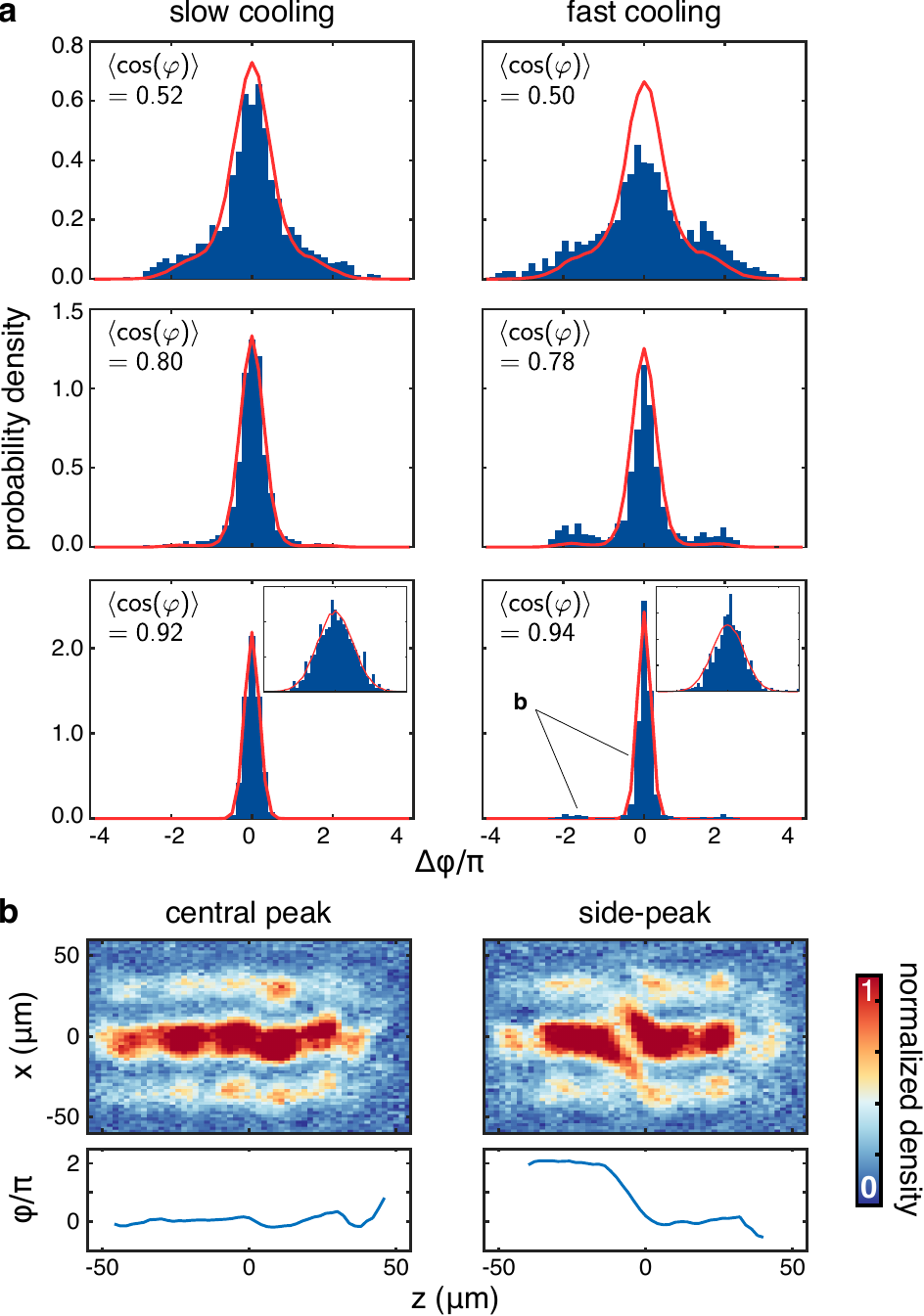}
	\caption{ \textbf{Full distribution functions and interference patterns of the phase.}
		\textbf{(a)} Full distribution (probability density) functions for the phase differences $\Delta\varphi = \varphi(z)-\varphi(z')$ for $z = -z' = 20\ \mu \mathrm{m}$ for different phase locking strengths and two different ways to prepare the quantum gas:  Slow cooling (left) and fast cooling (right). 
		The experimental data (blue bars) for the system prepared by slow cooling are in good agreement with the thermal sine-Gordon theory (red lines). 
		The rapidly cooled systems show substantial deviations, with especially pronounced peaks at $\pm 2 \pi$. 
		The temperature measured by density ripple thermometry (see \cref{sec:get_sG_param}) has been used for the theory calculations.
		\textbf{(b)}~Interference patterns of the two condensates (top panels) and the extracted phase profiles (bottom panels) contributing to the central and side peaks of the full distribution function (indicated in \textbf{a}) for the strongly coupled fast cooled system.		 
		Phase fluctuations are small in the central-peak sample, whereas a sine-Gordon soliton is clearly visible in the side-peak sample.
	}
	\label{fig:PhaseDist}
\end{figure}

To gain insight into the mechanisms leading to the difference between slow and fast cooling, we analyze the full distribution function of the phase differences $\Delta\varphi = \varphi(z)-\varphi(z')$ to which, in principle, all $N^{\mathrm{th}}$-order phase correlation functions contribute. 
\Cref{fig:PhaseDist}a shows the full distributions for one particular pair of coordinates $(z, z')$ chosen symmetrically around the center of the trap. 

For slow cooling and intermediate values of $\langle\cos(\varphi)\rangle$ the full distribution functions of the phase differences $\Delta\varphi$ are distinctly non-Gaussian. 
For strong phase locking ($\langle\cos(\varphi)\rangle \approx 1$) we find Gaussian full distribution functions, as anticipated from the observed validity of the Wick decomposition in this case. 
In contrast, for fast cooling all coupled cases show non-Gaussian distribution functions. 
With increasing phase locking, one can see the appearance of distinct side peaks at $\pm 2\pi$, becoming more localized, but at the same time more suppressed. 

For fast cooling and $\langle\cos(\varphi)\rangle = 0.94$ we observe a Gaussian central peak (see inset) as well as a few outliers at $\pm 2\pi$. 
Studying interference patterns for individual realizations corresponding to the side peaks reveals that the phase rotates through a full circle of $2\pi$ within a short distance (see \cref{fig:PhaseDist}b). 
These localized kinks represent transitions between different minima of the cosine potential and can be identified as solitons of the sine-Gordon model; they are topological excitations of $H_{\mathrm{SG}}$ \labelcref{eq:H_SG_corr}.

In the case of fast cooling these sine-Gordon solitons are frozen in, and the phase of the quantum field fluctuates around them. 
Such states may therefore be interpreted as topologically distinct, `false' vacua~\cite{Coleman77A} above which fluctuations are being excited. 
The energy of these false vacua increases with the number of sine-Gordon solitons.

Note that we also calculated correlation function for orders higher than four and presented the results in \rcite{Schweigler17}.
However, this becomes more challenging with increasing order and statistical as well as systematic uncertainties grow.

Moreover, we can also calculate higher-order correlation functions in momentum space.
Preliminary results for the slow cooled data show good agreement with the sine-Gordon theory for the first few modes.
However, more work needs to be done.


\section{Interpretation and robustness of the results}
\label{sec:corr_interpretation}

%
%

In this section, we will be analyzing simulated pictures in order to investigate the robustness of the experimental results.
The effect of the imaging procedure on the cosine-transformed second-order correlation functions has already been discussed in \cref{sec:width_of_Gaussian_psf}.
We will therefore focus on the fourth-order correlation functions as well as on explicitly discussing possible phase measurement errors here.

The details about the simulated pictures for the thermal sine-Gordon theory have been already discussed in \cref{sec:corr_sG_simu_pic}.
Here, we will also analyze pictures simulated from Gaussian phase fluctuations.
The details will be discussed in \cref{sec:corr_Gauss_simu_pic}.

\subsection{Simulating pictures from Gaussian fluctuations}
\label{sec:corr_Gauss_simu_pic}

In this section, we will discuss the details of how we simulate pictures for the Gaussian fluctuations following from two quadratically coupled Luttinger liquid Hamiltonians.
The simulation procedure for the sine-Gordon theory has already been discussed in \cref{sec:corr_sG_simu_pic}.
The actual procedure generating the simulated pictures is the same in the two cases, only the numerical realizations for the in-situ fields differ.
The discussion here will therefore only focus on this part. 

For simulating pictures from Gaussian fluctuation, both density and phase fluctuations are calculated for the Luttinger liquid Hamiltonian \labelcref{eq:H_LL_com_rel} plus quadratic coupling term \labelcref{eq:Ht_2}.
We use the classical fields approximation and assume thermal equilibrium.
The numerical realizations for the phase and density fluctuations are then obtained following the discussion in \cref{sec:bogo_matrix_vector}.
Note that we use the density dependent 1D interaction strength as given in \cref{eq:g1D_broadened}.   
The resulting numerical realizations for the density fluctuations are subsequently convolved by a Gaussian with a standard deviation of one micrometer in order to introduce an artificial cutoff.
Moreover, note that we use the background density for a  harmonic longitudinal trapping potential according to the discussion in \rcite{Gerbier_2004}.
For the harmonic trapping frequencies we choose $\omega_\perp = 2 \pi \times 1.35${\kHz} for the transverse directions and $\omega_\parallel = 2 \pi \times 6.7${\Hz} for the longitudinal direction.
The atom number is 5000 in one well.

\subsection{Phase measurement errors}
\label{sec:ph_err}

To investigate the reliability of the extraction of continuous phase profiles from the interference pictures, we compare the extracted phase profiles to the input phase profiles of simulated pictures.
We will do this by looking at the spatial mean of the squared error
\begin{equation}
E_\varphi = \frac{1}{N} \sum_{ n = 1 }^{N} \left( \varphi_\mathrm{in} (z_n) - \varphi_\mathrm{fit} (z_n) \right)^2 . \label{eq:ph_prof_mean_abs_err}
\end{equation}
Here $\varphi_\mathrm{fit} (z_n)$ represents the phase at pixel $n$ fitted from one simulated picture, and $\varphi_\mathrm{in} (z_n)$ is the corresponding phase used as the input for simulating that particular image.
To be more precise, $\varphi_\mathrm{in} (z_n)$ is the input phase profile convolved with a Gaussian with $\sigma_\mathrm{PSF} = 3${\um} (see discussion in \cref{sec:width_of_Gaussian_psf}).
For the presented results, the mean in \cref{eq:ph_prof_mean_abs_err} runs over $N = 25$ pixels, representing the central $50${\um} of the cloud. 

In general, one would expect \cref{eq:ph_prof_mean_abs_err} for be larger for phase profiles fluctuating more strongly. 
We therefore will plot $ E_\varphi $ in a scatter plot as a function of the spatial variance 
\begin{equation}
V_\varphi = \frac{1}{N-1} \sum_{ n = 1 }^{N} \left( \varphi_\mathrm{in} (z_n) - \frac{1}{N} \sum_{ n^\prime = 1 }^{N} \varphi_\mathrm{in} (z_{n^\prime}) \right)^2  \label{eq:ph_prof_spatial_var}
\end{equation}
of the input phase profiles.

\begin{figure}
	\centering
	\includegraphics{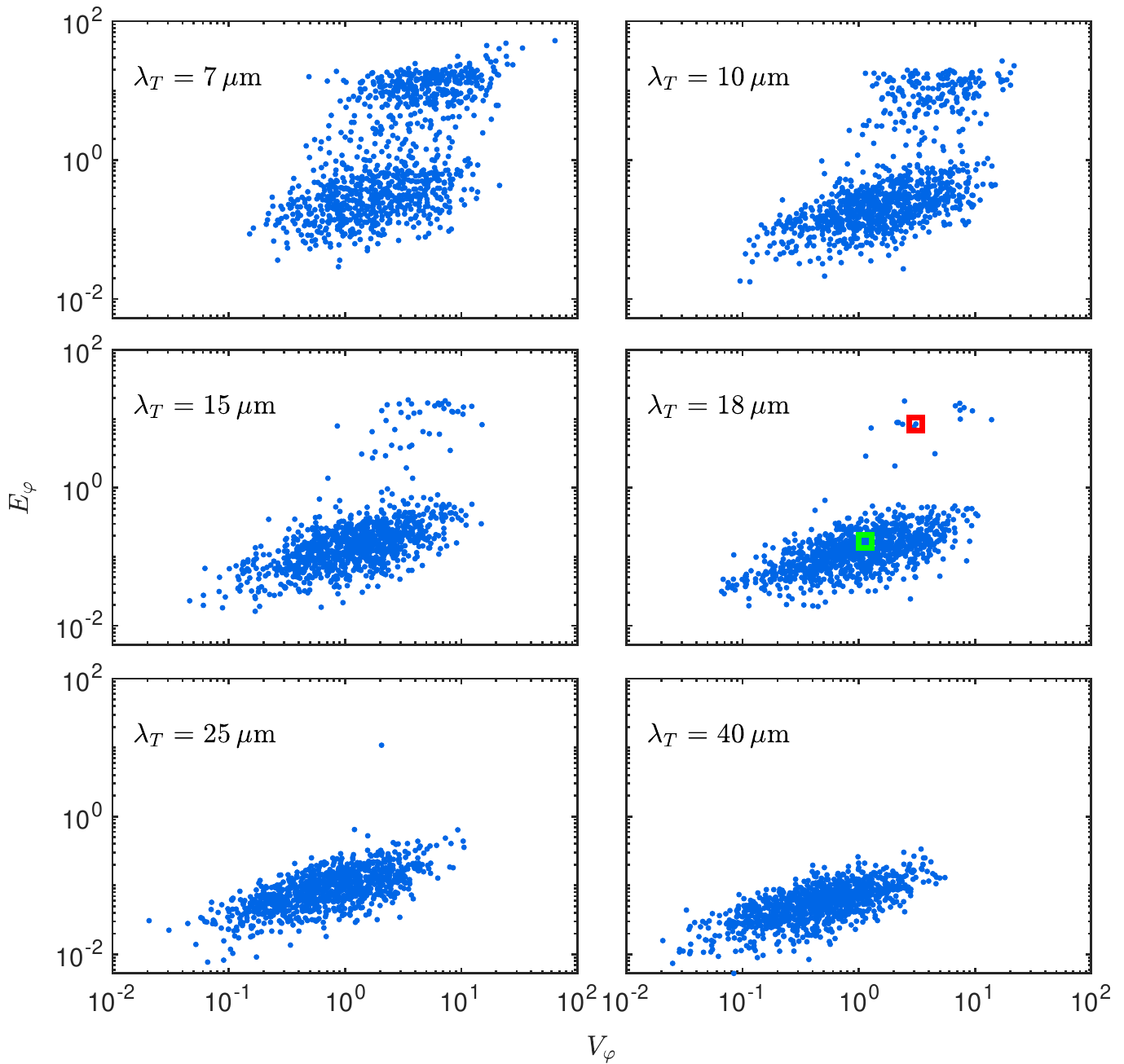}
	\caption{\textbf{Phase fitting error for zero tunnel coupling.}
		The vertical axis gives the spatial mean squared error \cref{eq:ph_prof_mean_abs_err}, the horizontal axis the spatial variance \cref{eq:ph_prof_spatial_var}.
		Each blue dot represents the result for one simulated image.
		For details about the simulation procedure see \cref{sec:corr_sG_simu_pic}.
		The different subplots represent the results for the different $\lambda_T$ indicated in the upper left corner of the subplots.
		With decreasing thermal coherence lengths, one sees the growth of a second cloud above the main cloud.
		Looking at the profiles of this upper cloud, one sees errors in the phase unwrapping procedure.
		As examples, the phase profiles marked with the red and green squares are plotted in \cref{fig:phaseproffromscatterq0}. 
	}
	\label{fig:scatterplotq0}
\end{figure}

\begin{figure}
	\centering
	\includegraphics{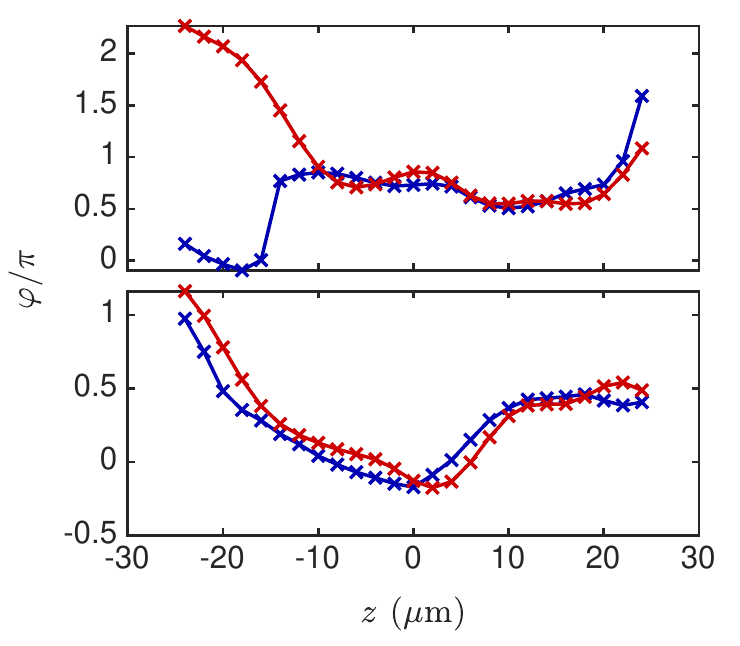}
	\caption{
		\textbf{Phase profiles fitted from simulated pictures.}
		The blue crosses represent the phases $\varphi_\mathrm{fit} (z_n)$  fitted from the simulated images, the red crosses the corresponding input for the simulation.
		The quantity $\varphi_\mathrm{in} (z_n)$ as discussed in the main text is plotted.
		Both plots represent phase profiles for the uncoupled case and $\lambda_T = 18${\um}.
		In the upper plot the phase unwrapping failed, it corresponds to the point marked by the red square in \cref{fig:scatterplotq0}.
		In the lower plot the phase extraction worked (green square in \cref{fig:scatterplotq0}).
	}
	\label{fig:phaseproffromscatterq0}
\end{figure}

\begin{figure}
	\centering
	\includegraphics{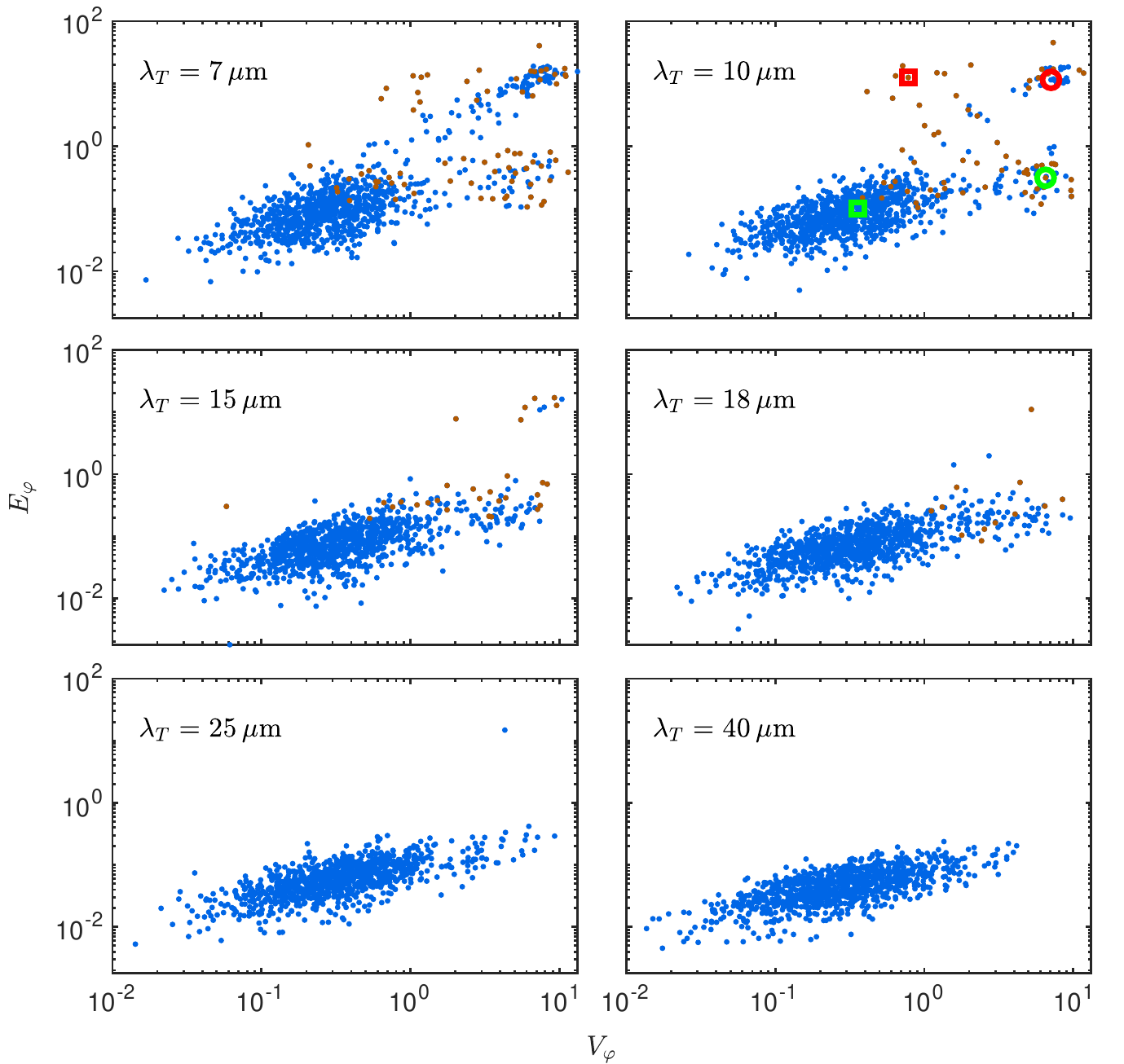}
	\caption{\textbf{Phase fitting error for intermediate phase locking.} 
		Like \cref{fig:scatterplotq0}, but for pictures simulated from the sine-Gordon theory with $q=2.9$.
		With decreasing thermal coherence length $\lambda_T$, one sees the appearance of three clouds separated from the main cloud.
		In the main cloud and the lower right cloud, the phase extraction more or less works.
		The difference between the two clouds is that in the right cloud a well captured physical phase slip occurs and in the main cloud not.
		In the upper left cloud the unwrapping introduces an unphysical phase slip and in the upper right cloud a physical phase slip is missed. 
		Note that most falsely introduced phase slips are filtered out by the condition \labelcref{eq:dphi_filt_cond} with $\varphi_\mathrm{lim} = \pi/2$. 
		The cases not fulfilling the condition are plotted in brown instead of blue.
		In the subplot for $\lambda_T = 10${\um}, the red/green squares/circles mark cases in each of the distinct clouds for which the phase profiles are plotted in \cref{fig:phaseproffromscatterq2p9}. 
	}
	\label{fig:scatterplotq2p9}
\end{figure}

\begin{figure}
	\centering
	\includegraphics{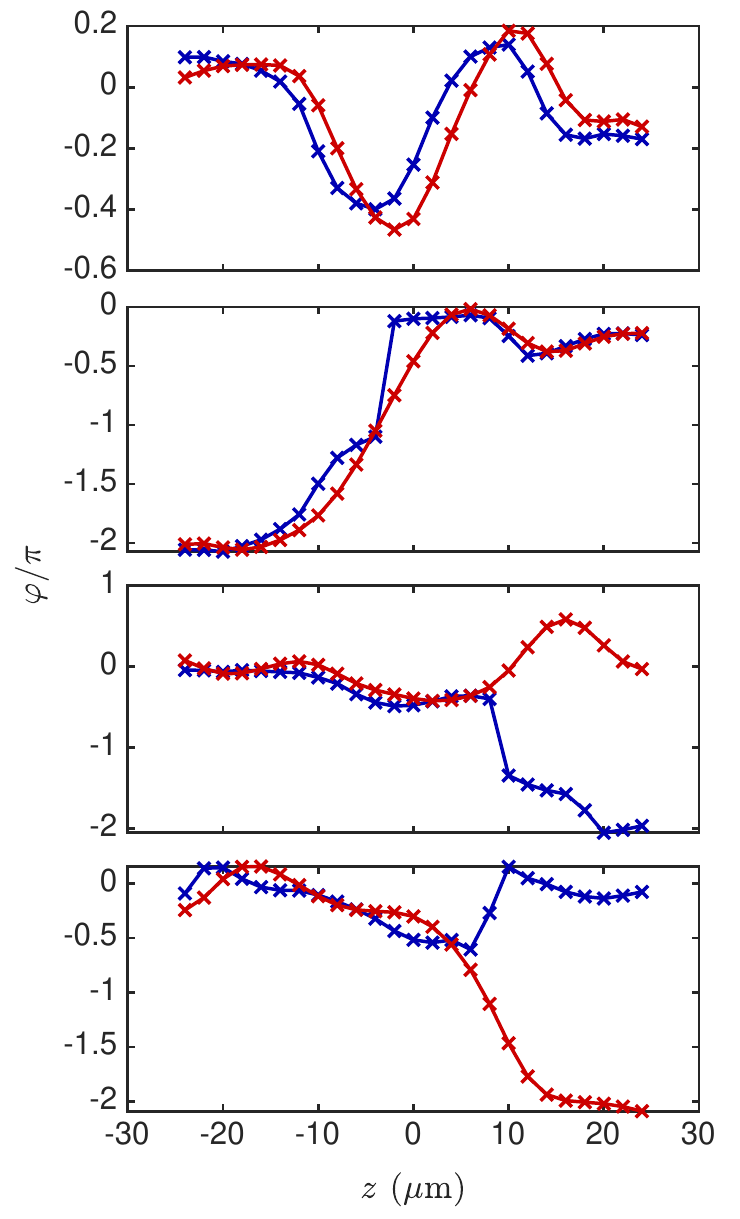}
	\caption{
		\textbf{Phase profiles fitted from simulated pictures.}
		Like \cref{fig:phaseproffromscatterq0}, but for sine-Gordon theory with $q = 2.9$ and $\lambda_T = 10${\um}.
		The uppermost plot corresponds to the point marked by the green square in \cref{fig:scatterplotq2p9}, the input phase profile doesn't contain a phase slip and none is introduced.
		In the next plot a phase slip is present and correctly fitted, the plot corresponds to the green circle in \cref{fig:scatterplotq2p9}.
		Going further down, we see that a phase slip is falsely introduced (red square in \cref{fig:scatterplotq2p9}), and in the lowermost plot a present phase slip is overlooked (red circle in \cref{fig:scatterplotq2p9}).
	}
	\label{fig:phaseproffromscatterq2p9}
\end{figure}

The case of having no tunnel coupling is presented in \cref{fig:scatterplotq0}.
One sees an increasing number of poorly fitted phase profiles with decreasing thermal coherence length.
Typical profiles for when the fit works and when it doesn't are shown in \cref{fig:phaseproffromscatterq0}.

The results for the pictures simulated from sine-Gordon fluctuations with tunnel coupling are shown in \cref{fig:scatterplotq2p9,fig:phaseproffromscatterq2p9}.
For details about the simulated pictures see the discussion in \cref{sec:corr_sG_simu_pic}.
Again, the number of poor fitting results increases with decreasing $\lambda_T$.
One sees the appearance of more and more profiles where a physical phase slip is missed or a non-existing phase slip is introduced by the extraction procedure.

In general, one would expect that erroneous phase slips are more likely when the phase fluctuates strongly, or at least, they should show up as a strong fluctuations in the fitted phase profiles.
We can therefore try to filter the phase profiles, discarding the ones in which phase differences between neighboring pixels are exceeding a certain value.
In other words, we will apply the condition
\begin{equation}
\left| \varphi_\mathrm{fit} (z_n) - \varphi_\mathrm{fit} (z_{n+1}) \right| < \varphi_\mathrm{lim} \ \text{  for all pixels } n. \label{eq:dphi_filt_cond}
\end{equation}
Using $\varphi_\mathrm{lim} = \pi/2$, most of the cases where phase slips were introduced by mistake are filtered. 
The filtered points are marked in brown in \cref{fig:scatterplotq2p9}.
Note that the condition \cref{eq:dphi_filt_cond} will be used in \cref{sec:corr_robust} to investigate the robustness of the experimental results.

\begin{figure}
	\centering
	\includegraphics{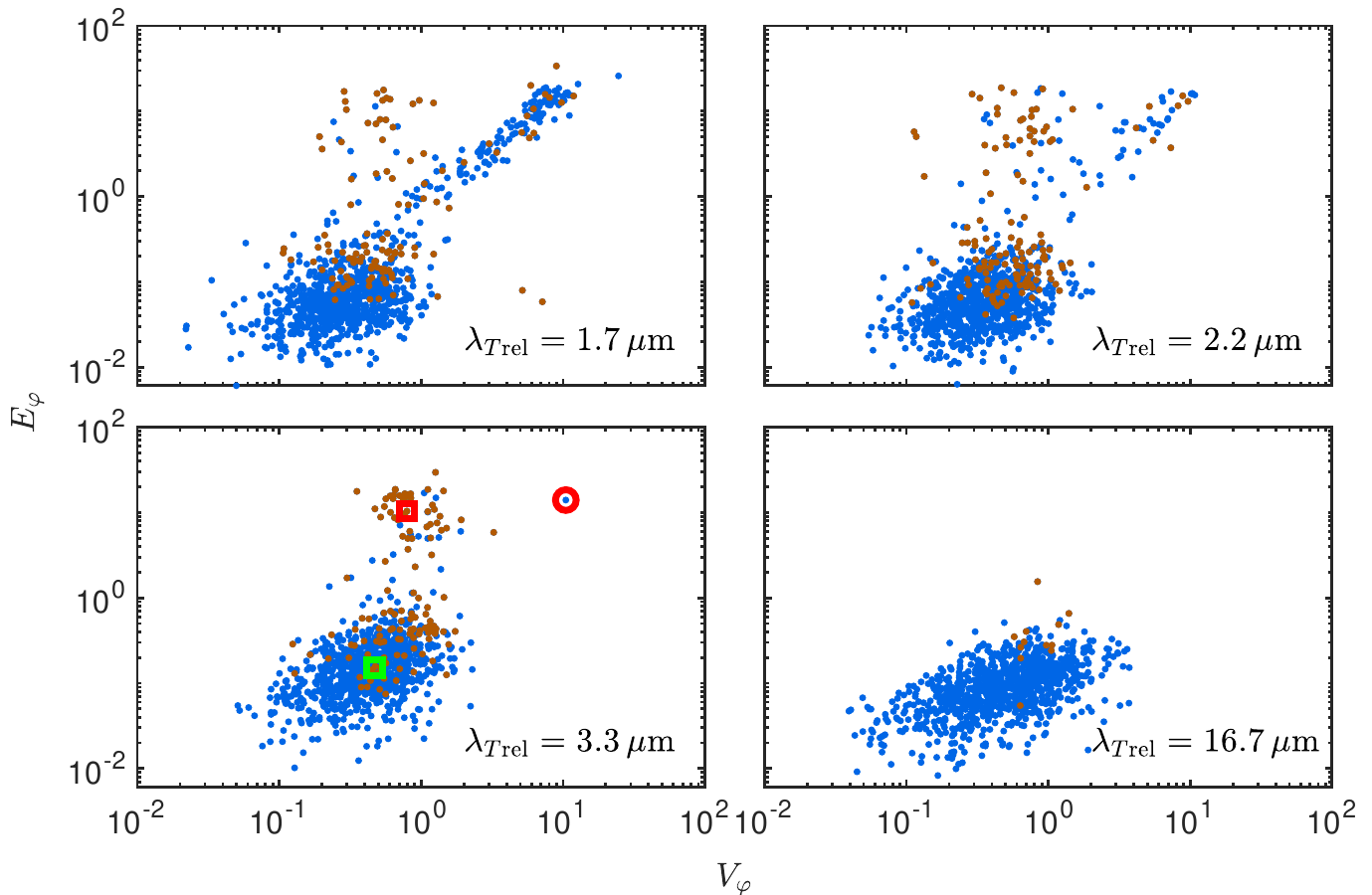}
	\caption{
		\textbf{Phase fitting error for pictures simulated from Gaussian fluctuations.}
		Same as in \cref{fig:scatterplotq2p9} but for a Gaussian theory with tunnel coupling.	
		The thermal coherence length $\lambda_{T \mathrm{rel}}$ for the relative degrees of freedom is indicated in the lower right corner of the subplots.
		All subplots have the same  $q = 1.40$ and the same thermal coherence length $\lambda_{T \mathrm{com}} = 16.7${\um} for the common degrees of freedom.
		Again one sees distinctive sub-clouds appearing.
		Compared to \cref{fig:scatterplotq2p9}, there is no lower right cloud as there are no phase slips in the Gaussian theory.
		The upper left cloud again represents pictures for which the phase unwrapping introduces a jump by mistake.
		In the upper right cloud, big Gaussian fluctuations are missed by the phase fitting.
		The phase profiles corresponding to the marked points in the lower left plot are shown in \cref{fig:phaseproffromscattergauss}.
	}
	\label{fig:scatterplotgauss}
\end{figure}

\begin{figure}
	\centering
	\includegraphics{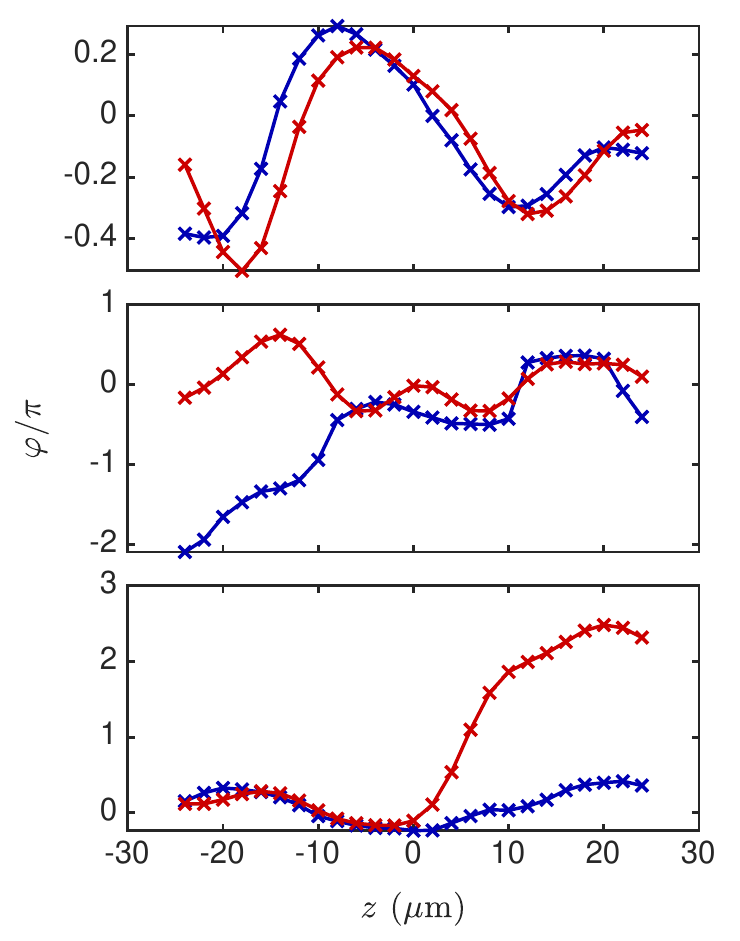}
	\caption{\textbf{Phase profiles fitted from Gaussian simulated pictures.}
		Like \cref{fig:phaseproffromscatterq0}, but for pictures simulated from Gaussian fluctuations.
		Red again represents the input of the simulated pictures and blue the extracted phase profile.
		The upper, middle and lower plot corresponds to the green square, red square and red circle in \cref{fig:phaseproffromscattergauss} respectively.
		In the upper plot, the fitting more or less works, in the middle plot, a phase slip is introduced by a mistake in the phase extraction procedure, and in the lower plot, a big Gaussian fluctuation is missed by the fitting procedure.		 
	}
	\label{fig:phaseproffromscattergauss}
\end{figure}

The results for pictures simulated from Gaussian fluctuations (according to \cref{sec:corr_Gauss_simu_pic}) are presented in \cref{fig:scatterplotgauss,fig:phaseproffromscattergauss}.
The particular value for the tunneling strength was chosen because it gives a rather large `fake' non-Gaussianity (see discussion in \cref{sec:corr_imaging}).
With increasing temperature, one again sees the appearance of more and more profiles where the extraction procedure introduces a non-existing phase slip.
And again, most of the erroneous phase slips are filtered out by using the criterion \cref{eq:dphi_filt_cond} with $\varphi_\mathrm{lim} = \pi/2$.

%

\subsection{Effect of the imaging on the fourth-order connected correlation functions}
\label{sec:corr_imaging}

\begin{figure}
	\centering
	\includegraphics[width=\linewidth]{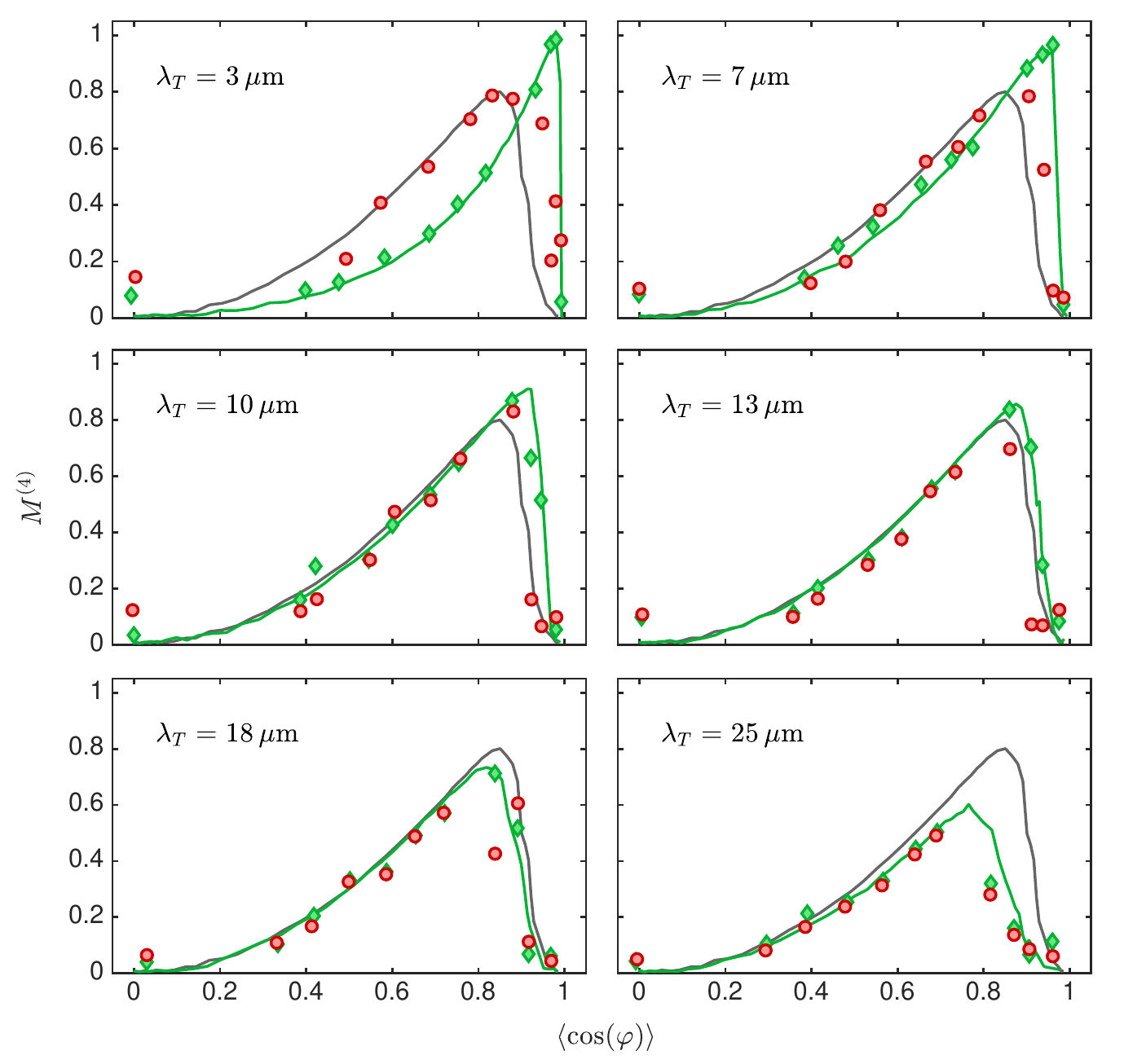}
	\caption{
		\textbf{Influence of the imaging system onto the measure \boldmath{$M^{(4)}$}.}
		 The red bullets represent the results calculated from 1000 phase profiles fitted from simulated images.
		 The green diamonds represent the quantities calculated from the same underlying data used to simulate the pictures.
		 The different subplots represent the results for the different $\lambda_T$ indicated in the upper left corner of the subplots.
		 The solid green lines represent the theory for the respective values of $\lambda_T$ calculated from $10^5$ numerical realizations.
		 Note that the only difference between the green diamonds and the solid green lines is the number of numerical realizations used to calculate the quantities.
		 The gray line acts as common reference curve, it is the same in each subplot and represents the theory for $\lambda_T = 15${\um}. 	 
		 For the green diamonds as well as the solid green and gray lines, the numerical phase profiles were convolved with a Gaussian with $\sigma_\mathrm{PSF} = 3${\um} before calculating the quantities.
	}
	\label{fig:diffnormvscohsgascans}
\end{figure}

In \cref{fig:diffnormvscohsgascans}, the results for the relative size of the connected fourth-order correlation functions extracted from simulated pictures are shown.
The pictures have been simulated from thermal phase fluctuations following the homogeneous sine-Gordon model. 
For details see the discussion in \cref{sec:corr_sG_simu_pic}.
As for the experimental results in \cref{fig:integrated_thermal}, $M^\mathrm{(4)}$ has been calculated from the central 25 pixels (50{\um}).

Note that the influence of the imaging system onto the measure $M^\mathrm{(4)}$ and the coherence factor $\cohfact$ is well described by the effective consideration through convolution of the phases with a Gaussian, as long as $\lambda_T$ is not too small.
For $\lambda_T < 10${\um}, the influence of the imaging system becomes more complicated, the results seem to partly coincide with the curves for bigger values of $\lambda_T$.

\begin{figure}
	\centering
	\includegraphics[width=\linewidth]{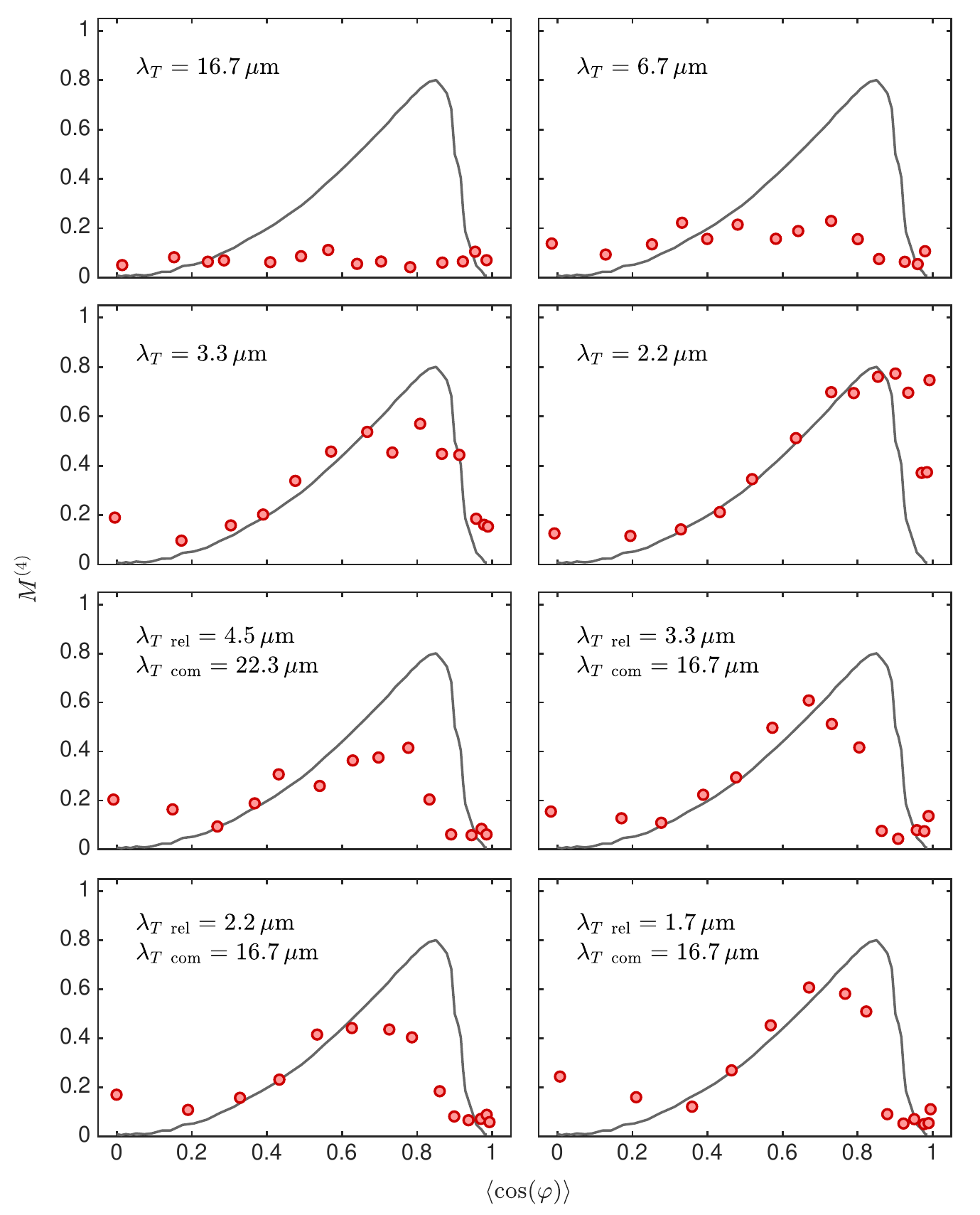}
	\caption{\textbf{Measure \boldmath{$ M^{(4)}$} following from pictures simulated for Gaussian fluctuations.}
	Same as \cref{fig:diffnormvscohsgascans}, but for pictures simulated from Gaussian phase fluctuations (see \cref{sec:corr_Gauss_simu_pic} for details).
	For very hot temperatures, the results partly coincide with the thermal sine-Gordon theory for $\lambda_T = 15${\um} (gray solid line).
	This is due to phase measurement errors (see discussion in main text).
	For the upper four subplots the thermal coherence length is the same for the relative and common phase and density fluctuations ($\lambda_T$ given in the upper left corner of the subplots). 
	For the lower four subplots, the thermal coherence length is given by $\lambda_{T \, \mathrm{rel}}$ for the relative and $\lambda_{T \, \mathrm{com}}$ for the common degrees of freedom.
	}
	\label{fig:diffnormvscohgaussianascans}
\end{figure}

Let us now move on to investigating the influence of the imaging system for Gaussian fluctuations.
Simulating pictures for a quadratic tunnel coupling term as discussed in \cref{sec:corr_Gauss_simu_pic}, we get the results presented in \cref{fig:diffnormvscohgaussianascans}.
Note that the values for $\lambda_T$ given in the figure represent mean values over the central 25 pixels (50{\um}), of the approximately 120{\um} long cloud.
This is the same range also used to calculate $M^\mathrm{(4)}$.
One can see that, for hot enough temperatures, also Gaussian fluctuations can partly mimic the non-Gaussian results predicted by the sine-Gordon theory.
This `fake' non-Gaussianity is due to the introduction of non-existent phase slips by mistakes in fitting and unwrapping the relative phase profiles (see discussion in \cref{sec:ph_err}).

\begin{figure}
	\centering
	\includegraphics{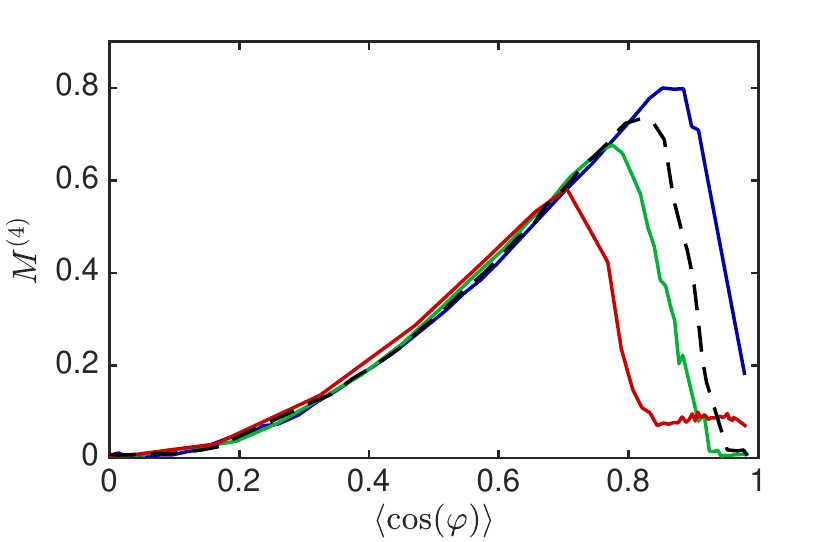}
	\caption{
		\textbf{Relative size of the fourth-order connected correlation function for sine-Gordon-like theories.}
		The measure $M^{(4)}$ as defined in \cref{eq:int_measure} is plotted as a function of the coherence factor $\cohfact$.
		The thermal results for sine-Gordon-like theories as defined in \cref{eq:H_SG_like} are shown.
		For the solid lines, the `potential' for the relative phase is given by \cref{eq:phi_n_pot} with $n = 0.5$, 2 and 4 (blue, green and red).
		The dashed black line represents the result for the sine-Gordon theory.	
		For all cases $\lambda_T = 18${\um} was used.
	}
	\label{fig:sgvsphin}
\end{figure}

However, it is quite puzzling why these erroneous phase slips lead to similar results as the thermal sine Gordon theory.
In order to understand this better, note that a class of theories similar to the sine-Gordon model leads to similar results when plotting the measure $M^{(4)}$ as a function of the coherence factor.
In the following, we will discuss the results calculated for classical thermal fluctuations governed by Hamiltonians of the form 
\begin{equation}
	H = \int d z \left[ g \, \delta \rho ^ { 2 }  + \frac { \hbar ^ { 2 } \nod } { 4 m } \left( \frac { \partial \varphi } { \partial z } \right) ^ { 2 } + V(\varphi) \right] . \label{eq:H_SG_like}
\end{equation}
This is basically the sine-Gordon Hamiltonian with a general `potential' $V$ for the relative phase $\varphi$.
In the case of the sine-Gordon theory we would simply have $V(\varphi) = - 2 \hbar J \nod \cos (\varphi)$.

In addition to the sine-Gordon model, we will also investigate the theory prediction for the $2 \pi$ periodic continuation of
\begin{equation}
V = 2 \hbar J \left| \varphi^n \right| \label{eq:phi_n_pot}
\end{equation}
defined on the interval $\varphi = \left[-\pi,\pi\right)$.
The results are shown in \cref{fig:sgvsphin}.
As can be seen, the results for the potential \labelcref{eq:phi_n_pot} with different exponents $n$ look quite similar to each other and the sine-Gordon theory.
The only real difference is the value of $\cohfact$ after which $M^{(4)}$ decreases again.
Note that the results differ with $\lambda_T$ and the periodicity of $V(\varphi)$.
For example, the $4\pi$ periodic continuation of \cref{eq:phi_n_pot} would give distinctively different results (not shown).

Phase-locked Gaussian fluctuations with phase slips introduced by error mimic a sine-Gordon-like theory \labelcref{eq:H_SG_like} with a $2 \pi$ periodic potential for the relative phase.
In addition, as also seen for the sine-Gordon case \cref{fig:diffnormvscohsgascans}, the imaging process makes very hot systems look colder.
Therefore, the results for the hot Gaussian fluctuations partly look like the results for the sine-Gordon theory with a much colder $\lambda_T$.
This gives an intuitive understanding of the results presented in \cref{fig:diffnormvscohgaussianascans}.

Note that the temperatures necessary to get this `fake' non-Gaussian fluctuations are way higher than what was measured by density ripple thermometry for the experimental data.
However, density ripple thermometry mostly gives the temperature of the common degrees of freedom (see \cref{fig:com_rel_g2}) and doesn't exclude a much higher temperature for the relative degrees of freedom. 
This is especially true for rather large tunnel couplings where the magnitude of the relative phase fluctuations is suppressed.
We show some results for simulated pictures with a large imbalance between the temperature of the common and the relative degrees of freedom in the four lowest plots of \cref{fig:diffnormvscohgaussianascans}.
One sees that the `fake' non-Gaussianity in this cases is not as high as when both the relative and common degrees of freedoms have a high temperature.
This suggests that the density ripple pattern emerging in TOF increases the error when fitting the phase profiles from the interference pictures.

\subsection{Robustness of the results}
\label{sec:corr_robust}

\begin{figure}
	\centering
	\includegraphics{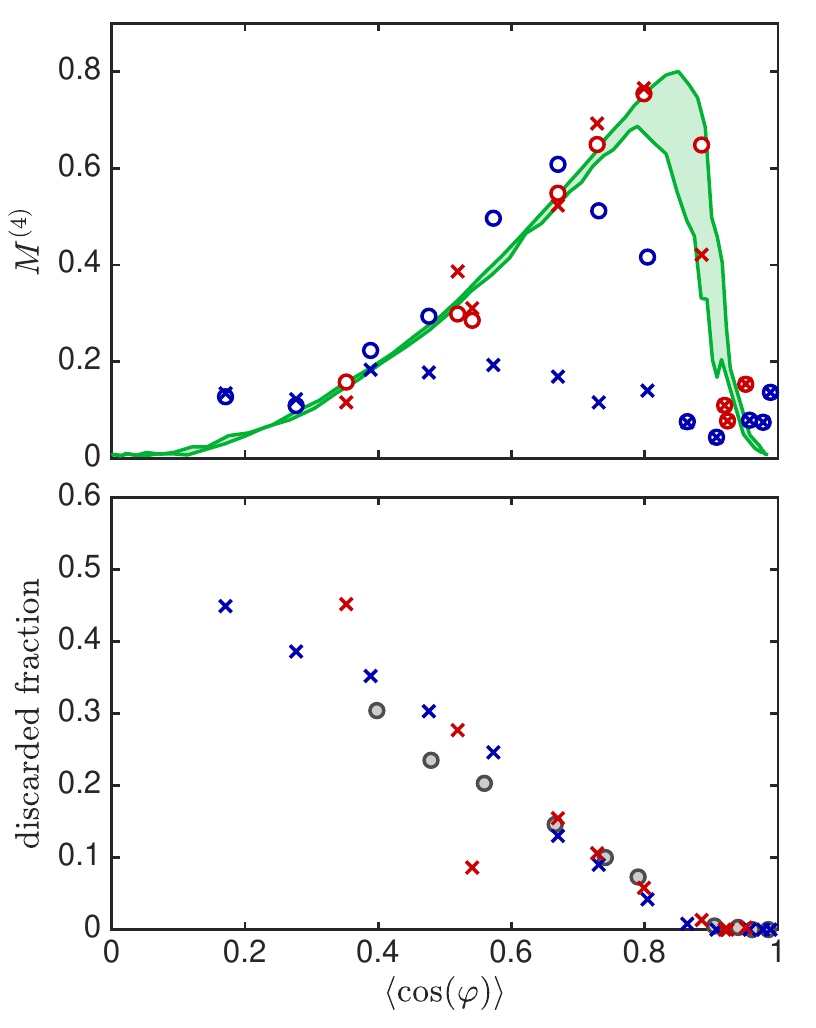}
	\caption{
		\textbf{Distinguishing genuine from fake non-Gaussianity.}
		In the upper subplot, the crosses show the measure $M^{(4)}$ calculated only from phase profiles fulfilling the condition \labelcref{eq:dphi_filt_cond} with $\varphi_\mathrm{lim} = \pi/2$.
		For the results marked by the bullets all phase profiles are used.
		The experimental results for slow cooling are marked in red, the results for the pictures simulated from Gaussian fluctuations are marked in blue.
		The thermal coherence length used to simulate the pictures is $\lambda_T = 3.3${\um} for the relative degrees of freedom and $\lambda_T = 16.7${\um} for the common degrees of freedom.
		Remember that the values measured with density ripple thermometry for the shown experimental data are in the range $\lambda_T = 15\dots20${\um}.
		The lower plot shows the fraction of phase profiles that were filtered out.
		The color-coding is the same as in the upper plot.
		In addition, we also show the results for the pictures simulated from the thermal sine-Gordon theory with $\lambda_T = 7${\um} (gray bullets).		
		Note that the coherence factor $\cohfact$ is always calculated from all phase profiles.
	}
	\label{fig:diffnormvscohdphilim}
\end{figure}

As already discussed in \cref{sec:corr_imaging}, we expect the `fake' non-Gaussianity seen in \cref{fig:diffnormvscohgaussianascans}, to be due to the introduction of erroneous phase slips during the fitting procedure.
We have seen in \cref{sec:ph_err} that we can filter out most of this cases by applying the condition \labelcref{eq:dphi_filt_cond} to the fitted phase profiles.
In \cref{fig:diffnormvscohdphilim}, the influence of this filtering on the measure $M^{(4)}$ is shown.

Using $\varphi_\mathrm{lim} = \pi/2$ for the filtering condition decreases the measure $M^{(4)}$ substantially for the pictures simulated from hot Gaussian fluctuations, but not for the slow cooled experimental data.
Also for the pictures simulated from the sine-Gordon theory and the fast cooled experimental data, no substantial reduction is visible for that value of $\varphi_\mathrm{lim}$ (not shown).
This supports that we see genuine non-Gaussian fluctuation in the experiment.

In the lower subplot of \cref{fig:diffnormvscohdphilim}, the fraction of phase profiles that don't fulfill the condition \cref{eq:dphi_filt_cond} is plotted.
Note that for the experimental data, the filtered fraction is much higher than what one expects form the sine-Gordon theory for the temperatures measured with density ripples ($\lambda_T = 15\dots20${\um}).
It coincides with the results for the sine-Gordon theory with the much shorter $\lambda_T = 7${\um}.
This again supports our suspicion of having a temperature imbalance between the relative and common degrees of freedom. 


\section{Conclusion}
\label{sec:corr_conclusion}

In this chapter we thoroughly analyzed the correlation functions of the relative phases between two quasicondensates in a double well trap with and without tunneling.
Many of the experimental results for the condensates prepared by slow evaporative cooling can be explained by the thermal sine-Gordon theory.
However, there are still many open questions, like the observed temperature imbalance between the relative and common degrees of freedom.
We realized, that we are working right on the edge of what is resolvable with our current imaging system.
The observed non-Gaussianity of the phase fluctuations appears to be genuine.
However, it is still not clear how much insight the higher-order connected correlation function can give into the occurring physical processes.
A large class of Hamiltonians seems to give similar results for the relative size of the fourth-order connected correlation functions in thermal equilibrium.
That said, we see very distinct differences between the results for slow and fast cooling as well as the non-equilibrium data presented in \cref{fig:scan5920diffnormvscoh}.
We have some intuitive understanding of how the differences between slow and fast cooling might arise.  
However, a quantitative description of the results for fast cooling is still lacking and will be a future objective. 


\chapter{Gaussification after switching off the tunnel coupling}
\label{chap:gaussification}

In this chapter, we will present experimental results showing how an initially non-Gaussian state becomes Gaussian after a quench to a free system.
Thereby, we give an experimental demonstration of a phenomenon often termed `Gaussification' in the literature \cite{Gogolin_2016,Gluza2016,Sotiriadis_2016,Sotiriadis_2017,murthy2018relaxation,gluza2018equilibration}.
In this chapter, we will first discuss the experimental procedure in \cref{sec:Gaussification_theory}.
We will then introduce a simple theoretical model in \cref{sec:Gaussification_theory} and subsequently present the experimental results in \cref{sec:Gaussification_experiment}.
Note that, throughout this chapter, we will denote the relative phase by $\varphi(z)$ (without the minus sign in subscript).

\section{Experimental procedure}
\label{sec:Gaussification_procedure}

As discussed in \cref{chap:corr}, cooling into a double well potential with tunneling can produce states with non-Gaussian phase fluctuations. 
Here, the slow cooling procedure discussed in \cref{sec:preparation} is used to prepare the initial state before the quench.\footnote{We also performed measurements where the initial state has been prepared by the fast cooling procedure outlined in \cref{sec:preparation}. The results are qualitatively similar to the ones with the slowly cooled initial state.}
Immediately after the evaporative cooling, the amplitude of the dressing fields is ramped up, increasing the separation of the two wells and consequently switching off the tunneling.
The ramp lasts for approximately 2{\ms}.\footnote{The ramp time varies slightly for the different measurements, it ranges from 1.6 to 2{\ms}.}
Afterwards, the cloud is held in the uncoupled double well for various times. 
The evolution of the system is investigated by recording interference pictures for these different times.
See the discussion in \cref{sec:extracting_time_evo} on how we generally investigate the time evolution of the physical system with our experimental apparatus.


In addition to measuring the interference pattern with the vertical imaging system, we also record density ripple patterns for the initial state using the transverse imaging system. 
Before starting each measurement, we made sure to have approximately equal atom number in the two clouds in the double well.
This was done by adjusting the ratio between the amplitudes of the currents creating the dressing fields (see \cref{sec:intro_atomchip}).
Some of the measurements have been performed in an harmonic trap, for others a box trap was used superimposed onto the harmonic trap.

\section{Theoretical model}
\label{sec:Gaussification_theory}

We have seen in \cref{chap:corr} that at least some aspects of the relative phase fluctuations of the slowly cooled clouds in the double well potential can be described by the thermal fluctuations of the sine-Gordon model in classical fields approximation.
We will therefore use this as the initial state in our theoretical model.
Both the relative number and phase fluctuations will be given by the thermal statistics of the sine-Gordon model \labelcref{eq:H_SG} in classical fields approximation.
For the evolution we assume the validity of the Luttinger liquid model \labelcref{eq:H_LL_com_rel}.
For comparing the theory predictions with the experimental results, we only need to calculate the relative phase fluctuations at the different times. 
The evolution with the Luttinger liquid model doesn't mix common and relative degrees of freedom (\cref{eq:def_rel_com}).
Therefore, we don't need to be concerned with the common phase and density fluctuations at all.
Of course, we still need to consider the relative density fluctuations as they rotate into the relative phase fluctuations during the evolution.

For simplicity, we will do the calculations for a large homogeneous system.
Note that the evolution times presented in this chapter are rather short compared to the system size.
To be more precise the evolution time multiplied with the speed of sound is smaller than the system size.
Therefore, the exact trapping geometry should not matter too much as long as we focus on the central part of the system.
In the experiment we will analyze the central $50${\um} of the cloud.

As discussed in \cref{sec:therm_sG}, the sine-Gordon stochastic process produces numerical realizations of the phase fluctuations for a homogeneous infinitely large system.
To be more precise, one can produce numerical realizations of a certain length, which represents a part of an infinite system.
The thermal density fluctuations for the sine-Gordon model in classical fields approximation don't depend on the length of the system.
Numerical realizations can easily be obtained following the discussion in \cref{sec:bogo_matrix_vector,sec:bogo_thermal_exp_val}.

For numerically evolving these realizations with the Luttinger liquid Hamiltonian, we have to assume a finite system with certain boundary conditions.
We will therefore just calculate the evolution for a large homogeneous system with Neumann boundary conditions.
In the central region, this will mimic the results for an infinite system, provided the evolution times are not too long.
Note that we used the density broadened 1D interaction strength \cref{eq:g1D_broadened} for calculating the initial density fluctuations and the evolution.
   
The results following from the theoretical model are discussed in comparison to the experimental results in the next section.
For all presented theory predictions, the effect of the imaging resolution was approximately considered by convolving the numerically obtained phase profiles with a Gaussian with a standard deviation of $\sigma_\mathrm{PSF} = 3${\um} (see discussion in \cref{sub_sec:eff_psf_fringes,sec:width_of_Gaussian_psf}).

\section{Experimental results and theoretical explanation}
\label{sec:Gaussification_experiment}

\begin{figure}
	\centering
	\includegraphics{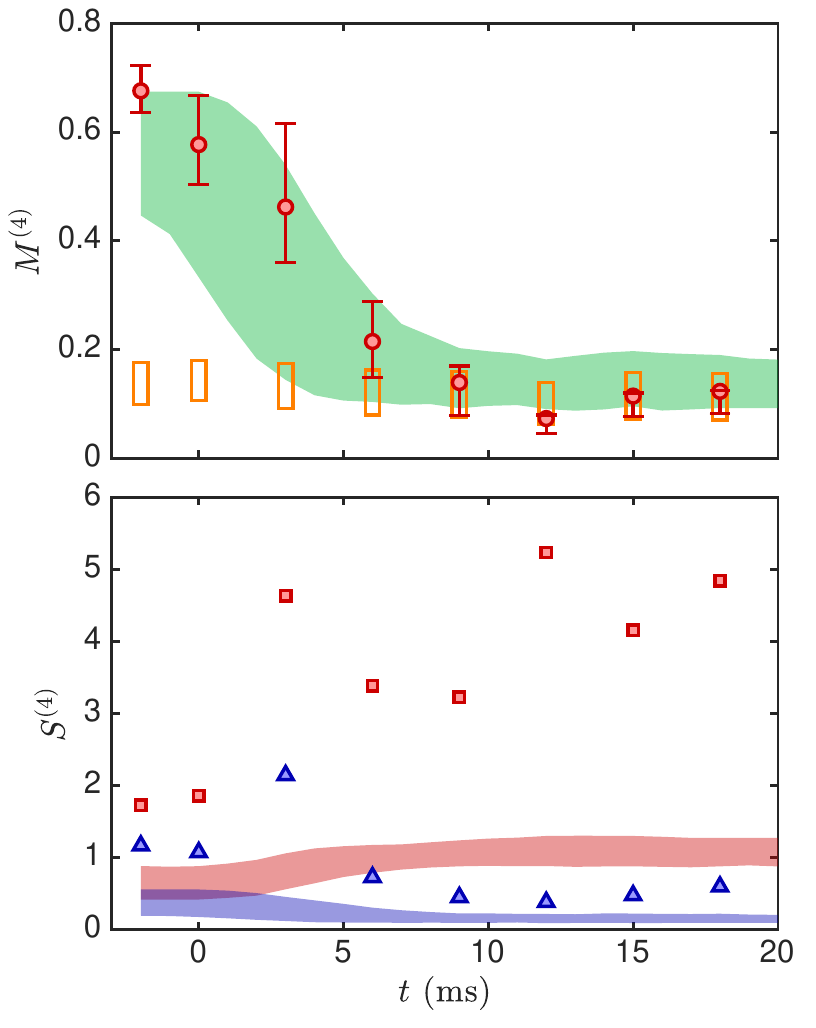}
	\caption{\textbf{Relative size of the fourth-order connected correlation functions.}
	In the upper plot, the red bullets represent the experimental results for the measure $M^{(4)}$ as a function of the evolution time $t$. 
	The errorbars represent 80\% confidence intervals calculated by using bootstrapping.
	The first point at $t = -2${\ms} represents the initial state prepared by slow evaporative cooling into the double well trap with tunneling, leading to phase locking with a coherence factor $\cohfact = 0.74$.
	Between $t = -2$ and $0${\ms}, the amplitude of the dressing fields is ramped up, leading to a decoupling of the two wells at some point during this time interval.
	The green shaded area represents the theory prediction, considering the finite statistics and uncertainty in the decoupling time point, but not the uncertainty in the intial $\lambda_T$ and $q$.
	The vertical extension of the shaded area gives the 80\% confidence intervals.
	The orange rectangles represent the predictions following from Gaussian fluctuations (see discussion in \cref{sec:exp_res_higher_order}). 
	Again, the vertical extension gives the 80\% confidence intervals, the horizontal extension was chosen arbitrarily.
	In the lower subplot, the red squares represent the experimental results for $S^{(4)}$ and the blue triangles the experimental results for $S^{(4)}_\mathrm{con}$ (see \cref{eq:M4_Gaussification}). The red and blue shaded areas show the corresponding theory predictions.  
	Note that the experimental results have been obtained for a harmonic longitudinal trapping potential. 
	  }
	\label{fig:scan5100integratedevol}
\end{figure}

In \cref{fig:scan5100integratedevol}, the evolution of the relative size of the fourth-order connected correlation function is shown.
The plot shows the measure (see discussion in \cref{chap:corr})
\begin{equation}
M^{(4)}=\frac{\sum_{\boldsymbol{z}}{\left|G^{(4)}_{\mathrm{con}}({\boldsymbol{z}},0)\right|}}{\sum_{\boldsymbol{z}}{\left|G^{(4)}({\boldsymbol{z}},0)\right|}} = \frac{S^{(4)}_\mathrm{con}}{S^{(4)}} \label{eq:M4_Gaussification}
\end{equation}
as a function of time.
One sees a fast decrease for the experimentally obtained values of $M^{(4)}$. 
For later times, the experimental results are not distinguishable from the predictions for Gaussian fluctuations when considering the finite statistics of the experimental sample (see discussion in \cref{sec:exp_res_higher_order}).
In \cref{fig:scan5100integratedevol}, the nominator $S^{(4)}_\mathrm{con}$ and denominator $S^{(4)}$ of \cref{eq:M4_Gaussification} are also plotted separately.
It seems like a big part of the reduction in $M^{(4)}$ is caused by an overall growth of the fluctuations.


Theoretically this can be explained by a rotating in of the Gaussian initial density fluctuations.
During the evolution with the Luttinger liquid model, the different non-interacting modes rotate and dephase with respect to each other.
The density quadrature rotates into the phase quadrature which is initially suppressed due to the finite tunnel coupling.
This is the same process as observed in \cite{Gring12,Kuhnert2013,Langen13b,Langen15,Rauereaan7938}.
Theses initial Gaussian density fluctuations are large and overshadow the initial phase fluctuations which are  suppressed due to the tunnel coupling. 
This makes the state appear Gaussian. 
Theoretically a little bit of non-Gaussianity should remain.
However, due to the limited experimental statistics and precision, it's not possible to distinguish such an almost Gaussian state from a fully Gaussian state.
Note that no evolution towards an apparently Gaussian state is predicted when assuming only initial phase but no density fluctuations in the theory calculation (not shown).

As can be seen from \cref{fig:scan5100integratedevol}, our theoretical model predicts the timescale for the decline of the measure $M^{(4)}$ quite well (upper plot).
However, the magnitude of the fluctuations grows much more than predicted by the theory.
A probable reason for this is that, for the initial state in the theoretical calculations, we use the temperatures measured by density ripple thermometry.
As already discussed in \cref{chap:corr}, this might not be the correct temperature for the relative degrees of freedom.
Also the thermal coherence length fitted from the relative phases according to the procedure discussed in \cref{sec:get_sG_param} seems to be still to cold (not shown).
Fitting with the differential entropy of the contrast distribution functions (see \cref{chap:contrast_dist}) looks more promising. 
However, more work needs to be done before this method can be used.

\begin{figure}
	\centering
	\includegraphics{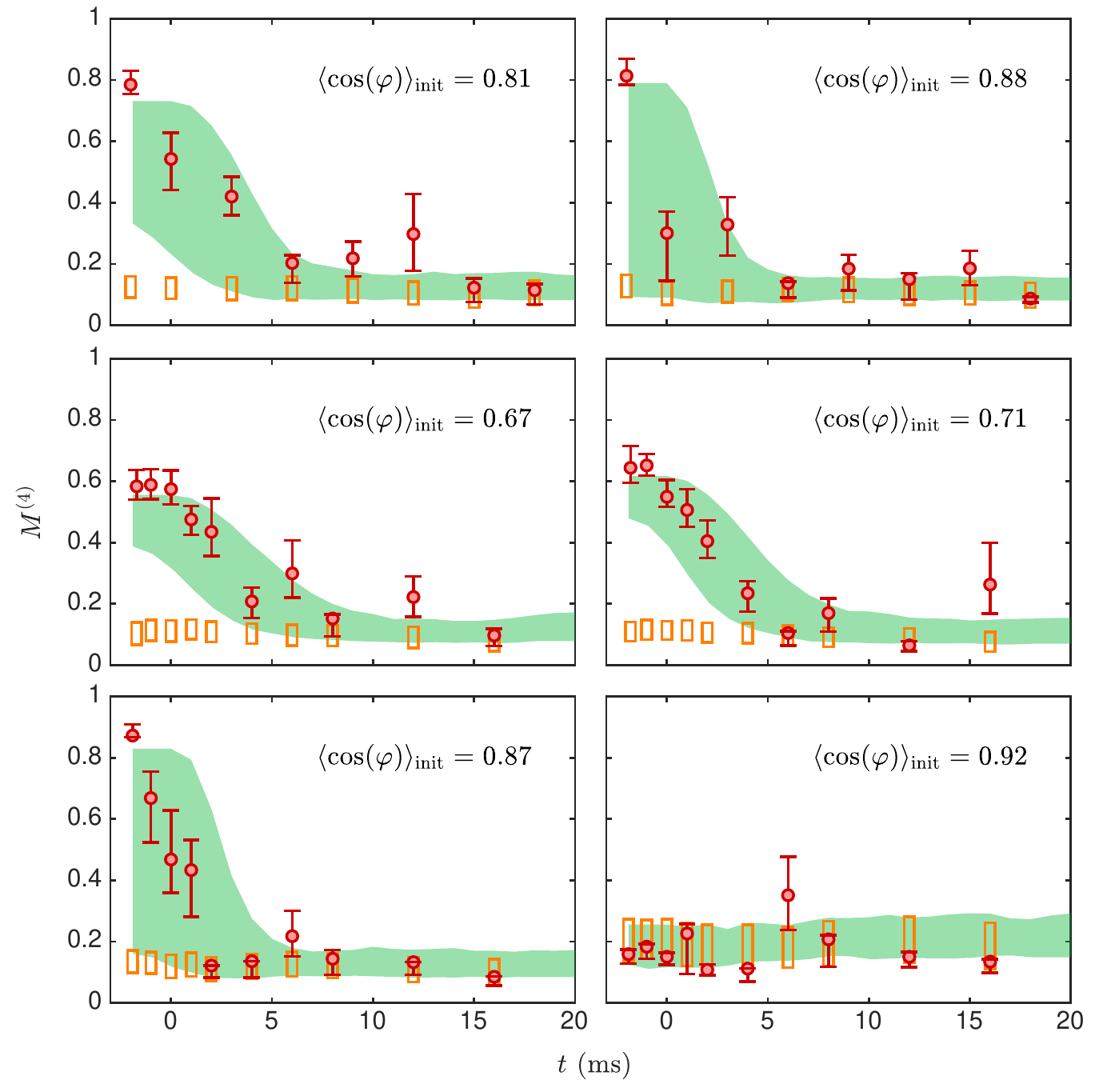}
	\caption{\textbf{Relative size of the fourth-order connected correlation function.}
	Same as the upper subplot in \cref{fig:scan5100integratedevol}, but for multiple different experimental measurements.
	The coherence factor of the initial states is given in the upper right corner of the subplots.
	The two upper subplots have been obtained with a harmonic longitudinal trapping potential. For the four lower ones, a 60{\um} long box trap has been superimposed.
	}
	\label{fig:multiscandiffnormevol}
\end{figure}

\begin{figure}
	\centering
	\begin{subfigure}{\textwidth}
		\centering
		\includegraphics{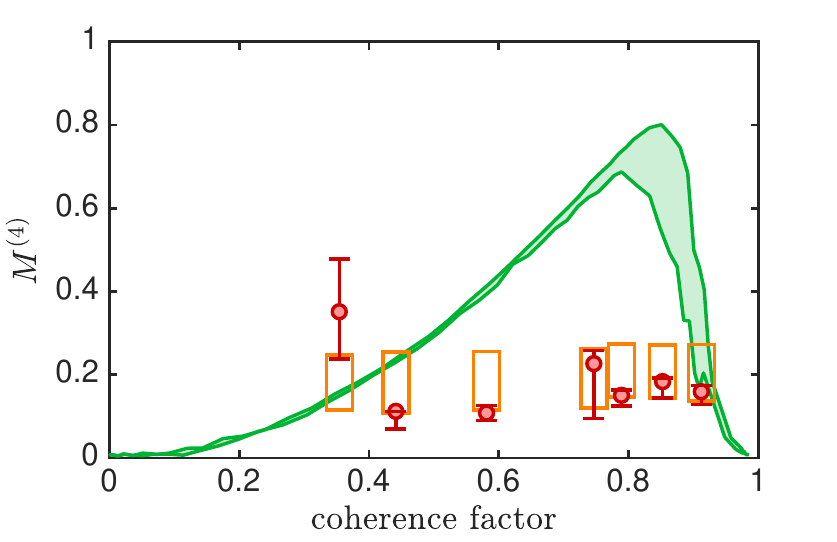}
		\caption{}
		\label{fig:scan5920diffnormvscoh}
	\end{subfigure}
	\\
	\begin{subfigure}{\textwidth}
		\centering
		\includegraphics{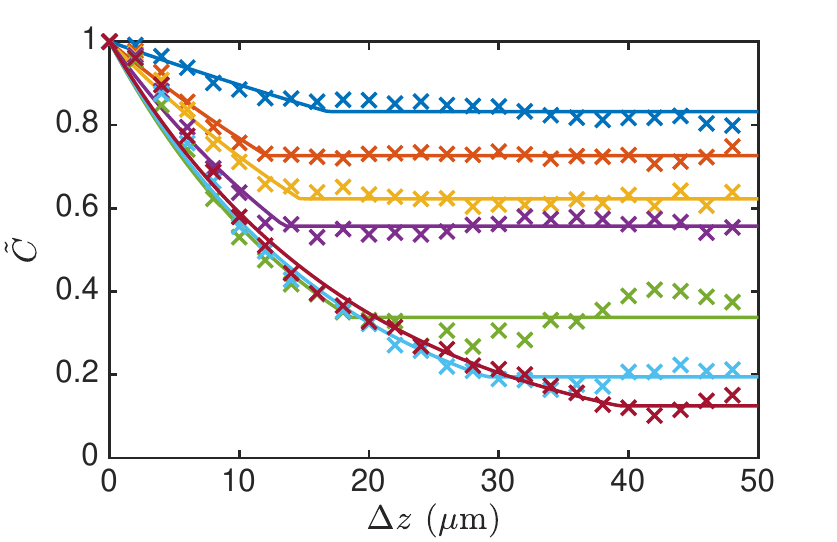}
		\caption{}
		\label{fig:scan5920pcfevo}
	\end{subfigure}
	\caption{
		\textbf{Quench from strong phase locking.}
		\textbf{(\subref{fig:scan5920diffnormvscoh})}~The red bullets show the experimental results for the measure $M^{(4)}$ for the fourth-order connected part as a function of the coherence factor (see discussion in the main text).
		The errorbars represent the 80\% confidence intervals calculated using bootstrapping.
		The different points correspond to different evolution times during the quench.
		From right to left, the points correspond to $t = -1.9$, $-1.0$, 0.0, 1.0, 2.0, 4.0 and 6.0{\ms}.
		The vertical extensions of the orange rectangles again represent the 80\% confidence intervals for the predictions from the Gaussian fluctuations.
		The green shaded area represents the sine-Gordon equilibrium theory for $\lambda_T = 15 \dots 20${\um}.
		This is the same theory as plotted in \cref{fig:integrated_thermal}.
		\textbf{(\subref{fig:scan5920pcfevo})}~The crosses represent the experimental results for the periodic phase correlation function \cref{eq:PCF_zbar}.
		The different colors represent the different evolution times.
		The times are the same as for subfigure \subref{fig:scan5920diffnormvscoh} with time increasing from top to bottom.
		The solid lines represent the corresponding fits used to extract the coherence factor (see main text).
		Note that one can see the light-cone like spreading of thermal correlations discussed in \rcite{Langen13b}.
	}
	\label{fig:scan5920}
\end{figure}

The evolution of the measure $M^{(4)}$ for several more experimental measurements is shown in \cref{fig:multiscandiffnormevol}.
A particularly interesting case is the quench from the strongly coupled system (lower right subplot).
The system basically stays Gaussian during the whole evolution.
In \cref{fig:scan5920diffnormvscoh}, the measure $M^{(4)}$ is plotted as a function of the coherence factor for these experimental data.
One can see a clear distinction between the Gaussian states during the evolution and the non-Gaussian states predicted by the sine-Gordon equilibrium theory.
Remember that the states produced by slow cooling follow this theory as was shown in \cref{fig:integrated_thermal}.

Note that for the time evolution, we cannot calculate the coherence factor directly from the measured phases using $\cohfact$.
It would be dominated by the accumulation of a global relative phase caused by random or systematic global atom number differences between the two clouds.
However, what we actually want to measure is the degree of phase locking.
We do this by looking at the periodic phase correlation function
\begin{equation}
C \left( z , z ^ { \prime } \right) = \left\langle \cos \left[ \varphi ( z ) - \varphi \left( z ^ { \prime } \right) \right] \right\rangle. \label{eq:PCF}
\end{equation} 
This can be rewritten as
\begin{equation}
 \left\langle \cos \left[ \varphi ( z ) \right] \, \cos \left[ \varphi \left( z ^ { \prime } \right) \right] \right\rangle + \left\langle \sin \left[ \varphi ( z ) \right] \, \sin \left[ \varphi \left( z ^ { \prime } \right) \right] \right\rangle.
\end{equation}
In equilibrium, for large spatial separations, $\cos \left[ \varphi ( z ) \right] $ and $\cos \left[ \varphi \left( z ^ { \prime } \right) \right] $ will become independent of each other.
The same is true for the sine functions.
Moreover, as the distribution of $\varphi ( z )$ is symmetric, the expectation value $ \left\langle \sin \left[ \varphi ( z ) \right] \right\rangle $ of the asymmetric sine-function will vanish.
Using this, we get $C \left( z , z ^ { \prime } \right) =  \left\langle \cos \left[ \varphi ( z ) \right] \right\rangle \left\langle \cos \left[ \varphi \left( z ^ { \prime } \right) \right] \right\rangle$ in the limit of large separation between $z$ and $z^\prime$.
For a homogeneous system this leads to the relation
\begin{equation}
	\lim_{\left| z - z^\prime \right| \to \infty} C \left( z , z ^ { \prime } \right) = \cohfact^2.\label{eq:pcf_coh_relation}
\end{equation} 
Of course, due to the global phase accumulation, the relation \labelcref{eq:pcf_coh_relation} doesn't hold for the non-equilibrium states during the evolution after the quench.
However the square root of the $C \left( z , z ^ { \prime } \right)$ for large separations of the coordinates is still the relevant measure for the phase locking strength.

In the experiment, we calculate \labelcref{eq:PCF} by averaging over many experimental realizations and choosing $z$ and $z^\prime$ symmetrically around the center of the trap.
With this particular choice of $z$ and $z^\prime$, we can define the function 
\begin{equation}
	\tilde C \left(\left| z - z^\prime \right|\right) = C \left( z , z ^ { \prime } \right) \label{eq:PCF_zbar}
\end{equation}
only depending on the distance $\Delta z = | z - z^\prime |$ between the coordinates.
As can be seen in \cref{fig:scan5920pcfevo}, $\tilde C(\Delta z)$ first falls off with $\Delta z$ and then more or less reaches a plateau.
We can extract the height of this plateau by fitting the piecewise function
\begin{equation}
	f(\Delta z) = 
	\left\{
	\begin{array}{ll}
		\exp\left[\frac{\Delta z}{z_0}  \log(p)\right]  & \Delta z < z_0 \\
		p & z_0 \leq \Delta z \\
	\end{array} 
	\right. . \label{eq:pcf_fit}
\end{equation}
The height $p$ of the plateau as well as the extension $z_0$ of the exponential decay will be fitted.
The dynamic coherence factor plotted in \cref{fig:scan5920diffnormvscoh} is then simply given by $\sqrt{p}$.
In \cref{fig:scan5920pcfevo} the evolution of the experimental $\tilde C(\Delta z)$ and the corresponding fits are plotted.
Note that the value of $z_0$ grows with time, which corresponds to the light-cone like spreading of thermal correlations observed in \rcite{Langen13b}.

In conclusion, we have experimentally observed Gaussification in a cold atom experiment and presented a possible theoretical explanation.
It is still unclear how the suspected mechanism fits in the more general theoretical frameworks for Gaussification discussed in the literature \cite{Gogolin_2016,Gluza2016,Sotiriadis_2016,Sotiriadis_2017,murthy2018relaxation,gluza2018equilibration}.
Future investigations might focus on possible revivals of non-Gaussianity, similar to the revivals of coherence observed in \rcite{Rauereaan7938}.

\chapter{Non-equilibrium physics in the coupled double well}
\label{chap:non_equi}

In this chapter, we will discuss the results following from two different experimental protocols. 
The physical situations are distinct, but related.
Note that, throughout this chapter, we will denote the relative phase by $\varphi(z)$ (without the minus sign in subscript).

We will first discuss the case of Josephson oscillations in \cref{sec:jos_osc}.
The experimental procedure will be described in \cref{sec:jos_osc_prep} and some simple theoretical models will be discussed in \cref{sec:jos_osc_model}.
Note that we used a very similar experimental procedure as was used in \rcite{Pigneur18}, where a fast damping of the oscillations has been observed.
We observe a similar damping in the experiment. 
The experimental results are presented in \cref{sec:jos_osc_exp}.
 



Compared to \rcite{Pigneur18}, the spatial resolution for our interference pictures is better, leading to a better longitudinal resolution of the measured relative phase profiles $\varphi(z)$.
On the other hand, we lack the atom number resolution of \cite{Pigneur18}.
Therefore, it is challenging for us to measure the evolution of the global atom number imbalance between the two clouds, a quantity which was observed in \rcite{Pigneur18}.
Also note that we work in a different parameter regime.
We are having longer condensates and smaller oscillation frequencies.
Our measurements should therefore be understood as being complementary to \cite{Pigneur18}, as we are looking at different parameter regimes and observables.
Unfortunately, no complete theoretical description of the measurements presented in \cite{Pigneur18}, or our measurements, exists yet.
Therefore, it is hard to say whether the same physical processes occur in the two cases.

In \cref{sec:inverse_quench}, we will discuss what happens, when we start with two independent condensates in thermal equilibrium and lower the separating barrier in order to introduce tunneling.
We think that in some respects this is similar to starting Josephson oscillations with a random initial phase.
Note that theoretical predictions for the investigated physical situation were made in \rcite{dalla_torre_13}.
Under certain conditions, universality was predicted for the evolution.
However, in the experiment, this proofed to be very difficult to observe.

Note that both the results presented in \cref{sec:jos_osc,sec:inverse_quench} are still somewhat preliminary and more work needs to be done to fully understand the experimental results.
In particular, we are lacking robust theoretical calculations necessary for a quantitative comparison between experiment and theory.

We will end the chapter by making concluding remarks and giving an outlook about possible next steps in \cref{sec:non_equi_outlook}. 
This includes a discussion about how we might be able to circumvent some of the problems encountered with the experimental procedures used so far by investigating other physical scenarios not discussed yet.

\section{Josephson oscillations}
\label{sec:jos_osc}

\subsection{Experimental procedure}
\label{sec:jos_osc_prep}


In order to start Josephson oscillations, we want to prepare two condensates with a global phase difference in a double well potential with tunneling.
The spatial phase fluctuations along the longitudinal $z$ direction should ideally be small.
In the following, we will discuss how we prepare such an initial state. 

The atomic cloud is cooled into a single well trap by forced evaporative cooling.
We use a single well trap with a slight radio frequency dressing.
The amplitude of the dressing fields is not big enough to create a double well potential, it only deforms the single well trap, giving it a more flat trap bottom in the transverse $x$ direction.
Starting from this slightly deformed trap when splitting one condensate into two seems to be beneficial in some respects~\cite{Langen13b, langen2013thesis}.

To get a cold cloud of atoms in this slightly dressed single well trap, we apply a procedure very similar to the one discussed in \cref{sec:preparation}.
We first pre-cool in the static trap on the atomchip.
We then switch off the cooling fields before the dressing amplitude is ramped up leading to the deformed single well trap.
Shortly after the ramp, the cooling fields are switched on again for a 470{\ms} long frequency ramp.
Subsequently, the cooling fields still stay on for 60{\ms} with the final frequency of the ramp.
After the cooling amplitude is switched off, we still wait for approximately $45${\ms} before ramping up the amplitude of the dressing fields, splitting the single cloud into a pair.

The first amplitude ramp takes 16{\ms} and leads to a double well potential with tunnel coupling.
The amplitude ratio between the two wires creating the dressing fields is chosen in such a way that the two wells have approximately the same total atom number right after this first ramp.
This criterion leads to slightly different voltage (and also current) amplitudes for the two wires. 
The second ramp takes 2{\ms} and increases the overall amplitude of the dressing fields further.
Simultaneously, the amplitude ratio for the two wires is ramped towards equal voltage amplitudes.
This leads to a tilt in the double well potential, i.e., the two wells are not only separated in the horizontal direction, but also in the vertical direction.
We, therefore, have a global energy difference between the two wells, mostly coming from a difference in the gravitational potential~\cite{Berrada2013}.
The condensates are held in the tilted double well potential for a varying amount of time, during which a global phase difference is accumulated due to the global energy difference.
Subsequently, the second ramp is performed in reverses, recoupling and untilting the double well in 2{\ms}.

The accumulated phase difference leads to the start of the Josephson oscillations after the recoupling.
Note that usually we don't hold the clouds in the tilted double well at all, the phase accumulation during the ramps are enough to start the oscillations. 
Only for the data presented in \cref{fig:contrastevolution} we used different non-zero holding times in order to compare the effects of different starting phases.
The data are also used in \cref{fig:omega0vsfrspacing}.
For all other presented data we did not use any hold time in the tilted double well.  

After the recoupling, the clouds are held for varying amount of time in order to get the time evolution of the investigated observables.
For a general discussion about how we can measure time evolution in our experiment, see \cref{sec:extracting_time_evo}.

\subsection{Theoretical model}
\label{sec:jos_osc_model}

As already mentioned above, we do not have a satisfactory theoretical understanding of the physical processes yet.
However we will still discuss a few simple theoretical models in this section.

Let us first discuss the theoretical predictions following from the sine-Gordon Hamiltonian \labelcref{eq:H_SG}.
In this case, the equation of motion for the phase field is given by
\begin{equation}
\frac{\partial^2 }{\partial t ^2} \varphi(z,t) = c^2(z) \frac{\partial^2 }{\partial z ^2}  \varphi(z,t) - \omega_0^2(z) \, \sin\left[ \varphi (z,t)\right]. \label{eq:sG_equation_of_motion}
\end{equation}
with
\begin{equation}
\omega_0(z) = \sqrt{4  J \, \frac{\g  \rho_0 (z)}{\hbar}}. \label{eq:local_omega_0}
\end{equation}
Here $c(z) = \sqrt{  \g  \rho_0(z) / m}$ is the local speed of sound. 
Please consult \cref{sec:coupled_DW} for an explanation of the other parameters.
Note that \cref{eq:sG_equation_of_motion} is valid for the quantum sine-Gordon model as well as its classical version.
However, in the following we will only be concerned with the classical fields approximation.
Therefore we didn't write any hats in \cref{eq:sG_equation_of_motion}.

Neglecting the spatial derivative in \cref{eq:sG_equation_of_motion}, the phases for the different positions $z$ evolve independent of each other.
For each position $z$ along the longitudinal direction of the clouds, we get the equation of motion
\begin{equation}
\frac{\partial^2 }{\partial t ^2} \varphi(z,t) + \omega_0^2(z) \, \sin\left[ \varphi (z,t)\right] = 0 \label{eq:sG_local_pendulum}
\end{equation}
for a simple pendulum.
For small amplitudes, \cref{eq:sG_local_pendulum} leads to harmonic oscillations with the angular frequency $\omega_0(z)$.

\begin{figure}
	\begin{subfigure}{\textwidth}
		\centering
		\includegraphics{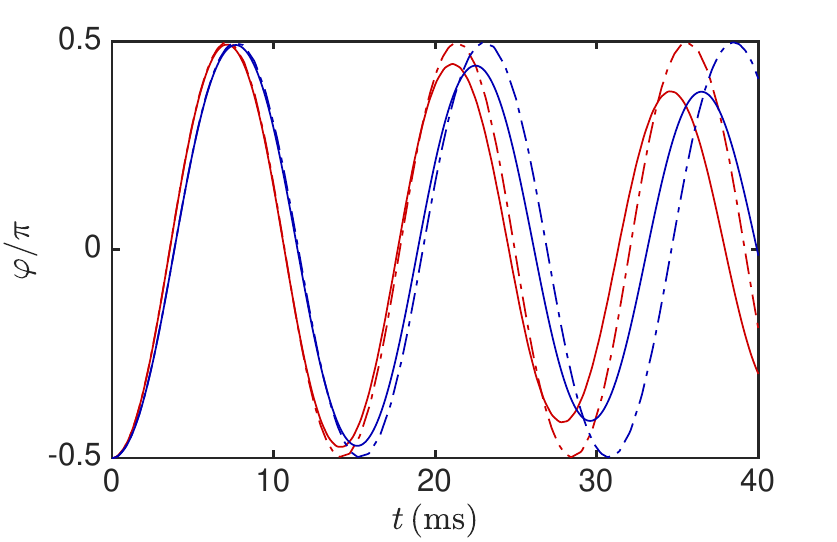}
		\caption{}
		\label{fig:sgnofluctosc}
	\end{subfigure}
	\\
	\begin{subfigure}{\textwidth}
		\centering
		\includegraphics{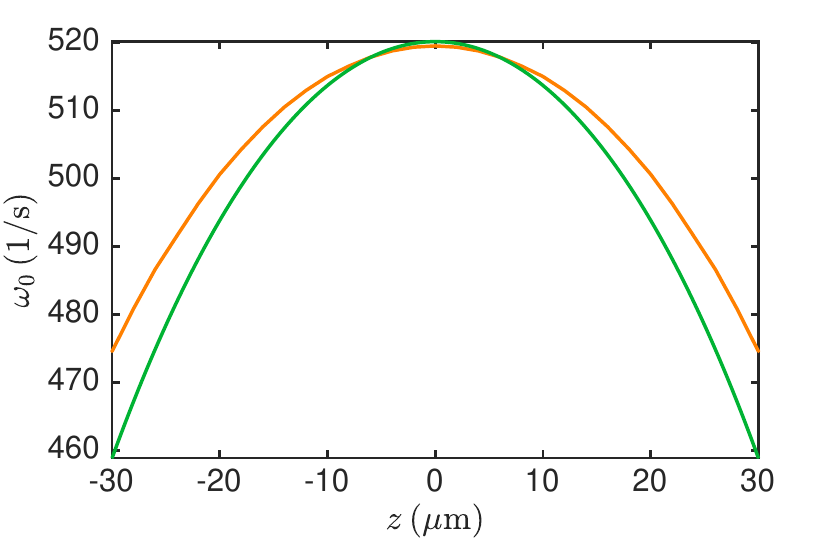}
		\caption{}
		\label{fig:sgnofluctomega}
	\end{subfigure}
	\caption{\textbf{Oscillations for the sine-Gordon equation without initial fluctuations.}
		\Cref{eq:sG_equation_of_motion} is solved for $\varphi (z, t=0) = \pi/2$ and $\left. \frac{\partial }{\partial t } \varphi(z,t) \right|_{t = 0} = 0$.
		\textbf{(\subref{fig:sgnofluctosc})} 
		Shows the time evolution of $\varphi (z,t)$ for $z = 0$ (solid red line) and $z = 24${\um} (solid blue line).
		The dash-dotted lines show the corresponding results of \cref{eq:sG_local_pendulum}.
		One sees that the solid and dashed lines coincide at the beginning, but then start to deviate as a phase gradient builds up.
		\textbf{(\subref{fig:sgnofluctomega})}
		The green curve shows the angular frequency $\omega_0$ given in \cref{eq:local_omega_0}, the orange line represents the results from fitting the solution of the damped pendulum \labelcref{eq:sG_local_pendulum_damped}.
		The first 40{\ms} as shown in (\subref{fig:sgnofluctosc}) are fitted.
		For a more detailed discussion, see the main text.
	}
	\label{fig:sG_no_init_fluct}
\end{figure}

Let us now discuss the implications of a non-homogeneous density profile $\rho_0 (z)$ onto the solutions of \cref{eq:sG_equation_of_motion}.
We will assume a parabolic density profile following from the Thomas-Fermi approximation (see \cref{sub_sec:bogo_single_cond}) for a harmonic trapping potential.
We choose typical trap frequencies and a typical total atom number with zero initial imbalance between the wells.
Moreover, we assume to initially have no spatial phase or density fluctuations.
The initial conditions are simply given by a global phase, i.e., 
\[
\varphi (z,t = 0) = \varphi_0 ,
\]
and zero relative density, i.e.,
\[
\left. \frac{\partial }{\partial t } \varphi(z,t) \right|_{t = 0} = 0 .
\]

Using this assumption, we solve \cref{eq:sG_equation_of_motion} numerically and show the results in \cref{fig:sgnofluctosc}.
As a comparison, we plot the corresponding solutions of \cref{eq:sG_local_pendulum}, i.e., of the undamped pendulum with local $\omega_0(z)$.
One can see clear deviations.
Apparently, the gradient term in \cref{eq:sG_equation_of_motion}, in combination with the non-homogeneous background density $\rho_0 (z)$ leads to a slight damping of the amplitude of the oscillations and a modified oscillation frequency.

However, the time evolution of the phase $\varphi(z,t)$ at a specific point $z$ seems to be quite well described by the solution of the damped pendulum
\begin{equation}
\frac{\partial^2 }{\partial t ^2} \varphi(z,t) + \eta(z) \frac{\partial }{\partial t} \varphi(z,t) + \omega_0^2(z) \, \sin\left[ \varphi (z,t)\right] = 0, \label{eq:sG_local_pendulum_damped}
\end{equation} 
with properly chosen parameters.
We will therefore do a fit with $\eta(z)$ and $\omega_0(z)$ being the free fitting parameters in addition to the initial conditions $\varphi (z, t=0)$ and $\left. \frac{\partial }{\partial t } \varphi(z,t) \right|_{t = 0}$.
The parameters for different points $z$ are fitted independently of each other. 

In \cref{fig:sgnofluctomega} the fitted\footnote{
	Note that in this case we fixed $\left. \frac{\partial }{\partial t } \varphi(z,t) \right|_{t = 0} = 0$ and only fitted the remaining three parameters. 
	If we wouldn't do that, we would basically get the same curve, only with a bit more noise. 
	For the experimental results presented in \cref{sec:jos_osc_exp}, we will also fit $\left. \frac{\partial }{\partial t } \varphi(z,t) \right|_{t = 0} = 0$.}
 $\omega_0 (z)$ is compared to \cref{eq:local_omega_0}.
One sees that the gradient term in \cref{eq:sG_equation_of_motion} leads to a slight slowing down of the oscillation in the middle of the cloud and a speed up for the parts closer to the edges.
This is compatible with the simple picture in which the middle of the cloud `drags' its edges behind.  
We will apply the same analysis to the experimental data in \cref{sec:jos_osc_exp} and will see a much stronger spatial dependence of the fitted $\omega_0 (z)$.
This points to a spatial variation of the tunneling strength $J$ as has been already observed in \cref{fig:cohspatialdependence}.

So far, we assumed that no spatial density or phase fluctuations are present in the initial state.
However, after splitting, we expect some spatial fluctuations of the relative density.
Unfortunately, the splitting process is very hard to model theoretically \cite{Langen15,langen2013thesis}.
A simple description assumes that every atom, independently from all the other atoms in the trap, has 50\% chance to go either to the right or left well, leading to shot-noise fluctuations in the relative density \cite{Kitagawa2011,Langen18}.
The relative phase after splitting will be very small, we can therefore neglect it in many cases.

This is a somewhat sensible assumption for the fluctuations right after splitting and was used for the theoretical description of the experimental results in \rcite{Gring12,Langen13b}.
For the results presented in \cref{sec:jos_osc_exp}, the initial splitting is followed by a further splitting and tilting as discussed in \cref{sec:jos_osc_prep}.
Ideally this should only imprint a global relative phase.
However, we expect that during this time already some dephasing occurs, i.e., some of the density fluctuation already rotate into the phase quadrature.
But at least in first approximation it would be sensible to assume shot noise for the initial density fluctuations as well as a uniform initial phase.
With this assumptions, one can draw stochastic initial conditions and subsequently calculate the time  evolution according to \cref{eq:sG_equation_of_motion}.
Expectation values are then calculated by averaging over the different stochastic realizations.
Preliminary results show a rather fast damping of the Josephson oscillations.
However it is still unclear whether quantitative agreement with the results presented in \cref{sec:jos_osc_exp} can be achieved.

Another similar approach would be to calculate the evolution with the 1D Gross-Pitaevskii equation with stochastic initial fluctuations.
One could assume shot noise in the relative degrees of freedom and thermal fluctuations in the common degrees of freedom.
In this approach, one could also, at least heuristically, consider the breathing motion introduced by the splitting process and observed in the experimental data.

A more complete theoretical description would use the 3D Gross-Pitaevskii equation.
Even though this approach doesn't give a correct description of the splitting process,  it should at least describe the breathing introduced by the splitting  correctly.


\subsection{Experimental results}
\label{sec:jos_osc_exp}

Let us start by discussing the spatially resolved time evolution of the circular mean of the relative phase $\varphi$.
It is defined as
\begin{equation}
	\bar \varphi = \arg \left( \left\langle \mathrm{e}^{ i \varphi } \right\rangle \right)
	\label{eq:circ_mean_def},
\end{equation}
where the expectation value is obtained by averaging over the different experimental realizations.

\begin{figure}
	\begin{subfigure}{\textwidth}
		\centering
		\includegraphics{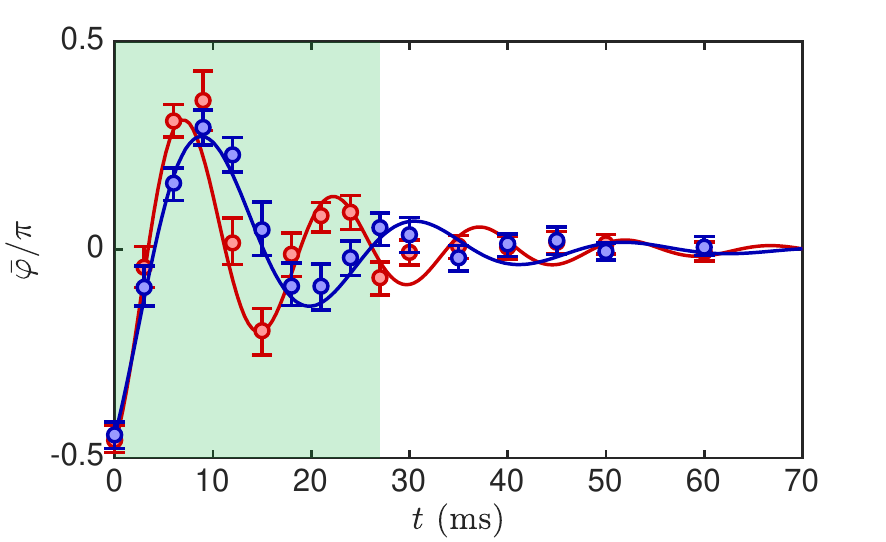}
		\caption{}
		\label{fig:scan4373circmeanph}
	\end{subfigure}
	\\
	\begin{subfigure}{\textwidth}
		\centering
	\includegraphics{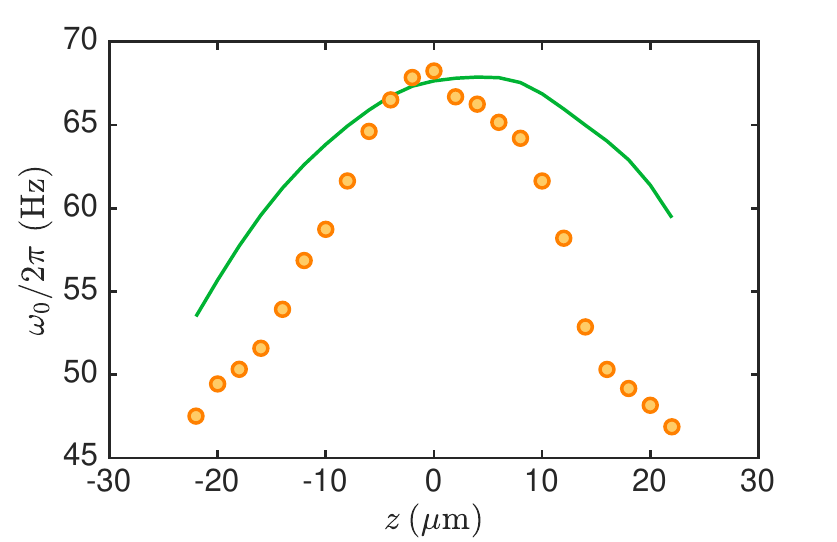}
		\caption{}
		\label{fig:scan4373localomega}
	\end{subfigure}
	\caption{\textbf{Circular mean phase and local oscillation frequency.}
		\textbf{(\subref{fig:scan4373circmeanph})} The circular mean $\bar \varphi$ of the phase is shown as a function of the time $t$. 
		The red and blue bullets show the experimental results for $z = 0$ and $z = 20${\um} respectively, with the errorbars giving the 80\% confidence intervals calculated using bootstrapping.
		Note that the oscillation for the point further towards the edge of the cloud ($z = 20${\um}) is clearly slower than the oscillation in the center ($z = 0$).
		The solid lines show corresponding fits with the solution of the damped pendulum.
		The green shaded area indicates which times we used as input for the fit.
		\textbf{(\subref{fig:scan4373localomega})} The orange bullets show the spatial variance of the fitted $\omega_0 (z)$.
		The solid green line shows \cref{eq:local_omega_0} with constant tunnel coupling $J$ and the experimentally obtained background density $\rho_0 (z)$.
		Note that the asymmetry of the green line is due to an asymmetry in the background density.  
	}
	\label{fig:circ_mean_local_omega}
\end{figure}

In \cref{fig:scan4373circmeanph}, we show the results for one particular experimental measurement.
The time evolution for two different points along the longitudinal direction is shown.
One sees a clear spatial dependence of the oscillation frequency.
Moreover, we observe a fast damping of the oscillations amplitude similar to what was observed in \cite{Pigneur18}.

The evolution of $\bar \varphi$ seems to be rather well described by the solution of \cref{eq:sG_local_pendulum_damped} with appropriate (spatially dependent) parameters.
To infer the parameters, a weighted least squares fit was performed.
As weights, we used the inverse variances obtained by bootstrapping.
The fitted functions are compared to the experimental results in the figure.

\Cref{fig:scan4373localomega} shows the spatial dependence of the fitted $\omega_0(z)$.
One sees that the experimentally obtained $\omega_0(z)$ cannot be explained by the inhomogeneous background density $\rho_0 (z)$ alone.  
We suspect a spatially depending tunnel coupling as already discussed in \cref{sec:phase_locking,fig:cohspatialdependence}.

\begin{figure}
	\centering
	\includegraphics{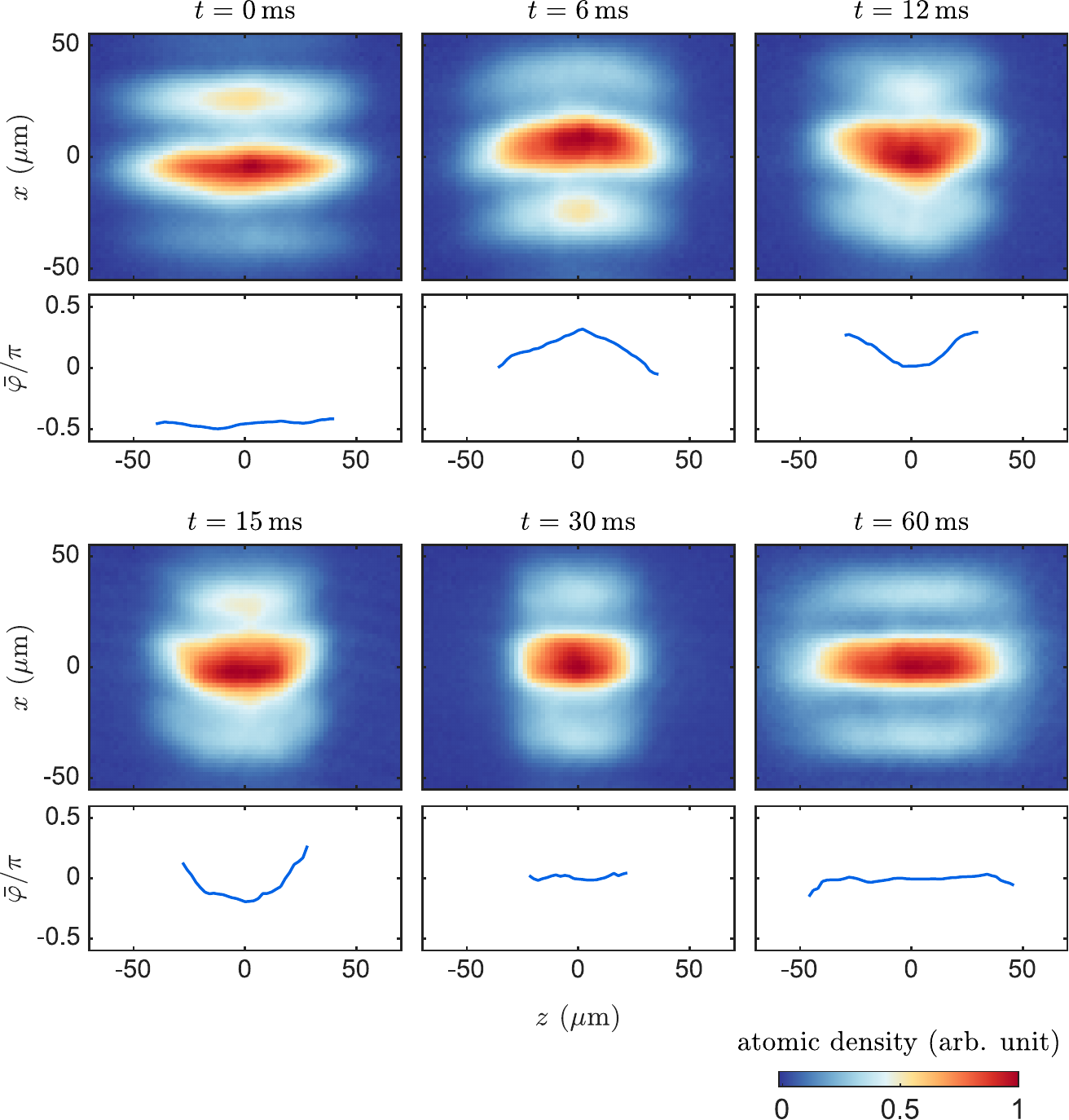}
	\caption{\textbf{Mean interference pictures,} for the same experimental measurement as presented in \cref{fig:circ_mean_local_omega}.
		The pictures show the 2D atomic density averaged over all experimental repetitions for the evolution time stated above the plots.
		One sees the bending of the interference fringes caused by the spatially dependent oscillation frequency.
		The spatially dependent circular mean $\bar \varphi$ is shown below the corresponding mean interference pictures. 
		Note the apparent longitudinal breathing which will be discussed later on in the main text.
		}
	\label{fig:scan4373meaninterferencepics}
\end{figure}	 

The spatial dependence of the oscillation frequency also manifests itself in a bending of the interference fringes.
To be more precise, we look at the atomic density integrated along the $y$ direction as recorded by the vertical imaging system.
Averaging the atomic density over the different experimental shots for the particular evolution times $t$ gives the results presented in \cref{fig:scan4373meaninterferencepics}.

\begin{figure}
	\centering
	\includegraphics{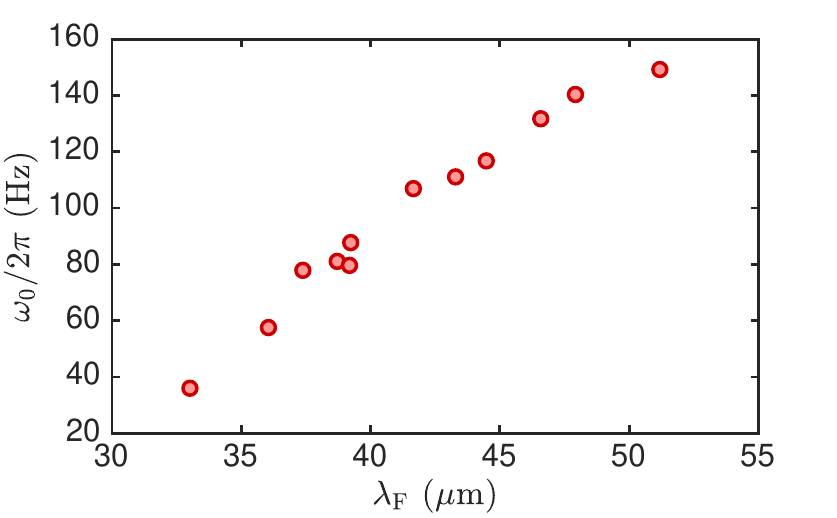}
	\caption{\textbf{Oscillation frequency as a function of the fringe spacing.}
		As for \cref{fig:scan4373localomega}, we obtain $\omega_0$ from fitting the circular mean phase with the solution of the damped pendulum.
		However, here we do not fit the local $\bar \varphi$ but the circular mean phase also averaged over the central $42${\um} of the cloud.
		We therefore get one value of $\omega_0$ per experimental parameters set, which is plotted here as a function of the fringe spacing.
	}
	\label{fig:omega0vsfrspacing}
\end{figure}

Note that \cref{fig:circ_mean_local_omega,fig:scan4373meaninterferencepics} only show the results for one particular experimental measurement.
However, similar results are obtained for a number of different experimental parameters (not shown).
In \cref{fig:omega0vsfrspacing}, we see the fitted $\omega_0$ for all the different experimental parameter sets plotted as a function of the fringe spacing.
As expected, the oscillation frequency grows with increasing fringe spacing (decreasing double well separation).

As already discussed in \cref{sec:jos_osc_model}, the spatial variation of the oscillation frequency can lead to a damping of the oscillations amplitude.
The mechanism is simply a dephasing between different points $z$ of the cloud.
Another source of dephasing would be the initial density and phase fluctuations, which change from shot to shot.
Global fluctuations, therefore, will lead to a dephasing between the different experimental realizations.
In addition, the spatially varying fluctuations again lead to different oscillations for the different values of $z$ and therefore to a dephasing between the different points along the longitudinal direction as well as the different experimental realization.

If the global fluctuations, leading to a dephasing between the different experimental realizations, would be the main reason for the damping, it should be evident from growing uncertainty for the circular mean phase.
However, no such growth of the errorbars is apparent in \cref{fig:scan4373circmeanph}.
On the other hand, the spatial dephasing would not necessarily lead to growing errorbars for $\bar \varphi$.
Instead, one should see a transfer of energy from the zero-mode (spatially constant) to higher modes (fluctuating with $z$).
This in turn might lead to a decrease in the integrated contrast of the interference fringes.
The decrease of contrast should depend on the amount of energy we put into the system, i.e., on the imprinted start phase.


\begin{figure}
	\centering
	\includegraphics{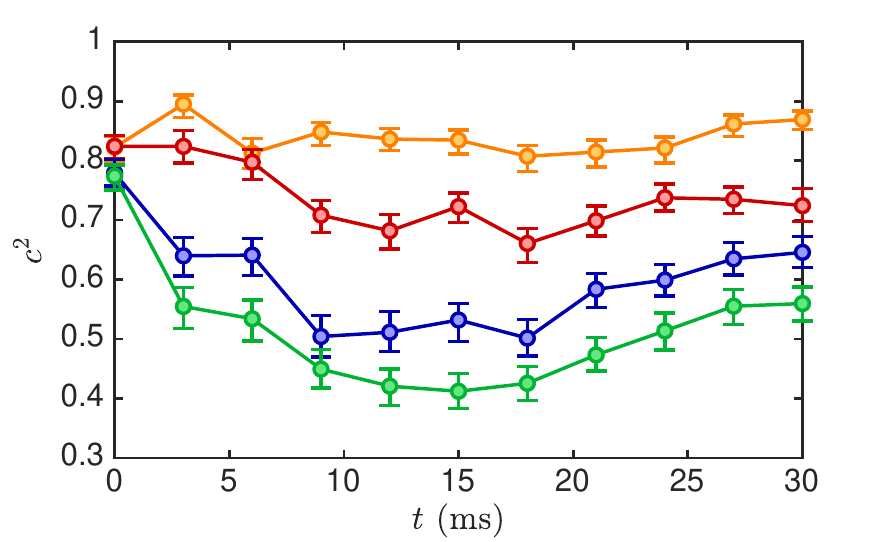}
	\caption{\textbf{Evolution of the integrated contrast for different start phases.}
	The squared integrated fringe contrast $c^2(L)$ calculated according to \cref{eq:pcf_contrast} is shown as a function of time.
	The integration length is $L = 42${\um}.
	The bullets represent the experimental results for different start phases.
	The measured circular mean phase (averaged over experimental shots and length $L$) at $t = 0$ is  $0.09 \, \pi$ (orange), $0.33 \, \pi$ (red), $0.48 \, \pi$ (blue) and $0.83 \, \pi$ (green).
	The errorbars represent 80\% confidence intervals obtained by using bootstrapping.
	The solid lines act as a guide to the eye.
} 
	\label{fig:contrastevolution}
\end{figure}

To test this hypothesis, we took several measurements in the same double well trap, varying the start phase.
The results for the evolution of the integrated fringe contrast $c (L)$ (integration length $L$) are shown in \cref{fig:contrastevolution}.
Note that we calculate $c (L)$ via~\cite{Langen18}
\begin{equation}
\left\langle c^2 (L) \right\rangle = \sum_{z,z ^ { \prime }} C \left( z , z ^ { \prime } \right) \label{eq:pcf_contrast}
\end{equation}
where $C \left( z , z ^ { \prime } \right)$ is the periodic phase correlation function defined in \cref{eq:PCF}.
The summation runs over all values $z$ and $z'$ in the interval $[-L/2,L/2]$.
This way of obtaining the integrated contrast seems to be experimentally more reliable than a direct measurement.

The results presented in \cref{fig:contrastevolution} show more or less what we expected.
Starting from a high initial contrast, we see a decay for approximately the first $15${\ms}. 
The decay is stronger for bigger initial phases.
The timespan of the contrast decay roughly corresponds to the timespan for the damping of the oscillations (not shown).
This is consistent with the picture of having an energy transfer from the zero-mode to the low lying non-zero modes.
After the zero mode is completely damped, there is no energy left to be transfered and the contrast decay stops.

Following the initial decay, the contrast grows again.
A possible explanation for this might be the transfer of the energy to even higher modes.
The higher modes store more energy for a given amplitude than the lower modes.
For the same amount of energy, they therefore don't decrease the contrast that strongly.
In addition, the more rapidly fluctuating higher modes are attenuated more strongly by the finite imaging resolution (see \cref{sub_sec:eff_psf_fringes}), which increases the measured contrast further.
Another possible explanation for the growth in contrast might be the transfer of energy from the relative to the common degrees of freedom.
However, as already mentioned above, we don't have a robust theoretical prediction yet.
A quantitative comparison between experiment and theory will be necessary to check for the occurrence and importance of the proposed mechanisms.

\begin{figure}
	\centering
	\includegraphics{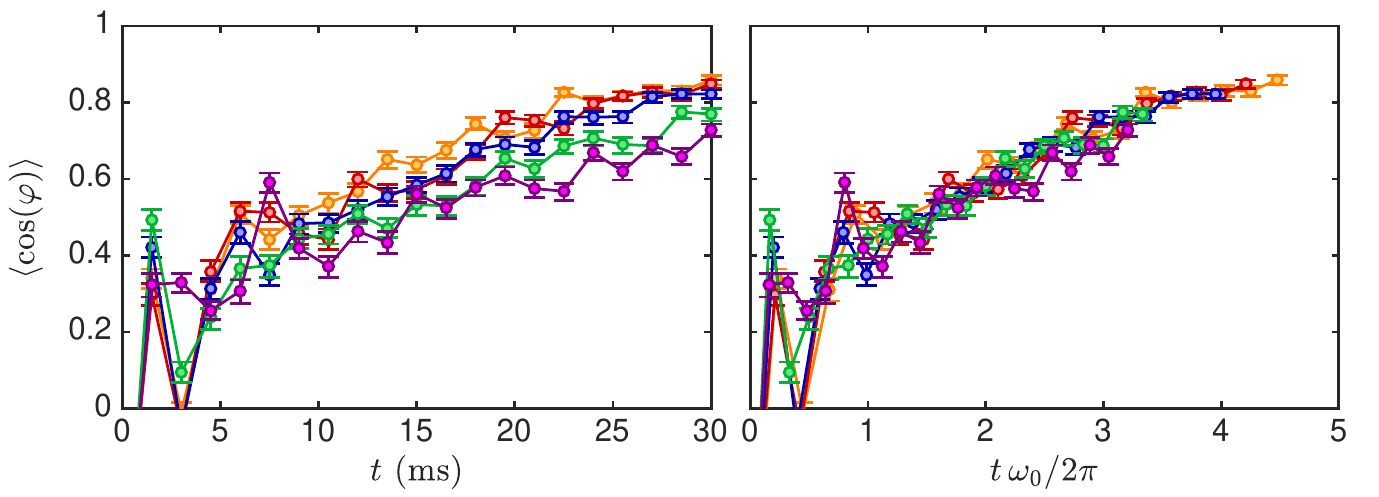}
	\caption{\textbf{Universal evolution of the coherence factor.}
	The bullets represent the experimental data for measurements with the same starting phase but different double well separations leading to different tunnel coupling strengths.
	The solid lines are a guide to the eye.
	On the vertical axis the coherence factor $\cohfact$ is shown. 
	Note that the expectation value is calculated by averaging over the experimental realizations as well as the central $42${\um} of the cloud.
	The errorbars represent 80\% confidence intervals obtained by using bootstrapping.
	On the horizontal axis we show the time in ms for the left subplot or rescaled with $\omega_0 / 2 \pi$ (the inverse oscillation period for small amplitudes) for the right subplot.
	The values for $\omega_0$ have been obtained the same way as in \cref{fig:omega0vsfrspacing}.
	The fitted  $\omega_0 / 2 \pi$ are $149${\Hz} (orange), $140${\Hz} (red), $132${\Hz} (blue), $111${\Hz} (green) and $107${\Hz} (purple).
	One sees that the rescaling more or less leads to a collapse of the curves after the damping of the initial oscillations.
	}
	\label{fig:cohfactevolution}
\end{figure}

The increase of the contrast after the damping of the oscillations can also be observed as an increase of the coherence factor $\cohfact$.
Looking at its evolution for measurements with the same starting phase, but different tunnel coupling, we found the curious universal behavior shown in \cref{fig:cohfactevolution}.
All the curves seem to collapse onto one when rescaling the time axis with the fitted $\omega_0$.
Let us state that it's not clear yet whether this collapse is exact and robust.
Further measurements with a wider range of $\omega_0$ as well as a theoretical prediction would be desirable.

Note that we also find the collapse to a single curve for the time evolution of the integrated contrast $c(L)$ when rescaling the time axis (not shown).
We just chose to show the coherence factor because we first observed the effect for it.
To check whether the universal scaling appears for further observables will be a task for the future.

The results presented in \cref{fig:cohfactevolution} have been obtained within a single experimental measurement.
There is another set of data which were obtained in different experimental measurement but have roughly the same starting phase.
For this data set we also observe the universal scaling behavior (not shown).
However, the rescaling does not work as well as for the results shown in \cref{fig:cohfactevolution}.
  
Let us now make some remarks about the longitudinal breathing motion.
We see such breathing for all experimental measurements.
Note that the common breathing of the two clouds is due to the splitting process and independent of the tunneling dynamics.
We checked this by investigating the breathing dynamics for the data presented in \cref{fig:cohfactevolution}.
No dependence on the tunnel strength was found.
However, we are still unsure whether the tunneling can introduce a relative breathing between the clouds, the cone-like shape for some of the mean interference pictures shown in \cref{fig:scan4373meaninterferencepics} might be an indication for that.

In conclusion, we triggered Josephson oscillations in a double well potential and saw fast damping similar to \rcite{Pigneur18}.
We have shown that this damping coincides with a decrease in the interference contrast, which is stronger for bigger amplitudes of the oscillations.
Moreover, we have observed a negative correlation between oscillation frequency and double well separation as well as confirmed the non-homogeneity of the tunneling strength already observed in \cref{sec:phase_locking}.
Future experiments might be performed in a box trap, starting from a double well trap with strong tunnel coupling.
This should eliminate the breathing motion observed for all measurement done so far.

\section{Tunnel coupling two independent condensates}
\label{sec:inverse_quench}

\subsection{Preparation of the system}
\label{sec:inverse_prep}

The initial preparation of the system is very similar to what is discussed in \cref{sec:preparation}, the only difference lies in the used timings and double well separations.
Here, we will always cool into a double well trap with large barrier, which completely separates the clouds in the two wells.
The frequency of the cooling fields is ramped in 470{\ms}, the final frequency of the ramp is then held for 60{\ms} more before switching off the cooling fields.
Subsequently, we wait for various time (ranging from 0 to 60 ms) before ramping down the amplitude of the dressing fields in order to introduce tunneling between the wells.
During this ramp, the amplitudes ratio between the two wires creating the dressing fields is ramped as well.
The used ratios are chosen so that one gets the same atom number in the two wells when evaporatively cooling into the initial/final double well potential.
After switching on the tunnel coupling, we wait for different hold times in order to investigate the time evolution.

\subsection{Experimental results}
\label{sec:inverse_exp}

\begin{figure}
	\centering
	\includegraphics{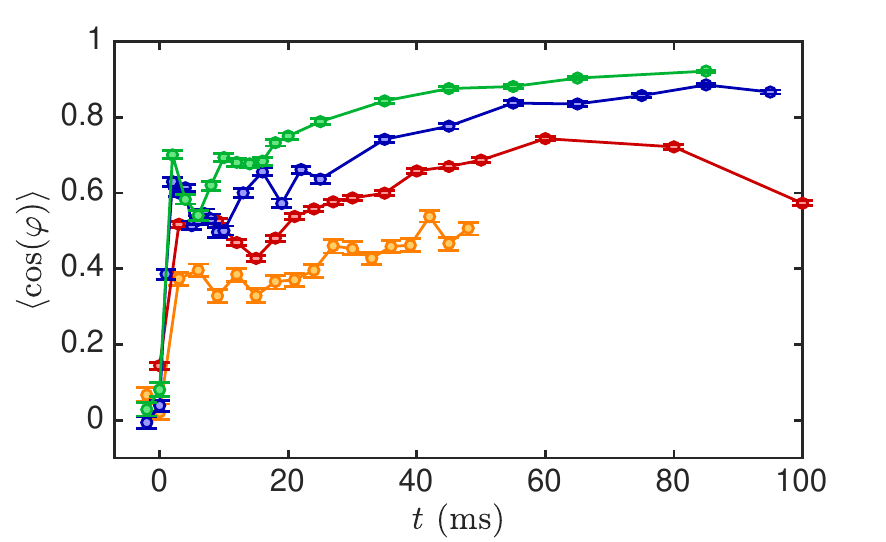}
	\caption{\textbf{Evolution of the coherence factor after recoupling.}
		Tunneling between two independent condensates is switched on by ramping down the double well barrier in 2{\ms}.
		The time $t = 0$ corresponds to the end of the ramp.
		The bullets represent the experimental results, with the different colors marking different double well separations.
		The errorbars represent 80\% confidence intervals obtained by using bootstrapping.
		The solid lines are a guide to the eye.
		The fitted fringe spacings $\lambda_\mathrm{F}$ are 35{\um} (orange), 36{\um} (red), 39{\um} (blue) and 46{\um} (green).
	}
	\label{fig:inverseqcohfact}
\end{figure}

We will start by discussing the evolution of the coherence factor $\cohfact$.
The experimental results for the evolution in double well traps with different barrier heights are presented in \cref{fig:inverseqcohfact}.
We see that starting from zero coherence factor (completely random phase), a phase locking between the condensates appears during the evolution.
The phase locking grows faster for higher values of the fitted fringe spacing (corresponding to smaller double well separations).
This is what one would naively expect and is in accordance with the results presented in \cref{fig:omega0vsfrspacing,fig:fringespacingvscoh}.

Note that the evolution in \cref{fig:inverseqcohfact} looks quite similar to the results presented in \cref{fig:cohfactevolution}. 
Moreover, note that in all cases, a fast initial increase of the coherence factor is followed by a slower growth.
These observations are somewhat consistent with the simple picture of the different experimental shots being interpreted as Josephson oscillations with random start phase.
The fast initial increase in $\cohfact$ corresponds to the damping of the oscillations.
The much slower subsequent growth is equivalent to the increase in contrast seen for later times in \cref{fig:contrastevolution}.

Using the relation between fringe spacing and oscillation frequency shown in \cref{fig:omega0vsfrspacing}, we can obtain a guess for $\omega_0$ for the data presented in \cref{fig:inverseqcohfact}.
Analogous to \cref{fig:cohfactevolution}, we can subsequently rescale the time axis and see if the curves collapse onto one.
The results of this are ambiguous, the curves definitely get closer to each other when rescaling, but don't collapse as nicely as in \cref{fig:cohfactevolution}.  
As we didn't control for the different parameters (e.g.\ temperature) and don't have a theoretical understanding of the processes, we decided not to show the rescaling here, but leave it for future works. 

Note that we are not sure why for the red curve presented in \cref{fig:inverseqcohfact}, the coherence factor decreases again for long times and whether that would also have happened for the other curves, for longer times than investigated.
The reason for the decrease might just simply be technical noise.

\begin{figure}
	\centering
	\includegraphics{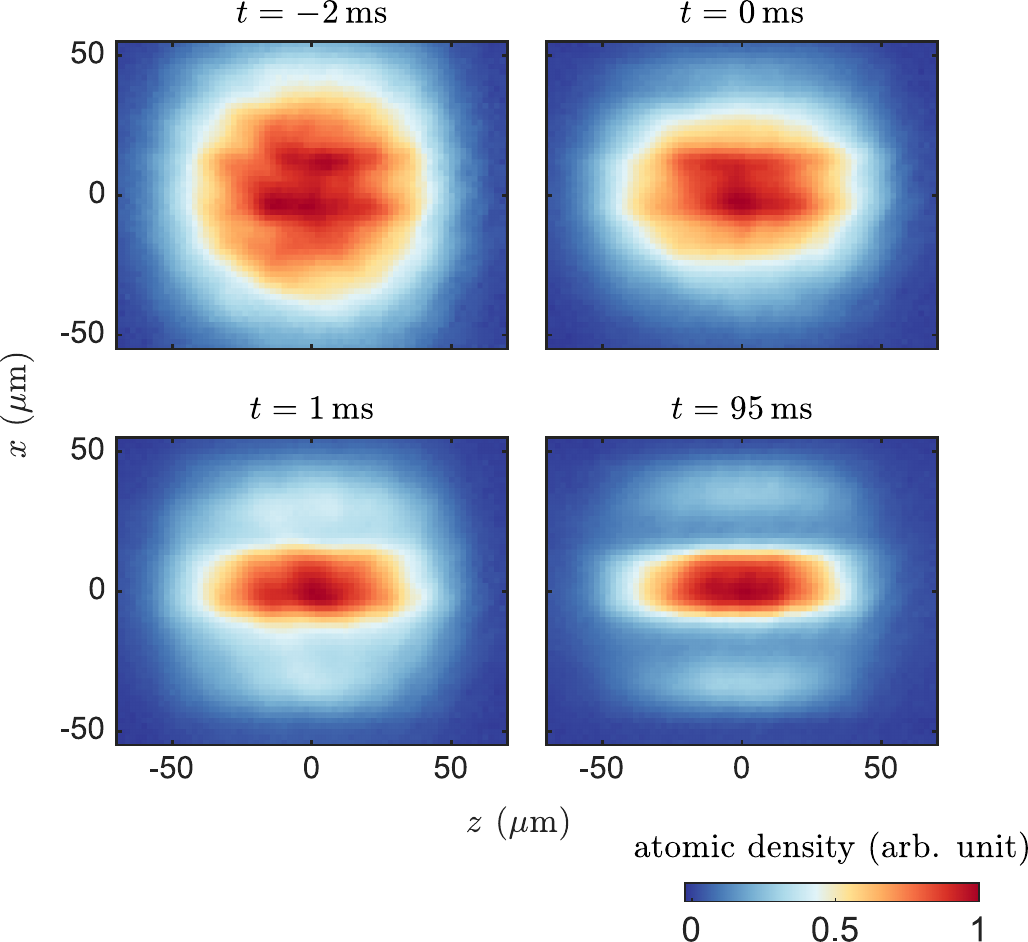}
	\caption{\textbf{Mean interference pictures.}
	Similar to the mean interference pictures in \cref{fig:scan4373meaninterferencepics}, but for the results corresponding to the blue curve in \cref{fig:inverseqcohfact}.
	One sees that at time $t = -2${\ms}, before the double well barrier is ramped down, the relative phase is completely random, leading to the blob seen in the mean interference picture.
	At $t = 0$, the ramp is finished.
	Already at $t = 1${\ms}, one can see the appearance of a central fringe, a clear sign for phase locking.
	}
	\label{fig:scan4328meaninterferencepics}
\end{figure}

Note that some mean interference pictures are presented in \cref{fig:scan4328meaninterferencepics}. They illustrate the rephasing process for one of the curves shown in \cref{fig:inverseqcohfact}.


\section{Outlook}
\label{sec:non_equi_outlook}

So far, for all the results presented in this chapter, we only looked at observables periodic in the relative phase.
It would be interesting to investigate the type of phase correlations discussed in \cref{chap:corr}.
Especially, it would be interesting to look at a possible build-up of non-Gaussianity.
However, we think that the phase profiles for the data presented in this chapter fluctuate too strongly for continues phase profiles to be extracted correctly with the procedure discussed in \cref{sec:phase_extraction}.
The reason for the strong fluctuations is that we add a lot of energy to the system by switching on the tunnel coupling for an initial state with global (\cref{sec:jos_osc}) or random (\cref{sec:inverse_quench}) phase difference. 

Other experimental protocols might lead to less strong fluctuations and, therefore, might be better suited for the investigation of higher-order correlation functions as defined in \cref{sec:calc_corr}.
One could for example start from the double well with strong phase locking (small Gaussian fluctuation) and subsequently decrease, but not completely switch off, the tunneling strength.
Thermalization (if it happens) should then lead to a build up of non-Gaussian fluctuations.
As no energy is added to the system by the change in tunnel coupling, but rather taken out of the system, we expect the phase profiles to fluctuate much less strongly.

\appendix

\chapter{Optical Bloch equations}
\label{chap:obe}

The optical Bloch equations~\cite{steck2015} describe the interaction of a quantum\hyp{}mechanical atom with a classical electromagnetic field. 
The ground and an excited state of the atom can each have several magnetic sub-states. The transition frequency between any combination of ground and excited magnetic sub-state is assumed to be near resonant with the electromagnetic field.\footnote{This 
	assumption is necessary to justify the rotating wave approximation used to obtain the optical Bloch equation.}
Spontaneous emission is considered through a decay term for the excited states.

The goal of the calculations using the optical Bloch equations is to obtain an effective cross-section for the absorption of light.
This is relevant for probing the atom cloud via absorption imaging (see \cref{chap:imaging}).
The numerically obtained results are presented in \secref{sec:obe_cross_sec} and also \cref{fig:TAndor_lineshape_width}.
Readers only interested in this results might skip the next \secref{sec:optical_bloch}, which is only about stating and explaining the equations.

 

\section{The equations}
\label{sec:optical_bloch}

In this section, we will state the full expressions of the optical bloch equations used to obtain the results presented in \secref{sec:obe_cross_sec}. 
Without loss of generality we will assume the quantization field to always be along the $z$ direction. Other field configurations can easily be investigated by performing a rotation of the coordinate system. 
We will discuss this in detail at the end of this section.

We image the cloud of \tss{87}Rb by using the D\tsub{2} (5\tss{2}S\tsub{1/2} $\rightarrow$ 5\tss{2}P\tsub{3/2}) transition.
The 5\tss{2}S\tsub{1/2} ground state has the quantum numbers $L_\mathrm{g} = 0$, $S = 1/2$, $J_\mathrm{g} = 1/2$ and $I = 3/2$.\footnote{
The standard nomenclature is used, denoting the orbital angular momentum quantum number by L, the spin quantum number by $S$, the total electronic angular momentum quantum number by $J$, the nuclear angular momentum quantum number by $I$ and the total atomic angular momentum quantum number by $F$.} 
The total angular momentum quantum number can therefore be either 1 or 2, in the case of our experiment we have $F_\mathrm{g} = 2$. 
The 5\tss{2}P\tsub{3/2} first excited state has $L_\mathrm{e} = 1$, $S = 1/2$, $J_\mathrm{e} = 3/2$ and $I = 3/2$. 
The total angular momentum quantum number can therefore be either 0, 1, 2 or 3, in the case of our experiment we have $F_\mathrm{e} = 3$. 
The energy difference between ground and excited state will be denoted by $\hbar \omega_0$.
Numerical values for the on-resonance transition frequency $\omega_0/2\pi$ can be found in \rcite{steck2015}.
Note that the first excited state could also have $J = 1/2$ which would lead to the D\tsub{1} line.

The equations will be stated for the density matrix in the basis of the ground and excited $m_F$-states:
\begin{align}
\begin{split}
\rho_{\mathrm{g} \, m_\mathrm{g},\, \mathrm{g} \, m_\mathrm{g}'} &= \langle \mathrm{g} \, m_\mathrm{g} | \, \hat{\rho} \, | \mathrm{g} \, m_\mathrm{g}'\rangle \\
\rho_{\mathrm{e} \, m_\mathrm{e},\, \mathrm{g} \, m_\mathrm{g}} &= \langle \mathrm{e} \, m_\mathrm{e} | \, \hat{\rho} \, | \mathrm{g} \, m_\mathrm{g}\rangle \\
\rho_{\mathrm{e} \, m_\mathrm{e},\, \mathrm{e} \, m_\mathrm{e}'} &= \langle \mathrm{e} \, m_\mathrm{e} | \, \hat{\rho} \, | \mathrm{e} \, m_\mathrm{e}'\rangle.
\end{split}
\end{align} 
Here $| \mathrm{g} \, m_\mathrm{g} \rangle$ represent the ground $m_F$-states with $m_\mathrm{g} = -2,\,\dots,\,2$ and $| \mathrm{e} \, m_\mathrm{e} \rangle$ the excited ones with $m_\mathrm{e} = -3,\,\dots,\,3$. 
In order to get simpler equations one defines
\begin{align}
\begin{split}
\tilde{\rho}_{\mathrm{e} \, m_\mathrm{e},\, \mathrm{g} \, m_\mathrm{g}} &= \rho_{\mathrm{e} \, m_\mathrm{e},\, \mathrm{g} \, m_\mathrm{g}} \, e^{i \omega_\mathrm{L} t} \\
\tilde{\rho}_{\mathrm{g} \, m_\mathrm{g},\, \mathrm{e} \, m_\mathrm{e}} &= \rho_{\mathrm{g} \, m_\mathrm{g},\, \mathrm{e} \, m_\mathrm{e}} \, e^{-i \omega_\mathrm{L} t}, 
\end{split}
\end{align}
where $\omega_\mathrm{L} = 2 \pi \nu_\mathrm{L}$, with $\nu_\mathrm{L}$ being the frequency of the laser light. 

Furthermore, we will need the Larmor frequencies for the ground and excited states. 
They are given by
\begin{equation}
\omega_\mathrm{l} = \frac{\mu_\mathrm{B} g_{F} B}{\hbar}, \label{eq:larmor}
\end{equation}
where $\mu_\mathrm{B}$ is the Bohr magneton, $g_{F}$ is the Land\'e factor for the total angular momentum $F$ of the ground or excited state and $B$ is the magnetic quantization field. 
The Land\'e factor $g_{F}$ can approximately be calculated using~\cite{steck2015}
\begin{equation}
g_{F} = g_{J} \, \frac{F (F+1)-I (I+1)+J (J+1)}{2 F (F+1)}.
\end{equation}
The values for $g_{J \mathrm{g}}$ and $g_{J \mathrm{e}}$ can be found in \rcite{steck2015}.
With this we get $g_{F \mathrm{g}} = 0.5006$ and $g_{F \mathrm{e}} = 0.6681$.
In the following, we will denote the Larmor frequencies of the ground and excited state by $\omega_\mathrm{lg}$ and $\omega_\mathrm{le}$ respectively. 

Moreover, we need to calculate the Rabi frequencies. 
In order to do so, we will first discuss the form of the electric field.
At fixed position, the electric field of a monochromatic light wave can be written as
\begin{align}
\begin{split}
\vec E (t) &= 
\begin{pmatrix}
E_x \, \cos(\omega_\mathrm{L} t + \phi_x)\\
E_y \, \cos(\omega_\mathrm{L} t + \phi_y)\\
E_z \, \cos(\omega_\mathrm{L} t + \phi_z)
\end{pmatrix}
= \frac{1}{2} \left( e^{-i \omega_\mathrm{L} t} \begin{pmatrix}
E_x \, e^{-i \phi_x}\\
E_y \, e^{-i \phi_y}\\
E_z \, e^{-i \phi_z}
\end{pmatrix} + \mathrm{c.c.}\right) \\
&= \frac 1 2 \left(\vec E ^{(+)} e^{-i \omega_\mathrm{L} t} + \mathrm{c.c.}  \right) ,
\end{split}
\end{align}
where $\vec E ^{(+)}$ is independent of time. 
We can write it as
\begin{equation}
\vec E ^{(+)} = \varepsilon \, \vec \epsilon_0 \quad \mathrm{with} \quad \left|\vec \epsilon_0 \right|^2 = \vec\epsilon_0 \cdot \vec{\epsilon_0}^{*} = 1.
\end{equation}
Here $\vec \epsilon_0$ represents the unit vector fixing the polarization and $\varepsilon$ is the amplitude of the electric field. 
Note that a total phase of $\vec E ^{(+)}$ corresponds to a shift in time which is unimportant for our calculations.

Let us give some useful examples for $\vec\epsilon_0$. For right-circularly polarized light propagating in the $y$ direction we have
\begin{equation}
\vec \epsilon_0 = \frac{1}{\sqrt{2}} 
\begin{pmatrix}
1\\
0\\
-i
\end{pmatrix},
\end{equation}
and for right-circularly polarized light propagating in the $z$ direction we get
\begin{equation}
\vec \epsilon_0 = \frac{1}{\sqrt{2}} 
\begin{pmatrix}
1\\
i\\
0
\end{pmatrix}.
\end{equation}

In the experiment we can measure the intensity $I$\footnote{
Note that we are using the same symbol $I$ for the nuclear angular momentum quantum number as well as the intensity. 
However, it should be known from context which quantity is meant.
} 
of the laser light rather than the field amplitude $\varepsilon$.
The two quantities can be connected via the relation 
\begin{equation}
I = \frac{1}{\mu_0 c} \overline{\,{| \vec{E} (t)|}^2}. \label{app_int}
\end{equation}
Here $\mu_0$ is the vacuum permeability and $c$ the speed of light. The overline in $\overline{\,{| \vec{E} (t)|}^2}$ stands for averaging over one period. We can write
\begin{equation}
\overline{\,{| \vec{E} (t)|}^2} = \frac{1}{2} \left|\varepsilon \right|^2 . \label{app_av}
\end{equation}
Remembering that a global phase is unimportant for us, we get from \cref{app_int,app_av}
\begin{equation}
\varepsilon = \sqrt{2 \mu_0 c I}.
\end{equation}

Using the introduced quantities, we can write the Rabi-frequencies as
\begin{equation}
\Omega(m_\mathrm{g},m_\mathrm{e}) = \langle F_\mathrm{g} \, m_\mathrm{g} | \, e \hat{\vec{r}} \, | F_\mathrm{e} \, m_\mathrm{e}\rangle \cdot \vec{\epsilon_0}^* \frac{\varepsilon^*}{\hbar},
\end{equation}
where $e$ is the elementary charge and $\hat{\vec{r}}$ is the position operator.
The dipole matrix elements 
\begin{equation}
\vec d = - \langle F_\mathrm{g} \, m_\mathrm{g} | \, e \hat{\vec{r}} \, | F_\mathrm{e} \, m_\mathrm{e}\rangle
\end{equation}
are most conveniently evaluated in the spherical basis
\begin{equation}
\left\{
\vec e_- = \frac{1}{\sqrt{2}}\begin{pmatrix}
1\\
-i\\
0
\end{pmatrix},
\vec e_0 = \begin{pmatrix}
0\\
0\\
1
\end{pmatrix},
\vec e_+ = \frac{1}{\sqrt{2}} \begin{pmatrix}
-1\\
-i\\
0
\end{pmatrix}
\right\}.
\end{equation}
Let us denote the coordinates for this basis by the subscript $q = -1,\,0\,,1$. 
Following\cite{steck2015} we can then evaluate
\begin{equation}
-d_q = \langle F_\mathrm{g} \, m_\mathrm{g} | \, e \hat r_q \, | F_\mathrm{e} \, m_\mathrm{e}\rangle = \langle F_\mathrm{g} \,||\, e \vec r \, || \, F_\mathrm{e} \rangle \, \langle F_\mathrm{g} \, m_\mathrm{g} \, |\, F_\mathrm{e} \, 1 \, m_\mathrm{e} \, q\rangle,
\end{equation}
where $ \langle F_\mathrm{g} \,||\, e \vec r \, || \, F_\mathrm{e} \rangle$ is the reduced matrix element and $\langle F_\mathrm{g} \, m_\mathrm{g} \, |\, F_\mathrm{e} \, 1 \, m_\mathrm{e} \, q\rangle$ are the Clebsch-Gordan coefficients. 
We can further simplify
\begin{equation}
\langle F_\mathrm{g} \,||\, e \vec r \, || \, F_\mathrm{e} \rangle = \langle J_\mathrm{g} || \, e \vec r \, || J_\mathrm{e} \rangle (-1)^{F_\mathrm{e}+J_\mathrm{g}+1+I} \sqrt{(2 F_\mathrm{e} + 1)(2 J_\mathrm{g} + 1)}
\begin{Bmatrix}
J_\mathrm{g} & J_\mathrm{e} & 1 \\
F_\mathrm{e} & F_\mathrm{g} & I
\end{Bmatrix}
\end{equation}
where
\begin{equation}
\begin{Bmatrix}
J_\mathrm{g} & J_\mathrm{e} & 1 \\
F_\mathrm{e} & F_\mathrm{g} & I
\end{Bmatrix}
\end{equation}
is the Wigner 6j-symbol. 
The reduced dipole matrix elements $\langle J_\mathrm{g} || e \vec r || J_\mathrm{e} \rangle$ can be found in \rcite{steck2015}. 

Note that different definitions (sign conventions etc.) for the Clebsch-Gordan coefficients and the 6j-symbols exist. 
To get the right dipole matrix elements with the above formulas, we have to use the definitions given in \rcite{brink1968angular}. 
The relevant equations in the cited edition are (2.34) on page 34, (3.15) on page 43 and (3.22) on page 44.

Transforming $\vec{\epsilon_0}$ also into the spherical basis (coordinates denoted by $\epsilon_{0 q}$) we can finally calculate the Rabi-frequencies as 
\begin{equation}
\Omega(m_\mathrm{g},m_\mathrm{e}) = \sum_{q} - d_q \, \epsilon_{0 q}^* \, \frac{\varepsilon^*}{\hbar}.
\end{equation}

With this we have everything to state the optical Bloch equations:
\begin{align}
\begin{split}
\frac{d}{dt}\rho_{\mathrm{g} \, m_\mathrm{g},\, \mathrm{g} \, m_\mathrm{g}'} = &- i \, \omega_\mathrm{lg} \left(m_\mathrm{g} - m_\mathrm{g}'\right) \rho_{\mathrm{g} \, m_\mathrm{g},\, \mathrm{g} \, m_\mathrm{g}'} \\ 
&- \frac{i}{2} \sum_{m_\mathrm{e}}{\left[\Omega(m_\mathrm{g},m_\mathrm{e}) \, \tilde\rho_{\mathrm{e} \, m_\mathrm{e},\, \mathrm{g} \, m_\mathrm{g}'} - \tilde\rho_{\mathrm{g} \, m_\mathrm{g},\, \mathrm{e} \, m_\mathrm{e}} \, \Omega^*(m_\mathrm{g}',m_\mathrm{e})\right]}  \\ &+ \Gamma \sum_{q = -1,0,1}{\rho_{\mathrm{e} \, (m_\mathrm{g} + q),\, \mathrm{e} \, (m_\mathrm{g}'+q)}} \\ & \qquad \qquad \qquad \langle F_\mathrm{e} \, (m_\mathrm{g} + q) \, |\, F_\mathrm{g} \, 1 \, m_\mathrm{g} \, q\rangle \, \langle F_\mathrm{e} \, (m_\mathrm{g}' + q) \, |\, F_\mathrm{g} \, 1 \, m_\mathrm{g}' \, q\rangle 
\end{split}
\end{align}
\begin{align}
\begin{split}
\frac{d}{dt}\tilde\rho_{\mathrm{e} \, m_\mathrm{e},\, \mathrm{g} \, m_\mathrm{g}} = 
& \, i \, \tilde\rho_{\mathrm{e} \, m_\mathrm{e},\, \mathrm{g} \, m_\mathrm{g}} \left[\left(\omega_\mathrm{L} - \omega_0\right)  - \left(\omega_\mathrm{le} m_\mathrm{e} - \omega_\mathrm{lg} m_\mathrm{g}\right) \right] \\ 
&- \frac{i}{2} \left(\sum_{m_\mathrm{g}'}{\Omega^*(m_\mathrm{g}',m_\mathrm{e}) \, \rho_{\mathrm{g} \, m_\mathrm{g}',\, \mathrm{g} \, m_\mathrm{g}}} - \sum_{m_\mathrm{e}'}{\Omega^*(m_\mathrm{g},m_\mathrm{e}') \, \rho_{\mathrm{e} \, m_\mathrm{e},\, \mathrm{e} \, m_\mathrm{e}'}}\right)  \\ 
&- \frac 1 2 \, \Gamma \,\tilde \rho_{\mathrm{e} \, m_\mathrm{e},\, \mathrm{g} \, m_\mathrm{g}}
\end{split}
\end{align}
\begin{align}
\begin{split}
\frac{d}{dt}\rho_{\mathrm{e} \, m_\mathrm{e},\, \mathrm{e} \, m_\mathrm{e}'} = &- i \, \omega_\mathrm{le} \left(m_\mathrm{e} - m_\mathrm{e}'\right) \rho_{\mathrm{e} \, m_\mathrm{e},\, \mathrm{e} \, m_\mathrm{e}'}\\  
&- \frac{i}{2} \sum_{m_\mathrm{g}}{\left[\Omega^*(m_\mathrm{g},m_\mathrm{e}) \, \tilde\rho_{\mathrm{g} \, m_\mathrm{g},\, \mathrm{e} \, m_\mathrm{e}'} - \tilde\rho_{\mathrm{e} \, m_\mathrm{e},\, \mathrm{g} \, m_\mathrm{g}} \, \Omega(m_\mathrm{g},m_\mathrm{e}')\right]}   \\ 
&- \, \Gamma \, \rho_{\mathrm{e} \, m_\mathrm{e},\, \mathrm{e} \, m_\mathrm{e}'}
\end{split}
\end{align} 

Let's now go back to discussing the case when we have a magnetic field 
\begin{equation}
\vec B = 
\begin{pmatrix}
B_x\\
B_y\\
B_z
\end{pmatrix},
\end{equation}
which is not aligned with the $z$ direction. Through a rotation we want to connect it with
\begin{equation}
\vec B' = |\vec B| 
\begin{pmatrix}
0\\
0\\
1
\end{pmatrix}.
\end{equation}
To be more precise, we want to find a rotation matrix $R$ so that 
\begin{equation}
\vec{B} = R \vec{B}'. \label{rot_cond}
\end{equation}
We can then simply use the equations as usual when replacing $\vec \epsilon_0$ with ${\vec{\epsilon_0}}^\prime = R^T \vec \epsilon_0$. Note that $R^T$ is $R^{-1}$ for any rotation matrix (orthogonal).

One can easily check that the matrix
\begin{equation}
R = 
\begingroup
\renewcommand*{\arraystretch}{1.7}
\begin{pmatrix}
\frac{B_y}{|\vec B_{\perp}|} \ & \frac{B_x}{|\vec B_{\perp}|} \, \frac{B_z}{|\vec B|} \ & \frac{B_x}{|\vec B|}  \\
-\frac{B_x}{|\vec B_{\perp}|} &  \frac{B_y}{|\vec B_{\perp}|} \, \frac{B_z}{|\vec B|} & \frac{B_y}{|\vec B|}\\
0 & -\frac{|\vec B_{\perp}|}{|\vec B|} & \frac{B_z}{|\vec B|}
\end{pmatrix}
\endgroup
\label{R_mat}
\end{equation} 
possesses the desired property. Here we have used $|\vec B_{\perp}| = \sqrt{B_x^2 + B_y^2}$. 
Note that this is not the only choice of $R$ fulfilling \cref{rot_cond}. 
Moreover, note that \cref{R_mat} is not defined for $|\vec B_{\perp}| = 0$, in this case we simply choose $R$ as the unit matrix.

\section{Calculating the effective cross-section}
\label{sec:obe_cross_sec}

The goal of this section is to find the effective absorption cross section for the different imaging situations. 
We will start by explaining the investigated quantities and how to calculate them and then present the result for the different imaging systems in section \labelcref{sec:LAndor_obe,sec:VAndor_obe,sec:TAndor_obe}.
Note that some results for the transverse imaging system are also presented in \cref{fig:TAndor_lineshape_width}.

The effective absorption cross section $\sigma_\mathrm{eff}$ is related to the number of scattered photons per atom $n_\mathrm{sc}$\footnote{
See the explanation below \cref{eq:alpha_formula} for an exact definition.	
} via
\begin{equation}
n_\mathrm{sc} = \sigma_\mathrm{eff} \frac{n_\mathrm{ph}}{A},
\end{equation}
where $n_\mathrm{ph}/A$ is the number of incident photons per unit area. The number of scattered photons can be inferred from the evolution of the atomic density matrix which we get as a result of solving the optical Bloch equations. Both $n_\mathrm{sc}$ and $n_\mathrm{ph}$ should be understood as the total numbers for the entire imaging process as this is what we have access to in the experiment.

In general, the cross-section will depend on the imaging duration and a number of parameters: 
It depends on the magnetic bias field, whose direction defines the quantization axis and whose magnitude determines the Zeeman splitting. 
It also depends on the polarization and the intensity of the incoming light. 
For an (effective) two-level system the dependence on the imaging intensity is simply
\begin{equation}
\sigma_\mathrm{eff} = \sigma \left(1 + \frac{I}{I_{\mathrm{sat}}} \right)^{-1} \label{eq:sigma_eff_2level}
\end{equation}
which is also used in \cref{eq:abs_imag_basic}.
Note that for the effective two level systems (see discussion in \cref{sec:im_cross_section}), \cref{eq:sigma_eff_2level} is only valid after the initial equilibration.
Using the definition \cref{eq:sigma_rescaled} and \cref{eq:isat_rescaled} following from it, we get
\begin{equation}
\sigma_\mathrm{eff} = \frac{\sigma_0}{\alpha} \left(1 + \frac{I}{\alpha \, I_{\mathrm{sat}}^0} \right)^{-1} . \label{eq:eff_sigma_w_alpha}
\end{equation}
Note that for a real two level transition one has $\alpha = 1$. 

In the following we will calculate the factor $\alpha$ from the results of the optical Bloch equations.
Naively applying\footnote{
As proposed in \rcite{Reinaudi2007}.
} 
\cref{eq:eff_sigma_w_alpha} to situations where one doesn't have an effective two level system, or where the initial equilibration is not negligible leads to an intensity dependent $\alpha$.
We will therefore investigate the dependence of $\alpha$ on all the parameters, the light polarization and intensity as well as the magnetic bias field and imaging duration. 
In particular, we will also investigate whether an intensity independent $\alpha$ is justified for the case of having perturbations to one of the effective two-level configurations. 
One example for such perturbations would be stray fields perturbing the quantization axis (see \secref{sec:influence-of-stray-fields}).

From the above equations one gets after some basic reformulations
\begin{equation}
\alpha = \frac{I}{I_{\mathrm{sat}}^0} \left(\frac{1}{2 \, n_\mathrm{sc}} \frac{t_\mathrm{exp}}{\tau} - 1 \right). \label{eq:alpha_formula}
\end{equation}
Here $t_\mathrm{exp}$ represents the imaging duration and $n_\mathrm{sc}$ is the total number of scattered photons per atom during this time. It consists of all photons absorbed and not re-emitted plus the ones absorbed and spontaneously emitted. 
These photons do not hit the CCD of the camera, in contrast to the photons absorbed and re-emitted by stimulated emission, which therefore do not contribute to $n_\mathrm{sc}$.

The number of absorbed and not re-emitted photons per atom can simply by calculated as
\begin{equation}
n_\mathrm{abs}(t) = p_\mathrm{exc} (t) - p_\mathrm{exc} (0),
\end{equation}  
where $p_\mathrm{exc}$ is the probability that the atom is in the excited state. We simply calculate it as
\begin{equation}
p_\mathrm{exc} = \sum_{m_\mathrm{e}} \left\langle \mathrm{e} \ m_\mathrm{e} \right| \hat{\rho}  \left| \mathrm{e} \ m_\mathrm{e} \right\rangle ,
\end{equation}
i.e., by tracing the density matrix over all excited $m_F$-states.
The number $n_\mathrm{sp}$ of absorbed and spontaneously re-emitted photons can be evaluated via the integral
\begin{equation}
n_\mathrm{sp}(t_\mathrm{exp}) = \int_{0}^{t_\mathrm{exp}} \frac{dt}{\tau} \ p_\mathrm{exc}(t) . \label{eq:n_sp}
\end{equation}

If we are only interested in the stationary state, i.e., if we neglect the initial equilibration, we can simplify \cref{eq:alpha_formula}. 
Assuming to be already stationary at $t = 0$, we get $n_\mathrm{abs}(t) = 0$ and can evaluate the integral in \cref{eq:n_sp} to $n_\mathrm{sp}(t_\mathrm{exp}) = {p_\mathrm{exc} \times t_\mathrm{exp}}/{\tau}$. \Cref{eq:alpha_formula} therefore simplifies to 
\begin{equation}
\alpha_\mathrm{stat} = \frac{I}{I_{\mathrm{sat}}^0} \left(\frac{1}{2 \, p_\mathrm{exc}} - 1 \right). \label{eq:alpha_formula_stat}
\end{equation}

\subsection{Longitudinal Imaging}
\label{sec:LAndor_obe}

In the longitudinal imaging system we use circularly polarized light.
The atomic quantization axis is aligned with the imaging direction. 
This is one of the two situations discussed in \cref{sec:im_cross_section}, which leads to an effective two level transition.
The bias field fixing the quantization axis has a magnitude of 1.5 Gauss. 
The imaging intensity is non-uniform, it's median value is about  $0.15 \, I^0_\mathrm{sat}$. 
The 5 and 95\% quantiles are typically $0.10 \, I^0_\mathrm{sat}$ and $0.20 \, I^0_\mathrm{sat}$, respectively. 
To get these values, the edges of the picture have not been taken into account. 
The values should be understood as an order of magnitude and not as exact values. 
They depend on the chosen region of interest and will fluctuate and drift in time. 
The imaging light is on for 75{\us}.

\begin{figure}
	\centering
	\includegraphics{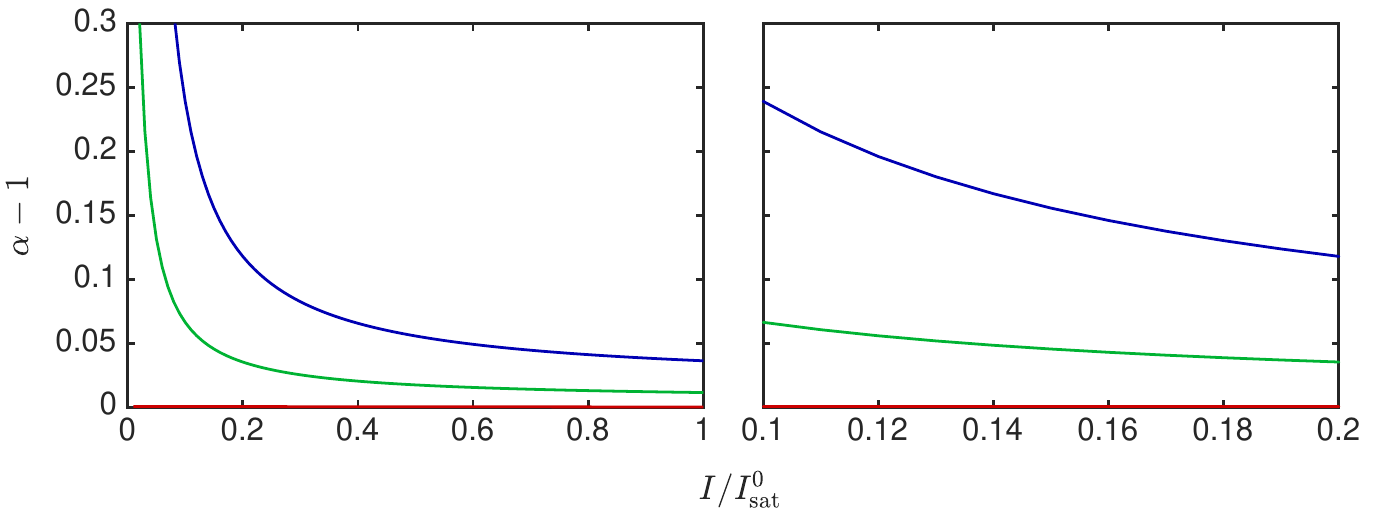}
	\caption{\textbf{ Effective cross section for the longitudinal imaging:} 
		Factor $\alpha$ calculated according to \cref{eq:alpha_formula} as a function of the imaging intensity $I$. 
		Initially all atoms are in the $m_F=2$ state (red), in an even mixture between all $m_F$-states (green) or in the $m_F=-2$ state (blue). 
		One sees clear differences depending on the initial $m_F$-state distribution. 
		For the median imaging intensity of typically $0.15 \, I^0_\mathrm{sat}$ the values are $\alpha = 1.000$, $1.045$ and $1.156$, respectively. 
		Starting from the $m_F=2$ state, 186 photons per atom are scattered for that intensity. 
		The detuning was chosen to be resonant with the transition between the $m_F = 2$ ground and the $m_{F'}=3$ excited state. 
		Note that due the initial equilibration phase this detuning will not exactly coincide with the maximum of the effective cross section (minimum of $\alpha$). 
		However, it was checked that the difference is small (not shown).
		Both the left and right subplot show the same data with different limits for the horizontal axis.
		The limits in the right subplot coincide with the typical range of imaging intensities occurring in the experiment (see discussion in the main text).
	}
	\label{fig:landoralphavsint}
\end{figure}

In \cref{fig:landoralphavsint} we see how the effective $\alpha$ depends on the initial (at the time the laser light is switched on) $m_F$-state distribution and the imaging intensity. 
For the longitudinal imaging system, the quantization field direction coincides with the Ioffe field direction during the magnetic trapping phase (both aligned with the longitudinal $z$ direction).
After switching off the trap, the field is simply ramped up from the Ioffe field value to the value desired for imaging. As the field is never off, we can expect the $m_F$-state distribution to be given by the distribution right after the switch-off of the magnetic trap. 

If we release the atoms from the rf-dressed trap, we can tune the phase of the rf-fields so that the majority (typically about 90\%) of the atoms will end up in the $m_F = 2$ state~\cite{langen2013thesis}. 
If the bias-field is positive and we use left circularly polarized light,\footnote{
	We think that this is used in the experiment. However, we never explicitly checked it.
} 
there will be no rearranging between the different $m_F$ ground states. 
Only the populations between the $m_F=2$ ground state and the $m_{F'} = 3$  excited state have to equilibrate, which happens very quickly. The expected value $\alpha = 1$ is reached even for very small exposure times or small imaging intensities.

To release the atoms out of the static trap (no rf-dressing) we usually switch-off all fields except the Ioffe-field abruptly (the same is also done in the rf-dressed trap). By doing so, we get a mixture of different $m_F$-states. The exact distribution between the $m_F$-states was to my knowledge never investigated in detail. By ramping the fields in a particular way, one can also manage to get most atoms in a certain mf-state after the static-field trap is switched off.

If we start from a state where all $m_F$ levels are occupied evenly, one sees substantial deviations from unity for the value of $\alpha$ (see \cref{fig:landoralphavsint}). 
This is due to the redistribution between the different $m_F$ groundstates. 
The deviations are even larger when initially all atoms are in the $m_F=-2$ state, as the redistribution takes even longer in this case. 

Note that for \cref{fig:landoralphavsint} the detuning from the zero-field resonance frequency was chosen to be resonant with the transition between the $m_F = 2$ ground and the $m_{F'}=3$ excited state. The detuning can be calculated as 
\begin{equation}
\Delta \omega = 3 \, \omega_\mathrm{le} - 2 \, \omega_\mathrm{lg}, \label{eq:landor_detuning_B_dep}
\end{equation}
where $\omega_\mathrm{l\, g,e}$ are the Larmor frequencies of the ground and excited state respectively. 
They are calculated by \cref{eq:larmor}.
Putting in numbers, we get
\begin{equation}
\Delta \omega = 2 \pi \, 1.404 \frac{\mathrm{MHz}}{\mathrm{G}} \times B.
\label{eq:landor_detuning_B_dep_num}
\end{equation}
Note that apart from the detuning, also the effective $\alpha$ depends on $B$ due to the equilibration phase.

As already discussed in \cref{sec:im_cross_section}, the scattering cross-section after the initial equilibration will follow a Lorentzian function with the natural line-width. 
As can be easily seen from \cref{eq:sigma_rescaled}, the quantity $1/\alpha$ will do the same when neglecting the initial equilibration. 
Not neglecting it, the line-shape will be broadened and slightly shifted depending on the magnetic field strength.

\subsection{Vertical Imaging}
\label{sec:VAndor_obe}

In the vertical imaging system we use circularly polarized light.
The atomic quantization axis is aligned with the imaging direction. 
This is one of the two situations discussed in \cref{sec:im_cross_section}, which leads to an effective two level transition.
The bias field fixing the quantization axis has a magnitude of 1.76 Gauss.
As for the two other imaging systems, the intensity is non-uniform. 
Considering a typical region of interest, the median is $0.22 \, I^0_\mathrm{sat}$, the 5 and 95\% quantiles are $0.07 \, I^0_\mathrm{sat}$ and $0.47 \, I^0_\mathrm{sat}$, respectively. 
The imaging intensity fluctuates quite a bit more than for the longitudinal and also the transverse imaging system. 
Again we state these values just to give an order of magnitude, the intensities fluctuate and drift with time. 
Due to the design of the imaging system~\cite{rauer2012mastersthesis}, the usable region of interest is strongly limited and usually closely resembles the choice for which we specified the median and quantiles. 
The imaging light is on for 50{\us}.

The vertical imaging system is basically only used to image matter-wave interference after releasing the atoms from a double well trap created by rf-dressing. 
All atoms should therefore be in the $m_F = 2$ state after the magnetic trap is switched off.
However, currently all fields are switched off when releasing the atoms.
Subsequently, the imaging field is turned on abruptly. 
It is therefore likely that we get a mixture of atoms in the different $m_F$-states after time of flight. 
In the future we are planning to slowly ramp down the Ioffe-field and at the same time slowly ramp up the imaging field so that most of the atoms stay in the $m_F = 2$ state. 

\begin{figure}
	\centering
	\includegraphics{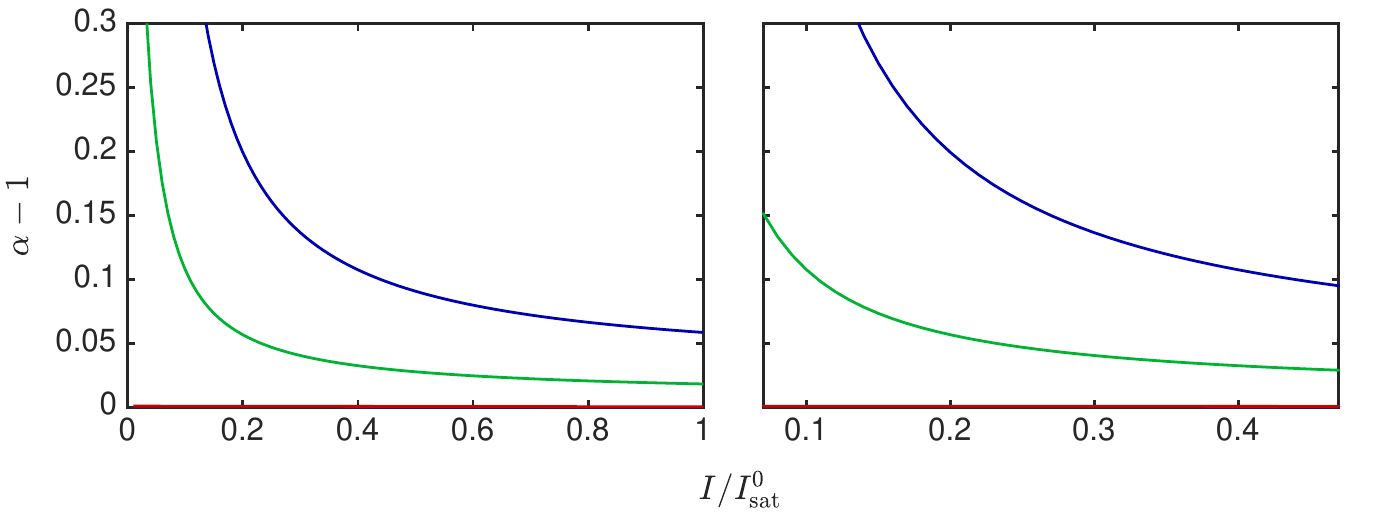}
	\caption{\textbf{ Effective cross section for the vertical imaging:} 
	Like \cref{fig:landoralphavsint}, but for the parameters of the vertical imaging system. 
	For the median imaging intensity of typically $0.22 \, I^0_\mathrm{sat}$ the values are $\alpha = 1.001$ when starting with all atoms initially in $m_F=2$, $\alpha = 1.052$ when starting with an equal mixture of all $m_F$-states and $\alpha = 1.182$ when all atoms are initially in the $m_F=-2$ state. 
	Starting from the $m_F=2$ state, 171 photons per atom are scattered for that intensity. 
	Again, the limits for the horizontal axis in the right subplot coincide with the typical range of imaging intensities occurring in the experiment (see discussion in the main text).
	}
	\label{fig:vandoralphavsint}
\end{figure}

As for the longitudinal imaging system we discuss the  dependence of $\alpha$ on the initial $m_F$-state distribution in \cref{fig:vandoralphavsint}. 
As it is basically the same imaging situation, the magnetic field dependence \cref{eq:landor_detuning_B_dep_num} of the detuning remains valid.

\subsection{Transversal Imaging}
\label{sec:TAndor_obe}

In the transverse imaging system we use linearly polarized light aligned with the atomic quantization axis. 
The imaging direction is then of course perpendicular to it. 
This is one of the two situations discussed in \cref{sec:im_cross_section}, which leads to an effective two level transition.
The bias field fixing the quantization axis is exactly the same (in direction and magnitude) as for the longitudinal imaging. 
It has a magnitude of 1.5 Gauss.
As for the two other imaging systems, the intensity is non-uniform.
Considering the whole picture, the median is $0.31 \, I^0_\mathrm{sat}$, the 5 and 95\% quantiles are 0.19 and 0.49, respectively.
Again we state these values just to give an order of magnitude, the intensities will fluctuate and drift in time. 
As for the longitudinal imaging, the light is on for 75{\us}.

\begin{figure}
	\centering
	\includegraphics{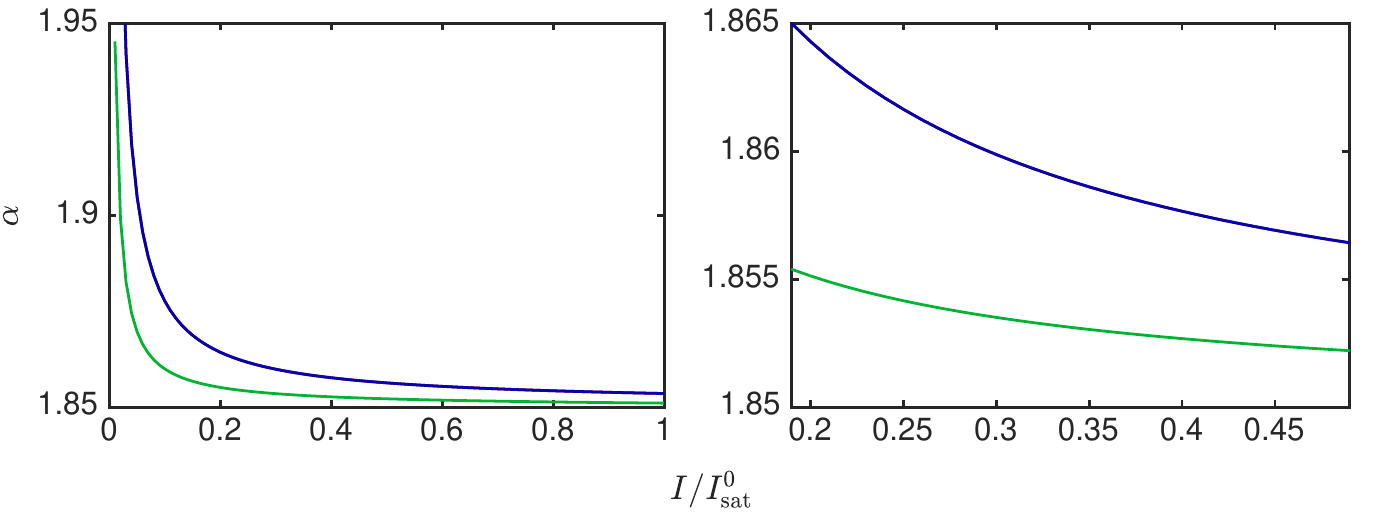}
	\caption{\textbf{ Effective cross section for the transverse imaging:} Like \cref{fig:landoralphavsint} but for the parameters of the transverse imaging system. Note that it does not matter whether one starts with initially all atoms in the $m_F = 2$ or the $m_F = -2$ state. For the median imaging intensity of typically $0.31 \, I^0_\mathrm{sat}$ the values are $\alpha = 1.860$ when starting with all atoms initially in $m_F=2$ or $m_F=-2$, and $\alpha = 1.853$ when starting with an equal mixture of all $m_F$-states. Starting from the $m_F=2$ state, 203 photons per atom are scattered for that intensity. The detuning is 0, which gives the largest scattering cross-section in this imaging situation once the stationary $m_F$-state distribution has been reached. Due to the initial equilibration, the minimal $\alpha$ value might be reached for a slightly different detuning. 
	Again, the limits for the horizontal axis in the right subplot coincide with the typical range of imaging intensities occurring in the experiment (see discussion in the main text).
	}
	\label{fig:tandoralphavsint}
\end{figure}

Also the switching-off of the trap and the ramping up of the quantization field happens exactly like for the longitudinal imaging, the whole discussion about the $m_F$-state distribution therefore also applies to the transverse imaging. Again we investigate the effective $\alpha$ value as a function of intensity for the different initial $m_F$-state distributions. The results are shown in \cref{fig:tandoralphavsint}. 

One sees that the dependence on the imaging intensity and initial $m_F$-state distribution is way less pronounced compared to the results for the longitudinal and vertical imaging system, despite the number of photons scattered per atom being roughly the same. 
The reason for this is that the stationary $m_F$ distribution is much broader and therefore achieved quicker. In the stationary distribution we have 43\% of the atoms in the $m_F=0$ state, 24\% in each of the $m_F = \pm 1$ state and 4\% in each of the $m_F = \pm 2$ state. These are the values for the typical median intensity of $0.31 \, I^0_\mathrm{sat}$. 
The populations change slightly with imaging intensity. 
They also change with magnetic field strength.

\begin{figure}
	\centering
	\includegraphics{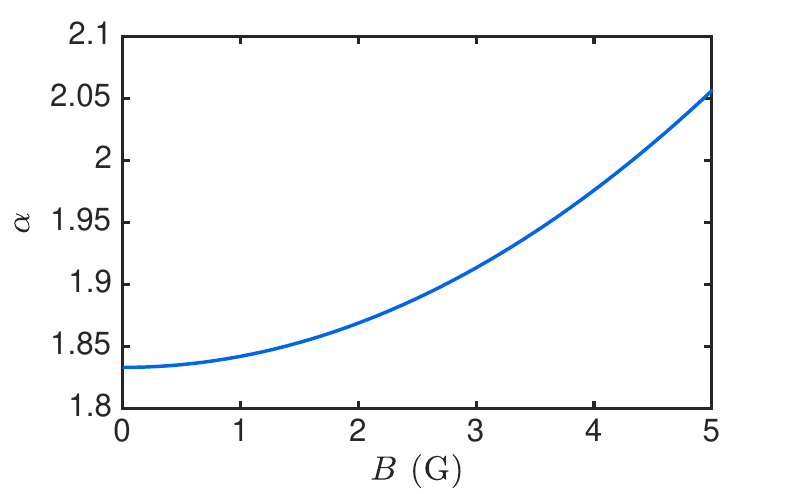}
	\caption{\textbf{ Effective cross section for the transverse imaging:} 
	The dependence of $\alpha$ on the magnetic field strength $B$ in Gauss is shown.
	Note that in the experiment $B = 1.5 \, \mathrm{G}$ is used.
	The imaging intensity is $0.31 \, I^0_\mathrm{sat}$ for this plot. Initially all atoms are in an equal superposition between the different $m_F$-states. The detuning is 0.
	}
	\label{fig:tandoralphavsb}
\end{figure}

\begin{figure}
	\centering
	\begin{subfigure}[b]{\textwidth}
		\centering
		\includegraphics{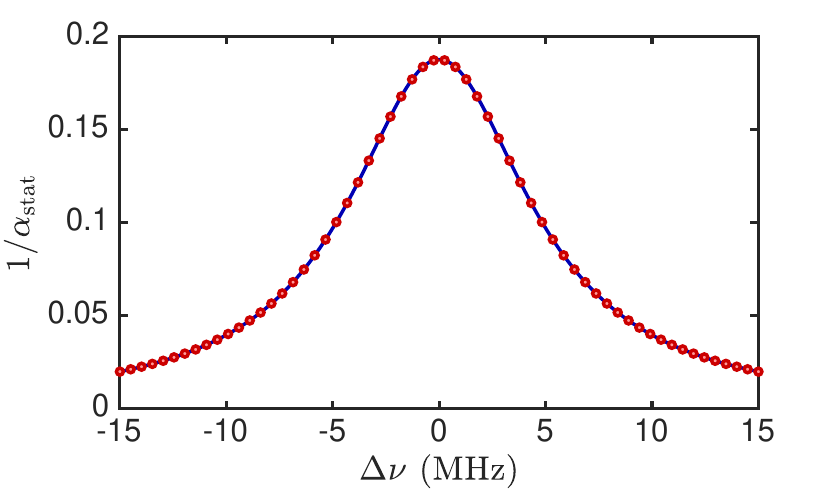}
		\caption{Lineshape for the effective two level-transition with linearly polarized light}
		\label{fig:lineshape20gtandor}
	\end{subfigure}%
	\\
	\begin{subfigure}[b]{\textwidth}
		\centering
		\includegraphics{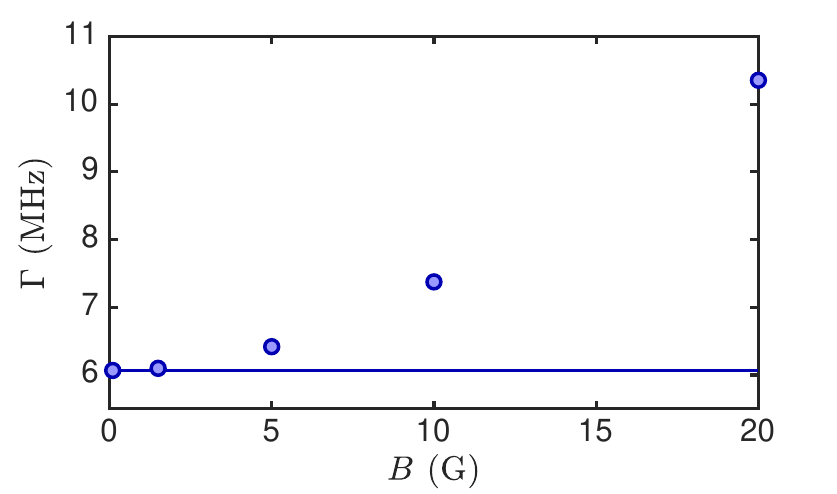}
		\caption{Linewidth increasing with the strength of the magnetic bias field}
		\label{fig:linewidthvsbtandor}
	\end{subfigure}
	\caption{ \textbf{ Effective cross section for the transverse imaging:}  
		As we have an effective two level-transition, the lineshape of the absorption cross-section stays Lorentzian for all bias fields. 
		In \textbf{(\subref{fig:lineshape20gtandor})}, the red bullets show the results of the optical Bloch equations for a bias field of $B = 20 \, \mathrm{G}$.
		Note that the stationary reduction factor $\alpha_\mathrm{stat}$ calculated according to \cref{eq:alpha_formula_stat} is used.
		The quantity $1 / \alpha_\mathrm{stat}$ proportional to the cross-section is shown as a function of the detuning $\Delta \nu$.
		The blue solid line represents a fitted Lorentzian. 
		The agreement with the numerical results is apparent. 
		In \textbf{(\subref{fig:linewidthvsbtandor})}, the linewidth $\Gamma$ (full width at half maximum (FWHM)) of the fitted Lorentzian is plotted as a function of the strength $B$ of the magnetic bias field (blue bullets). 
		For small bias fields it coincides with the natural linewidth represented by the solid blue line.
		Note that in the experiment $B = 1.5 \, \mathrm{G}$ is used.
	}
	\label{fig:TAndor_lineshape_width}
\end{figure}

After the initial equilibration phase, the largest possible scattering cross-section is achieved for 0 detuning, independent of the strength of the quantization field. 
The value for $\alpha$ however depends on the magnetic field strength even if neglecting the initial equilibration. 
The reason is that several $m_F$ states are populated also in the stationary distribution.
An increasing quantization field increases the Zeeman splitting, detuning the transitions between the different $m_F$ levels. 
As shown in \cref{fig:tandoralphavsb}, this leads to an increase in the value of $\alpha$, i.e., to a reduction in the scattering cross section, with magnetic field $B$.
Note that also the linewidth increases with field strength (\cref{fig:linewidthvsbtandor}), the lineshape stays Lorentzian (\cref{fig:linewidthvsbtandor}). 
The Lorentzian lineshape follows from the observation that we have an effective two-level transition.

\subsection{Influence of stray fields}
\label{sec:influence-of-stray-fields}

In the experiment, the ideal detuning, i.e., the detuning with the largest effective scattering cross section can be found by maximizing the measured atom number. 
As stated in \cref{eq:landor_detuning_B_dep_num}, the optimal detuning should depend on the magnetic field strength for the vertical and the longitudinal imaging. It should be independent of it for the transverse imaging.

To estimate the stray fields during imaging, we measured the ideal detuning for the longitudinal imaging system for three different quantization fields. 
The used values are $B = 0${\G}, 0.6{\G} and 1.5{\G}. As already mentioned above the ideal detuning for the transverse imaging system should be independent of the strength of the quantization field and coincide with the zero field resonance frequency for the atomic transition. 
To calibrate this zero line for the detuning, we also determined the ideal detuning for the transverse imaging system for the above stated values of the quantization field. 

Assuming \cref{eq:landor_detuning_B_dep_num} to be valid, the data for the longitudinal imaging should follow the equation
\begin{equation}
\Delta \omega = 2 \pi \, 1.404 \frac{\mathrm{MHz}}{\mathrm{G}} \times \sqrt{\left(B + B_{\mathrm{s}\parallel}\right)^2 + B_{\mathrm{s}\perp}^2 },
\end{equation}
where $B$ is the known applied quantization field. The stray fields parallel and perpendicular to it are denoted by $B_{\mathrm{s}\parallel}$ ans $B_{\mathrm{s}\perp}$, respectively. 
The best estimate for the stray fields was determined by a least squares fit giving $B_{\mathrm{s}\parallel} = 0.24${\G} and $B_{\mathrm{s}\perp} = 0.58${\G} when using all three values for the quantization field. When using only the data for non-zero quantization field, we get $B_{\mathrm{s}\parallel} = 0.21${\G} and $B_{\mathrm{s}\perp} = 0.63${\G}. Note that for such high values of the stray field it is not a priori clear that \cref{eq:landor_detuning_B_dep_num} remains valid for the longitudinal imaging system, especially in the case of vanishing quantization field. Therefore we checked for self consistency by solving the optical bloch equations with the obtained values of the stray fields. 

In the calculations for the longitudinal imaging parameters, we scanned the detuning in steps of 0.05{\MHz}. We found for the case of vanishing quantization field that the ideal detuning is approximately $-0.13${\MHz} instead of 0.
In the other two cases of $B = 0.6${\G} and 1.5{\G}, the deviation from the prediction of \cref{eq:landor_detuning_B_dep_num} was smaller than the detuning scanning step of 0.05{\MHz}.

For the transverse imaging system, we saw from the experimental data that the ideal detuning does not change with the bias field within the experimental uncertainty, meaning that the stray fields don't influence the ideal detuning very much. The optical Bloch equations predict a resonance shift of 0.11{\MHz} when having no quantization field, only the stray fields. In the other two cases of $B = 0.6${\G} and 1.5{\G} quantization fields, the shift was smaller than 0.03{\MHz}.

\begin{figure}
	\centering
	\includegraphics{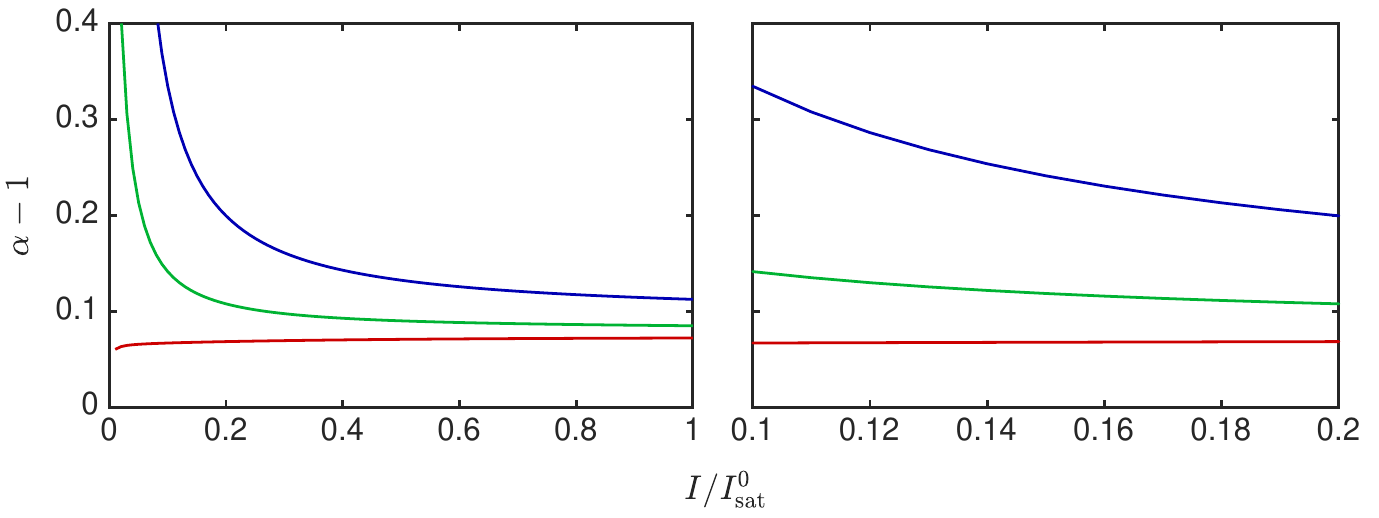}
	\caption{\textbf{ Effective cross section with stray fields for the longitudinal imaging:} Same as \cref{fig:landoralphavsint} but with the stray fields $B_{\mathrm{s}\parallel} = 0.21${\G} and $B_{\mathrm{s}\perp} = 0.63${\G} (see discussion in main text).}
	\label{fig:landoralphavsintstrayfields}
\end{figure}

\begin{figure}
	\centering
	\includegraphics{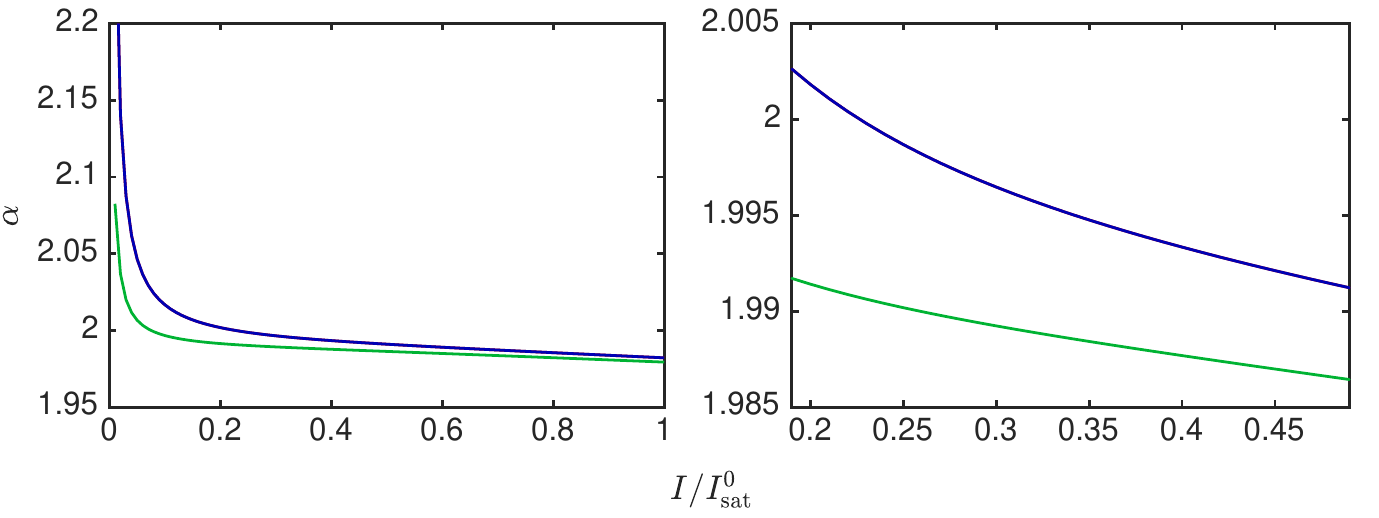}
	\caption{\textbf{ Effective cross section with stray fields for the transverse imaging:} Same as \cref{fig:tandoralphavsint} but with the stray fields $B_{\mathrm{s}\parallel} = 0.21${\G} and $B_{\mathrm{s}\perp} = 0.63${\G} (see discussion main text).}
	\label{fig:tandoralphavsintstrayfields}
\end{figure}

Bearing this in mind, we decided to trust the values $B_{\mathrm{s}\parallel} = 0.21${\G} and $B_{\mathrm{s}\perp} = 0.63${\G}, inferred from only the two data points with non-vanishing quantization field, more and will use them for the further discussion.
In \cref{fig:landoralphavsintstrayfields} the effective cross section in presence of the stray fields is plotted for the parameters of the longitudinal imaging. 
Comparing the plots to \cref{fig:landoralphavsint}, one sees that the stray-fields have a substantial influence. 
The same is true for the transverse imaging system as can be seen from \cref{fig:tandoralphavsintstrayfields} in comparison to \cref{fig:tandoralphavsint}. 
As in the case without stray field, the dependence on imaging intensity and initial $m_F$-state distribution is smaller than for the longitudinal imaging.
Note that we cannot do a similar analysis for the vertical imaging system as it uses a different quantization field and we don't know how much of the stray field is perpendicular and how much is parallel to this quantization field.

\chapter{Contrast distribution thermometry}
\label{chap:contrast_dist}

Due to the shortcomings of the fitting procedures discussed in \cref{sec:get_sG_param}, we also tried to fit a value for the parameters $\lambda_T$ and $q$ of the sine-Gordon model from the distribution functions of the integrated interference contrasts $\mathcal{C} (L)$.

The experimental values for $\mathcal{C} (L)$ are obtained from the interference patterns recorded with the vertical imaging system.
First, the experimentally obtained two dimensional atomic density is summed over a certain number of pixels (determining the integration length $L$) in the center of the cloud.
This summation only runs over the longitudinal $z$ direction, not over the transverse $x$ direction. 
The resulting 1D density depending on $x$ is then fitted with \cref{eq:fr_fit_func} in order to get the contrast.
The obtained value for the fit parameter $C$ gives the value of the integrated contrast.

To fit the parameters of the sine-Gordon model, we use the full distribution functions of the squared contrast normalized by its mean
\begin{equation}
\mathcal{C}^2 (L) / \left\langle \mathcal{C}^2 (L) \right\rangle. \label{eq:squared_contrast}
\end{equation}
One could then just fit the distribution functions for the different integration lengths with some theory predictions~\cite{rauer2018thesis}.
For long integration lengths, the effect of the imaging can be approximately considered in the theory predictions by Gaussian smearing of numerical realizations for $\mathrm{e}^{i \,\varphi(z)}$.
However, this seems to fail for short integration lengths.
For fitting measurements with weak or no tunnel coupling, the fit works well when using only the longer integrating lengths.
However, for coupled scans, we also need the information from the shorter integration lengths.
One therefore has to produce simulated images to consider the influence of the imaging process.

\begin{figure}
	\centering
	\begin{subfigure}{\textwidth}
		\centering
		\includegraphics{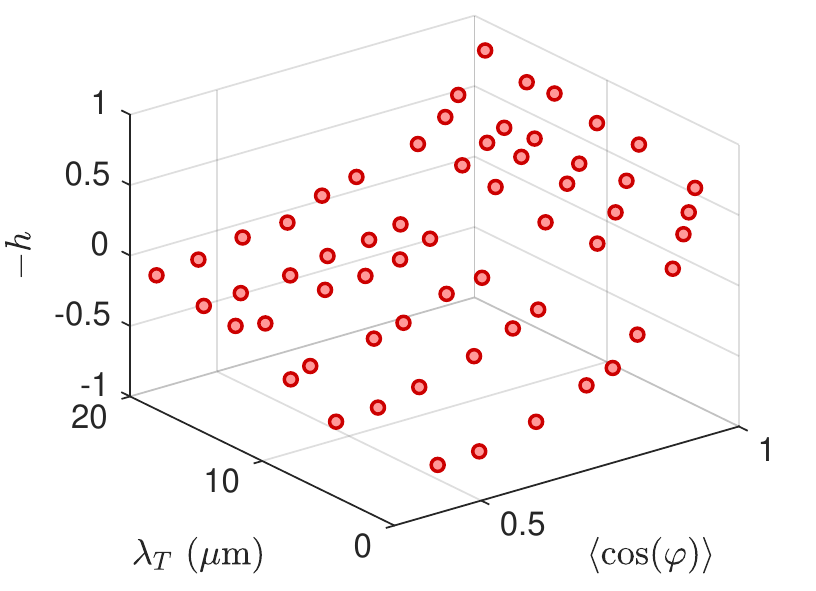}
		\caption{$L = 2${\um}}
		\label{fig:diffentrointlen2}
	\end{subfigure} 
	\\
	\begin{subfigure}{\textwidth}
		\centering
		\includegraphics{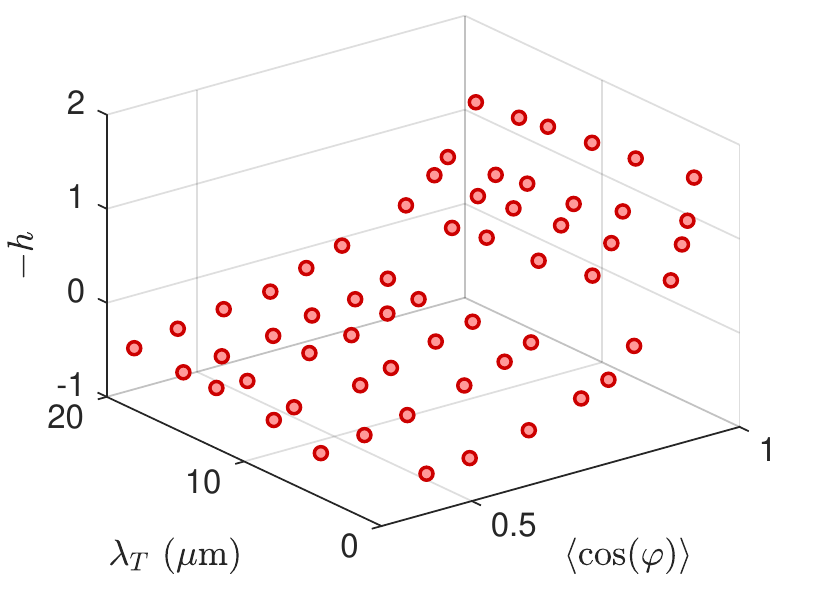}
		\caption{$L = 18${\um}}
		\label{fig:diffentrointlen18}
	\end{subfigure}
	\caption{\textbf{Differential entropy of the distribution functions for the normalized squared contrast.}
	The results following from simulated pictures for the sine-Gordon model (see section \cref{sec:corr_sG_simu_pic}) are shown.
	Each red bullet represents the result for a particular combination of the input parameters $\lambda_T$ and $q$ for the sine-Gordon theory used to simulate the pictures. 
	The left horizontal axis shows one of this input parameters, the thermal coherence length $\lambda_T$.
	On the right horizontal axis, we show the coherence factor $\cohfact$ extracted from the simulated pictures.
	On the vertical axis, the differential entropy $h$ for the distributions of the normalized squared integrated contrast \labelcref{eq:squared_contrast} is shown.
	The two subplots show the results for the two different integration lengths $L$ stated below the respective plots.
	}
	\label{fig:diffentrosin}
\end{figure}

For the direct fitting of the contrast distribution functions, one would have to simulate lots of images for a lot of different parameters.
This is not feasible as simulating images and subsequently analyzing them is computationally costly.
We will therefore focus on quantifying the peakedness of the contrast distributions.
This can be done by the differential entropy of the distribution.
For a general probability density function $f$, the differential entropy is defined as \cite{cover2012elements}
\begin{equation}
	h(f) = - \int \mathrm{d}x \  f(x) \, \log \left( f(x) \right) .
\end{equation}
Note that $h(f)$ is lower for higher peakedness.
The differential entropy for the distributions of the squared normalized contrast following from simulated pictures is shown in \cref{fig:diffentrosin}.


Comparing the experimentally obtained $h$ and $\cohfact$ to the results for the simulated pictures, we can infer the best fitting $\lambda_T$ and $q$.
One can imagine several different fitting procedures.
For the presented results, we start from an array of guesses for the thermal coherence length $\lambda_T$.
For each guess of $\lambda_T$ and the experimentally obtained coherence factor we calculate the `expected' differential entropy for every integration length.
We do this by interpolating between the data points obtained from the simulated pictures (using exactly the dependence shown in \cref{fig:diffentrosin}).
The squared deviation between experimental and `expected' $h$ for the different integration lengths is then summed up and the best guess for $\lambda_T$ found through minimization.
Note that by fitting with this procedure, 
we need to simulate far less pictures than when fitting the distribution functions directly.

\begin{figure}
	\centering
	\includegraphics{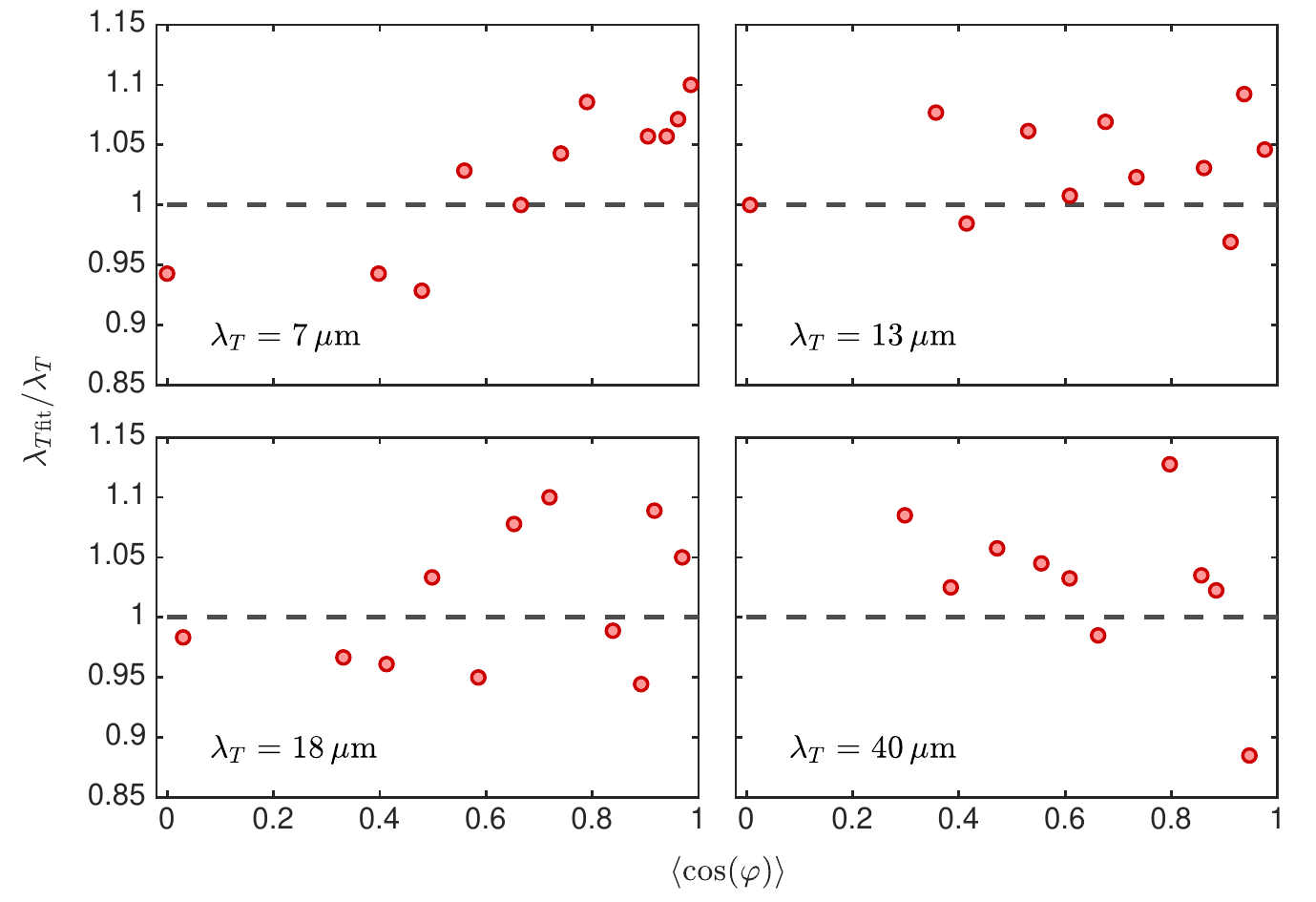}
	\caption{\textbf{Testing the temperature fit with simulated pictures.}
		The extraction of the thermal coherence length from the differential entropy of the contrast distributions (see discussion in main text) is tested by applying it to simulated pictures.	
		The red bullets present the ratio between the fitted thermal coherence lengths $\lambda_{T \mathrm{fit}}$ and the input values $\lambda_T$ used for simulating the pictures.
		The dashed gray line simply marks the value one.
		The different subplots show the results for the different $\lambda_T$ given in the lower left corner.
	}
	\label{fig:ltfitsimudiffentro}
\end{figure}

We obtained a simulated data set for a grid spanned by nine different values of $\lambda_T$ and eleven values of $q$. 
The fitting procedure was then tested by applying it to a different set of simulated pictures (with different values of $q$ and $\lambda_T$).
The results presented in \cref{fig:ltfitsimudiffentro} look quite promising.
However, one should note the influence of the fringe spacing (see also discussion in \cref{sec:width_of_Gaussian_psf}).
The fitting procedure only works well when choosing the same fringe spacing for the simulated pictures to fit as was used to obtain the data set.
Unfortunately, the fringe spacing $\lambda_\mathrm{F}$ varies quite a bit for experimental data.
We therefore need $\lambda_\mathrm{F}$ as an additional variable in the simulated data set and a modified fitting procedure.
This was not implemented yet and will be left to the future. 

One should also speak a further word of caution.
The whole method relies on simulating the pictures correctly. 
Mistakes in the simulation procedure automatically translate into errors for the fitted parameters.
Investigations concerning robustness, as well as consistency checks with experimental data should therefore be performed before applying the fitting procedure. 

After having obtained a value for $\lambda_T$, we can also fit a value for $q$.
In order to do so, we will use the connection between $\lambda_T$, $\cohfact$ and $q$ following from the simulated data.
Of course, we could also fit both, the values for $\lambda_T$ and $q$ in one step.
However, we are usually more interested in an exact value for $\lambda_T$ than $q$, that's why we chose the two step procedure for the discussion in this chapter.


\bibliography{biblio}

\chapter*{Acknowledgments}
\addcontentsline{toc}{chapter}{Acknowledgments}

Experimental physics is always a group effort, the result in this thesis wouldn't have been possible without the constant support from all members of the Atomchip group.
Being it in form of borrowed equipment, shared wisdom or codes, interesting discussions, constructive criticism or just by creating a nurturing work environment.
I therefore sincerely want to thank all current and past group members as well as all external collaborators.

\vspace{1em}

\noindent In particular, I would like to thank the following people:

\begin{itemize}
 
\item My supervisor J\"org Schmiedmayer for his support and guidance, but also for giving me freedom to work independently.

\item Bernhard Rauer, my main co-worker on the experiment, for being the bright and reliable person he is. Also for carefully proofreading most of this thesis. 

\item The former members of the KRb team: Micheal Gring, Max Kuhnert, Tim Langen and Remi Geiger for handing us the well running experimental apparatus with which we could do so many exciting experiments.

\item Sebsatian Erne, Igor Mazets, Marek Gluza and J\"urgen Berges for providing excellent theory support.

\item The next generation of KRb team members taking over the experiment: Federica Cataldini, Amin Tajik, Jo\~{a}o Sabino, Frederik M{\o}ller and SiCong Ji, may they continue to produce fantastic experimental results.
In particular, I would like to thank Frederik and Amin for proofreading parts of my thesis.

\item Selim Jochim and Ian Spielman for kindly agreeing to review my thesis.

\end{itemize}

\selectlanguage{ngerman}						

\noindent Zu guter Letzt möchte ich meinen Eltern für ihre fortwährende Unterstützung danken.

\selectlanguage{english}		

\chapter*{Curriculum Vitae}
\addcontentsline{toc}{chapter}{Curriculum Vitae}

\subsubsection{Personal data}

\begin{tabular}{p{0cm}p{4cm}p{5cm}}
\rule{0pt}{3ex} & Name:  & Thomas Schweigler \\
\rule{0pt}{3ex} & Date of birth:  & 14th October 1986 \\
\rule{0pt}{3ex} & Place of birth: & St.\ P{\"o}lten, Austria\\
\rule{0pt}{3ex} & Email: & thomas.schweigler@tuwien.ac.at \\
\end{tabular}

\subsubsection{Academic education}

\begin{tabular}{p{0cm}p{4cm}p{10cm}}
\rule{0pt}{3ex} & Since 12/2012  & TU Wien, Austria \\
& & Doctoral programme Technical Physics\\
\rule{0pt}{3ex} & 08/2009 - 10/2012 & TU Wien, Austria\\
& & Master's programme Technical Physics\\
\rule{0pt}{3ex} & 09/2009 - 08/2010 & University of Calgary, Canada\\
& & Exchange semesters\\
\rule{0pt}{3ex} & 03/2007 - 07/2009 & TU Wien, Austria\\
& & Bachelor's programme Technical Physics\\
\end{tabular}

\end{document}